%% file: main.tex
\documentclass[11pt,letterpaper]{article}
\usepackage[margin=1in]{geometry}
\input{command/commands.tex}

\newcommand{\ContentsGroup}[1]{%
  \par\addvspace{0.9\baselineskip}%
  \noindent\textbf{#1}\par\nobreak\smallskip
}

\begin{document}

\begin{titlepage}
  \begin{center}
    {\LARGE\bfseries Information Loss Is All You Need: $\TCZero$ Pseudorandomness from Worst-Case Lattices\par}
    \vspace{1.5em}
    {\large Chen Qian\par}
    \vspace{0.5em}
    {\small
      State Key Laboratory of Cryptography and Digital Economy Security,\\
      Shandong University, Qingdao 266237, China\par
      \vspace{0.5em}
      Key Laboratory of Cryptologic Technology  and Information Security of Ministry of Education,\\
      Shandong University, Qingdao 266237, China\par
      \vspace{0.5em}
      School of Cyber Science and Technology,\\
      Shandong University, Qingdao 266237, China\par
    }
  \end{center}
  \vspace{2em}
  \begin{center}
    \textbf{Abstract}
  \end{center}
  \input{chapters/abstract.tex}
\end{titlepage}

\pagenumbering{roman}
\setcounter{tocdepth}{1}
\begingroup
\makeatletter
\renewcommand*{\l@section}[2]{%
  \addvspace{0.35em}%
  \@dottedtocline{1}{0pt}{1.8em}{#1}{#2}%
}
\makeatother
\tableofcontents
\endgroup
\clearpage

\pagenumbering{arabic}
\addtocontents{toc}{\protect\ContentsGroup{Main text}}

\input{chapters/introduction.tex}
\input{chapters/preliminaries.tex}
\input{chapters/LAR.tex}
\input{chapters/uniform-zero-fiber-lwe-to-lar.tex}
\input{chapters/gaussian-lwe-to-uniform-zero-fiber-lwe.tex}
\input{chapters/t-aft-instantiation.tex}
\input{chapters/finite-dimensional-calibration.tex}
\input{chapters/probabilistic-transport.tex}

\input{chapters/direct-fourier-transport.tex}
\input{chapters/applications.tex}

\section*{AI Disclosure}
We used ChatGPT in preparing this paper. It assisted with
developing and checking the algorithm as well as drafting parts of the text and proofs, reviewing the
manuscript for correctness, consistency and presentation, and LaTeX typesetting.
The author checked the manuscript's statements and proofs and takes full
responsibility for its content.

The Prove2Me project\footnote{%
  \raggedright\url{https://prove2.me/missions/Information_Loss_Is_All_You_Need\%3A_TC\%5E0_Pseudorandomness_from_Worst-Case_Lattices}}
verifies four theorems, nine lemmas, and three corollaries from this paper in Lean~4.
Gaussian $\LWE$ reduces to uniform-zero-fiber $\LWE$
(\cref{thm:gaussian-lwe-to-uniform-zero-fiber-lwe}) and then to decisional $\LAR$
(\cref{thm:uniform-zero-fiber-lwe-to-dlar}), with the composed bound in
\cref{cor:gaussian-lwe-to-lar-composition}. Lean also verifies stable
finite-dimensional calibration (\cref{thm:stable-finite-dimensional-calibration}),
fixed-order Gaussian affine fiber transport for the explicit family (\cref{thm:df-constructive-aft}),
and the worst-case lattice reduction (\cref{cor:applications-worst-case}).

The algebraic results cover modular rank and image lattices (\cref{lem:rounding-smith-properties}),
quotient encoding, equivariance, and balanced fibers (\cref{lem:lar-properties}),
and zero-fiber decomposition (\cref{lem:zero-fiber-decomposition}).
Under affine transport the public matrix stays exactly uniform and independent of the
transported error (\cref{lem:t-aft-properties}). For the explicit matrix, the
proofs give its image lattice and Smith form (\cref{lem:rounding-image-lattice}),
the zero-fiber parametrization by zero-sum endpoints (\cref{lem:canonical-endpoint-pullback}),
and the endpoint and carry bound on transport error (\cref{lem:affine-transport-endpoint-carry-bound}).
The integral tree transport (\cref{cor:general-tree-incidence-transport}) uses
the tree incidence basis (\cref{lem:tree-incidence-basis}) and the determinant
formula for a one-column extension (\cref{lem:tree-incidence-determinant}).

AI generated all project-specific code for the author's earlier Lean formalization of
\cref{thm:df-constructive-aft} in the counted binary-operation model. The author
checked its main statement against the paper, compiled the code, and verified
that the proof uses only Lean's standard logical axioms.

\bibliographystyle{alpha}
\AddToHookNext{env/thebibliography/begin}{%
  \phantomsection
  \addcontentsline{toc}{section}{\protect\numberline{}\refname}%
}
\bibliography{cryptobib/abbrev0,cryptobib/crypto,references}

\appendix
\addtocontents{toc}{\protect\ContentsGroup{Appendices}}
\input{chapters/appendix-foundations-proofs.tex}
\input{chapters/appendix-incidence-background.tex}
\input{chapters/appendix-affine-proofs.tex}

\input{chapters/appendix-bounded-sampling.tex}

\input{chapters/appendix-direct-fourier-reference.tex}

\input{chapters/appendix-direct-fourier-calibration.tex}

\input{chapters/appendix-direct-fourier-tail.tex}
\input{chapters/appendix-direct-fourier-charge.tex}
\input{chapters/appendix-applications-proofs.tex}

\end{document}

%% file: command/commands.tex
\input{command/packages.tex}
\input{command/font.tex}

\input{command/common.tex}

\input{command/pseudocode.tex}
\input{command/theorems.tex}

%% file: command/packages.tex
\usepackage[T1]{fontenc}
\usepackage{lmodern}
\usepackage{amsmath,amsfonts,amssymb,amsthm}
\usepackage{mathrsfs}
\usepackage{xcolor,xspace,graphicx,booktabs}
\ExplSyntaxOn
\cs_new_eq:NN \lar_cryptocode_saved_newcounter:n \newcounter
\cs_set_protected:Npn \newcounter #1
  {
    \str_case:nn {#1}
      {
        {Hpclinenumber}  {\cs_undefine:c {theHpclinenumber}}
        {H@pclinenumber} {\cs_undefine:c {theH@pclinenumber}}
        {Hpcgamecounter} {\cs_undefine:c {theHpcgamecounter}}
        {Hpcrlinenumber} {\cs_undefine:c {theHpcrlinenumber}}
      }
    \lar_cryptocode_saved_newcounter:n {#1}
  }
\ExplSyntaxOff
\usepackage[
  operators,adversary,landau,probability,sets,notions,logic,
  ff,mm,advantage,primitives,oracles,events,complexity,
  asymptotics,keys,n,lambda
]{cryptocode}
\ExplSyntaxOn
\cs_set_eq:NN \newcounter \lar_cryptocode_saved_newcounter:n
\cs_undefine:N \lar_cryptocode_saved_newcounter:n
\ExplSyntaxOff
\usepackage{nicodemus}
\usepackage{pseudocodemacros}
\usepackage{algorithm}
\usepackage{algpseudocode}
\usepackage{mdframed}
\usepackage[hidelinks,hypertexnames=false,pagebackref]{hyperref}
\usepackage[capitalize,noabbrev]{cleveref}
\usepackage{macros}

%% file: command/font.tex
\newcommand{\varfont}[1]{\mathsf{#1}}
\newcommand{\algfont}[1]{\mathsf{#1}}\newcommand{\notionfont}[1]{\mathsf{#1}}

\DeclareFontFamily{U}{lardistgreek}{}
\DeclareFontShape{U}{lardistgreek}{m}{n}{<-> fcsrgr}{}
\DeclareFontShape{U}{lardistgreek}{b}{n}{<-> fcsbgr}{}
\DeclareSymbolFont{distgreek}{U}{lardistgreek}{m}{n}
\SetSymbolFont{distgreek}{bold}{U}{lardistgreek}{b}{n}
\makeatletter
\DeclareMathSymbol{\dist@mu}{\mathord}{distgreek}{`m}
\DeclareMathSymbol{\dist@pi}{\mathord}{distgreek}{`p}
\DeclareMathSymbol{\dist@nu}{\mathord}{distgreek}{`n}
\DeclareMathSymbol{\dist@chi}{\mathord}{distgreek}{`q}
\DeclareMathSymbol{\dist@delta}{\mathord}{distgreek}{`d}
\DeclareMathSymbol{\dist@upsilon}{\mathord}{distgreek}{`u}
\DeclareMathSymbol{\dist@Pi}{\mathord}{distgreek}{`P}
\DeclareMathSymbol{\dist@tau}{\mathord}{distgreek}{`t}
\DeclareMathSymbol{\dist@psi}{\mathord}{distgreek}{`y}
\DeclareMathSymbol{\dist@rho}{\mathord}{distgreek}{`r}
\newcommand{\distfont}[1]{{%
    \let\mu\dist@mu%
    \let\pi\dist@pi%
    \let\nu\dist@nu%
    \let\chi\dist@chi%
    \let\delta\dist@delta%
    \let\upsilon\dist@upsilon%
    \let\Pi\dist@Pi%
    \let\tau\dist@tau%
    \let\psi\dist@psi%
    \let\rho\dist@rho%
\mathsf{#1}}}
\makeatother
\newcommand{\polyfont}[1]{\mathsf{#1}}
\newcommand{\polyvarfont}[1]{\mathsf{#1}}
\newcommand{\polyvecfont}[1]{\boldsymbol{\polyvarfont{#1}}}
\newcommand{\mat}[1]{\vec{#1}}

\newcommand{\setfont}[1]{\mathcal{#1}}

%% file: command/common.tex
\newcommand{\pparam}{\varfont{pp}} 

\providecommand{\llbracket}{\lbrack\!\lbrack}
\providecommand{\rrbracket}{\rbrack\!\rbrack}

\renewcommand{\sample}{\hookleftarrow}
\newcommand{\Unif}[1]{\distfont{U}\left(#1\right)}
\newcommand{\BernoulliLaw}[1]{\distfont{Bernoulli}\left(#1\right)}

\renewcommand{\Gaussian}[2]{\ensuremath{\distfont{D}_{#1,#2}}}
\renewcommand{\GaussianCenter}[3]{\ensuremath{\distfont{D}_{#1,#2,#3}}}

\newcommand{\trans}{\intercal}

\makeatletter
\newcommand{\ubar}[1]{%
  \mathpalette\@ubar{#1}%
}
\newcommand{\@ubar}[2]{%
  \vtop{%
    \offinterlineskip
    \halign{%
      \hfil##\hfil\cr
      $\m@th#1#2$\cr
      \noalign{\kern0.3ex}
      \rule{0.8ex}{0.4pt}\cr
    }%
  }%
}
\makeatother

\newcommand{\Span}{\mathsf{Span}}

\newcommand{\LatticeRank}{n}
\newcommand{\AmbientDim}{m}
\newcommand{\LatticeBasis}{\mat{B}}
\newcommand{\BasisVec}{\vec{b}}
\newcommand{\BasisIndex}{i}
\newcommand{\LatticeCoeff}{z}
\newcommand{\LatticeCoeffVec}{\vec{\LatticeCoeff}}
\newcommand{\LatticeGen}[1]{\mathcal{L}(#1)}
\newcommand{\ZeroVec}{\vec{0}}
\newcommand{\ZeroSumIntegerLattice}[1]{A_{#1-1}}

\newcommand{\LatticeModulus}{q}

\newcommand{\GaussianWidth}{\sigma}
\newcommand{\GaussianCenterVec}{\vec{c}}
\newcommand{\GaussianPoint}{\vec{x}}
\newcommand{\GaussianSumPoint}{\vec{y}}
\newcommand{\GaussianSupport}{S}
\newcommand{\GaussianWeight}{\rho}

\newcommand{\LAR}{\mathsf{LAR}}
\newcommand{\LARModulus}{q}
\newcommand{\LARField}{\mathbb{F}_{\LARModulus}}
\newcommand{\BlockDim}{d}
\newcommand{\RoundingRank}{r}
\newcommand{\OutputDim}{\ell}
\newcommand{\SmithLeft}{\mat{U}}
\newcommand{\SmithRight}{\mat{V}}
\newcommand{\IdentityMatrix}{\mat{I}}
\newcommand{\ZeroMatrix}{\mat{0}}
\newcommand{\QuotientSelector}{\mat{E}}
\newcommand{\QuotientProjection}{\mat{P}}
\newcommand{\OnesVec}{\vec{1}}
\newcommand{\SecretDim}{N}
\newcommand{\CoordinateIndex}{j}
\newcommand{\StandardBasisVec}{\vec{e}}
\newcommand{\RoundingColumn}{\vec{v}_0}
\newcommand{\RoundingMatrix}{\mat{T}}
\newcommand{\IntegerEntry}{z}
\newcommand{\IntegerVec}{\vec{\IntegerEntry}}
\newcommand{\CanonicalLift}[1]{[#1]_{\LARModulus}}
\newcommand{\RoundingInput}{\vec{x}}
\newcommand{\RoundingMap}{\algfont{AR}_{\RoundingMatrix}}
\newcommand{\LARLabelMap}{f_{\RoundingMatrix}}
\newcommand{\RoundingLabel}{\vec{y}}
\newcommand{\RoundingDirection}{\mat{J}}
\newcommand{\RoundingDirectionInt}{\RoundingDirection_{\mathrm{int}}}
\newcommand{\RoundingOffset}{\vec{a}}
\newcommand{\LARSecret}{\vec{s}}
\newcommand{\LARPublicMatrix}{\mat{M}}
\newcommand{\LARUniformLabel}{\vec{u}}
\newcommand{\SampleCount}{t}
\newcommand{\SampleIndex}{i}
\newcommand{\LARReal}{\distfont{Real}^{\LAR}}
\newcommand{\LARIdeal}{\distfont{Ideal}^{\LAR}}
\newcommand{\LARDistinguisher}{\adversary{D}}
\newcommand{\DecisionalLAR}{\notionfont{dLAR}}
\newcommand{\LARDecisionalAdv}[1]{\Advantage^{\DecisionalLAR}_{#1}}
\newcommand{\SecurityDecayExponent}{c}
\newcommand{\LARSecretSeparator}{\vec{a}}
\newcommand{\LARLabelShift}{\vec{u}}
\newcommand{\LARRecoverySlack}{k}

\newcommand{\LWR}{\mathsf{LWR}}
\newcommand{\LWE}{\mathsf{LWE}}
\newcommand{\LWESecret}{\vec{s}}
\newcommand{\LWEPublicMatrix}{\mat{A}}
\newcommand{\LWEError}{\vec{e}}
\newcommand{\LWEErrorLaw}{\distfont{\chi}}
\newcommand{\LWEUniformResponse}{\vec{u}}
\newcommand{\LWEReal}{\distfont{Real}^{\LWE}}
\newcommand{\LWEIdeal}{\distfont{Ideal}^{\LWE}}
\newcommand{\LWEDistinguisher}{\adversary{D}}
\newcommand{\ZeroFiberFunction}{g}
\newcommand{\FunctionZeroFiber}{\mathcal{F}_{\ZeroFiberFunction,0}}
\newcommand{\LWEZeroFiber}{\mathcal{F}_{\RoundingMatrix,0}}
\newcommand{\ZeroFiberDecomposition}{\Phi_{\RoundingMatrix}}

\newcommand{\LWEResponse}{\vec{b}}
\newcommand{\LWEReduction}{\adversary{B}}
\newcommand{\IdealLWEReduction}{\LWEReduction^{\mathrm{id}}}

\newcommand{\ReductionFieldSamplingError}{\varepsilon_{\mathrm{field}}}
\newcommand{\LWEAdvantage}[1]{\Advantage^{\LWE}_{#1}}
\newcommand{\ReductionVector}{\vec{a}_0}
\newcommand{\ZeroSecretBadEvent}{\mathsf{E}_{0}}
\newcommand{\ReductionBadEvent}{\mathsf{E}}
\newcommand{\LARNonzeroReal}{\distfont{Real}^{\LAR,\ne0}}

\newcommand{\TransportMatrix}{\mat{Q}}
\newcommand{\TransportOffset}{\vec{\beta}}
\newcommand{\TransportAccuracy}{\delta}
\newcommand{\TransportAccuracyOrder}{c}

\newcommand{\TransportAccuracyThreshold}{\lambda_0}
\newcommand{\SampleCountExponent}{b}
\newcommand{\AdvantageExponent}{{c_{\mathrm{adv}}}}
\newcommand{\ModulusExponent}{a}
\newcommand{\AffineFiberTransport}{\mathsf{AFT}_{\RoundingMatrix}}
\newcommand{\TransportSampler}{\mathsf{Sample}}
\newcommand{\TransportEvaluator}{\mathsf{Eval}}
\newcommand{\TransportInput}{\vec{x}}
\newcommand{\TransportOutput}{\TransportInput'}
\newcommand{\StatisticalDistance}[2]{\Delta\!\left(#1,#2\right)}
\newcommand{\DistributionLaw}[1]{\mathcal{L}\!\left(#1\right)}
\newcommand{\FiniteSampleSpace}{\Omega}
\newcommand{\FiniteSamplePoint}{\omega}
\newcommand{\FiniteProbabilityLaw}{\distfont{P}}
\newcommand{\FiniteReferenceLaw}{\distfont{Q}}
\newcommand{\FiniteDensity}{\distfont{f}}
\newcommand{\ProbabilityEvent}{A}
\newcommand{\EventIndicator}[1]{\mathbf{1}_{#1}}
\newcommand{\ProbabilityRandomVariable}{X}
\newcommand{\FiniteSecondTestFunction}{g}
\newcommand{\ProbCalLaw}{\distfont{\mu}}
\newcommand{\ProbCalSpace}{V}
\newcommand{\ProbCalOutputSpace}{Y}
\newcommand{\ProbCalSynthesis}{\Psi}
\newcommand{\ProbCalResponse}{\mathcal M}
\newcommand{\ProbCalEvaluation}{\mathcal E}
\newcommand{\ProbCalOperator}{\mathcal A}
\newcommand{\ProbCalDefect}{\mathcal K}
\newcommand{\ProbCalIdentity}{I}
\newcommand{\ProbCalTarget}{\tau}
\newcommand{\ProbCalCoefficient}{z}
\newcommand{\ProbCalReference}{z_0}
\newcommand{\ProbCalVector}{v}
\newcommand{\ProbCalCorrectionSpace}{V_{\mathrm{corr}}}
\newcommand{\ProbCalNorm}[1]{\lVert #1\rVert_V}
\newcommand{\ProbCalWeight}{w}
\newcommand{\ProbCalTest}{h}
\newcommand{\ProbCalContraction}{\kappa}
\newcommand{\ProbCalPower}{k}

\newcommand{\ProbCalExponent}{p}
\newcommand{\ProbCalBridgeAt}[1]{C_{#1}}
\newcommand{\ProbCalFunction}{f}
\newcommand{\ProbCalUpperDefect}{a}
\newcommand{\ProbCalAbsoluteDefect}{b}
\newcommand{\ProbCalSupport}{S}
\newcommand{\ProbCalSignedWeight}{W}
\newcommand{\ProbCalClippedWeight}{V_{\mathrm{clip}}}
\newcommand{\ProbCalClipMap}{\operatorname{clip}_{[0,2]}}
\newcommand{\ProbCalClipError}{\delta_{\mathrm{clip}}}
\newcommand{\GeneralLinearGroup}{\operatorname{GL}}
\newcommand{\GaussianErrorLaw}{\LWEErrorLaw_{\mathrm{G}}}
\newcommand{\GaussianLWEAdvantage}[1]{\Advantage^{\mathsf{Gauss\text{-}LWE}}_{#1}}
\newcommand{\ZeroFiberLWEAdvantage}[1]{\Advantage^{\mathsf{UZF\text{-}LWE}}_{#1}}
\newcommand{\AdvantagePolynomial}{\polyfont{p}}

\newcommand{\GraphEdges}{E}
\newcommand{\GraphEdge}{e}
\newcommand{\HyperedgeSupport}[1]{\operatorname{supp}(#1)}
\newcommand{\EdgeTail}[1]{\operatorname{tail}(#1)}
\newcommand{\EdgeHead}[1]{\operatorname{head}(#1)}
\newcommand{\GraphIncidenceMatrix}{\mat{D}_{\TransportGraph}}
\newcommand{\WeightedIncidenceMatrix}{\mat{D}_{\TransportGraph,\mathrm{wt}}}

\newcommand{\HypergraphIncidenceMatrix}{\mat{D}_{\RoundingMatrix}}
\newcommand{\InitialVertexConfiguration}{\vec{z}_0}
\newcommand{\EndpointResidue}{\bar{\EndpointLoad}}
\newcommand{\ZeroSumEndpointHyperplane}{H_0}
\newcommand{\CanonicalEndpointPullback}{c_{\RoundingMatrix}}

\newcommand{\PathIncidenceMatrix}{\mat{D}_{\BlockDim}}

\newcommand{\RootLattice}{\ZeroSumIntegerLattice{\BlockDim}}
\newcommand{\TransportGraph}{G}
\newcommand{\TransportTree}{\mathscr{T}}
\newcommand{\TransportVertices}{V}
\newcommand{\InteriorVertices}{\setfont{B}}
\newcommand{\InteriorVertexCount}{n_{\InteriorVertices}}
\newcommand{\BodyIncidenceMatrix}{\mat{D}_{\mathrm B}}

\newcommand{\BodyAffineOffset}{\vec{a}_{\mathrm B}}
\newcommand{\BodyGaussianInput}{\BodySourceCoordinates}
\newcommand{\BodyLoadIncrement}{\vec\Delta_{\mathrm B}^{\mathrm{loc}}}

\newcommand{\NumericalSamplingError}{\nu_{\mathrm{num}}}

\newcommand{\TreeIncidenceMatrix}{\mat{D}_{\TransportTree}}
\newcommand{\ExtendedTreeIncidenceMatrix}{\overline{\mat{D}}_{\TransportTree}}
\newcommand{\ExtendedEdgeCoordinateMatrix}{\overline{\mat{R}}}
\newcommand{\TreeExtensionColumn}{\vec{v}}

\newcommand{\EdgeCoordinateMatrix}{\mat{R}_{\mathrm{edge}}}
\newcommand{\GadgetCount}{g}
\newcommand{\GadgetIndex}{j}

\newcommand{\GadgetScale}{c}
\newcommand{\GadgetShear}{C}
\newcommand{\ChargeCluster}{S_{\mathrm Q}}
\newcommand{\QueueIncidenceMatrix}{\mat{D}_{\mathrm Q}}
\newcommand{\ChargeColumn}{\QueueIncidenceMatrix}
\newcommand{\VertexIndex}{v}

\newcommand{\IntegerCharge}{k}
\newcommand{\EndpointLoad}{\vec{w}}
\newcommand{\EndpointCarry}{\vec{k}}

\newcommand{\EndpointMixingError}{\varepsilon_{\mathrm{end}}}
\newcommand{\CarryFailureError}{\varepsilon_{\mathrm{carry}}}

\newcommand{\LocalSlice}{I}

\newcommand{\LegInputCount}{{r_{\mathrm L}}}

\newcommand{\LocalFirstError}{u}
\newcommand{\LocalSecondError}{t}
\newcommand{\LocalChartBase}{{a_0}}

\newcommand{\LocalFieldValue}{b}

\newcommand{\LocalTestFunction}{f}

\newcommand{\LocalAtomWeight}{\alpha}

\newcommand{\LocalReservoirBase}{{z_{\mathrm R}}}
\newcommand{\LocalPointMass}[1]{\distfont{\delta}_{#1}}
\newcommand{\LocalFirstEndpoint}{X}
\newcommand{\LocalSecondEndpoint}{Y}

\newcommand{\PaddingSource}{\TransportInput}
\newcommand{\BodySourceCoordinates}{\TransportInput_{\mathrm B}}
\newcommand{\LegSourceCoordinates}[1]{\TransportInput_{\mathrm L,#1}}
\newcommand{\QueueSourceCoordinate}{x_{\mathrm Q}}
\newcommand{\PreRelabelIncidenceMatrix}{\HypergraphIncidenceMatrix^{\mathrm{pre}}}
\newcommand{\PreRelabelOffset}{\InitialVertexConfiguration^{\mathrm{pre}}}
\newcommand{\PreRelabelBaseline}{\BaselineShift^{\mathrm{pre}}}

\newcommand{\GlobalRelabelingMatrix}{\mat P_{\mathrm{glob}}}
\newcommand{\PaddingVector}{\vec W}

\newcommand{\PaddingInteriorCutoff}{K_{\mathrm B}}
\newcommand{\PaddingChargeCutoff}{K_{\mathrm Q}}
\newcommand{\PaddingSafeMargin}{K_{\mathrm{BS}}}

\newcommand{\PaddingInteriorRootLattice}{{\ZeroSumIntegerLattice{\InteriorVertexCount}}}

\newcommand{\BaselineShift}{\vec\Delta_{\mathrm{BS}}}
\newcommand{\BaselineCenters}{b}
\newcommand{\BaselineMarks}{m}
\newcommand{\BaselineCoreLaw}{\distfont{D}_{\mathrm{BS}}}
\newcommand{\BaselineImplementedCoreLaw}{\widehat{\distfont{D}}_{\mathrm{BS}}}
\newcommand{\BaselineJointSamplingError}{\varepsilon_{\mathrm{BS,joint}}}
\newcommand{\BaselineDataLaw}{\distfont{P}_{\mathrm{BS}}}
\newcommand{\BaselineImplementedDataLaw}{\widehat{\distfont{P}}_{\mathrm{BS}}}

\newcommand{\DiscreteReferenceQueueIncrement}{\PlacedQueueIncrement^{\mathrm{ref}}}
\newcommand{\PlacedBodyIncrement}{\vec\Delta_{\mathrm B}}
\newcommand{\PlacedLegIncrement}{\vec\Delta_{\mathrm L}}
\newcommand{\PlacedQueueIncrement}{\vec\Delta_{\mathrm Q}}

\newcommand{\PaddingChargeError}{\QueueSourceCoordinate}
\newcommand{\PaddingProtectedEvent}{{\mathsf E_{\mathrm{safe}}}}
\newcommand{\LocalIdentityOperator}{I}

\newcommand{\CountExpectation}{\mathbb{E}}

\newcommand{\AssemblyAcceptanceProbability}{{p_{\mathrm{acc}}}}

\newcommand{\Advantage}{\mathsf{Adv}}

\newcommand{\LocalComponentFunction}{g}

\newcommand{\AssemblyLiftDifference}{\vec z_{\mathrm{lift}}}

\newcommand{\LocalChartPoint}{\omega}

\newcommand{\ChartLaw}{{\distfont{\Pi}_{\mathrm{iso}}}}

\newcommand{\LocalSamplingDenominator}{Q_{\mathrm{samp}}}
\newcommand{\LocalSamplingInteger}{J_{\mathrm{samp}}}
\newcommand{\LocalSamplingBits}{b_{\mathrm s}}

\newcommand{\GaussianLaplaceParameter}{t}
\newcommand{\GaussianTestDirection}{\vec u_{\mathrm{test}}}

\newcommand{\LegLeftVertex}{v_{\mathrm L}}
\newcommand{\LegMiddleVertex}{v_{\mathrm M}}
\newcommand{\LegRightVertex}{v_{\mathrm R}}
\newcommand{\LegBaselineShift}{\zeta}
\newcommand{\LegSumBound}{B_{\mathrm L}}
\newcommand{\LegCutoff}{K_{\mathrm L}}
\newcommand{\LegBaselineFirst}{a}
\newcommand{\LegScaleLift}{\widetilde{\GadgetScale}}
\newcommand{\LegShearLift}{\widetilde{\GadgetShear}}
\newcommand{\LegInjectionMatrix}{\mat M_{\mathrm L,\LocalChartPoint}}
\newcommand{\LegBaseLoad}{\vec\Delta_{\mathrm{BS},\mathrm L}^{\mathrm{loc}}}
\newcommand{\LegGaussianInput}{\TransportInput_{\mathrm L}}
\newcommand{\LegPaddingVector}{\PaddingVector_{\mathrm L}}
\newcommand{\LegAffineOffset}{\vec a_{\mathrm L}}
\newcommand{\LegLoadIncrement}{\vec\Delta_{\mathrm L}^{\mathrm{loc}}}
\newcommand{\LegAttachmentLoad}{Z}
\newcommand{\LegSliceShift}{\xi}

\newcommand{\LegProtectedInputs}{\mathcal E}

\newcommand{\LegMarkedFirst}{a_*}
\newcommand{\LegTrimMargin}{h}
\newcommand{\LegInteriorColors}{C_{\LegTrimMargin}}
\newcommand{\LegBoundaryColors}{E_{\partial,\LegTrimMargin}}
\newcommand{\LegTrimmedSlice}[1]{J_{#1,\LegTrimMargin}}
\newcommand{\LegTrimmedLaw}[1]{\distfont{L}_{#1,\LegTrimMargin}^{\mathrm{trunc}}}
\newcommand{\LegMarkedSecond}{b_*}
\newcommand{\LegMarkFlag}{{B_{\mathrm{mark}}}}
\newcommand{\LegRoundedAtomWeight}{\widehat{\LocalAtomWeight}}
\newcommand{\LegFlagChartLaw}{\distfont{P}_{\mathrm{fc}}}
\newcommand{\LegJointReferenceLaw}{\distfont{Q}_{\LegBaselineShift,\LegMarkedFirst,\LocalAtomWeight,\LegTrimMargin}}
\newcommand{\ChartInteriorSupport}{S}

\newcommand{\ChartUnrestrictedSpace}{\Omega}
\newcommand{\ChartRegularDensity}{\distfont{Q}_{\mathrm{reg}}}
\newcommand{\ChartMarkedDensity}{\distfont{Q}_*}

\newcommand{\ChartDeletionOperator}{D}
\newcommand{\ChartDeletionBound}{\delta_{\mathrm{aff}}}
\newcommand{\ChartOperatorBound}{\kappa_{\mathrm{aff}}}
\newcommand{\ChartMarkDefect}{D_{\mathrm{mark}}}
\newcommand{\ChartMarkOdds}{s_{\LocalAtomWeight}}
\newcommand{\ChartMarkedSurvival}{p_*}

\newcommand{\ChartLineSupport}{S_{\mathrm{line}}}
\newcommand{\ChartLineIndices}{I_{0}}
\newcommand{\ChartOffLineIndices}{I_{\ne0}}
\newcommand{\ChartLocalPositivity}{A_{\mathrm{loc}}}

\newcommand{\LegClippedLaw}{\distfont{\mu}_{\mathrm L}}

\newcommand{\LegActivationEvent}{\mathsf A_{\mathrm L}}
\newcommand{\LegTailProbability}{\theta_{\mathrm L}}

\newcommand{\LegChartError}{\varepsilon_{\mathrm c}}
\newcommand{\LegApproximationError}{\varepsilon_{\mathrm{leg}}}
\newcommand{\LegTableRoutine}{\ensuremath{\algfont{LegTables}}}
\newcommand{\LegDyadicRoutine}{\ensuremath{\algfont{DyadicSample}}}
\newcommand{\LegExactSolve}{\ensuremath{\algfont{RationalSolve}}}
\newcommand{\LegUniformBits}{\ensuremath{\algfont{UniformBits}}}
\newcommand{\LegPrimeTest}{\ensuremath{\algfont{IsPrime}}}
\newcommand{\LegReferenceColor}{b^{\mathrm{ref}}}
\newcommand{\LegInputIndex}{i}
\newcommand{\LegSecondInputIndex}{i'}
\newcommand{\LegAffineEvaluation}{\ell}
\newcommand{\LegAllowedColors}{J}
\newcommand{\LegSecondColor}{b'}

\newcommand{\LegAveragedFunction}{h}
\newcommand{\LegColorMean}{m}
\newcommand{\LegColorImbalance}{\mathcal C}
\newcommand{\LegCountReconstruction}{\mathcal C}
\newcommand{\LegBasisIndices}{\mathcal J}
\newcommand{\LegBasisIndex}{j}
\newcommand{\LegSecondBasisIndex}{k}
\newcommand{\LegBasisFunction}{\varphi}
\newcommand{\LegGramMatrix}{\mat G_{\mathrm L}}
\newcommand{\LegSolveRHS}{\vec d_{\mathrm L}}
\newcommand{\LegSolveCoefficients}{\vec z_{\mathrm L}}
\newcommand{\LegSamplingSupport}{T}
\newcommand{\LegSamplingLaw}{\distfont p}
\newcommand{\LegSamplingCounts}{n_{\mathrm s}}
\newcommand{\LegStoredMap}{\omega_*}
\newcommand{\LegSecondChartPoint}{\omega'}
\newcommand{\LegSignalWeight}{\gamma}
\newcommand{\LegUnmarkedReferenceLaw}{\distfont{Q}_{\mathrm L}}
\newcommand{\LegMarkedReferenceLaw}{\distfont{Q}_{\mathrm L,*}}

\newcommand{\LegIncidenceMatrix}{\mat{D}_{\mathrm L}}
\newcommand{\LegCoordinateMatrix}{\mat{R}_{\mathrm L,\LocalChartPoint}}
\newcommand{\LegCoordinateBlock}[1]{\mat{R}_{\mathrm L,\LocalChartPoint_{#1}}}
\newcommand{\LegEdgeCoordinates}{\vec r_{\mathrm L,\LocalChartPoint}}

\newcommand{\QueueClippedLaw}{\distfont{\mu}_{\mathrm Q}}
\newcommand{\QueueTailProbability}{\theta_{\mathrm Q}}

\newcommand{\HistogramSafeCenters}{\setfont{I}_{\mathrm{safe}}}

\newcommand{\AssemblyPermutationBits}{b_{\mathrm{perm}}}

\newcommand{\HoeffdingDegree}{r}
\newcommand{\HoeffdingOrder}{R}

\newcommand{\HoeffdingScalarValue}{x}

\newcommand{\HoeffdingAttachment}{R}

\newcommand{\HoeffdingParameterBase}{L}
\newcommand{\DFPublicScale}{t}

\newcommand{\HoeffdingEndpointVector}{\vec Y}

\newcommand{\FockArgument}{\polyvecfont{t}}
\newcommand{\FockPolynomial}{\polyfont{P}}
\newcommand{\FockTensor}{T}
\newcommand{\FockRadius}{\rho}
\newcommand{\FockNorm}[2]{\left\|#1\right\|_{\mathcal F,#2}}
\newcommand{\FockLinear}{\polyfont{\ell}}
\newcommand{\FockNonlinear}{\polyfont{h}}
\newcommand{\FockLinearActivity}{\eta}
\newcommand{\FockQuadraticActivity}{{A_2}}
\newcommand{\FockCubicActivity}{{A_3}}
\newcommand{\FockMean}{M}
\newcommand{\FockConstant}{C}
\newcommand{\FockSmallConstant}{c}
\newcommand{\FockGaussian}{Z}
\newcommand{\FockGram}{S}
\newcommand{\FockQuadraticCount}{a}
\newcommand{\FockCubicCount}{b}
\newcommand{\FockSelectedCount}{j}
\newcommand{\FockSelectedDegree}{k}

\newcommand{\FockWeight}{W}

\newcommand{\FockMomentOrder}{p}

\newcommand{\CalibAnchorWidth}{{a_{\mathrm{anc}}}}

\newcommand{\CalibAnchorLaw}{\distfont{A}}

\newcommand{\CalibActivation}{{\gamma_{\mathrm{cal}}}}
\newcommand{\CalibCompensation}{{\varepsilon_{\mathrm{anc}}}}

\newcommand{\CalibUniform}{\distfont{u}}
\newcommand{\CalibSafeUniform}{{\distfont{u}_{\mathrm{safe}}}}

\newcommand{\CalibSliceMean}{\distfont{v}}

\newcommand{\CalibSignResponse}{\distfont{E}}

\newcommand{\CalibColor}{c}
\newcommand{\CalibOtherColor}{a}

\newcommand{\CalibPublicBlur}{{z_{\mathrm{pub}}}}

\newcommand{\CalibProfile}{{\distfont{v}_{\mathrm{in}}}}

\newcommand{\CalibRootValue}{z}
\newcommand{\CalibContext}{\xi}

\newcommand{\CalibProposal}{\distfont{\mu}}
\newcommand{\CalibCore}{\theta}

\newcommand{\CalibChargeEvent}{\mathcal E}

\newcommand{\ActivityTranslation}[1]{T_{#1}}

\newcommand{\ActivityTailArgument}{{t_{\mathrm H}}}

\newcommand{\ActivityPrimaryDegree}{{a_{\mathrm P}}}
\newcommand{\ActivityBodyDegree}{{b_{\mathrm B}}}
\newcommand{\ActivityPrimaryPool}{{m_{\mathrm P}}}
\newcommand{\ActivityBodyPool}{{m_{\mathrm B}}}

\newcommand{\ActivityCoreScore}{{f_{\mathrm{in}}}}
\newcommand{\ActivityCoreLift}{{J_{\mathrm{in}}}}

\newcommand{\CorePrimaryFixed}{{I_{\mathrm P}}}
\newcommand{\CoreBodyFixed}{{I_{\mathrm B}}}

\newcommand{\CoreUStatistic}{\mathsf U}
\newcommand{\CoreDisjointProduct}{\diamond}

\newcommand{\CoreHilbertBound}{{B_2}}

\newcommand{\BoundaryInverseOrder}{{R}}

\newcommand{\GapSVP}{\mathsf{GapSVP}}
\newcommand{\SIVP}{\mathsf{SIVP}}
\newcommand{\LatticeApproxFactor}{\gamma}
\newcommand{\LWENoiseRate}{\alpha}
\newcommand{\TCZero}{\mathrm{TC}^{0}}
\newcommand{\NCOne}{\mathrm{NC}^{1}}
\newcommand{\WeakPRFFunction}{F}
\newcommand{\PRFKeySpace}{\mathcal{K}}
\newcommand{\PRFKeyError}{\delta_{\mathrm{key}}}
\newcommand{\WeakPRFAdvantage}{\operatorname{Adv}_{\mathrm{wPRF}}}
\newcommand{\LearningInputDim}{D}
\newcommand{\LearningGrid}{\mathcal X}
\newcommand{\LearningFunction}{g}
\newcommand{\LearningHypothesis}{\widehat{g}}
\newcommand{\LearningRationalHypothesis}{\widetilde{g}}
\newcommand{\LearningValidationCount}{m_{\mathrm{val}}}
\newcommand{\LearningEmpiricalRisk}{\widehat{\mathcal R}}

\newcommand{\LearningVariance}{v_{\LARModulus}}
\newcommand{\ReLU}{\operatorname{ReLU}}
\newcommand{\LearningIntegerScore}{H}
\newcommand{\LearningOuterMap}{\psi}
\newcommand{\LearningRange}{R}
\newcommand{\LearningScalar}{z}
\newcommand{\LearningThresholdIndex}{u}
\newcommand{\LearningSplineSlope}{a}

\newcommand{\SignedEndpointLaw}{\distfont{P}}

\newcommand{\BodyClosureGaussian}{x_{\mathrm{cl}}}
\newcommand{\BodyClosureSamplingError}{\nu_{\mathrm G}}
\newcommand{\BodyPermutationSamplingError}{\nu_{\mathrm{perm}}}
\newcommand{\BodyTailProbability}{\theta_{\mathrm B}}
\newcommand{\CycleGaussianScalar}{x}
\newcommand{\CycleReferenceScalar}{z}
\newcommand{\CycleScalarLaw}{\distfont{\mu}_{\GaussianWidth}}

\newcommand{\BodyPlacementMatrix}{\mat P_{\mathrm B}}
\newcommand{\CycleIndex}{i}

\newcommand{\ClosureSamplingRadius}{R_{\mathrm G}}
\newcommand{\ClosureSamplingCutoff}{H_{\mathrm G}}
\newcommand{\ClosureProposalSize}{M_{\mathrm G}}
\newcommand{\ClosureTrialCap}{T_{\mathrm G}}
\newcommand{\ClosureProposalBits}{b_{\mathrm{G,prop}}}
\newcommand{\ClosureAcceptanceBits}{b_{\mathrm{G,acc}}}
\newcommand{\ClosureProposal}{z_{\mathrm G}}

\newcommand{\DFFrequency}{\vec a}
\newcommand{\DFFrequencyEntry}{a}
\newcommand{\DFCharacter}[1]{\chi_{#1}}
\newcommand{\DFTransform}{\widehat{\distfont{P}}}
\newcommand{\DFCharacterPolynomial}{\polyfont{P}_{\mathrm{char}}}
\newcommand{\DFPolynomialVariable}{\polyvarfont{z}}
\newcommand{\DFPolynomialVariables}{\polyvecfont{z}}
\newcommand{\DFFourierTensor}{T}
\newcommand{\DFFourierNorm}{\mathfrak F}
\newcommand{\DFEnergy}{\mathcal E}
\newcommand{\DFChargeRadius}{s_{\mathrm F}}
\newcommand{\DFPositiveMass}{b_{\mathrm{ch}}}
\newcommand{\DFContextChargeEvent}{\mathcal E_{\CalibContext}}
\newcommand{\DFContextPositiveEndpointLaw}{\distfont{P}_{V,\CalibContext}}
\newcommand{\DFProposalError}{\delta_{\mathrm{prop}}}
\newcommand{\DFWeightError}{\delta_W}
\newcommand{\DFUnnormalizedError}{\delta_{\mathrm{unn}}}
\newcommand{\DFNumericalBudget}{\epsilon_{\mathrm{num}}}
\newcommand{\DFApproxWeight}{\widetilde W}
\newcommand{\DFClipMap}{\operatorname{clip}_{[0,2]}}
\newcommand{\DFClippedWeight}{V}
\newcommand{\DFTableClippedWeight}{V_{\HoeffdingOrder}}
\newcommand{\DFClippingError}{\delta_{\mathrm{clip}}}
\newcommand{\DFAmbientError}{\delta_{\mathrm{amb}}}
\newcommand{\DFConditioningError}{\epsilon_{\mathrm{ch}}}
\newcommand{\DFClippedEndpointMeasure}{\distfont{P}_{V}}
\newcommand{\DFNegligibleError}{\operatorname{negl}}
\newcommand{\DFTrialCap}{T_R}
\newcommand{\DFTrialError}{\epsilon_{\mathrm{cap}}}
\newcommand{\DFRationalProposal}{\distfont{\nu}_{0,\mathrm{num}}}
\newcommand{\DFIdealProposal}{\distfont{\nu}_0}
\newcommand{\DFTolerance}{\epsilon}
\newcommand{\DFLocalTolerance}{\delta_{\mathrm{tab}}}
\newcommand{\DFTableSize}{S_R}
\newcommand{\DFStabilityFactor}{\kappa_{\mathrm{num},\BoundaryInverseOrder}}
\newcommand{\DFDiagonalInverseBound}{\kappa_{\mathrm{diag}}}
\newcommand{\DFDiagonalEntryError}{\delta_{\mathrm{diag}}}
\newcommand{\DFRationalPrimary}{\distfont{\pi}_{0,\mathrm{num}}}
\newcommand{\DFRationalCenter}{\distfont{p}_{0,\mathrm{num}}}
\newcommand{\DFRationalAnchor}{\mat A_{\mathrm{anc},\mathrm{num}}}
\newcommand{\DFRationalActivation}{\gamma_{\mathrm{num}}}
\newcommand{\DFRationalLift}{J_{\mathrm{in},\mathrm{num}}}
\newcommand{\DFNonzeroField}{\LARField^\times}

\newcommand{\DFProtectedRootMarginal}{\distfont{R}_{\mathrm B}^{\mathrm{prot}}}
\newcommand{\DFEdgeCutoff}{H_{\mathrm e}}
\newcommand{\DFProtectedEdgeLaw}{\distfont{\mu}_{\mathrm e}}

\newcommand{\DFBodyClippingError}{\varepsilon_{\mathrm{B,clip}}}
\newcommand{\DFRootTagSize}{\eta_{\mathrm R}}
\newcommand{\DFCoarseLocalScale}{Q_{\mathrm{loc}}}
\newcommand{\DFRootNewSlot}{\tau_{\mathrm R}}
\newcommand{\DFRootOutputFactor}{\omega_{\mathrm R}}
\newcommand{\DFRootPopulation}{N_{\mathrm R}}
\newcommand{\DFRootInputFactor}{M_{\mathrm R}}
\newcommand{\DFRootDeviation}{\delta_{\mathrm R}}
\newcommand{\DFRootTransferBudget}{\varepsilon_{\mathrm R}}
\newcommand{\DFPathLength}{k}
\newcommand{\DFRootOrder}{r}

\newcommand{\DFClippedPathLaw}{\distfont{p}_{\mathrm{path},H}}

\newcommand{\DFClippedIntegerMarginal}{\distfont{\nu}_{H}}
\newcommand{\DFPathDifference}{\vec y}
\newcommand{\DFPathDifferenceEntry}{y}
\newcommand{\DFPathBoundaryVector}{\vec u_{\mathrm{path}}}
\newcommand{\DFPathTransferMatrix}{\mat K_{\mathrm{path}}}

\newcommand{\DFProofConstant}{C}
\newcommand{\DFJointRootLaw}{\distfont{P}_{\mathrm R}}
\newcommand{\DFJointRootDensity}{\distfont{k}_{\mathrm R}}
\newcommand{\DFCenteredRootDensity}{\distfont{H}}
\newcommand{\DFPlacementForestEdges}{h}
\newcommand{\DFRootPushforward}{\phi}
\newcommand{\DFAttachmentSet}{A_{\mathrm{att}}}
\newcommand{\DFQueueSet}{S_{\mathrm Q}}
\newcommand{\DFFreeSet}{F_{\mathrm{free}}}
\newcommand{\DFFreeCount}{n_{\mathrm F}}
\newcommand{\JointPlacementMatrix}{\mat P_{\mathrm{place}}}
\newcommand{\ReferenceQueueSet}{S_{\mathrm Q}^{(0)}}
\newcommand{\ReferenceAttachmentSet}{A_{\mathrm{att}}^{(0)}}
\newcommand{\ReferenceAttachmentVertex}{R^{(0)}}
\newcommand{\PlacementQueueSet}{S}
\newcommand{\PlacementAttachmentVertex}{a}

\newcommand{\DFBaselinePrimary}{\distfont{\pi}_0}
\newcommand{\DFBaselineCenter}{\distfont{p}_0}
\newcommand{\DFBodyProfile}{\distfont{b}}
\newcommand{\DFAnchorIndex}{j_{\mathrm{anc}}}
\newcommand{\DFAnchorSign}{\chi_{\mathrm{sign}}}
\newcommand{\DFSignChoice}{\varsigma}
\newcommand{\DFMidpoint}{c_{\mathrm{mid}}}
\newcommand{\DFAnchorPoint}{A_{\CalibColor}}
\newcommand{\DFSmoothedAnchor}{\widetilde{\CalibAnchorLaw}}
\newcommand{\DFReflection}{\iota}
\newcommand{\DFSliceDeletion}{D_{\mathrm{sl}}}
\newcommand{\DFBaselineLoss}{\delta_{\mathrm{base}}}
\newcommand{\DFCenterLoss}{\delta_{\mathrm{ctr}}}
\newcommand{\DFPrimaryDeviation}{\delta_{\mathrm{prim}}}
\newcommand{\DFFirstResponse}{\mat T_1}
\newcommand{\DFDensityInput}{\FiniteDensity}

\newcommand{\DFTable}{{\FiniteDensity}}
\newcommand{\DFOtherTable}{{\FiniteDensity_{\mathrm{aux}}}}
\newcommand{\DFTableNorm}{{\mathcal C_{\FockRadius}}}
\newcommand{\DFTableEvaluation}{{\mathcal E}}
\newcommand{\DFTableSym}{{\operatorname{Sym}}}
\newcommand{\DFTableConsumedPrimary}{{u}}
\newcommand{\DFTableConsumedBody}{{v}}
\newcommand{\DFTableNewPrimary}{{m}}
\newcommand{\DFTableQueue}{{y}}
\newcommand{\DFTableScalar}{{t}}
\newcommand{\DFTableCentering}[1]{\mathsf C_{#1}}
\newcommand{\DFTableSafeProjection}{{\mathsf B_{\mathrm{safe}}}}
\newcommand{\DFTableTranslate}{{\mathsf T}}
\newcommand{\DFTableSafeHeight}{{h_{\mathrm{safe}}}}
\newcommand{\DFTableFringe}{{\distfont{F}_{\partial}}}
\newcommand{\DFTableShiftDifference}{{\distfont{d}_{\mathrm{sh}}}}
\newcommand{\DFTableUnsafePart}{{\distfont{e}_{\mathrm{sh}}}}
\newcommand{\DFTableUnsafeMass}{{\beta_{\mathrm{sh}}}}
\newcommand{\DFTableSafeSeed}{{\distfont{k}_{\mathrm{sh}}}}
\newcommand{\DFTableShiftResidual}{{\distfont{r}_{\mathrm{sh}}}}
\newcommand{\DFTableDiffusePair}{{\distfont{P}_{\mathrm{diff}}}}
\newcommand{\DFTableDiffuseSeed}{{\distfont{S}_{\mathrm{diff}}}}
\newcommand{\DFTableDiffuseResidual}{{\distfont{E}_{\mathrm{diff}}}}
\newcommand{\DFTableMarkedPair}{{\distfont{P}_{\mathrm{mark}}}}
\newcommand{\DFTableMarkedSeed}{{\distfont{K}_{\mathrm{mark}}}}
\newcommand{\DFTableAnchorAtScalar}{{\distfont{A}_{\mathrm{sh}}}}
\newcommand{\DFTableFreeCertificate}{{\mathsf F}}
\newcommand{\DFTableTranslationCertificate}{{\mathsf J}}
\newcommand{\DFTableLegLinear}{{\mathsf l}}
\newcommand{\DFTableLegQuadratic}{{\mathsf h_2}}
\newcommand{\DFTableLegCubic}{{\mathsf h_3}}
\newcommand{\DFTableMeanCertificate}{{\mathsf M}}
\newcommand{\DFTableBackground}{{\mathsf Q}}
\newcommand{\DFTableLog}{{\mathsf H}}
\newcommand{\DFTablePower}{{k}}
\newcommand{\DFTableIndependentPolynomial}{{\polyfont F^0}}
\newcommand{\DFTableCyclePolynomial}{{\polyfont F^{\mathrm{cyc}}}}
\newcommand{\DFTableSignedPolynomial}{{\polyfont S}}
\newcommand{\DFTableUpper}{{\mathsf U_{\mathrm{raise}}}}
\newcommand{\DFTableRootDefect}{{\mathsf V}}
\newcommand{\DFTableWeightedDefect}{{\mathsf B_{\mathrm{root}}}}
\newcommand{\DFTableTotalBackground}{{\mathsf B_{\mathrm{bg}}}}
\newcommand{\DFTableCorrection}{{\FiniteDensity_{\BoundaryInverseOrder}}}
\newcommand{\DFTableWeight}{{W_{\BoundaryInverseOrder}}}

\newcommand{\DFTailFluctuation}{F_{\mathrm{cyc}}}
\newcommand{\DFTailLocalDeviation}{\delta_0}
\newcommand{\DFTailDifferenceBound}{c_{\mathrm{bd}}}
\newcommand{\DFTailVarianceProxy}{v_{\mathrm{bd}}}
\newcommand{\DFTailLaplace}{\alpha_{\mathrm{mgf}}}
\newcommand{\DFTailBlockLaw}{\distfont{p}}
\newcommand{\DFTailSlot}{j_{\mathrm{slot}}}
\newcommand{\DFTailSlotSize}{{n_{\mathrm{slot}}}}
\newcommand{\DFMomentPrimaryAtom}{\alpha_{\mathrm P}}
\newcommand{\DFMomentBodyAtom}{\alpha_{\mathrm B}}
\newcommand{\DFMomentOccupancy}{c_{\mathrm{at}}}
\newcommand{\DFMomentAtom}{\alpha_{\mathrm{at}}}
\newcommand{\DFMomentAverage}{\mathcal T}
\newcommand{\DFMomentBlockCount}{k_{\mathrm{eq}}}
\newcommand{\DFTailCompensatorScore}{r_{\mathrm{cmp}}}
\newcommand{\DFTailComplexValue}{z_{\mathrm{ev}}}
\newcommand{\DFTailVectorFirst}{x_{\mathrm{aux}}}
\newcommand{\DFTailVectorSecond}{y_{\mathrm{aux}}}

\newcommand{\DFTableTranslationAverage}{{\overline{\mathsf J}}}

\newcommand{\DFMatrixCoord}{\operatorname{coord}}
\newcommand{\DFMatrixVector}{\vec f}
\newcommand{\DFMatrixOtherVector}{\vec g}
\newcommand{\DFMatrixZero}{\vec e_0}
\newcommand{\DFMatrixMultiply}{\mat M_{\BoundaryInverseOrder}}
\newcommand{\DFMatrixBackground}{\mat M_{\mathrm{bg}}}
\newcommand{\DFMatrixBackgroundVector}{\vec b}
\newcommand{\DFMatrixNormalized}{\widetilde{\mat A}_{\BoundaryInverseOrder}}
\newcommand{\DFMatrixRaising}{\mat U_{\BoundaryInverseOrder}}
\newcommand{\DFMatrixRoot}{\mat Z_{\BoundaryInverseOrder}}
\newcommand{\DFMatrixLifted}{\mat A_{\BoundaryInverseOrder}}
\newcommand{\DFMatrixPhysical}{\vec\Psi_{\BoundaryInverseOrder}}
\newcommand{\DFMatrixLeading}{\mat E_{\BoundaryInverseOrder}}
\newcommand{\DFMatrixCorrection}{\vec c}
\newcommand{\DFMatrixSolution}{\vec z}
\newcommand{\DFCalibrationBackgroundSize}{\varepsilon_{\mathrm{bg}}}
\newcommand{\DFCalibrationResponseSize}{\eta_{\mathrm{resp}}}

\newcommand{\DFMatrixNorm}[1]{\lVert #1\rVert_{\mathrm{simp},\FockRadius}}
\newcommand{\DFMatrixDimension}{m_{\BoundaryInverseOrder}}
\newcommand{\DFMatrixPrimarySpace}{V_{\mathrm P}}
\newcommand{\DFMatrixBodySpace}{V_{\mathrm B}}
\newcommand{\DFMatrixDegreeSpace}{\mathcal V}
\newcommand{\DFMatrixSpace}{\mathcal V_{\leq\BoundaryInverseOrder}}
\newcommand{\DFMatrixReferenceColor}{c_*}
\newcommand{\DFMatrixReferenceCenter}{a_*}
\newcommand{\DFMatrixPrimaryBasis}{\varphi^{\mathrm P}}
\newcommand{\DFMatrixBodyBasis}{\varphi^{\mathrm B}}
\newcommand{\DFMatrixBasis}{\varphi}
\newcommand{\DFMatrixBasisIndex}{i_{\mathrm{bas}}}
\newcommand{\DFMatrixSecondIndex}{j_{\mathrm{bas}}}
\newcommand{\DFMatrixThirdIndex}{k_{\mathrm{bas}}}
\newcommand{\DFMatrixStructure}{s^{\mathrm{alg}}}
\newcommand{\DFMatrixOrbitSize}{o_{\mathrm{sym}}}
\newcommand{\DFMatrixDiagonalBlock}{\mat A}
\newcommand{\DFMatrixDegreeIndex}{k}

\newcommand{\BaseMatrixBodyConvolution}{\mat C_{\mathrm B}}
\newcommand{\BaseMatrixLegConvolution}{\mat C_{\mathrm L}}
\newcommand{\BaseMatrixAnchor}{\mat A_{\mathrm{anc}}}
\newcommand{\BaseMatrixPrimaryDiagonal}{\mat D_{\pi}}

\newcommand{\BaseMatrixTarget}{\vec h_{\mathrm{base}}}
\newcommand{\BaseMatrixSignVector}{\vec e}
\newcommand{\BaseMatrixSignFunctional}{\vec\chi}

\newcommand{\AppMatrixQueueBackground}{{\vec q_{\DFTableQueue}}}

\newcommand{\AppMatrixQueueDefect}{{\vec v_{\DFTableQueue}}}

\newcommand{\AppMatrixQueueRoot}{{\mat B_{\mathrm{root},\DFTableQueue}}}
\newcommand{\AppMatrixQueueConnected}{{\mat C_{\DFTableQueue}}}

\newcommand{\RealMatrixBodyEmbedding}{\mat E_{\mathrm B}}
\newcommand{\RealMatrixLegEmbedding}[1]{\mat E_{#1}}
\newcommand{\RealMatrixIncidence}{\mat D_{\CalibContext}}
\newcommand{\RealMatrixSourceMixer}{\mat R_{\mathrm{src}}}

\newcommand{\DFMatrixWeightedPolynomial}{\polyfont{P}_{\mathrm{char},\DFTableWeight}}

\newcommand{\DFContextPositiveMeasure}{\distfont{\nu}_{V,\CalibContext}}
\newcommand{\DFContextIdealProposal}{\distfont{\nu}_{0,\CalibContext}}

\newcommand{\LegFunctionSpace}{\mathcal H}
\newcommand{\LegWeightSynthesis}{\Psi_{\mathrm{leg}}}
\newcommand{\LegCalibrationResponse}{\mathcal A_{\mathrm{leg}}}
\newcommand{\LegCalibrationTarget}{\tau_{\mathrm{leg}}}

\newcommand{\DFFallbackAdjustment}{\delta_{\mathrm{fb}}}
\newcommand{\DFFallbackLoad}{\vec{b}_{\mathrm{fb}}}

%% file: command/pseudocode.tex
\renewcommand{\gamedes}[1]{%
  \par\noindent\underline{\textbf{Game $\GAME_{#1}$:}}%
}

%% file: command/theorems.tex
\makeatletter
\patchcmd{\mdf@put@frame@i}
  {\advance\dimen@.8\pageshrink}
  {}
  {}{\PackageError{LAR}{Cannot repair the mdframed split height}{}}
\patchcmd{\mdf@put@frame@i}
  {\ifdim\wd\mdf@splitbox@two=\wd\mdf@splitbox@one\relax}
  {\edef\lar@mdf@firstwidth{\the\wd\mdf@splitbox@two}%
   \ifdim\wd\mdf@splitbox@two=\wd\mdf@splitbox@one\relax}
  {}{\PackageError{LAR}{Cannot inspect the mdframed first fragment}{}}
\patchcmd{\mdf@put@frame@i}
  {\ifdimequal{\ht\mdf@splitbox@two}{0pt}}
  {\ifboolexpr{
     test {\ifdimequal{\ht\mdf@splitbox@two}{0pt}}
     or test {\ifdimequal{\lar@mdf@firstwidth}{0pt}}}}
  {}{\PackageError{LAR}{Cannot repair empty mdframed first fragments}{}}
\makeatother

\makeatletter
\patchcmd{\mdf@putbox@single}
  {\mdf@frame@topline@single\mdf@frame@background@single}
  {\mdf@frame@background@single\mdf@frame@topline@single}
  {}{\PackageError{LAR}{Cannot preserve the mdframed top border}{}}
\makeatother
\theoremstyle{definition}
\newtheorem{theorem}{Theorem}[section]
\surroundwithmdframed[
  linewidth=0.5pt,
  linecolor=black,
  backgroundcolor=white,
  roundcorner=0pt,
  topline=true,
  innerleftmargin=8pt,
  innerrightmargin=8pt,
  innertopmargin=6pt,
  innerbottommargin=6pt,
  skipabove=10pt,
  skipbelow=10pt
]{theorem}
\theoremstyle{definition}
\newtheorem{lemma}[theorem]{Lemma}
\surroundwithmdframed[
  linewidth=0.5pt,
  linecolor=black,
  backgroundcolor=white,
  roundcorner=0pt,
  topline=true,
  innerleftmargin=8pt,
  innerrightmargin=8pt,
  innertopmargin=6pt,
  innerbottommargin=6pt,
  skipabove=10pt,
  skipbelow=10pt
]{lemma}
\newtheorem{corollary}[theorem]{Corollary}
\surroundwithmdframed[
  linewidth=0.5pt,
  linecolor=black,
  backgroundcolor=white,
  roundcorner=0pt,
  topline=true,
  innerleftmargin=8pt,
  innerrightmargin=8pt,
  innertopmargin=6pt,
  innerbottommargin=6pt,
  skipabove=10pt,
  skipbelow=10pt
]{corollary}
\theoremstyle{definition}
\newtheorem{proposition}[theorem]{Proposition}
\surroundwithmdframed[
  linewidth=0.5pt,
  linecolor=black,
  backgroundcolor=white,
  roundcorner=0pt,
  topline=true,
  innerleftmargin=8pt,
  innerrightmargin=8pt,
  innertopmargin=6pt,
  innerbottommargin=6pt,
  skipabove=10pt,
  skipbelow=10pt
]{proposition}
\theoremstyle{definition}

\theoremstyle{definition}
\newtheorem{definition}[theorem]{Definition}
\surroundwithmdframed[
  linewidth=0.5pt,
  linecolor=black,
  backgroundcolor=white,
  roundcorner=0pt,
  innerleftmargin=8pt,
  innerrightmargin=8pt,
  innertopmargin=6pt,
  innerbottommargin=6pt,
  skipabove=10pt,
  skipbelow=10pt
]{definition}

\surroundwithmdframed[
  linewidth=0.5pt,
  linecolor=black,
  backgroundcolor=white,
  roundcorner=0pt,
  topline=true,
  innerleftmargin=8pt,
  innerrightmargin=8pt,
  innertopmargin=6pt,
  innerbottommargin=6pt,
  skipabove=10pt,
  skipbelow=10pt
]{requirement}
\crefname{requirement}{requirement}{requirements}
\Crefname{requirement}{Requirement}{Requirements}

\theoremstyle{remark}

%% file: chapters/abstract.tex
Pseudorandomness plays a central role in algorithms, complexity theory,
and cryptography. We ask whether deterministic information loss alone
can yield post-quantum pseudorandomness with $\TCZero$ evaluation.
Learning with errors ($\LWE$) hides linear measurements with random noise.
Learning with rounding ($\LWR$) replaces noise with deterministic
rounding, but a standard polynomial-modulus hardness reduction requires
parameters tailored to a predetermined sample bound. Decisional
Diffie--Hellman offers a noiseless algebraic route to pseudorandomness,
but its standard instantiations are vulnerable to quantum algorithms.
We show that deterministic rounding achieves this under quantum
worst-case lattice hardness, with fixed polynomial parameters and no
predetermined sample bound.

We introduce the \emph{learning with algebraic rounding} ($\LAR$) assumption, which
applies an explicit integer rounding map to blocks of secret-dependent
linear measurements and encodes the result in a lattice quotient.
For one fixed family with polynomially bounded modulus and dimensions,
its deterministic labels on exactly uniform, secret-independent public
matrices are computationally indistinguishable from independent uniform
labels for every polynomial number of classical random examples.
Our classical reduction from Gaussian $\LWE$, combined with the quantum
worst-case reduction to $\LWE$, bases $\LAR$ on $\GapSVP$ and $\SIVP$
over general lattices at approximation factor $\widetilde O(\LARModulus)$,
where $\LARModulus$ is the modulus.

To prove this, we develop a \emph{source-independent affine transport
framework} that may be of independent interest. Using public affine maps
sampled independently of the hidden errors, it transports the Gaussian
error law close to uniform on the rounding map's zero-label fiber while
preserving exactly uniform public matrices. For every fixed
inverse-polynomial accuracy, we give a polynomial-time sampler for the
same public parameter family.

This yields the first post-quantum weak pseudorandom function with joint
key-and-input evaluation in $\TCZero$, secure against uniform quantum
polynomial-time distinguishers given any polynomial number of classical
uniform random examples. We also establish the hardness of learning polynomial-size two-hidden-layer ReLU networks from noiseless uniform random examples, even for quantum polynomial-time improper learners.

%% file: chapters/introduction.tex
\section{Introduction}
\label{sec:introduction}

Pseudorandomness plays a central role in algorithms, complexity theory,
and cryptography, providing efficiently computable objects that appear
random to computationally bounded observers. We ask the following natural question:

\begin{quotation}\bfseries
  Can deterministic information loss alone yield post-quantum
  pseudorandomness with $\TCZero$ evaluation?
\end{quotation}

\paragraph{Pseudorandom functions.}
Weak pseudorandom functions (weak PRFs) give a concrete formulation of
this goal. A weak PRF is a keyed family of deterministic functions.
Under one uniformly random
key, its evaluations at independently sampled uniform inputs are
computationally indistinguishable from evaluations of a truly random
function at the same inputs.

Pseudorandom functions were introduced by Goldreich, Goldwasser, and
Micali~\cite{GolGolMic86}. Their strong notion requires computational
indistinguishability even under adaptively chosen oracle queries.
The original construction builds PRFs from pseudorandom generators
and hence from one-way functions. Naor and Reingold~\cite{NaoRei04} gave more efficient
algebraic constructions based on factoring and the decisional
Diffie--Hellman assumption, with evaluation in
$\TCZero$. However, Shor~\cite{FOCS:Shor94} gave quantum polynomial-time
algorithms for the underlying factoring and discrete-logarithm problems.

Banerjee, Peikert, and Rosen~\cite{EC:BanPeiRos12} initiated the study of direct, highly
parallelizable lattice-based PRFs by introducing learning with rounding
($\LWR$) as a deterministic counterpart of learning with errors
($\LWE$). Their constructions include low-depth PRFs with $\TCZero$
evaluation under suitable parameters.
Subsequent work developed key-homomorphic PRFs from
$\LWE$~\cite{C:BLMR13,C:BanPei14} and constrained key-homomorphic PRFs
from standard lattice assumptions~\cite{TCC:BraVai15}. Further work
addressed efficiency, reduction tightness, and modulus
size~\cite{AC:JagKurPan18,INDOCRYPT:Montgomery18,EC:Kim20}.

A central issue for deterministic rounding is to establish hardness
with a polynomial modulus and no a priori bound on the number of
samples. Alwen, Krenn, Pietrzak, and Wichs~\cite[Theorem~4.1]{C:AKPW13}
proved $\LWE$-based hardness
of $\LWR$ with polynomial modulus, but their reduction requires
parameters that depend explicitly on the sample
bound. Newton and Richelson~\cite{C:NewRic23} established
limitations for non-aborting pointwise reductions that preserve the
source dimension and modulus in a specified parameter
regime. These limitations apply to the stated reduction
model and do not establish a general impossibility of
polynomial-modulus $\LWR$ hardness.

Another line of work studies PRFs from learning parity with noise (LPN).
Ding, Jain, and Komargodski~\cite{STOC:DinJaiKom25} constructed $\NCOne$ weak PRFs from standard
LPN and stronger variants under structured LPN
assumptions. The same authors~\cite{ARXIV:DinJaiKom26} subsequently gave a generic
depth-preserving transformation from weak to strong PRFs and obtained
$\NCOne$ strong PRFs from standard $\LWE$ with a polynomial
modulus-to-noise ratio, as well as from standard LPN and computational
Diffie--Hellman.

\paragraph{Computational hardness of learning.}
Weak PRFs with low-complexity evaluation also give hardness of
noise-free learning from uniform random examples. An efficient learner
that predicts a fresh label with a nonnegligible advantage over random
guessing would distinguish pseudorandom labels from independent uniform
ones. In Valiant's
probably approximately correct (PAC) framework~\cite{STOC:Valiant84}, a
concept class can have polynomial sample complexity and efficiently
computable functions, yet admit no polynomial-time learner under suitable
computational assumptions. Such hardness is especially significant for
\emph{improper learning}, where the learner may output any efficiently
evaluable hypothesis, including one outside the target class. Kearns and
Valiant~\cite{KearnsValiant1994} established the cryptographic approach to
this representation-independent hardness.

The input distribution is central to this separation. Distribution-free
hardness rules out a learner that succeeds for every input distribution;
it need not imply hardness from uniformly random examples. A reduction
establishing the latter must preserve the prescribed input law.
Kharitonov~\cite{Kharitonov1995} established cryptographic hardness under
uniform inputs. For distribution-free learning, Klivans and
Sherstov~\cite{KlivansSherstov2009} based the hardness of learning
intersections of halfspaces on worst-case lattice problems.
Low-depth circuits make the role of computational resources concrete:
Linial, Mansour, and Nisan~\cite{LinialMansourNisan1993} gave a Fourier-based
quasipolynomial-time uniform-distribution learning algorithm for
constant-depth AND/OR/NOT circuits ($\mathrm{AC}^{0}$). For constant-depth
threshold circuits ($\TCZero$), Arunachalam, Grilo, and
Sundaram~\cite{ArunachalamGriloSundaram2021} ruled out polynomial-time
quantum learning under the uniform distribution, even with quantum
membership queries, assuming quantum hardness of ring learning with errors.

Shallow neural networks can be hard to learn even without label noise.
Daniely and Vardi~\cite{DanielyVardi2021} obtained
hardness for shallow rectified linear unit (ReLU) networks under Gaussian
inputs from local pseudorandom generators. Chen, Gollakota, Klivans, and
Meka~\cite{NeuroIPS:CGKM22} proved superpolynomial statistical-query lower
bounds for noise-free two-hidden-layer ReLU learning under Gaussian inputs.
They also established cryptographic hardness results based on learning
with rounding. These are distinct types of lower bounds: statistical-query
hardness alone does not rule out general efficient learners.

We seek this separation under exactly uniform inputs, with deterministic
$\TCZero$ evaluation and a worst-case lattice foundation. We consider
uniform quantum polynomial-time learners receiving classical noiseless
random examples, with possibly improper hypotheses given by classical
descriptions and admitting efficient deterministic classical evaluation
as specified in \cref{sec:applications-learning}. Our neural-network
application uses a uniform finite grid. The same fixed family of
polynomial parameters must support every polynomial sample budget for
both pseudorandomness and this learning consequence.

\subsection{Our Results}
\label{sec:introduction-results}

Under quantum worst-case hardness of general-lattice problems, we construct
an explicit deterministic block-rounding family whose labels on exactly
uniform, secret-independent public matrices are indistinguishable from
independent uniform labels.

Learning with algebraic rounding ($\LAR$) mixes a block of linear
measurements using an integer matrix
$\RoundingMatrix\in\ZZ^{\BlockDim\times\BlockDim}$, divides by an odd
prime $\LARModulus$, and rounds down coordinatewise. The label is its coset modulo $\RoundingMatrix\ZZ^{\BlockDim}$,
invariant under changes of integer representatives.
For the rounding matrices of \cref{def:rounding-matrix}, Smith
coordinates encode these cosets in
$\LARField^{\OutputDim}$, where $\LARField=\ZZ/\LARModulus\ZZ$ and
$\OutputDim=\BlockDim-\RoundingRank\geq1$ for modular rank
$\RoundingRank$. These coordinates also give translations prescribing label changes
(\cref{lem:lar-properties}), as used below.
With the quotient coordinate matrix $\QuotientProjection$ of
\cref{def:quotient-coordinates}, the label map $\LARLabelMap$ of
\cref{def:lar-label-map}, and one uniformly random secret
$\LARSecret\in\LARField^{\SecretDim}$, the examples are
\[
  \begin{aligned}
    \LARPublicMatrix_{\SampleIndex}&\sample
    \Unif{\LARField^{\BlockDim\times\SecretDim}},
    \qquad \SampleIndex=1,\ldots,\SampleCount,\\
    \RoundingLabel_{\SampleIndex}
    &=\LARLabelMap(\LARPublicMatrix_{\SampleIndex}\LARSecret)
    =\QuotientProjection
    \left\lfloor
    \frac{\RoundingMatrix
    \CanonicalLift{\LARPublicMatrix_{\SampleIndex}\LARSecret}}
    {\LARModulus}
    \right\rfloor\bmod\LARModulus
    \in\LARField^{\OutputDim}.
  \end{aligned}
\]
Here the matrix-secret product is over $\LARField$,
$\CanonicalLift{\cdot}$ chooses representatives in
$\set{0,\ldots,\LARModulus-1}$. The subsequent matrix product is over
the integers, and the floor acts coordinatewise. Labels use no fresh randomness.
The public matrices are mutually independent and secret-independent;
ideal labels are independent and uniform in $\LARField^{\OutputDim}$.
Exact public-input uniformity gives a random-example learning problem
on the full matrix space.

\paragraph{A fixed family for every polynomial sample budget.}
Our hardness result uses one explicit, polynomial-time computable
family with $\SecretDim=\secpar$, an odd prime $\LARModulus$, and
$\BlockDim\geq\LARModulus$, all polynomially bounded in $\secpar$, and
\[
  \RoundingMatrix=
  \left[\StandardBasisVec_1-\StandardBasisVec_2\mid\cdots\mid
    \StandardBasisVec_{\BlockDim-1}-\StandardBasisVec_{\BlockDim}\mid
    \sum_{\CoordinateIndex=1}^{\LARModulus}\StandardBasisVec_{\CoordinateIndex}
  \right],
  \qquad \QuotientProjection=\OnesVec^{\trans},
  \qquad \OutputDim=1,
\]
where the $\StandardBasisVec_{\CoordinateIndex}$ are standard basis
vectors and $\OnesVec$ is the all-ones vector. Thus each label sums the
rounded coordinates modulo $\LARModulus$. The modulus, dimensions, and
rounding map are fixed before the adversary and its polynomial sample
budget are chosen.

\begingroup
\mdfsetup{topline=true,everyline=true}
\begin{theorem}[Main reduction, informal]
  \label{thm:introduction-main}
  For this fixed public family, any uniform QPT distinguisher that
  separates real and ideal $\LAR$ distributions on polynomially many
  classical random examples with nonnegligible advantage yields
  uniform quantum polynomial-time algorithms for worst-case
  $\GapSVP_{\LatticeApproxFactor}$ and $\SIVP_{\LatticeApproxFactor}$
  on general lattices of dimension $\SecretDim$, with
  $\LatticeApproxFactor=\widetilde O(\LARModulus)$.
\end{theorem}
\endgroup

Quantum worst-case hardness therefore gives pseudorandom labels for every polynomial sample budget
under the same fixed public family.
The reduction passes through Gaussian $\LWE$: the worst-case lattice
step is quantum, while the subsequent sample transformations are
classical and remain valid when composed with a QPT distinguisher.

\paragraph{From Gaussian noise to deterministic information loss.}
The zero fiber consists of inputs with label zero. Translation equivariance
combines a uniform fiber residual and an independent uniform label into a
uniform ambient vector, reducing uniform-zero-fiber $\LWE$ to $\LAR$.
Our \emph{source-independent affine transport} reduces Gaussian $\LWE$ to
uniform-zero-fiber $\LWE$ by making Gaussian errors nearly uniform on this
fiber using public randomness.

Applying this public affine map to both matrix and noisy response preserves
exact matrix uniformity and independence from the transported error through
invertibility modulo $\LARModulus$. The map observes neither hidden errors
nor $\LWE$ data. Its accuracy is chosen after fixing the adversary and
polynomial sample budget, without changing the public family. Transport is
used only in the security reduction; labels remain deterministic.

\paragraph{Calibrating public randomness.}
Local calibration prescribes color probabilities at every protected input
through one positive public law of safe maps. Balanced expected counts still
leave endpoints correlated through shared source coordinates.

Global calibration reweights public choices to cancel bounded-support Fourier
tests; a full-frequency tail bound controls remaining joint error. The solve
temporarily omits the integrality congruence and may yield negative weights. Moment bounds
justify bounded nonnegative clipping; restoring the congruence yields
efficient, source-independent sampling.

\paragraph{Consequences for pseudorandomness and noiseless learning.}
Under the same worst-case hardness condition, restricting the secret
to nonzero keys gives a field-valued weak PRF on
$\LARField^{\BlockDim\times\SecretDim}$ with joint key-and-input
$\TCZero$ evaluation. Security holds against uniform QPT distinguishers
receiving any polynomial number of classical uniform random examples.
The normalized labels also have polynomial-size two-hidden-layer ReLU
representations on a uniform finite grid, with rational weights and
biases of polynomially bounded magnitude and bit length.

The same assumption rules out uniform QPT PAC learning of these
networks in squared loss from classical noiseless random examples,
including improper learning with classically described hypotheses
whose rational predictions admit uniform deterministic polynomial-time
classical evaluation. The separation is computational: exhaustive
search identifies the secret from $O(\SecretDim)$ examples at fixed
confidence.

\paragraph{Formal verification.}
The main theorems and lemmas listed in the Prove2Me project\footnote{%
  \raggedright\url{https://prove2.me/missions/Information_Loss_Is_All_You_Need\%3A_TC\%5E0_Pseudorandomness_from_Worst-Case_Lattices}}
have all been verified in Lean~4. These include the fixed-order Gaussian
transport of \cref{thm:df-constructive-aft} and the reduction from worst-case
lattice problems to decisional $\LAR$ in \cref{cor:applications-worst-case}.

\subsection{Technical Overview}
\label{sec:introduction-techniques}

Gaussian $\LWE$ hides a linear function of the secret using random
errors; a $\LAR$ label is deterministic. Simply rounding an $\LWE$
response can leave its label dependent on the hidden error.
We instead transport Gaussian errors to a law that rounding absorbs
exactly, while preserving exactly uniform public inputs. The transport
uses randomness independent of the samples in the security reduction;
label evaluation remains deterministic.

The resulting reduction chain is
\[
  \begin{aligned}
    \{\GapSVP,\SIVP\}
    &\xrightarrow{\text{\cite{STOC:Regev05}; \cref{sec:applications-worst-case-proof}}}
    \text{Gaussian }\LWE
    \xrightarrow{\text{\cref{sec:gaussian-lwe-to-uniform-zero-fiber-lwe}}}
    \text{uniform-zero-fiber }\LWE\\
    &\xrightarrow{\text{\cref{sec:uniform-zero-fiber-lwe-to-lar}}}\DecisionalLAR
    \xrightarrow{\text{\cref{sec:applications-weak-prf,sec:applications-learning}}}
    \text{weak PRFs and noiseless learning hardness}.
  \end{aligned}
\]
The first arrow supplies the quantum worst-case lattice foundation;
the last gives consequences for classical random-example access.
For the middle arrows, we identify the required error law, construct
affine maps that reach its support except on a Gaussian tail event,
and choose a positive public law of maps giving joint uniformity up
to controlled error.

\paragraph{I. Finding the error law that rounding can absorb.}
For the label map $\LARLabelMap$ recalled in
\cref{sec:introduction-results}, the zero fiber $\LWEZeroFiber$
consists of inputs with label zero. Its role follows from
\emph{translation equivariance}: the matrix $\RoundingDirection$ in
\cref{lem:lar-properties} converts any desired label change
$\RoundingLabel\in\LARField^{\OutputDim}$ into an input translation.
Adding $\RoundingDirection\RoundingLabel$ adds precisely
$\RoundingLabel$ to the label. Thus every input has a unique
description by a zero-fiber residual and a label. The decomposition
of \cref{lem:zero-fiber-decomposition} is the bijection
\[
  (\LWEError,\RoundingLabel)
  \longmapsto\LWEError+\RoundingDirection\RoundingLabel,
  \qquad
  \LWEZeroFiber\times\LARField^{\OutputDim}
  \xrightarrow{\ \sim\ }\LARField^{\BlockDim}.
\]
A uniform residual and an independent uniform label reconstruct an
exactly uniform input whose rounded label is the chosen label.
Thus fiber membership removes the error from the label, while
uniformity supplies the required input law. The residual need not
be coordinatewise small: zero-fiber membership concerns integer
rounding and quotient encoding, not the modular kernel of
$\RoundingMatrix$. Coordinatewise error bounds alone do not solve
this joint distributional problem.

\Cref{thm:uniform-zero-fiber-lwe-to-dlar} samples one uniform auxiliary
vector $\ReductionVector\in\LARField^{\SecretDim}$ shared across blocks
and independent uniform labels, all independently of the input table.
On real uniform-zero-fiber $\LWE$ input with hidden secret $\LWESecret$,
its rank-one update makes the matrix-secret product equal the zero-fiber
residual plus the chosen label direction when $\ReductionVector^{\trans}\LWESecret=1$.
With exact auxiliary draws, this gives real $\LAR$ labels on exactly
uniform matrices; otherwise the update gives the ideal product law.
For each nonzero secret, alignment has probability $1/\LARModulus$,
explaining the advantage loss. The shared vector selects one branch
for the whole table. The theorem also accounts for the zero-secret
event and bounded sampling.
It remains to reach the zero-fiber error law from Gaussian samples.

\paragraph{II. Moving errors that the reduction cannot see.}
An $\LWE$ reduction receives a public matrix $\LWEPublicMatrix$ and
a noisy response $\LWEResponse$, but cannot extract the hidden
Gaussian error $\LWEError$. An error-dependent transport would thus
be unusable. We instead sample an integer matrix $\TransportMatrix$
and offset $\TransportOffset$ independently of the secret and the
entire sample table. Their affine action can be applied directly to
the observed block:
\[
  (\LWEPublicMatrix,\LWEResponse)
  \longmapsto
  (\TransportMatrix\LWEPublicMatrix,
  \TransportMatrix\LWEResponse+\TransportOffset),
  \qquad
  \LWEError\longmapsto
  \TransportMatrix\LWEError+\TransportOffset\pmod{\LARModulus}.
\]
Source independence makes this operation available to the reduction.
If every sampled $\TransportMatrix$ is invertible modulo
$\LARModulus$, the transformed public matrix is exactly uniform even
conditional on the affine map and error. It is therefore independent
of the transformed error. An ideal uniform response also remains
uniform. It suffices to make the transformed error close to
$\Unif{\LWEZeroFiber}$, averaging over the independently drawn map
and source; an individual fixed map need not achieve this closeness.
\Cref{lem:t-aft-properties,thm:gaussian-lwe-to-uniform-zero-fiber-lwe}
then give the reduction, with statistical loss at most the number of
blocks times the single-block transport error.

\paragraph{Why incidence geometry is useful.}
We first design integer vertex \emph{loads} $\EndpointLoad$, then
recover the affine output through $\RoundingMatrix^{-1}$. Their sum is the
\emph{charge}, their residues are endpoint \emph{colors}, and their
integer quotients on division by $\LARModulus$ are \emph{carries}.
For the matrix in \cref{sec:introduction-results}, two distinct
conditions govern a load's preimage:
\[
  \begin{aligned}
    \sum_{\CoordinateIndex=1}^{\BlockDim}\EndpointLoad_{\CoordinateIndex}
    &\equiv0\pmod{\LARModulus}
    &&\text{(integral preimage)},\\
    \sum_{\CoordinateIndex=1}^{\BlockDim}
    \left\lfloor\frac{\EndpointLoad_{\CoordinateIndex}}{\LARModulus}\right\rfloor
    &\equiv0\pmod{\LARModulus}
    &&\text{(zero label)}.
  \end{aligned}
\]
The first says $\EndpointLoad\in\RoundingMatrix\ZZ^{\BlockDim}$,
so $\RoundingMatrix^{-1}\EndpointLoad$ is integral
(\cref{lem:rounding-image-lattice}). Once it holds, the second
ensures that this preimage modulo $\LARModulus$ belongs to the zero
fiber (\cref{lem:affine-transport-endpoint-carry-bound}). Individual
carries may be nonzero; only their sum must vanish modulo $\LARModulus$.

The charge condition places the colors in the zero-sum hyperplane
$\ZeroSumEndpointHyperplane$. Taking their canonical representatives,
applying $\RoundingMatrix^{-1}$, and reducing modulo $\LARModulus$
gives a bijection to the zero fiber
(\cref{lem:canonical-endpoint-pullback}). We must therefore make the
colors jointly uniform while enforcing compatible carries.
For the affine output $\TransportOutput$ and its colors
$\EndpointResidue$, the bound of
\cref{lem:affine-transport-endpoint-carry-bound} reads
\[
  \StatisticalDistance{\DistributionLaw{\TransportOutput}}
  {\Unif{\LWEZeroFiber}}
  \leq
  \StatisticalDistance{\DistributionLaw{\EndpointResidue}}
  {\Unif{\ZeroSumEndpointHyperplane}}
  +\Pr[\text{incompatible carry}],
\]
where $\DistributionLaw{\cdot}$ denotes a probability law and
$\StatisticalDistance{\cdot}{\cdot}$ statistical distance.

A directed edge has an incidence column with entries $+1$ and $-1$
at its endpoints. Multiplying by the edge's source coordinate
transfers load without changing the charge. A tree's columns form
an integer basis of all zero-sum loads; adjoining a column of charge
$\LARModulus$ gives a basis of $\RoundingMatrix\ZZ^{\BlockDim}$.
Applying $\RoundingMatrix^{-1}$ produces an integer matrix of
determinant $\pm1$. Source-coordinate changes invertible modulo
$\LARModulus$ preserve this modular invertibility
(\cref{cor:general-tree-incidence-transport}). The remaining task is
to choose a tree and local maps with safe carries and adjustable colors.

Safety is analyzed on a \emph{protected source}, obtained by conditioning
each underlying Gaussian on a prescribed bounded interval. Protection
includes the independent public Gaussian used below to close the body
path into a cycle. The sampler never tests hidden inputs.
Coupling the full and protected experiments fails only when a Gaussian
leaves its interval. For our parameter choice, the total probability is
exponentially small (\cref{sec:joint-injection-protection}).

\paragraph{A geometry with safe carries and adjustable colors.}
The \emph{chameleon lobster} in
\cref{fig:introduction-transport-tree} supplies public parameters
that adjust colors while ensuring compatible carries throughout
this protected experiment. We call these parameters the public \emph{controls}.

\begin{figure}[H]
  \centering
  \begin{tikzpicture}[
      x=1cm,y=1cm,font=\footnotesize,line cap=round,
      vertex/.style={circle,draw=black,fill=white,line width=.7pt,
      minimum size=4.2mm,inner sep=0pt},
      queue vertex/.style={vertex,draw=blue!55!black,fill=blue!12},
      control vertex/.style={vertex,fill=black!15},
      tree edge/.style={line width=.85pt},
      closure/.style={draw=blue!55!black,densely dotted,line width=1pt},
      pairing/.style={draw=black!65,dashed,line width=.8pt},
      note/.style={align=center,inner sep=2pt}
    ]
    \draw[draw=blue!55!black,fill=blue!3,dashed,rounded corners=7pt]
    (9.55,-.43) rectangle (13.65,.43);

    \node[vertex] (first) at (0,0) {};
    \node[control vertex] (attachment) at (1.6,0) {};
    \node[vertex] (body-two) at (3.2,0) {};
    \node[vertex] (other-attachment) at (4.8,0) {};
    \node[control vertex] (compensator) at (6.5,0) {};
    \node[vertex] (free) at (8.2,0) {};
    \node[queue vertex] (queue-first) at (10,0) {};
    \node[queue vertex] (queue-middle) at (11.2,0) {};
    \node[queue vertex] (last) at (13.2,0) {};
    \draw[tree edge] (first) -- (attachment) -- (body-two)
    -- (other-attachment) -- (compensator) -- (free)
    -- (queue-first) -- (queue-middle) -- (11.7,0);
    \node[inner sep=0pt] at (12.2,0) {$\cdots$};
    \draw[tree edge] (12.7,0) -- (last);
    \node[note,anchor=south] at (6.5,.38)
    {body path: hidden Gaussian coordinates};

    \draw[closure] (first) -- (0,1.22) -- (13.2,1.22) -- (last);
    \node[note,fill=white,text=blue!55!black] at (6.6,1.22)
    {public Gaussian closure: opposite offsets};

    \node[vertex] (companion) at (1.6,-1.05) {};
    \node[control vertex] (primary) at (1.6,-2.1) {};
    \node[vertex] (other-companion) at (4.8,-1.05) {};
    \node[vertex] (other-primary) at (4.8,-2.1) {};
    \draw[tree edge] (attachment) -- (companion) -- (primary);
    \draw[tree edge] (other-attachment) -- (other-companion) -- (other-primary);
    \node[note,anchor=south] at (1.6,.38) {attachment};
    \node[note,anchor=west] at (1.85,-.33) {$+\LocalSecondError$};
    \node[note,anchor=east] at (1.3,-1.05) {companion};
    \node[note,anchor=east] at (1.3,-2.1) {primary};
    \node[note,anchor=west] at (1.95,-1.55)
    {one leg uses\\hidden $(\LocalFirstError,\LocalSecondError)$};
    \node[note,anchor=north] at (4.8,-2.32) {other legs};
    \node[note,anchor=north west,align=left] at (6.8,-.27)
    {free body\\compensator};

    \draw[pairing] (primary) -- (1.6,-2.95) -- (6.5,-2.95)
    -- (compensator.south);
    \node[note,anchor=north] at (4.05,-3.04)
    {public control pairing (one distinct compensator per leg)};
    \node[note,text=blue!55!black,anchor=north] at (11.6,-.58)
    {queue: $\LARModulus$ body vertices\\
      one shared hidden Gaussian\\
    added at every queue site};
  \end{tikzpicture}
  \caption{The chameleon lobster and its public controls.
    Solid segments are tree edges. The dotted closure supplies opposite
    public offsets; the dashed link pairs a primary with a separate
    body vertex whose distribution compensates for changes in the companion.
    Neither adds a hidden-source edge. The queue enclosure represents
  one shared source coordinate, with sites drawn together for clarity.}
  \label{fig:introduction-transport-tree}
\end{figure}

The \emph{body} is a path of hidden Gaussian edge values, with
vertex increments given by neighboring differences. An independent
public Gaussian adds opposite endpoint offsets, closing these into
a cycle law through $\TransportOffset$. All increments then have
the same marginal law, but neighboring differences share a Gaussian
and all increments sum to zero. Public initial loads, called
\emph{centers}, shift the body colors.

The \emph{legs} provide more targeted color control. A leg runs from
an outer \emph{primary} through a \emph{companion} to an
\emph{attachment} on the body. A public affine map acts on two
hidden Gaussian coordinates $(\LocalFirstError,\LocalSecondError)$.
For a public shift $\LegBaselineShift$, the primary and companion
loads sum to
$\LARModulus-1+\LegBaselineShift-\LocalSecondError$;
the attachment receives $+\LocalSecondError$ from the very same
coordinate. This shared input makes the leg's total contribution
independent of $\LocalSecondError$. Safe affine maps also make the
two off-body residue representatives have the same integer sum as
their loads, so their carries cancel even when either load wraps.
Body centers leave enough room for protected increments that body
carries vanish. \Cref{prop:matrix-carry-safety} proves the resulting
carry compatibility.

The \emph{queue} supplies the direction missing from tree incidence.
It adds one hidden Gaussian scalar at all $\LARModulus$ queue sites,
giving a column of total charge $\LARModulus$ and hence full rank.
This shared scalar creates a global dependence. A congruence on public
centers and leg shifts ensures that the initial loads also have
charge divisible by $\LARModulus$, making the affine offset integral.
This is the \emph{public charge condition}.

A leg's control changes both its primary and companion colors.
To isolate the primary response, each leg is assigned a distinct
free body vertex, its \emph{compensator}, whose center law can offset
the companion's change in expected color counts. These vertices lie
outside the attachments and queue. The initial assignment of
attachment and queue sites, which also determines the compensators,
is the public \emph{placement}. Two subsequent permutations have
separate purposes.
A uniform body permutation distributes the cycle increments among
body roles, making selected positions unlikely to be neighbors.
An independent final permutation relabels all endpoints, so symmetric
histogram and Fourier information applies to arbitrary endpoint
positions. Neither permutation makes the endpoint colors independent.

\paragraph{Marking a color without observing the hidden input.}
At a fixed protected input, the leg's two canonical color
representatives must sum to
$\LARModulus-1+\LegBaselineShift-\LocalSecondError$.
The diffuse target law is uniform on such pairs with both colors in a
trimmed interior (\cref{def:leg-trimmed-pair-law}); choosing the
primary fixes its companion. This law misses boundary primary
colors. A \emph{marked color} is a public target, chosen independently
of hidden inputs, that lets us add probability at any prescribed primary.

An affine map can create this bias without testing the hidden input.
Fix its constant term to the marked color and its coefficient on
$\LocalSecondError$ to zero. The primary then equals the marked
color whenever $\LocalFirstError=0$, regardless of
$\LocalSecondError$. Mixing these \emph{marked maps} with a diffuse
branch gives a controllable contribution whose source-averaged mass
is the marking probability times $\Pr[\LocalFirstError=0]$ under
the protected Gaussian law. Because activation depends only on the
first coordinate, it preserves the second coordinate's law at the
attachment.

The difficulty is to retain only this intended bias. Marked maps
also affect inputs off the line $\LocalFirstError=0$, and deleting
unsafe maps biases the diffuse branch. Repairing the output after
seeing the input would violate source independence. We must repair
\emph{one public distribution of maps} so that every protected input
has its prescribed color law simultaneously.

\paragraph{Local calibration gives a positive law.}
For this repair, begin with uniform unrestricted affine maps. Their
evaluations at distinct protected inputs are pairwise independent.
A centered function of one evaluation therefore changes that input's
color marginal and averages to zero at every other input. Summing
such functions gives an exact reference response to the desired
marginal adjustments.

We then exclude zero first-coordinate coefficients and unsafe maps.
The excluded probability is small even conditional on an allowed
evaluation, so deletion only slightly perturbs this response.
A finite linear solve reweights the remaining, \emph{regular} maps
to restore the marginals and cancel the marked branch's off-line
bias. A separate one-sided pointwise bound proves positivity on the
regular support (\cref{sec:leg-calibration-proof,cor:calibration-pointwise-positivity}).
Under its feasibility and marked-color conditions,
\cref{thm:leg-constructive-sampler} gives an exact public map law whose
output at every protected input is uniform on the interior pairs above when
$\LocalFirstError\ne0$, and, for our parameter choice, an equal
mixture of that law and the pair with the marked primary and the
companion fixed by the same integer sum when
$\LocalFirstError=0$.
Every supported map is safe at every protected input, with cancelling
off-body carries, and has the nonzero first-coordinate coefficient
required for modular invertibility.
Bounded sampling approximates the complete flag--map law, including
activation probabilities, while preserving this support.

\paragraph{One calibration principle, used twice.}
Both calibrations solve for weights on public choices. Local
constraints specify color probabilities at each protected input;
global constraints specify Fourier coefficients averaged over the
protected source. The weight itself uses only public data.

The interface in \cref{sec:finite-dimensional-calibration} separates
coefficients, weights, and responses. Synthesis $\ProbCalSynthesis$
maps coefficients $\ProbCalCoefficient$ to weights relative to a
reference law; response $\ProbCalResponse$ records unnormalized
constraints, including mass as the weight's expectation. Reference
evaluation $\ProbCalEvaluation$ converts coefficients into ideal
responses; $\ProbCalOperator$ accounts for the actual construction:
\[
  \ProbCalResponse\ProbCalSynthesis
  =\ProbCalEvaluation\ProbCalOperator.
\]
Solving $\ProbCalOperator\ProbCalCoefficient=\ProbCalTarget$ for target
coefficients $\ProbCalTarget$ gives response $\ProbCalEvaluation\ProbCalTarget$. Only
$\ProbCalOperator$ must be invertible; $\ProbCalEvaluation$ need not be.
Both applications prove
$\|\ProbCalOperator-\ProbCalIdentity\|\leq\ProbCalContraction<1$,
with $\ProbCalIdentity$ the identity, using the supremum norm locally
and a weighted coefficient norm globally. Neumann inversion gives
inverse norm at most $1/(1-\ProbCalContraction)$ and a controlled
coefficient correction (\cref{thm:stable-finite-dimensional-calibration}).
The construction-specific work is to build this stable response and
realize a probability law: positive directly in the local case,
initially signed in the joint case.

\paragraph{III. From balanced colors to a uniform joint law.}
The local law gives safe, adjustable color marginals; joint uniformity
still requires choosing the marked-color and body-center laws.
Uniform marginals alone do not suffice: coordinates that all equal
the same uniform field element
have uniform marginals and a highly nonuniform joint law. Here body
increments share Gaussians, each companion shares
$\LocalSecondError$ with its attachment, and all queue sites share
one scalar.

Charge conditioning would also couple the public controls. To keep
an independent reference for these controls, we first omit charge and
target the uniform law on the entire color space
$\LARField^{\BlockDim}$, in which coordinates are independently uniform.
This is an auxiliary target: the candidate integer loads need not yet
have integral affine preimages under $\RoundingMatrix$.
Conditioning the uniform target on zero color sum gives exactly
$\Unif{\ZeroSumEndpointHyperplane}$. We will impose public charge
only after controlling the ambient error, restoring integral affine data.

\paragraph{A positive proposal that balances first moments.}
We first choose a positive public reference law, the \emph{proposal},
that balances expected color counts. Marking can supply missing
primary mass, but also moves the companion. To compensate using
body centers, their effects must undergo the same random-shift
averaging. We place the companion's reference color in a safe interval
and add an independent public offset with the marginal law of one
protected cycle difference. This offset copies a law, never an actual
hidden increment. Companion and body-center responses then share
the required averaging.

For the chosen public family and sufficiently large $\secpar$,
these laws balance the histogram in expectation over the proposal
controls, protected cycle, leg inputs, and shared queue scalar, at every
fixed placement (\cref{thm:df-positive-baseline}).
Independent uniform final relabeling gives uniform coordinate marginals.
The proposal's control groups are independent at this placement;
their endpoints remain dependent through the hidden inputs.

\paragraph{Pairing makes the first response invertible.}
Reweight a leg's marked color and its distinct compensator. Common
shift averaging cancels the marked companion contribution. The diffuse
target uses only the mark's side of the midpoint, mixing fairly at
equality. Its change is therefore one scalar times a fixed profile
difference. Dividing by the source-averaged activation probability
leaves the primary adjustment plus this rank-one residual.
\Cref{lem:df-first-inverse} bounds this perturbation and supplies a
stable inverse for the normalized \emph{first response}. The attachment
has zero first response, but its shared hidden coordinate remains in joint tests.

\paragraph{Hoeffding-type corrections give a triangular response.}
Joint uniformity requires every nonconstant product of Fourier
characters to have zero expectation. Its \emph{support} is the set of
endpoints tested. We use a Hoeffding-type grading of public
reweightings: degree counts how many control groups interact in each
term. This orders the correction of pairwise, triple, and higher biases.

A degree-$\HoeffdingDegree$ correction averages terms selecting
$\HoeffdingDegree$ distinct control groups, each a primary with its
compensator or an ordinary free body center. Every term is centered
in each selected group: averaging that group under the proposal
gives zero with the others fixed. Different groups affect disjoint
endpoint sets. A test on fewer than $\HoeffdingDegree$ endpoints
misses at least one selected group, whose averaging kills that term,
even with the protected cycle fixed. This holds for every selection
and hence their average, so higher-degree corrections preserve
lower-degree constraints.

\paragraph{A small perturbation makes the response invertible.}
Fix a positive integer calibration order $\BoundaryInverseOrder$ and
write $\HoeffdingParameterBase=\secpar+2$.
The square coefficient response $\DFMatrixLifted$ of
\cref{sec:df-actual-inversion} realizes $\ProbCalOperator$ through
degree $\BoundaryInverseOrder$. Triangularity orders the equations
but does not bound their diagonal inverses. We compare the response
with the identity after normalizing its leading terms.

Three effects remain. Unselected controls supply background
correlations after first-moment balancing; selected groups can emit
more endpoint factors than their leading response. Local profile
bounds and control-pool normalization make these effects small, with
only boundedly many pattern types at fixed order.

The third source is cycle dependence. Compare distinct body increments
with independent copies of their protected marginal. Actual increments
share Gaussians only at neighboring positions; the body permutation
makes these overlaps rare. \Cref{thm:df-cycle-density,lem:df-actual-cycle-defect}
bound this defect in a weighted $L^2$ coefficient norm matching
distinct-group averaging. Each complete leg retains its shared hidden
input, and complete expressions use one common queue scalar, averaged
before inversion. Together the estimates give
$\|\DFMatrixLifted-\IdentityMatrix\|
=O_{\BoundaryInverseOrder}(\HoeffdingParameterBase^{-3})$
(\cref{sec:df-graded-application}). For each fixed order and
sufficiently large $\secpar$, the common Neumann bound therefore
gives a bounded inverse and a small coefficient correction.

\paragraph{Second application: exact signed joint calibration.}
We solve the public coefficient equation
$\DFMatrixLifted\DFMatrixSolution=\DFMatrixZero$, where
$\DFMatrixZero$ is the constant-one target and
$\DFMatrixSolution$ is the coefficient vector to be chosen.
At each degree, subtract the response of the coefficients already
determined and solve the current diagonal block. The proposal has
zero degree-one part, so corrections start at degree two
(\cref{sec:df-canonical-first-degree}), retaining all mixtures of
paired and body controls.
\Cref{thm:df-actual-table-inverse} gives a
public weight $\DFTableWeight$ with expectation one under the
fixed-placement proposal.
After independent uniform endpoint relabeling, its
\emph{signed, uncharged, protected-source} endpoint law has every
nontrivial Fourier coefficient of support at most
$\BoundaryInverseOrder$ exactly zero. This is the global counterpart
of the local marginal repair, but the weight may be negative.

At fixed order this polynomial-size system averages the known protected-source
law in advance; its weight reads only public controls and placement.

\paragraph{Finite cancellation needs a complete Fourier tail bound.}
Finite cancellation leaves too many frequencies for individual bounds
to suffice. We control their total energy using the \emph{complete}
Fourier norm $\DFFourierNorm_{\FockRadius}$ of \cref{sec:df-target}.
Its square sums squared coefficients over all frequencies, weighted
by support size. The analysis radius $\FockRadius$ penalizes large supports.

Set $\FockRadius=\HoeffdingParameterBase^{1/4}$, as in
\cref{sec:df-parameter-choice}. For the public correction coefficients
$\DFMatrixCorrection$ and the endpoint law
$\SignedEndpointLaw_{\DFTableWeight}$ weighted by $\DFTableWeight$,
\cref{thm:df-actual-table-inverse,thm:df-weighted-tail,prop:df-corrected-weight-bounds}
give
\[
  \DFMatrixNorm{\DFMatrixCorrection}
  =O_{\BoundaryInverseOrder}(\HoeffdingParameterBase^{-3}),
  \qquad
  \DFFourierNorm_{\FockRadius}(\SignedEndpointLaw_{\DFTableWeight})
  =O_{\BoundaryInverseOrder}(1).
\]
By \cref{lem:df-fourier-isometry}, cancellation through
$\BoundaryInverseOrder$ bounds the square root of the remaining Fourier
energy by $O_{\BoundaryInverseOrder}(\FockRadius^{-(\BoundaryInverseOrder+1)})$.
Parseval and Cauchy--Schwarz give the same half-$L^1$ distance to
uniformity on $\LARField^{\BlockDim}$ (\cref{lem:df-ambient-target}),
hence decay $\HoeffdingParameterBase^{-(\BoundaryInverseOrder+1)/4}$
before charge conditioning. The radius affects only the analysis.

Fix placement, protected cycle, body permutation, and shared queue.
The unweighted proposal factors into independent singletons and
complete leg triples, so their character polynomials multiply.
An all-degree product bound combines local coefficient bounds with
exponential moments of the normalized imbalance of the conditional mean
histogram. Changing one latent Gaussian affects only two cycle
differences, controlling fluctuations; the queue adds uniformly small
drift and is averaged after multiplication. For each summand in the
correction average, local operators handle its bounded set of selected
groups, including compensators; the product bound controls the remaining endpoints.

\paragraph{High moments turn the signed solution into a sampler.}
Mass one and Fourier cancellation do not ensure positivity,
and pointwise closeness to one is unavailable. For
each fixed calibration order $\BoundaryInverseOrder$ and each fixed
positive integer moment order $\FockMomentOrder$,
\cref{prop:df-corrected-weight-bounds} gives, for sufficiently large
$\HoeffdingParameterBase$ depending on these orders,
\[
  \|\DFTableWeight-1\|_{2\FockMomentOrder}
  \leq\DFProofConstant_{\BoundaryInverseOrder,\FockMomentOrder}
  \HoeffdingParameterBase^{-3}.
\]
Here $\DFProofConstant_{\BoundaryInverseOrder,\FockMomentOrder}$
depends only on the two fixed orders. The norm uses the uncharged
public proposal, uniformly over fixed placements.
In the moment expansion, centering kills terms with a control index
used only once. The large independent control pools bound the
remaining overlaps, transferring coefficient control to high moments
even for paired controls (\cref{lem:df-rectangular-moments,cor:df-weight-moments}).

Clip $\DFTableWeight$ to $[0,2]$. Changed weights satisfy
$|\DFTableWeight-1|\geq1$, so the $L^1$ loss is bounded by the
corresponding even moment. All fixed moment orders apply to the same
weight at fixed calibration order, making clipping loss negligible
uniformly over placements (\cref{lem:df-positive-repair}). The protected
endpoint channel contracts $L^1$, giving a nonnegative, unnormalized
ambient measure negligibly close to the signed law.

\paragraph{Charge and the passage to the zero fiber.}
Charge is public: cycle increments sum
to zero, each leg's second input cancels at its attachment, and the
queue contributes a multiple of $\LARModulus$. Conditioning on the
public charge congruence therefore selects
$\ZeroSumEndpointHyperplane$ and restores integral affine offsets
without inspecting the source.
On this hyperplane, frequencies differing by a multiple of the all-ones
vector define the same character.

We control the restricted law through the ambient approximation.
The ambient uniform law assigns the hyperplane mass exactly $1/\LARModulus$,
so restricting and normalizing
the nonnegative measure amplifies ambient error by only $O(\LARModulus)$.
For sufficiently large fixed calibration order, the ambient error
is $o(1/\LARModulus)$. \Cref{thm:df-positive-conditioning} then gives
clipped charge mass at least $1/(4\LARModulus)$ and conditional
error
$O_{\BoundaryInverseOrder}
(\LARModulus\HoeffdingParameterBase^{-(\BoundaryInverseOrder+1)/4})$
plus a negligible term. Polynomially bounded $\LARModulus$ preserves
the negligibility of clipping.

Rejection with the clipped weight and public charge test samples
this law. Retaining placement through retries preserves its marginal.
Acceptance of order $1/\LARModulus$ and a deterministic retry cap give
worst-case polynomial time with controlled failure
(\cref{lem:df-bounded-rejection}). Numerical approximations and
fallbacks preserve integrality and modular invertibility on every
execution (\cref{sec:df-numerical-verification}). Conditional local
maps, independent closure, and permutations complete the
source-independent transport.

\paragraph{One public family for every polynomial sample budget.}
Use the public family and Gaussian source specified below.
For each sufficiently large fixed calibration order
$\BoundaryInverseOrder$, an implementation with numerical budget
$\DFNumericalBudget$ satisfies the bound of
\cref{thm:df-constructive-aft}, for all sufficiently large
$\secpar$, with $\DFProofConstant_{\BoundaryInverseOrder}$ depending
only on the fixed calibration order:
\[
  \StatisticalDistance{\DistributionLaw{
  \TransportMatrix\LWEError+\TransportOffset\bmod\LARModulus}}
  {\Unif{\LWEZeroFiber}}
  \leq
  \DFProofConstant_{\BoundaryInverseOrder}
  \LARModulus\HoeffdingParameterBase^{-(\BoundaryInverseOrder+1)/4}
  +\DFNegligibleError_{\BoundaryInverseOrder}(\HoeffdingParameterBase)
  +\DFNumericalBudget.
\]
The terms are the charge-amplified Fourier remainder, the negligible
sum of charge-amplified clipping loss and source-protection error,
and numerical error.
At fixed calibration order, the negligible bound is uniform over
placements. One source coupling under the positive public law covers
both endpoint disagreement and unsafe carries.

For the family below,
$\LARModulus=\Theta(\HoeffdingParameterBase^{64})$.
Fix an integer accuracy order $\TransportAccuracyOrder\geq1$
and choose a fixed calibration order
$\BoundaryInverseOrder\geq4\TransportAccuracyOrder+267$.
The Fourier term is then
$O_{\BoundaryInverseOrder}
(\HoeffdingParameterBase^{-\TransportAccuracyOrder-3})$.
With $\DFNumericalBudget\leq
\tfrac14\HoeffdingParameterBase^{-(\TransportAccuracyOrder+2)}$
the total error is at most $\secpar^{-\TransportAccuracyOrder}$ eventually.
The calibration order determines $\DFTableWeight$ and remains fixed as
$\secpar$ grows; the numerical budget may shrink. Both choices leave
the public family and source unchanged.

For security, first fix this public family and Gaussian source.
Given a uniform QPT distinguisher with classical sample access and
nonnegligible advantage, fix its polynomial block-count bound
$\SampleCount$ and an
inverse-polynomial advantage lower bound on an infinite set of security
parameters. By \cref{cor:gaussian-lwe-to-lar-composition}, the
statistical loss is at most
$\LARModulus\SampleCount\secpar^{-\TransportAccuracyOrder}
+\LARModulus^{-\SecretDim}$.
Choose one fixed accuracy order making this loss smaller than half
that advantage for all sufficiently large parameters in that set.
The corresponding uniform PPT sampler's running-time polynomial may
depend on the order. Thus the same LAR family supports every polynomial
sample budget; this asserts neither uniform polynomial complexity
in inverse accuracy nor negligible error for one fixed-order sampler.
The quantum worst-case lattice foundation then yields the $\TCZero$
weak PRF and noise-free finite-grid learning hardness under classical
random-example access in \cref{sec:introduction-applications}.

These guarantees hold for the explicit family in
\cref{sec:df-parameter-choice}, which takes $\LARModulus$ to be the
least prime at least $2^{16}\HoeffdingParameterBase^{64}$ and sets
\[
  \SecretDim=\secpar,\qquad
  \GaussianWidth=\HoeffdingParameterBase,\qquad \OutputDim=1,\qquad
  \BlockDim=\LARModulus\HoeffdingParameterBase^{96}
  +2\HoeffdingParameterBase^{128}.
\]
Thus $\BlockDim=\Theta(\HoeffdingParameterBase^{160})$, with the remaining
choices verified in \cref{sec:df-parameter-verification}.
These conservative exponents reflect coarse bounds chosen to simplify
the proof. We believe that a more refined analysis could reduce the
modulus and block dimension.

\subsection{Applications and Relation to Prior Work}
\label{sec:introduction-applications}

\paragraph{Uniform inputs and nonstandard rounding.}
The Nearby Learning with Lattice Rounding (NLWLR) problem of
Montgomery~\cite{PQCRYPTO:Montgomery18}
provides fully deterministic rounding with polynomial modulus and
no predetermined polynomial bound on the number of
samples.
Its real sampler first chooses a uniform lattice-coefficient vector
and a small perturbation, then chooses a matrix constrained so that
its product with the secret equals that lattice vector plus the
perturbation. Deterministic lattice inversion recovers the coefficient
vector, whose reduction modulo two is the label
\cite[full version, Section~3.1 and Definition~4.2]{PQCRYPTO:Montgomery18}.
The ideal sampler instead uses uniform matrices and independent
uniform binary-vector labels. Thus the real-input interface uses a
secret-dependent nearby-lattice sampler; it is not an experiment
that draws real public matrices independently and exactly uniformly.

Our contribution combines an explicit deterministic
quotient-rounding map, exactly uniform secret-independent public
matrices, and fixed polynomial parameters supporting every polynomial
sample budget. Blockwise rounding records a lattice coset, and the
source-independent affine transport preserves the public-matrix law
exactly. These structural properties supply the uniform random-example
interface used in the weak-PRF and finite-grid learning consequences.

\paragraph{Compatibility with lower bounds for ordinary LWR.}
Newton and Richelson~\cite[Definition~5 and Theorem~2]{C:NewRic23}
analyze non-aborting pointwise reductions from
$\LWE$ to ordinary scalar LWR: a fixed deterministic map acts on
each sample separately, with the same dimension and source modulus.
For prime polynomial source modulus and their specified range of
rounding moduli, such a reduction with polynomial security loss yields
an efficient attack on the source $\LWE$ problem.
Our reduction changes the target to quotient-valued block rounding,
uses a source-independent randomized affine transport, and shares
randomness across the sample table in the zero-fiber step.
It therefore lies outside that reduction model. We do not establish
hardness of ordinary scalar LWR with polynomial modulus and unrestricted
polynomial sample counts.

\paragraph{Weak pseudorandom functions over the field.}
Rounding-based weak PRFs with $\TCZero$ evaluation appear
in~\cite{EC:BanPeiRos12,C:AKPW13}. Under the worst-case hardness
condition above, the direct family in
\cref{sec:applications-weak-prf}, with a uniformly sampled nonzero key,
is a weak PRF on
$\LARField^{\BlockDim\times\SecretDim}$ with range $\LARField$.
Joint key-and-input evaluation is in $\TCZero$, with polynomial-time
constructible circuits, and each fixed nonzero key has exactly uniform
output marginals under uniform inputs.
For this same fixed public family, security holds against every uniform
QPT adversary receiving classical random examples.
\Cref{sec:applications-weak-prf} states the security and circuit guarantees;
evaluation does not invoke the transport sampler.

Ding, Jain, and Komargodski~\cite{ARXIV:DinJaiKom26}
obtain weak and strong PRFs in $\NCOne$
from standard $\LWE$ with polynomial modulus-to-noise
ratio. Our direct quotient-rounding construction
is a weak PRF with $\TCZero$ evaluation and security for random examples.

\paragraph{Noise-free learning.}
A learner that predicts fresh field labels with an inverse-polynomial
improvement over success probability $1/\LARModulus$ gives a
decisional $\LAR$ distinguisher. We also give a concrete representation
of the normalized labels by two-hidden-layer ReLU networks on a
uniform finite grid, with polynomial size and polynomially bounded
rational weights and biases of logarithmic bit length in the input
dimension. A uniform QPT learner that, for every target, attains
squared loss at most $1/100$ with probability at least $2/3$ yields
the worst-case lattice algorithm above. These guarantees use independent
classical noiseless random examples and permit improper hypotheses
given by classical descriptions with deterministic classical
finite-precision evaluation, as in \cref{sec:applications-learning}.

The learning-to-distinguishing argument follows the rounding-based
approach of Chen, Gollakota, Klivans, and
Meka~\cite{NeuroIPS:CGKM22}.
\Cref{sec:applications-learning} states the precise guarantee, which
does not cover arbitrary adaptive label queries.

A complementary route to improper learning hardness uses average-case
assumptions about refuting random constraint satisfaction
problems~\cite{STOC:DanLinSha14}.
Within lattice-based learning hardness, Li, Zadik, and
Zampetakis~\cite{LiZadikZampetakis2025} study one-hidden-layer ReLU networks
under standard Gaussian inputs with polynomially small additive Gaussian
label noise. Their setting differs from our uniform finite-grid inputs
and noiseless labels.

\paragraph{Organization.}
\Cref{sec:preliminaries,sec:LAR} give the preliminaries and the formal
rounding definitions.
\Cref{sec:uniform-zero-fiber-lwe-to-lar,sec:gaussian-lwe-to-uniform-zero-fiber-lwe}
state the generic reductions and explain their distributional mechanisms.
\Cref{sec:aft-instantiation} develops the integer affine geometry.
\Cref{sec:finite-dimensional-calibration} gives the common calibration
principle; \cref{sec:probabilistic-transport,sec:direct-fourier-transport}
apply it within the safe affine construction and joint endpoint analysis.
\Cref{sec:applications} states the lattice, pseudorandomness, and
learning consequences. Each section points to the appendix containing
its complete proofs and technical verification.

%% file: chapters/preliminaries.tex
\section{Preliminaries}
\label{sec:preliminaries}

Throughout, residues modulo an integer $\LatticeModulus\geq2$ are
represented by $\set{0,\ldots,\LatticeModulus-1}$. For vectors and matrices,
this convention is applied entrywise. We use $\norm{\cdot}_2$ for the
Euclidean norm, and $\StandardBasisVec_{\VertexIndex}$ for the standard
unit vector at coordinate $\VertexIndex$ in the ambient space under
discussion.
Sample tables are written as indexed families in braces, retaining
repeated entries and the order specified by their indices.

\subsection{Lattices}
\label{sec:prelim-lattices}

We use the standard lattice terminology of~\cite[Section~2.1]{EPRINT:Peikert15}.

\begin{definition}[Lattice]
  \label[definition]{def:lattice}
  Let $\AmbientDim\geq1$ be an integer.
  A \emph{lattice} $\Lattice\subseteq\RR^{\AmbientDim}$ is a discrete
  additive subgroup of $\RR^{\AmbientDim}$.

  Equivalently, for some integer $0\leq\LatticeRank\leq\AmbientDim$,
  there are vectors
  $\BasisVec_1,\ldots,\BasisVec_{\LatticeRank}\in\RR^{\AmbientDim}$
  that are linearly independent over $\RR$ and generate $\Lattice$
  by integer linear combinations.
  Write
  $\LatticeBasis=\left[\BasisVec_1\mid\cdots\mid\BasisVec_{\LatticeRank}\right]
  \in\RR^{\AmbientDim\times\LatticeRank}$.
  For integer coefficients $\LatticeCoeff_1,\ldots,\LatticeCoeff_{\LatticeRank}$,
  write
  $\LatticeCoeffVec=(\LatticeCoeff_1,\ldots,\LatticeCoeff_{\LatticeRank})^{\trans}$.
  Then
  \[
    \Lattice = \LatticeGen{\LatticeBasis}
    := \set{\LatticeBasis\LatticeCoeffVec : \LatticeCoeffVec \in \ZZ^{\LatticeRank}}
    = \set{\sum_{\BasisIndex=1}^{\LatticeRank}
      \LatticeCoeff_{\BasisIndex}\BasisVec_{\BasisIndex} :
    \LatticeCoeff_1,\ldots,\LatticeCoeff_{\LatticeRank} \in \ZZ}.
  \]
  The matrix $\LatticeBasis$ is a \emph{basis matrix} of $\Lattice$,
  and its columns form a \emph{basis}: each point of $\Lattice$ has
  a unique coefficient vector in $\ZZ^{\LatticeRank}$.
  The \emph{rank} of $\Lattice$ is
  $\LatticeRank=\dim(\Span_{\RR}(\Lattice))$, the dimension of its
  real linear span, and $\AmbientDim$ is its \emph{ambient dimension}.
  The lattice is \emph{full rank} if $\LatticeRank=\AmbientDim$.
\end{definition}

\begin{definition}[Zero-sum integer lattice]
  \label[definition]{def:zero-sum-integer-lattice}
  For an integer $\AmbientDim\geq1$, let
  $\OnesVec\in\ZZ^{\AmbientDim}$ be the column vector whose entries
  are all $1$. Define
  \[
    \ZeroSumIntegerLattice{\AmbientDim}
    :=\set{\IntegerVec\in\ZZ^{\AmbientDim}:
    \OnesVec^{\trans}\IntegerVec=0}.
  \]
\end{definition}

\subsection{Rounding Matrices}
\label{sec:rounding-matrices}

We use the integer Smith normal form and its description of quotient
groups as in~\cite[Section~2]{Stanley2016Smith}.
A square integer matrix is \emph{unimodular} if its determinant is
$\pm1$, equivalently if it has an integer inverse.

\begingroup
\mdfsetup{topline=true}
\begin{definition}[Rounding Matrix]
  \label[definition]{def:rounding-matrix}
  Let $\LARModulus$ be an odd prime and let $\BlockDim\geq1$ be an integer.
  Write $\LARField:=\ZZ/\LARModulus\ZZ$.
  Fix an integer $0\leq\RoundingRank\leq\BlockDim$ and set
  $\OutputDim:=\BlockDim-\RoundingRank$.
  An integer matrix
  $\RoundingMatrix\in\ZZ^{\BlockDim\times\BlockDim}$ is a
  \emph{rounding matrix of modular rank $\RoundingRank$} if there exist
  unimodular matrices
  $\SmithLeft,\SmithRight\in\ZZ^{\BlockDim\times\BlockDim}$ such that
  \[
    \SmithLeft\RoundingMatrix\SmithRight
    =\operatorname{diag}(\IdentityMatrix_{\RoundingRank},
    \LARModulus\IdentityMatrix_{\OutputDim}).
  \]
\end{definition}
\endgroup

The Smith form connects modular rank to the integer quotient in which
rounding outputs are encoded.
\begin{lemma}[Modular rank and image lattice]
  \label[lemma]{lem:rounding-smith-properties}
  A rounding matrix as in \cref{def:rounding-matrix} satisfies
  \[
    \begin{gathered}
      \operatorname{rank}_{\LARField}(\RoundingMatrix\bmod\LARModulus)
      =\RoundingRank,\qquad
      \dim_{\LARField}\ker(\RoundingMatrix\bmod\LARModulus)=\OutputDim,\\
      |\det\RoundingMatrix|=\LARModulus^{\OutputDim}.
    \end{gathered}
  \]
  Its image lattice obeys
  \[
    \LARModulus\ZZ^{\BlockDim}\subseteq\LatticeGen{\RoundingMatrix}
    \subseteq\ZZ^{\BlockDim},\qquad
    \ZZ^{\BlockDim}/\LatticeGen{\RoundingMatrix}
    \cong\LARField^{\OutputDim}.
  \]
  For the fixed Smith matrix $\SmithLeft$, the isomorphism takes the
  last $\OutputDim$ coordinates of $\SmithLeft\IntegerVec$ modulo
  $\LARModulus$. In particular the quotient has
  $\LARModulus^{\OutputDim}$ elements and is nontrivial exactly when
  $\OutputDim\geq1$.
\end{lemma}
The proof, including both endpoint ranks, is in
Appendix~\ref{sec:foundations-smith}.

\input{chapters/prelim-probability}
\input{chapters/prelim-affine-maps}

\subsection{Error Distributions and Learning with Errors}
\label{sec:lwe}

For Gaussian errors, we use the discrete Gaussian convention
of~\cite{STOC:GenPeiVai08}; see also
\cite[Section~2.3]{EPRINT:Peikert15}.
Fix an integer $\AmbientDim\geq1$.
For a width parameter $\GaussianWidth>0$ and a center
$\GaussianCenterVec\in\RR^{\AmbientDim}$, define
\[
  \GaussianWeight_{\GaussianWidth,\GaussianCenterVec}(\GaussianPoint)
  :=\exp\!\left(-\pi
    \frac{\norm{\GaussianPoint-\GaussianCenterVec}_2^2}
  {\GaussianWidth^2}\right),
  \qquad \GaussianPoint\in\RR^{\AmbientDim}.
\]
For a nonempty countable set $\GaussianSupport\subseteq\RR^{\AmbientDim}$,
write
\[
  \GaussianWeight_{\GaussianWidth,\GaussianCenterVec}(\GaussianSupport)
  :=\sum_{\GaussianSumPoint\in\GaussianSupport}
  \GaussianWeight_{\GaussianWidth,\GaussianCenterVec}(\GaussianSumPoint).
\]

\begin{definition}[Discrete Gaussian distribution]
  \label[definition]{def:discrete-gaussian}
  If
  $0<\GaussianWeight_{\GaussianWidth,\GaussianCenterVec}
  (\GaussianSupport)<\infty$, the
  \emph{discrete Gaussian distribution}
  $\GaussianCenter{\GaussianSupport}{\GaussianWidth}{\GaussianCenterVec}$
  assigns to each $\GaussianPoint\in\GaussianSupport$ probability
  \[
    \GaussianCenter{\GaussianSupport}{\GaussianWidth}{\GaussianCenterVec}
    (\GaussianPoint)
    =\frac{
    \GaussianWeight_{\GaussianWidth,\GaussianCenterVec}(\GaussianPoint)}{
    \GaussianWeight_{\GaussianWidth,\GaussianCenterVec}(\GaussianSupport)}.
  \]
  It assigns zero probability outside $\GaussianSupport$.
\end{definition}

The strict positivity of the normalizing sum follows from
$\GaussianSupport\ne\varnothing$.
We omit the center when it is zero, writing
$\Gaussian{\GaussianSupport}{\GaussianWidth}
:=\GaussianCenter{\GaussianSupport}{\GaussianWidth}{\ZeroVec}$ and
$\GaussianWeight_{\GaussianWidth}
:=\GaussianWeight_{\GaussianWidth,\ZeroVec}$.
Every lattice and every translate of a lattice has finite Gaussian
mass for all $\GaussianWidth>0$ and all centers, so these sets are
valid choices of $\GaussianSupport$.
For the integer lattice, the centered distribution factors coordinatewise:
\[
  \Gaussian{\ZZ^{\AmbientDim}}{\GaussianWidth}
  =\bigl(\Gaussian{\ZZ}{\GaussianWidth}\bigr)^{\otimes\AmbientDim}.
\]
The continuous Gaussian with Lebesgue density
$\GaussianWidth^{-\AmbientDim}
\GaussianWeight_{\GaussianWidth,\GaussianCenterVec}(\GaussianPoint)$
on $\RR^{\AmbientDim}$ has mean $\GaussianCenterVec$ and covariance
$\GaussianWidth^2\IdentityMatrix_{\AmbientDim}/(2\pi)$.
In one dimension its standard deviation is
$\GaussianWidth/\sqrt{2\pi}$.

The next inequality bounds the exponential moment of each real
linear functional of a discrete Gaussian whose weight is centered
at zero~\cite[full version, Lemma~2.8]{EC:MicPei12}.
It uses the Gaussian-width convention above and requires no lower
bound on the width beyond positivity.

\begingroup
\mdfsetup{topline=true}
\begin{lemma}[Exponential moments of a centered discrete Gaussian]
  \label[lemma]{lem:gaussian-linear-mgf}
  Let $\Lattice\subseteq\RR^{\AmbientDim}$ be a lattice,
  let $\GaussianWidth>0$, and sample
  $\GaussianPoint\sample\Gaussian{\Lattice}{\GaussianWidth}$.
  For $\GaussianTestDirection\in\Span_{\RR}(\Lattice)$
  and $\GaussianLaplaceParameter\in\RR$,
  \begin{equation*}
    \mathbb E\!\left[
      e^{\GaussianLaplaceParameter
    \langle\GaussianTestDirection,\GaussianPoint\rangle}\right]
    \leq e^{\GaussianWidth^2\GaussianLaplaceParameter^2
    \norm{\GaussianTestDirection}_2^2/(4\pi)}.
  \end{equation*}
\end{lemma}
\endgroup

\paragraph{Learning with errors.}
The $\LWE$ problem was introduced by Regev~\cite{STOC:Regev05}.
We use a shared-secret block formulation: one secret is sampled for
the entire collection of blocks. When each error block consists of
independent draws from the same scalar error law, this is the usual
$\LWE$ experiment~\cite[Section~4.2.1]{EPRINT:Peikert15} with scalar
samples grouped into blocks.

\begingroup
\mdfsetup{everyline=true}
\begin{definition}[Shared-secret block $\LWE$]
  \label[definition]{def:block-lwe}
  Let $\LARModulus$ be an odd prime, let $\SecretDim,\BlockDim\geq1$
  be integers, and let $\LWEErrorLaw$ be a distribution on
  $\LARField^{\BlockDim}$. For an integer block count $\SampleCount\geq1$,
  draw one secret $\LWESecret\sample\Unif{\LARField^{\SecretDim}}$ and,
  independently for $\SampleIndex=1,\ldots,\SampleCount$, draw
  \[
    \LWEPublicMatrix_{\SampleIndex}\sample
    \Unif{\LARField^{\BlockDim\times\SecretDim}},
    \qquad
    \LWEError_{\SampleIndex}\sample\LWEErrorLaw,
    \qquad
    \LWEUniformResponse_{\SampleIndex}\sample
    \Unif{\LARField^{\BlockDim}}.
  \]
  All these choices are mutually independent and independent of the secret.
  The real and ideal laws of indexed families of $\SampleCount$ blocks are
  \[
    \begin{aligned}
      \LWEReal_{\LWEErrorLaw,\SampleCount}
      &:=\DistributionLaw{\set{(
            \LWEPublicMatrix_{\SampleIndex},
            \LWEPublicMatrix_{\SampleIndex}\LWESecret
            +\LWEError_{\SampleIndex}
      )}_{\SampleIndex=1}^{\SampleCount}},\\
      \LWEIdeal_{\SampleCount}
      &:=\DistributionLaw{\set{(
            \LWEPublicMatrix_{\SampleIndex},
            \LWEUniformResponse_{\SampleIndex}
      )}_{\SampleIndex=1}^{\SampleCount}},
    \end{aligned}
  \]
  where responses are computed in $\LARField^{\BlockDim}$.

  A \emph{search $\LWE$} algorithm receives a draw from the real
  experiment and succeeds exactly when its output is the sampled
  vector $\LWESecret$.
  \emph{Decisional $\LWE$} asks to distinguish the real and ideal laws.
  For a randomized algorithm $\LWEDistinguisher$ returning a bit, its
  advantage is
  \[
    \left|
    \Pr[\LWEDistinguisher(\LWEReal_{\LWEErrorLaw,\SampleCount})=1]
    -\Pr[\LWEDistinguisher(\LWEIdeal_{\SampleCount})=1]
    \right|.
  \]
\end{definition}

\begin{definition}[Gaussian $\LWE$]
  \label[definition]{def:gaussian-lwe}
  For $\GaussianWidth>0$, \emph{Gaussian $\LWE$} is the specialization of
  \cref{def:block-lwe} to
  \[
    \LWEErrorLaw
    =\Gaussian{\ZZ^{\BlockDim}}{\GaussianWidth}\bmod\LARModulus
    =\bigl(\Gaussian{\ZZ}{\GaussianWidth}\bmod\LARModulus
    \bigr)^{\otimes\BlockDim}.
  \]
  The notation $\bmod\LARModulus$ on a law means its image under
  coordinatewise reduction from $\ZZ^{\BlockDim}$ to
  $\LARField^{\BlockDim}$ (or from $\ZZ$ to $\LARField$ in the
  scalar factor). Equivalently, for every
  $\LWEError\in\LARField^{\BlockDim}$,
  \[
    \LWEErrorLaw(\LWEError)
    =\frac{\GaussianWeight_{\GaussianWidth}
    (\CanonicalLift{\LWEError}+\LARModulus\ZZ^{\BlockDim})}
    {\GaussianWeight_{\GaussianWidth}(\ZZ^{\BlockDim})}.
  \]
\end{definition}
\endgroup

Here $\SampleCount$ blocks contain $\SampleCount\BlockDim$ scalar
$\LWE$ samples with independent errors. We use this discrete Gaussian
convention throughout, rather than rounding a continuous Gaussian to
the nearest integer; see \cite[Section~4.2.1]{EPRINT:Peikert15}.

We also consider errors drawn uniformly from the inputs on which a
public function is zero.

\begingroup
\mdfsetup{topline=true}
\begin{definition}[Uniform-zero-fiber $\LWE$]
  \label[definition]{def:uniform-zero-fiber-lwe}
  In the setting of \cref{def:block-lwe}, fix an integer
  $\OutputDim\geq1$ and a public function
  $\ZeroFiberFunction\colon\LARField^{\BlockDim}\to\LARField^{\OutputDim}$,
  whose \emph{zero fiber}
  \[
    \FunctionZeroFiber
    :=\ZeroFiberFunction^{-1}(\ZeroVec)
    =\set{\LWEError\in\LARField^{\BlockDim}:
    \ZeroFiberFunction(\LWEError)=\ZeroVec}
  \]
  is nonempty. \emph{Uniform-zero-fiber $\LWE$} associated with
  $\ZeroFiberFunction$ is the specialization of \cref{def:block-lwe}
  to $\LWEErrorLaw=\Unif{\FunctionZeroFiber}$.
\end{definition}
\endgroup

Uniformity is over the entire fiber, which need not be a linear
subspace. Thus every fiber point has probability
$|\FunctionZeroFiber|^{-1}$.

%% file: chapters/prelim-probability.tex
\subsection{Probability Laws and Conditioning}
\label{sec:probability-notation}

For a random variable $\ProbabilityRandomVariable$ taking values in a
space $\FiniteSampleSpace$, write
$\DistributionLaw{\ProbabilityRandomVariable}$ for its probability law:
$\DistributionLaw{\ProbabilityRandomVariable}(\ProbabilityEvent)
=\Pr[\ProbabilityRandomVariable\in\ProbabilityEvent]$ for every
measurable set $\ProbabilityEvent\subseteq\FiniteSampleSpace$.
On a finite or countable space all subsets are measurable; on a
Euclidean space we use the Borel sets.
For a law $\FiniteProbabilityLaw$ on a finite or countable space, write
$\FiniteProbabilityLaw(\FiniteSamplePoint)
:=\FiniteProbabilityLaw(\set{\FiniteSamplePoint})$ for its point probability.
Thus
$\FiniteProbabilityLaw(\ProbabilityEvent)
=\sum_{\FiniteSamplePoint\in\ProbabilityEvent}
\FiniteProbabilityLaw(\FiniteSamplePoint)$.
The point-mass law $\LocalPointMass{\FiniteSamplePoint}$ assigns
probability one to $\FiniteSamplePoint$.
For laws $\FiniteProbabilityLaw$ and $\FiniteReferenceLaw$, possibly
on different spaces, their product
$\FiniteProbabilityLaw\otimes\FiniteReferenceLaw$ is the joint law of
independent variables $\ProbabilityRandomVariable_1\sim\FiniteProbabilityLaw$
and $\ProbabilityRandomVariable_2\sim\FiniteReferenceLaw$.

\begin{definition}[Convolution of probability laws]
  \label[definition]{def:probability-convolution}
  Let $\FiniteSampleSpace$ be a finite or countable abelian group
  written additively, or $\RR^{\AmbientDim}$ for an integer
  $\AmbientDim\geq1$. For probability laws
  $\FiniteProbabilityLaw,\FiniteReferenceLaw$ on $\FiniteSampleSpace$,
  their \emph{convolution} is
  \[
    \FiniteProbabilityLaw*\FiniteReferenceLaw
    :=\DistributionLaw{\ProbabilityRandomVariable_1
      +\ProbabilityRandomVariable_2},
  \]
  where $\ProbabilityRandomVariable_1\sim\FiniteProbabilityLaw$ and
  $\ProbabilityRandomVariable_2\sim\FiniteReferenceLaw$ are independent.
  In the finite or countable case, the point probabilities satisfy
  \[
    (\FiniteProbabilityLaw*\FiniteReferenceLaw)(\FiniteSamplePoint)
    =\sum_{\FiniteSamplePoint'\in\FiniteSampleSpace}
      \FiniteProbabilityLaw(\FiniteSamplePoint')
      \FiniteReferenceLaw(\FiniteSamplePoint-\FiniteSamplePoint'),
    \qquad \FiniteSamplePoint\in\FiniteSampleSpace.
  \]
  Addition and subtraction are the group operations in
  $\FiniteSampleSpace$. In particular, on $\ZZ/\LARModulus\ZZ$ for
  an integer $\LARModulus\geq2$, they are taken modulo $\LARModulus$.
\end{definition}

For the remainder of this subsection, all state spaces are finite or
countable. For laws $\FiniteProbabilityLaw$ and $\FiniteReferenceLaw$
on the same space $\FiniteSampleSpace$, their \emph{statistical distance}
(SD), or total variation distance, is
\[
  \StatisticalDistance{\FiniteProbabilityLaw}{\FiniteReferenceLaw}
  :=\frac12\sum_{\FiniteSamplePoint\in\FiniteSampleSpace}
  |\FiniteProbabilityLaw(\FiniteSamplePoint)
   -\FiniteReferenceLaw(\FiniteSamplePoint)|.
\]
A randomized transformation assigns to each input point a probability
law on a fixed output space. Applying the same assignment to inputs
with laws $\FiniteProbabilityLaw$ and $\FiniteReferenceLaw$ produces
output laws whose SD is at most
$\StatisticalDistance{\FiniteProbabilityLaw}{\FiniteReferenceLaw}$.
A deterministic function is the special case in which each assigned
law is a point mass.

\begingroup
\mdfsetup{topline=true}
\begin{definition}[Coupling]
  \label[definition]{def:probability-coupling}
  Let $\FiniteProbabilityLaw$ and $\FiniteReferenceLaw$ be probability
  laws on the same finite or countable space $\FiniteSampleSpace$.
  A \emph{coupling} of these laws is a joint law of a pair
  $(\ProbabilityRandomVariable_1,\ProbabilityRandomVariable_2)$ in
  $\FiniteSampleSpace^2$ with marginal laws
  $\FiniteProbabilityLaw$ and $\FiniteReferenceLaw$, respectively.
  The coordinates need not be independent.
\end{definition}
\endgroup

Under any such coupling
\cite[Proposition~4.7]{LevinPeres2017},
\[
  \StatisticalDistance{\FiniteProbabilityLaw}{\FiniteReferenceLaw}
  \leq\Pr[\ProbabilityRandomVariable_1\ne\ProbabilityRandomVariable_2].
\]
For laws $\FiniteProbabilityLaw_1,\FiniteReferenceLaw_1$ on one
space and $\FiniteProbabilityLaw_2,\FiniteReferenceLaw_2$ on another,
the product bound is~\cite[Exercise~4.4]{LevinPeres2017}
\[
  \StatisticalDistance{\FiniteProbabilityLaw_1\otimes\FiniteProbabilityLaw_2}
    {\FiniteReferenceLaw_1\otimes\FiniteReferenceLaw_2}
  \leq
  \StatisticalDistance{\FiniteProbabilityLaw_1}{\FiniteReferenceLaw_1}
  +\StatisticalDistance{\FiniteProbabilityLaw_2}{\FiniteReferenceLaw_2}.
\]
Iterating this inequality bounds the SD between products of any
finite number of laws by the sum of the SDs of corresponding factors.
These bounds also apply on countable spaces; the verification is in
\cref{sec:foundations-conditioning}.

For $\ProbabilityEvent\subseteq\FiniteSampleSpace$ with
$\FiniteProbabilityLaw(\ProbabilityEvent)>0$, the conditional law
$\FiniteProbabilityLaw(\cdot\mid\ProbabilityEvent)$ is a law on the
same space $\FiniteSampleSpace$, defined by
\[
  \FiniteProbabilityLaw(\ProbabilityEvent'\mid\ProbabilityEvent)
  :=\frac{\FiniteProbabilityLaw(\ProbabilityEvent'\cap\ProbabilityEvent)}
    {\FiniteProbabilityLaw(\ProbabilityEvent)},
  \qquad \ProbabilityEvent'\subseteq\FiniteSampleSpace.
\]
It assigns zero probability to
$\ProbabilityEvent^c:=\FiniteSampleSpace\setminus\ProbabilityEvent$.
The cost of conditioning is exactly the discarded probability mass:
\begin{equation*}
  \StatisticalDistance{\FiniteProbabilityLaw}
    {\FiniteProbabilityLaw(\cdot\mid\ProbabilityEvent)}
  =\FiniteProbabilityLaw(\ProbabilityEvent^c).
\end{equation*}
We call $\FiniteProbabilityLaw(\cdot\mid\ProbabilityEvent)$ the
\emph{truncation} of $\FiniteProbabilityLaw$ to $\ProbabilityEvent$;
it includes normalization by $\FiniteProbabilityLaw(\ProbabilityEvent)$.
If independent variables
$\ProbabilityRandomVariable_1\sim\FiniteProbabilityLaw$ and
$\ProbabilityRandomVariable_2\sim\FiniteReferenceLaw$ are jointly
conditioned on
$\set{\ProbabilityRandomVariable_1\in\ProbabilityEvent_1,\,
\ProbabilityRandomVariable_2\in\ProbabilityEvent_2}$, where
$\FiniteProbabilityLaw(\ProbabilityEvent_1)>0$ and
$\FiniteReferenceLaw(\ProbabilityEvent_2)>0$, their conditional joint
law is
$\FiniteProbabilityLaw(\cdot\mid\ProbabilityEvent_1)\otimes
\FiniteReferenceLaw(\cdot\mid\ProbabilityEvent_2)$.
Thus they remain independent under this conditional law. The same
statement holds for any finite number of independent variables, each
restricted to a set of positive probability in its own coordinate.
These conditioning identities are proved in
\cref{sec:foundations-conditioning}. They let us analyze protected
sources by truncation without making the sampler inspect a source
realization.

\subsection{Finite-Space Densities}
\label{sec:finite-space-operators}

Let $\FiniteSampleSpace$ be a nonempty finite set, equipped with the
reference law $\Unif{\FiniteSampleSpace}$.
Write $\RR^{\FiniteSampleSpace}$ for the vector space of real
functions on $\FiniteSampleSpace$, and $1$ for the constant function
with value one.
Unless another law is explicitly specified, probabilities and
expectations in this subsection refer to this law. In particular,
for a real function $\LocalTestFunction:\FiniteSampleSpace\to\RR$,
\[
  \mathbb E\LocalTestFunction
  :=\frac1{|\FiniteSampleSpace|}
  \sum_{\FiniteSamplePoint\in\FiniteSampleSpace}
  \LocalTestFunction(\FiniteSamplePoint).
\]
For $\ProbabilityEvent\subseteq\FiniteSampleSpace$, superscript $c$
denotes its complement in $\FiniteSampleSpace$, and
$\EventIndicator{\ProbabilityEvent}(\FiniteSamplePoint)$ is one
when $\FiniteSamplePoint\in\ProbabilityEvent$ and zero otherwise.

A function $\FiniteDensity:\FiniteSampleSpace\to\RR_{\geq0}$
with $\mathbb E\FiniteDensity=1$ is a \emph{probability density
relative to the uniform law}. It defines a probability law whose
point mass at $\FiniteSamplePoint$ is
$\FiniteDensity(\FiniteSamplePoint)/|\FiniteSampleSpace|$.
More generally, a real function $\LocalTestFunction$ defines a signed
measure with point mass
$\LocalTestFunction(\FiniteSamplePoint)/|\FiniteSampleSpace|$.
Its total mass is $\mathbb E\LocalTestFunction$, which need not be one
or nonnegative.
For real functions $\LocalTestFunction,\FiniteSecondTestFunction$ on
$\FiniteSampleSpace$, define the inner product and supremum norm by
\[
  \langle\LocalTestFunction,\FiniteSecondTestFunction\rangle
  :=\mathbb E[\LocalTestFunction\FiniteSecondTestFunction],
  \qquad
  \|\LocalTestFunction\|_\infty
  :=\max_{\FiniteSamplePoint\in\FiniteSampleSpace}
  |\LocalTestFunction(\FiniteSamplePoint)|.
\]
For a linear operator between subspaces of real functions on
$\FiniteSampleSpace$, its $\infty\to\infty$ norm is the supremum of
the output supremum norm over input functions of supremum norm at
most one.

%% file: chapters/prelim-affine-maps.tex
\subsection{Canonical Representatives and Carries}
\label{sec:prelim-residues}

Use the prime field of \cref{def:rounding-matrix}. The following
notation makes the canonical residue convention explicit.

\begingroup
\mdfsetup{topline=true}
\begin{definition}[Canonical lift]
  \label[definition]{def:canonical-lift}
  For an integer $\BlockDim\geq1$ and
  $\RoundingInput\in\LARField^{\BlockDim}$, its \emph{canonical integer
  lift} $\CanonicalLift{\RoundingInput}$ is the unique vector in
  $\ZZ^{\BlockDim}$ whose reduction modulo $\LARModulus$ equals
  $\RoundingInput$ and whose entries lie in
  $\set{0,\ldots,\LARModulus-1}$.
\end{definition}
\endgroup

The same notation applies to scalars and vectors of any dimension.
For an integer $\IntegerCharge$, its canonical residue is
$\CanonicalLift{\IntegerCharge\bmod\LARModulus}$ and its \emph{carry}
is $\floor{\IntegerCharge/\LARModulus}$. They satisfy
\[
  \IntegerCharge
  =\CanonicalLift{\IntegerCharge\bmod\LARModulus}
   +\LARModulus\floor{\IntegerCharge/\LARModulus}.
\]
The carry is the integer quotient in this decomposition, including
when $\IntegerCharge<0$. These conventions apply coordinatewise to
integer vectors. We call an integer in
$\set{0,\ldots,\LARModulus-1}$ a \emph{color}.

%% file: chapters/LAR.tex
\section{Learning with Algebraic Rounding}
\label{sec:LAR}

\begingroup
\mdfsetup{topline=true}
\begin{definition}[Algebraic Rounding]
  \label[definition]{def:lar}
  Let $\LARModulus$ be an odd prime, let $\BlockDim\geq1$ be an integer,
  and let $\RoundingMatrix\in\ZZ^{\BlockDim\times\BlockDim}$ be a rounding
  matrix of modular rank $\RoundingRank$, as in
  \cref{def:rounding-matrix}, with output dimension $\OutputDim$.
  The associated \emph{algebraic rounding function} is
  \begin{equation*}
    \begin{aligned}
      \RoundingMap\colon \ZZ^{\BlockDim}
      &\longrightarrow \ZZ^{\BlockDim},\\
      \RoundingInput
      &\longmapsto \floor{\RoundingMatrix\RoundingInput/\LARModulus}.
    \end{aligned}
  \end{equation*}
  The product and division inside the floor are computed over $\QQ$.
\end{definition}
\endgroup

An integer rounded output determines a coset in
$\ZZ^{\BlockDim}/\LatticeGen{\RoundingMatrix}$ that depends only on
the input modulo $\LARModulus$. Smith coordinates encode this coset
as a label in $\LARField^{\OutputDim}$. The corresponding rounding
directions translate labels bijectively, giving fibers of equal size.
We use these coordinates and directions to define the label function.

Fix public Smith matrices $\SmithLeft,\SmithRight$ as in
\cref{def:rounding-matrix}. The quotient isomorphism of
\cref{lem:rounding-smith-properties} gives the following coordinate maps.

\begin{definition}[Quotient coordinates]
  \label[definition]{def:quotient-coordinates}
  Define the \emph{coordinate embedding} $\QuotientSelector$ and the
  \emph{quotient coordinate matrix} $\QuotientProjection$ by
  \[
    \QuotientSelector:=
    \begin{pmatrix}
      \ZeroMatrix_{\RoundingRank\times\OutputDim}\\
      \IdentityMatrix_{\OutputDim}
    \end{pmatrix}\in\ZZ^{\BlockDim\times\OutputDim},
    \qquad
    \QuotientProjection:=\QuotientSelector^{\trans}\SmithLeft
    \in\ZZ^{\OutputDim\times\BlockDim}.
  \]
\end{definition}

For $\IntegerVec\in\ZZ^{\BlockDim}$,
$\QuotientProjection\IntegerVec$ extracts the last $\OutputDim$
coordinates of $\SmithLeft\IntegerVec$. Reducing them modulo
$\LARModulus$ labels its quotient coset. The matching columns of
$\SmithRight$ define the rounding directions.

\begingroup
\mdfsetup{topline=true}
\begin{definition}[Rounding directions]
  \label[definition]{def:rounding-directions}
  Define the \emph{integer rounding direction matrix}
  $\RoundingDirectionInt:=\SmithRight\QuotientSelector
  \in\ZZ^{\BlockDim\times\OutputDim}$ and its
  \emph{modular reduction}
  $\RoundingDirection:=\RoundingDirectionInt\bmod\LARModulus
  \in\LARField^{\BlockDim\times\OutputDim}$.
\end{definition}
\endgroup

These directions shift the argument of the floor by integer vectors.
Its coordinatewise fractional parts therefore remain unchanged, while
the rounded output shifts by the same integer vector. The quotient
encoding turns this shift into the prescribed label change
(\cref{lem:lar-properties}).

For an input in $\LARField^{\BlockDim}$, first use the canonical integer
lift of \cref{def:canonical-lift}, then apply the integer rounding map
and reduce its quotient coordinates modulo $\LARModulus$.

\begin{definition}[Encoded label function]
  \label[definition]{def:lar-label-map}
  The \emph{encoded label function} associated with
  $\RoundingMatrix$ and $\QuotientProjection$ is defined by
  \begin{equation*}
    \begin{aligned}
      \LARLabelMap\colon \LARField^{\BlockDim}
      &\longrightarrow \LARField^{\OutputDim},\\
      \RoundingInput
      &\longmapsto \QuotientProjection\RoundingMap(\CanonicalLift{\RoundingInput})
      \bmod\LARModulus.
    \end{aligned}
  \end{equation*}
\end{definition}

The notation $\LARLabelMap$ suppresses the fixed choice of
$\QuotientProjection$.
When $\OutputDim=0$, the label space $\LARField^0$ consists of the
single empty vector; the definitions and properties below include
this case.

\begingroup
\mdfsetup{topline=true,everyline=true}
\begin{lemma}[Basic properties of $\LAR$]
  \label[lemma]{lem:lar-properties}
  Under the definitions above, the following properties hold.
  \begin{description}
    \item[Well-defined quotient encoding.]
      For all $\RoundingInput,\IntegerVec\in\ZZ^{\BlockDim}$,
      \[
        \RoundingMap(\RoundingInput+\LARModulus\IntegerVec)
        =\RoundingMap(\RoundingInput)+\RoundingMatrix\IntegerVec.
      \]
      The map
      $\IntegerVec\mapsto\QuotientProjection\IntegerVec\bmod\LARModulus$
      induces an isomorphism
      $\ZZ^{\BlockDim}/\LatticeGen{\RoundingMatrix}
      \cong\LARField^{\OutputDim}$.
      Consequently, for $\RoundingInput\in\LARField^{\BlockDim}$,
      $\LARLabelMap(\RoundingInput)$ encodes the quotient output
      $\RoundingMap(\CanonicalLift{\RoundingInput})
      +\LatticeGen{\RoundingMatrix}$ and is unchanged if another integer
      lift is used.
    \item[Translation equivariance.]
      The matrix $\RoundingDirection$ has full column rank over
      $\LARField$, and its columns form a basis of
      $\ker(\RoundingMatrix\bmod\LARModulus)$.
      For all $\RoundingInput\in\LARField^{\BlockDim}$ and
      $\RoundingOffset\in\LARField^{\OutputDim}$,
      \begin{equation*}
        \LARLabelMap(\RoundingInput+\RoundingDirection\RoundingOffset)
        =\LARLabelMap(\RoundingInput)+\RoundingOffset.
      \end{equation*}
      All operations in this identity are over $\LARField$.
    \item[Balanced fibers.]
      For every label $\RoundingLabel\in\LARField^{\OutputDim}$, its
      \emph{fiber}, the set of inputs mapped to that label, satisfies
      \[
        \left|\set{\RoundingInput\in\LARField^{\BlockDim}:
        \LARLabelMap(\RoundingInput)=\RoundingLabel}\right|
        =\LARModulus^{\RoundingRank}.
      \]
  \end{description}
\end{lemma}
\endgroup

The complete proof is in Appendix~\ref{sec:foundations-lar}.
The Smith coordinates make integral changes of lift disappear in the
quotient. Their rounding directions then translate each label once
along every kernel coset, which gives balanced fibers.

If $\OutputDim=1$ and the chosen Smith decomposition satisfies
$\QuotientProjection=\OnesVec^{\trans}$, where
$\OnesVec\in\ZZ^{\BlockDim}$ is the all-ones vector, then
\[
  \LARLabelMap(\RoundingInput)
  =\sum_{\CoordinateIndex=1}^{\BlockDim}
  \left\lfloor
  \frac{(\RoundingMatrix\CanonicalLift{\RoundingInput})_{\CoordinateIndex}}
  {\LARModulus}
  \right\rfloor\bmod\LARModulus
  \qquad(\RoundingInput\in\LARField^{\BlockDim}).
\]
Thus in this case the label is a single field element obtained by summing
the integer rounded coordinates.

\subsection{Decisional \texorpdfstring{$\LAR$}{LAR} assumption}
\label{sec:lar-assumptions}

Fix a family of public parameters $\pparam=\pparam(\secpar)$ indexed
by an integer security parameter $\secpar\geq1$. A fixed deterministic
algorithm computes $\pparam(\secpar)$ from the unary input $\secparam$
in time polynomial in $\secpar$.
These parameters specify the prime $\LARModulus$, the integer dimensions
$\SecretDim,\BlockDim,\RoundingRank,\OutputDim$, the rounding matrix
$\RoundingMatrix$, and the chosen Smith matrices
$\SmithLeft,\SmithRight$, hence also $\QuotientProjection$ and
$\LARLabelMap$. Require $\SecretDim\geq1$ and $\OutputDim\geq1$.
The dimensions and the binary encoding length of $\pparam$ are
polynomially bounded in $\secpar$. The parameter family is fixed
before the sample-count function and the adversary are chosen.

A nonnegative function of $\secpar$ is \emph{negligible} if, for
every constant $\SecurityDecayExponent>0$, it is at most
$\secpar^{-\SecurityDecayExponent}$ for all sufficiently large
$\secpar$.

We first define the real and ideal $\LAR$ distributions.

\begin{definition}[Real and ideal $\LAR$ distributions]
  \label[definition]{def:lar-distributions}
  For an integer sample count $\SampleCount\geq1$, draw one secret
  $\LARSecret\sample\Unif{\LARField^{\SecretDim}}$.
  Independently of that secret and independently for each
  $\SampleIndex=1,\ldots,\SampleCount$, draw
  \[
    \LARPublicMatrix_{\SampleIndex}\sample
    \Unif{\LARField^{\BlockDim\times\SecretDim}},
    \qquad
    \LARUniformLabel_{\SampleIndex}\sample\Unif{\LARField^{\OutputDim}}
  \]
  with the matrices and uniform labels mutually independent.
  The \emph{real $\LAR$ distribution} and the
  \emph{ideal $\LAR$ distribution} are the following laws of indexed
  families of $\SampleCount$ samples:
  \[
    \begin{aligned}
      \LARReal_{\SampleCount}
      &:=\DistributionLaw{\set{(\LARPublicMatrix_{\SampleIndex},
            \LARLabelMap(\LARPublicMatrix_{\SampleIndex}\LARSecret)
      )}_{\SampleIndex=1}^{\SampleCount}},\\
      \LARIdeal_{\SampleCount}
      &:=\DistributionLaw{\set{(
            \LARPublicMatrix_{\SampleIndex},\LARUniformLabel_{\SampleIndex}
      )}_{\SampleIndex=1}^{\SampleCount}}.
    \end{aligned}
  \]
\end{definition}

The products $\LARPublicMatrix_{\SampleIndex}\LARSecret$ are computed
over $\LARField$ before applying $\LARLabelMap$.
Conditional on the single sampled secret, the real samples are
independent; their unconditional joint law averages over that same
secret. In the ideal law, the matrices and labels are all independent.

\begin{definition}[Decisional $\LAR$ assumption]
  \label[definition]{def:decisional-lar-assumption}
  For a uniform QPT distinguisher $\LARDistinguisher$ on classical inputs,
  define its decisional advantage by
  \[
    \begin{aligned}
      \LARDecisionalAdv{\LARDistinguisher}(\secpar,\SampleCount)
      :=\bigl|&\Pr[\LARDistinguisher(\secparam,\pparam,
      \LARReal_{\SampleCount})=1]\\
      &-\Pr[\LARDistinguisher(\secparam,\pparam,
      \LARIdeal_{\SampleCount})=1]\bigr|.
    \end{aligned}
  \]
  The parameter family satisfies the \emph{decisional $\LAR$ assumption}
  if this advantage is negligible in $\secpar$ for every such
  $\LARDistinguisher$ and every positive integer-valued sample-count
  function $\SampleCount=\SampleCount(\secpar)$ bounded by a polynomial
  in $\secpar$.
\end{definition}

This is the decisional $\LAR$ assumption against QPT adversaries.
Restricting the same definition to PPT adversaries gives its classical
version.

\begin{lemma}[Information-theoretic identifiability]
  \label[lemma]{lem:lar-information-theoretic-identifiability}
  Fix public parameters as in \cref{def:lar-distributions}, with
  $\OutputDim\geq1$, and let $\SampleCount\geq1$ be an integer.
  Let
  $\LARPublicMatrix_1,\ldots,\LARPublicMatrix_{\SampleCount}$ be
  independent and uniform on $\LARField^{\BlockDim\times\SecretDim}$.
  For every fixed $\LARSecret\in\LARField^{\SecretDim}$, the examples
  $\set{(\LARPublicMatrix_{\SampleIndex},
  \LARLabelMap(\LARPublicMatrix_{\SampleIndex}\LARSecret))}
  _{\SampleIndex=1}^{\SampleCount}$ uniquely determine $\LARSecret$
  with probability at least
  $1-(\LARModulus^{\SecretDim}-1)\LARModulus^{-\OutputDim\SampleCount}$.
\end{lemma}

The proof and recovery procedure are in
\cref{sec:foundations-lar-identifiability}.

%% file: chapters/uniform-zero-fiber-lwe-to-lar.tex
\section{From Uniform-Zero-Fiber \texorpdfstring{$\LWE$}{LWE} to Decisional \texorpdfstring{$\LAR$}{LAR}}
\label{sec:uniform-zero-fiber-lwe-to-lar}

To establish the decisional $\LAR$ assumption from uniform-zero-fiber
$\LWE$ hardness, we use a decomposition of each input into its encoded
label and a residual in the zero fiber. Fix a public parameter family
as in \cref{sec:lar-assumptions}. For each security parameter, write
\[
  \LWEZeroFiber:=\LARLabelMap^{-1}(\ZeroVec)
  =\set{\LWEError\in\LARField^{\BlockDim}:
    \LARLabelMap(\LWEError)=\ZeroVec}.
\]
By the balanced-fiber property in \cref{lem:lar-properties},
$|\LWEZeroFiber|=\LARModulus^{\RoundingRank}
=\LARModulus^{\BlockDim-\OutputDim}$, so each point has probability
$\LARModulus^{-\RoundingRank}$ under $\Unif{\LWEZeroFiber}$.
In particular, this set is nonempty. Set
$\LWEErrorLaw:=\Unif{\LWEZeroFiber}$, the error law in
\cref{def:uniform-zero-fiber-lwe} for
$\ZeroFiberFunction=\LARLabelMap$.
Membership in this zero fiber means that the encoded label is zero;
it does not require
$(\RoundingMatrix\bmod\LARModulus)\LWEError=\ZeroVec$.

Define the \emph{zero-fiber decomposition map} by
\begin{equation*}
  \begin{aligned}
    \ZeroFiberDecomposition\colon
    \LWEZeroFiber\times\LARField^{\OutputDim}
    &\longrightarrow\LARField^{\BlockDim},\\
    (\LWEError,\RoundingLabel)
    &\longmapsto\LWEError+\RoundingDirection\RoundingLabel.
  \end{aligned}
\end{equation*}
All operations are over $\LARField$.

\begin{lemma}[Zero-fiber decomposition]
  \label[lemma]{lem:zero-fiber-decomposition}
  The zero-fiber decomposition map $\ZeroFiberDecomposition$ defined above
  is a bijection. For every $\RoundingInput\in\LARField^{\BlockDim}$,
  its inverse is given by
  \[
    \ZeroFiberDecomposition^{-1}(\RoundingInput)
    =\bigl(\RoundingInput-\RoundingDirection\LARLabelMap(\RoundingInput),
    \LARLabelMap(\RoundingInput)\bigr).
  \]
\end{lemma}

\begin{proof}
  For every $\LWEError\in\LWEZeroFiber$ and
  $\RoundingLabel\in\LARField^{\OutputDim}$, translation equivariance
  in \cref{lem:lar-properties} gives
  \[
    \LARLabelMap\bigl(\ZeroFiberDecomposition(\LWEError,\RoundingLabel)\bigr)
    =\LARLabelMap(\LWEError+\RoundingDirection\RoundingLabel)
    =\LARLabelMap(\LWEError)+\RoundingLabel
    =\RoundingLabel.
  \]
  Thus the image uniquely determines $\RoundingLabel$; subtracting
  $\RoundingDirection\RoundingLabel$ then recovers $\LWEError$.
  Hence $\ZeroFiberDecomposition$ is injective.

  Conversely, fix any $\RoundingInput\in\LARField^{\BlockDim}$ and set
  $\RoundingLabel:=\LARLabelMap(\RoundingInput)$ and
  $\LWEError:=\RoundingInput-\RoundingDirection\RoundingLabel$.
  Applying the translation equivariance of \cref{lem:lar-properties}
  with offset $-\RoundingLabel$ yields
  \[
    \LARLabelMap(\LWEError)
    =\LARLabelMap(\RoundingInput-\RoundingDirection\RoundingLabel)
    =\LARLabelMap(\RoundingInput)-\RoundingLabel
    =\ZeroVec.
  \]
  Therefore $\LWEError\in\LWEZeroFiber$, and
  $\ZeroFiberDecomposition(\LWEError,\RoundingLabel)=\RoundingInput$.
  This proves surjectivity and the stated inverse formula.
\end{proof}

By \cref{lem:zero-fiber-decomposition}, if
$\RoundingInput\sample\Unif{\LARField^{\BlockDim}}$, then
$\ZeroFiberDecomposition^{-1}(\RoundingInput)$ is uniform over
$\LWEZeroFiber\times\LARField^{\OutputDim}$. Consequently, its residual
$\RoundingInput-\RoundingDirection\LARLabelMap(\RoundingInput)$ and label
$\LARLabelMap(\RoundingInput)$ are independent and uniformly distributed
over $\LWEZeroFiber$ and $\LARField^{\OutputDim}$, respectively.
Conversely, independent draws
$\LWEError\sample\Unif{\LWEZeroFiber}$ and
$\RoundingLabel\sample\Unif{\LARField^{\OutputDim}}$ give a uniform
vector $\LWEError+\RoundingDirection\RoundingLabel$ in
$\LARField^{\BlockDim}$ whose label is $\RoundingLabel$.

For any classical or quantum algorithm $\LWEReduction$ with classical
input and a classical output bit, define
its advantage on these block $\LWE$ experiments by
\[
  \begin{aligned}
    \LWEAdvantage{\LWEReduction}(\secpar,\SampleCount)
    :=\bigl|&
      \Pr[\LWEReduction(\secparam,\pparam,
        \LWEReal_{\LWEErrorLaw,\SampleCount})=1]\\
      &-\Pr[\LWEReduction(\secparam,\pparam,
        \LWEIdeal_{\SampleCount})=1]\bigr|.
  \end{aligned}
\]
\begin{theorem}[From uniform-zero-fiber $\LWE$ to decisional $\LAR$]
  \label{thm:uniform-zero-fiber-lwe-to-dlar}
  Fix an efficiently computable public parameter family as in
  \cref{sec:lar-assumptions}, and set
  $\LWEErrorLaw=\Unif{\LWEZeroFiber}$.
  For every uniform QPT distinguisher $\LARDistinguisher$ on classical
  sample tables and every positive integer-valued sample-count function
  $\SampleCount=\SampleCount(\secpar)$ bounded by a polynomial in
  $\secpar$, there is a uniform QPT distinguisher $\LWEReduction$
  using $\SampleCount$ input blocks and one call to $\LARDistinguisher$ such that
  \begin{equation*}
    \LARDecisionalAdv{\LARDistinguisher}(\secpar,\SampleCount)
    \leq \LARModulus\,
    \LWEAdvantage{\LWEReduction}(\secpar,\SampleCount)
    +\LARModulus^{-\SecretDim}.
  \end{equation*}
\end{theorem}

\begin{proof}[Proof sketch]
  First consider exact uniform auxiliary choices. The reduction samples
  one uniform vector $\ReductionVector\in\LARField^{\SecretDim}$ shared
  across blocks and independent uniform labels
  $\RoundingLabel_{\SampleIndex}\in\LARField^{\OutputDim}$, all
  independently of the input. Given input blocks
  $(\LWEPublicMatrix_{\SampleIndex},\LWEResponse_{\SampleIndex})$, it
  outputs $(\LARPublicMatrix_{\SampleIndex},\RoundingLabel_{\SampleIndex})$
  with the rank-one update
  \[
    \LARPublicMatrix_{\SampleIndex}:=
    -\LWEPublicMatrix_{\SampleIndex}
    +(\LWEResponse_{\SampleIndex}
      +\RoundingDirection\RoundingLabel_{\SampleIndex})
      \ReductionVector^{\trans}.
  \]
  All arithmetic is over $\LARField$. For real input,
  $\LWEResponse_{\SampleIndex}
  =\LWEPublicMatrix_{\SampleIndex}\LWESecret+\LWEError_{\SampleIndex}$.
  Set $\RoundingInput_{\SampleIndex}:=\LWEError_{\SampleIndex}
  +\RoundingDirection\RoundingLabel_{\SampleIndex}$. By
  \cref{lem:zero-fiber-decomposition}, these vectors are independent
  and uniform over $\LARField^{\BlockDim}$, independent of the input
  matrices, secret, and $\ReductionVector$, with
  $\LARLabelMap(\RoundingInput_{\SampleIndex})
  =\RoundingLabel_{\SampleIndex}$. Substitution gives the identity
  \[
    \LARPublicMatrix_{\SampleIndex}
    =-\LWEPublicMatrix_{\SampleIndex}
      (\IdentityMatrix_{\SecretDim}
       -\LWESecret\ReductionVector^{\trans})
      +\RoundingInput_{\SampleIndex}\ReductionVector^{\trans}.
  \]
  Hence, whenever $\ReductionVector^{\trans}\LWESecret=1$,
  \[
    \LARPublicMatrix_{\SampleIndex}\LWESecret
    =\RoundingInput_{\SampleIndex},\qquad
    \LARLabelMap(\LARPublicMatrix_{\SampleIndex}\LWESecret)
    =\RoundingLabel_{\SampleIndex}.
  \]
  For each fixed secret and auxiliary vector satisfying this alignment
  condition, the update is a surjective linear map of the independent
  uniform input matrix and vector, so the output matrices are independent
  and uniform. Averaging over aligned choices yields the real $\LAR$
  table conditioned on a nonzero secret. Otherwise,
  $\det(\IdentityMatrix_{\SecretDim}
  -\LWESecret\ReductionVector^{\trans})
  =1-\ReductionVector^{\trans}\LWESecret\ne0$.
  Conditional on the vectors $\RoundingInput_{\SampleIndex}$, the update
  is then an affine bijection in each input matrix, so the output
  matrices and labels have exactly the ideal product law.

  For every nonzero secret, the alignment event
  $\ReductionVector^{\trans}\LWESecret=1$ has probability
  $1/\LARModulus$. Thus the transformed real $\LWE$ table mixes the
  nonzero-secret real $\LAR$ law with the ideal law, with real-branch
  weight $(1-\LARModulus^{-\SecretDim})/\LARModulus$.
  Ideal $\LWE$ input maps exactly to the ideal $\LAR$ law.
  Comparing these mixtures gives the factor $\LARModulus$ in the
  advantage bound; the zero-secret event accounts for the additive
  $\LARModulus^{-\SecretDim}$ allowance.

  Capped rejection sampling implements the auxiliary choices in
  worst-case polynomial time. Its error fits within the unused portion
  of the zero-secret allowance, retaining the stated bound. The reduction
  prepares a classical table and calls $\LARDistinguisher$ once, so it
  remains uniform QPT. The full game-based proof, including the exact
  mixture identities and sampling-error budget, is deferred to
  Appendix~\ref{sec:zero-fiber-lar-reduction-proof}.
\end{proof}

%% file: chapters/gaussian-lwe-to-uniform-zero-fiber-lwe.tex
\section{From Gaussian \texorpdfstring{$\LWE$}{LWE} to Uniform-Zero-Fiber \texorpdfstring{$\LWE$}{LWE}}
\label{sec:gaussian-lwe-to-uniform-zero-fiber-lwe}

A sufficiently accurate source-independent affine transport reduces
Gaussian $\LWE$ to uniform-zero-fiber $\LWE$. We state the transport
property and prove the generic reduction under that hypothesis.
\Cref{sec:direct-fourier-transport} constructs the transport for the
public rounding family used in this paper.

\subsection{\texorpdfstring{$\RoundingMatrix$}{T}-adapted affine fiber transport}
\label{sec:t-adapted-affine-fiber-transport}

Retain the public rounding matrix $\RoundingMatrix$, its quotient encoding,
and the zero fiber $\LWEZeroFiber$ from
\cref{sec:uniform-zero-fiber-lwe-to-lar}, with $\OutputDim\geq1$.
Let $\LWEErrorLaw$ be a source distribution on $\LARField^{\BlockDim}$.

\begingroup
\mdfsetup{everyline=true,topline=true}
\begin{definition}[$\RoundingMatrix$-adapted affine fiber transport]
  \label[definition]{def:t-aft}
  Fix an accuracy $\TransportAccuracy=\TransportAccuracy(\secpar)\in[0,1]$.
  A \emph{$\TransportAccuracy$-accurate $\RoundingMatrix$-adapted affine
  fiber transport} ($\RoundingMatrix$-AFT) for $\LWEErrorLaw$ is a
  pair $\AffineFiberTransport=(\TransportSampler,\TransportEvaluator)$
  of uniform classical PPT algorithms with access to the fixed public
  modulus, dimensions, and quotient encoding. The evaluator
  $\TransportEvaluator$ is deterministic. Their input/output interfaces are:
  \begin{itemize}
    \item $\TransportSampler(\secparam,\RoundingMatrix,\LWEErrorLaw,\TransportAccuracy)
      \to(\TransportMatrix,\TransportOffset)$:
      uses only public descriptions and fresh randomness to return
      $\TransportMatrix\in\ZZ^{\BlockDim\times\BlockDim}$ and
      $\TransportOffset\in\ZZ^{\BlockDim}$. Its code and worst-case
      running-time polynomial in $\secpar$ may depend on the designated
      accuracy.
    \item $\TransportEvaluator(\TransportMatrix,\TransportOffset,\TransportInput)
      \to\TransportOutput$:
      takes the integer matrix and offset and an input
      $\TransportInput\in\LARField^{\BlockDim}$, and returns
      $\TransportOutput\in\LARField^{\BlockDim}$.
  \end{itemize}
  The algorithms satisfy:
  \begin{description}
    \item[Affine evaluation.]
      For all
      $\TransportMatrix\in\ZZ^{\BlockDim\times\BlockDim}$,
      $\TransportOffset\in\ZZ^{\BlockDim}$, and
      $\TransportInput\in\LARField^{\BlockDim}$,
      \begin{equation*}
        \TransportEvaluator(\TransportMatrix,\TransportOffset,\TransportInput)
        =\TransportMatrix\TransportInput+\TransportOffset
        \bmod\LARModulus\in\LARField^{\BlockDim}.
      \end{equation*}
    \item[Invertibility.]
      Every output of $\TransportSampler$ satisfies
      $\TransportMatrix\bmod\LARModulus\in
      \GeneralLinearGroup_{\BlockDim}(\LARField)$.
    \item[Fiber transport.]
      For independently sampled
      \[
        (\TransportMatrix,\TransportOffset)\sample
        \TransportSampler(\secparam,\RoundingMatrix,\LWEErrorLaw,\TransportAccuracy),
        \qquad \LWEError\sample\LWEErrorLaw,
      \]
      \begin{equation*}
        \StatisticalDistance{
          \DistributionLaw{\TransportEvaluator(
        \TransportMatrix,\TransportOffset,\LWEError)}}{\Unif{\LWEZeroFiber}}
        \leq\TransportAccuracy.
      \end{equation*}
  \end{description}
  The transport is \emph{exact} if $\TransportAccuracy=0$.
\end{definition}
\endgroup

The accuracy is designated before choosing the algorithm; the
notation does not require uniform complexity across all accuracies.
Exact transport obeys the same worst-case running-time bound.
The sampler may equivalently return the matrix and offset modulo
$\LARModulus$. We retain integer representatives for the construction,
while the following $\LWE$ experiments operate over $\LARField$.

\begin{lemma}[Basic properties of $\RoundingMatrix$-AFT]
  \label[lemma]{lem:t-aft-properties}
  Let $\AffineFiberTransport=(\TransportSampler,\TransportEvaluator)$
  be a $\TransportAccuracy$-accurate
  $\RoundingMatrix$-AFT for $\LWEErrorLaw$. Sample
  \[
    \begin{gathered}
      \LWEPublicMatrix\sample\Unif{\LARField^{\BlockDim\times\SecretDim}},
      \qquad \LWEError\sample\LWEErrorLaw,\\
      (\TransportMatrix,\TransportOffset)\sample
      \TransportSampler(\secparam,\RoundingMatrix,\LWEErrorLaw,\TransportAccuracy)
    \end{gathered}
  \]
  independently, and set
  $\LWEPublicMatrix':=\TransportMatrix\LWEPublicMatrix$ and
  $\LWEError':=\TransportEvaluator(\TransportMatrix,\TransportOffset,\LWEError)$.
  Then $\LWEPublicMatrix'$ is exactly uniform over
  $\LARField^{\BlockDim\times\SecretDim}$ and independent of
  $\LWEError'$, and
  \[
    \StatisticalDistance{\DistributionLaw{\LWEError'}}{
    \Unif{\LWEZeroFiber}}\leq\TransportAccuracy.
  \]
  If $\LWEUniformResponse\sample\Unif{\LARField^{\BlockDim}}$
  is independent of the joint choice of $\LWEPublicMatrix$ and
  $(\TransportMatrix,\TransportOffset)$, then
  \[
    \DistributionLaw{\TransportMatrix\LWEPublicMatrix,
    \TransportEvaluator(\TransportMatrix,\TransportOffset,\LWEUniformResponse)}
    =\Unif{\LARField^{\BlockDim\times\SecretDim}}
    \times\Unif{\LARField^{\BlockDim}}.
  \]
\end{lemma}

\begin{proof}
  Condition on $(\TransportMatrix,\TransportOffset,\LWEError)$.
  Multiplication by the fixed invertible matrix $\TransportMatrix$ is a
  bijection on $\LARField^{\BlockDim\times\SecretDim}$. Thus the
  conditional law of $\LWEPublicMatrix'$ is uniform and does not depend
  on the conditioning variables. Averaging shows that
  $\LWEPublicMatrix'$ is uniform and independent of
  $(\TransportMatrix,\TransportOffset,\LWEError)$, hence also of
  $\LWEError'$. The error bound is the transport guarantee of \cref{def:t-aft}.

  For the last property, condition on
  $(\TransportMatrix,\TransportOffset)$. The two maps
  $\LWEPublicMatrix\mapsto\TransportMatrix\LWEPublicMatrix$ and
  $\LWEUniformResponse\mapsto
  \TransportMatrix\LWEUniformResponse+\TransportOffset$ are bijections.
  They act separately on independent uniform inputs, so their outputs
  have the stated product law, independently of the conditioned pair.
  Averaging proves the claim even when $\TransportMatrix$ and
  $\TransportOffset$ are correlated.
\end{proof}

\subsection{Gaussian errors and the reduction}
\label{sec:gaussian-t-aft-reduction}

Write $\GaussianErrorLaw$ for the error law of
\cref{def:gaussian-lwe}. A \emph{Gaussian $\RoundingMatrix$-AFT} is a
$\RoundingMatrix$-AFT for this source.

\begingroup
\mdfsetup{everyline=true,topline=true}
\begin{definition}[Fixed-order Gaussian $\RoundingMatrix$-AFT]
  \label[definition]{def:gaussian-t-aft}
  Fix an efficiently computable public parameter family as in
  \cref{sec:lar-assumptions}, augmented by a positive Gaussian width
  $\GaussianWidth=\GaussianWidth(\secpar)$ included in $\pparam$.
  This fixed family admits \emph{fixed-order Gaussian
  $\RoundingMatrix$-AFTs} if, for every fixed integer
  $\TransportAccuracyOrder\geq1$, there is a uniform classical PPT sampler
  $\TransportSampler_{\TransportAccuracyOrder}$ satisfying the
  algebraic requirements of \cref{def:t-aft}
  at every security parameter, with
  \[
    \StatisticalDistance{
      \DistributionLaw{\TransportEvaluator(
        \TransportMatrix,\TransportOffset,\LWEError)}}{
      \Unif{\LWEZeroFiber}}
    \leq\secpar^{-\TransportAccuracyOrder}
  \]
  for all $\secpar\geq\TransportAccuracyThreshold(\TransportAccuracyOrder)$,
  where $(\TransportMatrix,\TransportOffset)\sample
  \TransportSampler_{\TransportAccuracyOrder}(\secparam,\pparam)$ and
  $\LWEError\sample\GaussianErrorLaw$ independently.
  The running-time polynomial and threshold may depend on
  $\TransportAccuracyOrder$. We also call this property
  \emph{arbitrary fixed inverse-polynomial accuracy}.
\end{definition}
\endgroup

\paragraph{Choosing the accuracy order.}
\phantomsection
\label{rem:gaussian-t-aft-existence}
After fixing a hypothetical uniform QPT distinguisher and polynomial
bounds on its block count and inverse advantage, choose one fixed
$\TransportAccuracyOrder$ so that the accumulated transport error is
smaller than that advantage. This chooses a sampler
$\TransportSampler_{\TransportAccuracyOrder}$, not a new public
parameter family or Gaussian source. No single sampler taking the
order as input is required. The reduction runs at every security
parameter without computing
$\TransportAccuracyThreshold(\TransportAccuracyOrder)$.

\begingroup
\mdfsetup{everyline=true,topline=true}
\begin{theorem}[Gaussian $\LWE$ to uniform-zero-fiber $\LWE$]
  \label{thm:gaussian-lwe-to-uniform-zero-fiber-lwe}
  Fix an efficiently computable public parameter family as above.
  Suppose the family admits fixed-order Gaussian
  $\RoundingMatrix$-AFTs as in \cref{def:gaussian-t-aft}.
  For every fixed positive integer $\TransportAccuracyOrder$, every
  uniform QPT distinguisher $\LWEDistinguisher$ on classical sample
  tables, and every
  polynomially bounded block count $\SampleCount=\SampleCount(\secpar)$,
  using $\TransportSampler_{\TransportAccuracyOrder}$ yields a uniform QPT
  reduction $\LWEReduction_{\TransportAccuracyOrder}$ with
  \[
    \ZeroFiberLWEAdvantage{\LWEDistinguisher}(\secpar,\SampleCount)
    \leq
    \GaussianLWEAdvantage{\LWEReduction_{\TransportAccuracyOrder}}
    (\secpar,\SampleCount)
    +\SampleCount\secpar^{-\TransportAccuracyOrder}
  \]
  for all sufficiently large $\secpar$.
  The reduction uses $\SampleCount$ input blocks and one call to
  $\LWEDistinguisher$.

  Since the family admits such transports for every fixed accuracy
  order, Gaussian decisional $\LWE$ hardness against uniform QPT
  distinguishers on classical inputs for every polynomial block count
  implies the same hardness for uniform-zero-fiber $\LWE$ with this
  fixed public family.
\end{theorem}
\endgroup

\begin{proof}[Proof sketch]
  Fix $\TransportAccuracyOrder$ and its sampler
  $\TransportSampler_{\TransportAccuracyOrder}$. For each input block,
  sample a fresh transport pair independently of the entire input
  experiment and apply the transformation
  \[
    (\LWEPublicMatrix_{\SampleIndex},\LWEResponse_{\SampleIndex})
    \longmapsto
    (\TransportMatrix_{\SampleIndex}\LWEPublicMatrix_{\SampleIndex},
      \TransportMatrix_{\SampleIndex}\LWEResponse_{\SampleIndex}
      +\TransportOffset_{\SampleIndex})
  \]
  over $\LARField$. On real Gaussian input this preserves the shared
  secret and replaces each error by its affine transport. Conditional
  on the shared secret, \cref{lem:t-aft-properties} makes the entire
  transformed matrix table exactly uniform and independent of the
  transported errors. The error law in each block, averaged over both
  the Gaussian source and its independently sampled transport pair,
  is within $\secpar^{-\TransportAccuracyOrder}$ of
  $\Unif{\LWEZeroFiber}$. Independence across blocks gives total
  statistical distance at most
  $\SampleCount\secpar^{-\TransportAccuracyOrder}$ from the real
  uniform-zero-fiber table, also after averaging over the shared secret.

  On ideal input, modular invertibility preserves exactly the product
  law of independent uniform matrices and responses. The reduction
  prepares the transformed classical table and invokes
  $\LWEDistinguisher$ once. Statistical distance bounds the resulting
  acceptance gap even for quantum postprocessing, giving the stated
  inequality. For fixed $\TransportAccuracyOrder$, this procedure is
  uniform QPT at every security parameter; the distance bound holds
  for all sufficiently large $\secpar$ as in \cref{def:gaussian-t-aft}.
  The full game-based proof is deferred to
  Appendix~\ref{sec:gaussian-zero-fiber-reduction-proof}.
\end{proof}

\paragraph{Composition with the rounding reduction.}
The two reductions preserve the sample count and the fixed public
family. Their quantitative losses combine as follows.
\begin{corollary}[Gaussian $\LWE$ to decisional $\LAR$]
\label[corollary]{cor:gaussian-lwe-to-lar-composition}
Under the hypotheses of
\cref{thm:gaussian-lwe-to-uniform-zero-fiber-lwe}, for every fixed
positive integer $\TransportAccuracyOrder$, every polynomially bounded
$\SampleCount$, and every uniform QPT distinguisher $\LARDistinguisher$
on classical sample tables, there is a uniform QPT reduction
$\LWEReduction_{\TransportAccuracyOrder}$
such that, for all sufficiently large $\secpar$,
\begin{equation*}
 \LARDecisionalAdv{\LARDistinguisher}(\secpar,\SampleCount)
 \leq\LARModulus\,
 \GaussianLWEAdvantage{\LWEReduction_{\TransportAccuracyOrder}}
 (\secpar,\SampleCount)
 +\LARModulus\SampleCount\secpar^{-\TransportAccuracyOrder}
 +\LARModulus^{-\SecretDim}.
\end{equation*}
The reduction uses $\SampleCount$ source blocks and invokes
$\LARDistinguisher$ once. If $\LARModulus$ is polynomially bounded,
$\LARModulus^{-\SecretDim}$ is negligible, and Gaussian decisional
$\LWE$ is hard against uniform QPT distinguishers on classical inputs
for every polynomial block count, then decisional $\LAR$ is hard
against such distinguishers for every polynomial sample count for
the same family.
\end{corollary}
\begin{proof}
Apply \cref{thm:uniform-zero-fiber-lwe-to-dlar} to
$\LARDistinguisher$, then apply
\cref{thm:gaussian-lwe-to-uniform-zero-fiber-lwe} to the resulting
uniform-zero-fiber distinguisher. Substitution yields the advantage bound of
\cref{cor:gaussian-lwe-to-lar-composition}. Each reduction preserves the
number of blocks and makes one call to its downstream distinguisher.
Their composition is uniform QPT, with classical preprocessing and
one invocation of $\LARDistinguisher$ on a classical table.

If the $\LAR$ advantage is at least
$1/\AdvantagePolynomial(\secpar)$ infinitely often, choose fixed
$\ModulusExponent,\SampleCountExponent,\AdvantageExponent$ bounding
$\LARModulus,\SampleCount,\AdvantagePolynomial$ by the corresponding
powers of $\secpar$. Taking
$\TransportAccuracyOrder>
\ModulusExponent+\SampleCountExponent+\AdvantageExponent+2$
makes the statistical term at most $1/(4\AdvantagePolynomial(\secpar))$
eventually, while negligibility gives the same bound for
$\LARModulus^{-\SecretDim}$. The Gaussian advantage is then at least
$1/(2\LARModulus\AdvantagePolynomial(\secpar))$ infinitely often,
contradicting Gaussian $\LWE$ hardness. The parameter family is fixed
first, followed by the hypothetical distinguisher and its polynomial
bounds, then by one fixed accuracy order and its sampler. The
reduction runs at every security parameter and need not compute the
accuracy threshold.
\end{proof}

For the concrete family, \cref{thm:df-constructive-aft} supplies the
transport hypothesis. The following sections develop that construction.
Both reductions remain uniform classical PPT when their downstream
distinguisher is classical PPT.

%% file: chapters/t-aft-instantiation.tex
\section{Zero-Fiber Geometry and Affine Matrix Realization}
\label{sec:aft-instantiation}

We express affine transport as an integer incidence map and reduce
its distributional guarantee to endpoint uniformity and carry safety.
A fixed rounding matrix identifies the relevant image lattice;
a tree incidence basis then supplies integral affine maps with
invertible linear parts modulo $\LARModulus$.

\subsection{Graph Interpretation}
\label{sec:graph-interpretation}

All vectors are columns, and a matrix acts from its input columns to its
output rows. We represent affine transport in the image lattice as integer loads on
the vertex set $\TransportVertices:=\set{1,\ldots,\BlockDim}$.
Rows of an integer matrix correspond to vertices, and each nonzero
column represents a signed, integer-weighted hyperedge: positive entries
are tail incidences, negative entries are head incidences, and their
absolute values are the incidence weights. These columns need not have
zero sum, and either the tail or head set may be empty.
Appendix~\ref{sec:incidence-background} records the full graph conventions.

\begingroup
\mdfsetup{everyline=true,topline=true}

\begin{definition}[Incidence representation of affine transport]
  \label[definition]{def:affine-transport-incidence-representation}
  Fix the rounding matrix $\RoundingMatrix$ and a transport pair
  $(\TransportMatrix,\TransportOffset)$ satisfying the exact algebraic
  requirements of \cref{def:gaussian-t-aft}.
  Its \emph{incidence representation} is the pair
  \[
    \HypergraphIncidenceMatrix:=\RoundingMatrix\TransportMatrix
    \in\ZZ^{\BlockDim\times\BlockDim},
    \qquad
    \InitialVertexConfiguration:=\RoundingMatrix\TransportOffset
    \in\ZZ^{\BlockDim}.
  \]
  For an integer representative $\TransportInput\in\ZZ^{\BlockDim}$
  of the source input, the associated \emph{vertex load} is
  \[
    \EndpointLoad:=\HypergraphIncidenceMatrix\TransportInput
    +\InitialVertexConfiguration
    =\RoundingMatrix(\TransportMatrix\TransportInput+\TransportOffset).
  \]
\end{definition}

Each source coordinate gives the coefficient of one hyperedge, and
$\InitialVertexConfiguration$ is the initial vertex load.
For the construction, we choose a tree incidence basis, mix the source
coordinates before applying it, and append a charge-injection column.
The image-lattice constraints ensure that solving these incidence
equations gives integral affine data. For the fixed rounding matrix
below, \cref{cor:general-tree-incidence-transport} reduces these
constraints to total charge and the mixer determinant.

Ordinary directed edges supply charge-neutral updates: their incidence
columns have $+1$ at the tail and $-1$ at the head, so each column sums
to zero. Let $\OnesVec\in\ZZ^{\BlockDim}$ denote the all-ones vector,
and write $\RootLattice$ for the zero-sum integer lattice of
\cref{def:zero-sum-integer-lattice} in dimension $\BlockDim$.
The following standard fact lets us parametrize every such update
using a tree.

\begin{lemma}[Tree incidence basis]
  \label[lemma]{lem:tree-incidence-basis}
  Let $\TransportTree$ be a tree on $\BlockDim$ vertices with an
  arbitrary orientation. It has $\BlockDim-1$ edges, and the columns
  of its signed incidence matrix $\TreeIncidenceMatrix$ form an
  integral basis of $\RootLattice$. In particular,
  \[
    \TreeIncidenceMatrix\ZZ^{\BlockDim-1}=\RootLattice,
    \qquad
    \operatorname{rank}_{\RR}(\TreeIncidenceMatrix)=\BlockDim-1.
  \]
\end{lemma}

The leaf-elimination proof is given in
Appendix~\ref{sec:tree-incidence-basis-proof}.

\endgroup

\subsection{Zero-Fiber Geometry}
\label{sec:zero-fiber-geometric-structure}

For a load $\EndpointLoad\in\LatticeGen{\RoundingMatrix}$, define its
canonical carry by
\begin{equation*}
  \EndpointCarry:=\floor{\frac{\EndpointLoad}{\LARModulus}}
  \in\ZZ^{\BlockDim}.
\end{equation*}
The load's membership in the image lattice guarantees an integral
preimage. Reduction modulo $\LARModulus$ retains the load's remainders
as endpoint colors and discards its integer quotients, which form the
carry. The label of the preimage modulo $\LARModulus$ is
$\QuotientProjection\EndpointCarry\bmod\LARModulus$; the zero fiber
requires this label to vanish. The criterion in
\cref{lem:affine-transport-endpoint-carry-bound} below permits
nonzero carries, which is essential when off-body loads wrap around.
Canonical loads in the box $[0,\LARModulus)^{\BlockDim}$ have zero
carry and supply the endpoint parametrization.

We now specialize to a rounding matrix built from a path incidence
matrix and a column that injects charge at $\LARModulus$ vertices.
This rounding matrix remains fixed throughout the subsequent
tree-incidence transport construction.

\begin{definition}[Charge distribution and simultaneous injection]
  \label[definition]{def:explicit-rounding-matrix}
  Let $\LARModulus$ be an odd prime and let $\BlockDim\geq\LARModulus$
  be an integer. Write $\StandardBasisVec_1,\ldots,\StandardBasisVec_{\BlockDim}$
  for the standard basis of $\ZZ^{\BlockDim}$. Define
  \[
    \begin{aligned}
      \PathIncidenceMatrix
      &:=[\StandardBasisVec_1-\StandardBasisVec_2\mid\cdots\mid
      \StandardBasisVec_{\BlockDim-1}-\StandardBasisVec_{\BlockDim}],\\
      \RoundingColumn
      &:=\sum_{\CoordinateIndex=1}^{\LARModulus}
      \StandardBasisVec_{\CoordinateIndex},
      \qquad
      \RoundingMatrix:=[\PathIncidenceMatrix\mid\RoundingColumn]
      \in\ZZ^{\BlockDim\times\BlockDim}.
    \end{aligned}
  \]
  We use the scalar quotient encoding
  \begin{equation*}
    \QuotientProjection:=\OnesVec^{\trans}.
  \end{equation*}
\end{definition}

The path columns redistribute charge, while the last column injects
one unit at each of the first $\LARModulus$ vertices:
\[
  \OnesVec^{\trans}\PathIncidenceMatrix=\ZeroVec^{\trans},\qquad
  \OnesVec^{\trans}\RoundingColumn=\LARModulus.
\]
The image lattice of $\RoundingMatrix$ consists precisely of the integer
vectors whose coordinate sum is divisible by $\LARModulus$. Equivalently,
it is the preimage of the zero-sum hyperplane in $\LARField^{\BlockDim}$
under reduction modulo $\LARModulus$.

\begin{lemma}[Image lattice and Smith normal form]
  \label[lemma]{lem:rounding-image-lattice}
  Let $\RoundingMatrix$ be the matrix from
  \cref{def:explicit-rounding-matrix}. Its image lattice is
  \begin{equation*}
    \LatticeGen{\RoundingMatrix}
    =\set{\IntegerVec\in\ZZ^{\BlockDim}:
      \sum_{\CoordinateIndex=1}^{\BlockDim}\IntegerEntry_{\CoordinateIndex}
    \equiv0\pmod{\LARModulus}}.
  \end{equation*}
  This lattice has index $\LARModulus$ in $\ZZ^{\BlockDim}$, and the
  Smith normal form of $\RoundingMatrix$ is
  $\operatorname{diag}(\IdentityMatrix_{\BlockDim-1},\LARModulus)$.
  Consequently, $\RoundingMatrix$ is a rounding matrix with
  $\RoundingRank=\BlockDim-1$ and $\OutputDim=1$ in the sense of
  \cref{def:rounding-matrix}.
\end{lemma}

The proof is in Appendix~\ref{sec:incidence-image-lattice-proof}.
The coordinate-sum map identifies the quotient and its scalar encoding.

For this rounding matrix, define the zero-sum hyperplane
\begin{equation*}
  \ZeroSumEndpointHyperplane
  :=\set{\EndpointResidue\in\LARField^{\BlockDim}:
  \OnesVec^{\trans}\EndpointResidue=0}.
\end{equation*}
By \cref{lem:rounding-image-lattice}, the image lattice
$\LatticeGen{\RoundingMatrix}=\RoundingMatrix\ZZ^{\BlockDim}$ is precisely
the integer preimage of $\ZeroSumEndpointHyperplane$ under reduction
modulo $\LARModulus$.
In particular, canonical lifting identifies
$\ZeroSumEndpointHyperplane$ with
$\LatticeGen{\RoundingMatrix}\cap[0,\LARModulus)^{\BlockDim}$.
The preceding observation therefore defines the \emph{canonical endpoint
pullback}
\[
  \begin{aligned}
    \CanonicalEndpointPullback\colon\ZeroSumEndpointHyperplane
    &\longrightarrow\LWEZeroFiber,\\
    \EndpointResidue
    &\longmapsto\RoundingMatrix^{-1}\CanonicalLift{\EndpointResidue}
    \bmod\LARModulus.
  \end{aligned}
\]
Here $\RoundingMatrix^{-1}$ is taken over $\QQ$ before reduction modulo
$\LARModulus$; its product with
$\CanonicalLift{\EndpointResidue}\in\LatticeGen{\RoundingMatrix}$ is integral.

\begin{lemma}[Canonical parametrization of the zero fiber]
  \label[lemma]{lem:canonical-endpoint-pullback}
  The map $\CanonicalEndpointPullback$ is a bijection from
  $\ZeroSumEndpointHyperplane$ onto $\LWEZeroFiber$, with inverse
  \[
    \CanonicalEndpointPullback^{-1}(\RoundingInput)
    =\RoundingMatrix\CanonicalLift{\RoundingInput}\bmod\LARModulus,
    \qquad \RoundingInput\in\LWEZeroFiber.
  \]
\end{lemma}

The proof in Appendix~\ref{sec:incidence-pullback-proof} verifies the
two inverse maps. Canonical representatives identify the finite
hyperplane with the zero fiber.

The bijection in \cref{lem:canonical-endpoint-pullback} transfers
uniformity and statistical distance, but need not preserve affine
structure. The following lemma bounds the distance of an affine output
to uniform on $\LWEZeroFiber$ by its endpoint distance to uniform and
the probability that its carry lies outside the image lattice.

\begingroup
\mdfsetup{everyline=true,topline=true}
\begin{lemma}[Endpoint and carry bound for affine transport]
  \label[lemma]{lem:affine-transport-endpoint-carry-bound}
  Let $\RoundingMatrix$ be the matrix from
  \cref{def:explicit-rounding-matrix}, and let
  $\TransportInput\sample\GaussianErrorLaw$. Sample
  $(\TransportMatrix,\TransportOffset)\in
  \ZZ^{\BlockDim\times\BlockDim}\times\ZZ^{\BlockDim}$ independently of
  $\TransportInput$, with $\TransportMatrix\bmod\LARModulus\in
  \GeneralLinearGroup_{\BlockDim}(\LARField)$.
  Use the incidence load of
  \cref{def:affine-transport-incidence-representation} at the canonical
  source lift $\CanonicalLift{\TransportInput}$, and its carry as defined in
  \cref{sec:zero-fiber-geometric-structure}. Define the output and endpoint by
  \[
    \begin{aligned}
      \TransportOutput
        &:=(\TransportMatrix\CanonicalLift{\TransportInput}
           +\TransportOffset)\bmod\LARModulus,\\
      \EndpointResidue
        &:=\EndpointLoad\bmod\LARModulus
         =\RoundingMatrix\TransportOutput\bmod\LARModulus
         \in\ZeroSumEndpointHyperplane.
    \end{aligned}
  \]
  Then
  \begin{equation*}
    \StatisticalDistance{\DistributionLaw{\TransportOutput}}
    {\Unif{\LWEZeroFiber}}
    \leq
    \StatisticalDistance{\DistributionLaw{\EndpointResidue}}
    {\Unif{\ZeroSumEndpointHyperplane}}
    +\Pr[\EndpointCarry\notin\LatticeGen{\RoundingMatrix}].
  \end{equation*}
  The distributions and probability are over both the transport data and
  the Gaussian source.
  The exact pointwise criterion is
  \begin{equation*}
    \TransportOutput=\CanonicalEndpointPullback(\EndpointResidue)
    \quad\Longleftrightarrow\quad
    \EndpointCarry\in\LatticeGen{\RoundingMatrix}
    \quad\Longleftrightarrow\quad
    \OnesVec^{\trans}\EndpointCarry\equiv0\pmod{\LARModulus}.
  \end{equation*}
\end{lemma}
\endgroup

The proof is in Appendix~\ref{sec:incidence-carry-proof}.
The carry measures exactly the discrepancy between an affine output and
the canonical pullback of its endpoint. Coupling these two outputs
reduces the distributional question to the displayed endpoint and carry
terms.

\subsection{General Tree-Incidence Transport}
\label{sec:general-tree-incidence-transport}

For the rest of this paper, fix the rounding matrix $\RoundingMatrix$
of \cref{def:explicit-rounding-matrix}. Let $\TransportTree$ be any
oriented spanning tree on $\TransportVertices$. Its signed incidence
matrix is an integral basis for charge-neutral motion by
\cref{lem:tree-incidence-basis}. The following extension determines the effect of adjoining a
column of nonzero total charge, independently of the shape of the tree.

\begin{lemma}[Determinant of an extended tree incidence matrix]
  \label[lemma]{lem:tree-incidence-determinant}
  For any oriented tree $\TransportTree$ on $\BlockDim$ vertices with
  signed incidence matrix
  $\TreeIncidenceMatrix\in\ZZ^{\BlockDim\times(\BlockDim-1)}$ and any
  integer column $\TreeExtensionColumn\in\ZZ^{\BlockDim}$,
  \begin{equation*}
    \left|\det[\TreeIncidenceMatrix\mid\TreeExtensionColumn]\right|
    =\left|\OnesVec^{\trans}\TreeExtensionColumn\right|.
  \end{equation*}
\end{lemma}

The proof is in Appendix~\ref{sec:incidence-extended-determinant-proof}.
Appending one charge column measures the index of the resulting lattice
in $\ZZ^{\BlockDim}$. This gives the determinant formula independently
of the tree shape.

\begin{corollary}[General tree-incidence transport]
  \label[corollary]{cor:general-tree-incidence-transport}
  Let $\TransportTree$ be any oriented spanning tree, let
  $\EdgeCoordinateMatrix\in\ZZ^{(\BlockDim-1)\times(\BlockDim-1)}$, and let
  $\TreeExtensionColumn\in\ZZ^{\BlockDim}$ satisfy
  $\OnesVec^{\trans}\TreeExtensionColumn=\LARModulus$.
  Define the extended incidence basis and source-coordinate mixer by
  \[
    \begin{aligned}
      \ExtendedTreeIncidenceMatrix
      &:=[\TreeIncidenceMatrix\mid\TreeExtensionColumn],
      &
      \ExtendedEdgeCoordinateMatrix
      &:=\operatorname{diag}(\EdgeCoordinateMatrix,1),\\
      \HypergraphIncidenceMatrix
      &:=\ExtendedTreeIncidenceMatrix\ExtendedEdgeCoordinateMatrix
      \in\ZZ^{\BlockDim\times\BlockDim}.
    \end{aligned}
  \]
  Let the integer affine baseline and public closure offset together
  form $\InitialVertexConfiguration\in\ZZ^{\BlockDim}$ with
  $\OnesVec^{\trans}\InitialVertexConfiguration\equiv0\pmod{\LARModulus}$,
  and set
  \[
    \TransportMatrix:=\RoundingMatrix^{-1}\HypergraphIncidenceMatrix,
    \qquad
    \TransportOffset:=\RoundingMatrix^{-1}\InitialVertexConfiguration.
  \]
  Then $|\det\ExtendedTreeIncidenceMatrix|=\LARModulus$,
  $\HypergraphIncidenceMatrix\ZZ^{\BlockDim}
  \subseteq\LatticeGen{\RoundingMatrix}$, and
  $\InitialVertexConfiguration\in\LatticeGen{\RoundingMatrix}$.
  Consequently $\TransportMatrix\in\ZZ^{\BlockDim\times\BlockDim}$ and
  $\TransportOffset\in\ZZ^{\BlockDim}$, with
  \begin{equation*}
    \begin{aligned}
      \det\HypergraphIncidenceMatrix
      &=\pm\LARModulus\det\EdgeCoordinateMatrix,\\
      \det\TransportMatrix
      &=\frac{\det\ExtendedTreeIncidenceMatrix
      \det\ExtendedEdgeCoordinateMatrix}{\det\RoundingMatrix}
      =\pm\det\EdgeCoordinateMatrix.
    \end{aligned}
  \end{equation*}
  Every integer input $\TransportInput\in\ZZ^{\BlockDim}$ gives
  \[
    \EndpointLoad:=\ExtendedTreeIncidenceMatrix
    \ExtendedEdgeCoordinateMatrix\TransportInput
    +\InitialVertexConfiguration\in\LatticeGen{\RoundingMatrix},
    \qquad
    \EndpointLoad\bmod\LARModulus\in\ZeroSumEndpointHyperplane.
  \]
  If $\det\EdgeCoordinateMatrix\not\equiv0\pmod{\LARModulus}$, then
  $\TransportMatrix\bmod\LARModulus\in
  \GeneralLinearGroup_{\BlockDim}(\LARField)$.
\end{corollary}

The tree columns have total charge zero, while the extra column has
total charge $\LARModulus$. The extended incidence basis therefore has
determinant of absolute value $\LARModulus$, matching that of
$\RoundingMatrix$. Pulling back through $\RoundingMatrix^{-1}$ cancels
this common factor and leaves the mixer determinant, up to sign.
Appendix~\ref{sec:incidence-general-transport-proof}
verifies integrality and this determinant identity.

The extended incidence basis is singular modulo $\LARModulus$.
The invertibility conclusion concerns the integral quotient
$\TransportMatrix=\RoundingMatrix^{-1}\HypergraphIncidenceMatrix$:
compute its determinant over the integers before reducing modulo
$\LARModulus$.

\subsection{Probabilistic Requirements for Fiber Transport}
\label{sec:probabilistic-fiber-transport-requirements}

\Cref{cor:general-tree-incidence-transport} supplies the algebraic part
of fiber transport. Fix a public parameter family and, for each fixed
positive integer $\TransportAccuracyOrder$, an admissible sampler
$\TransportSampler_{\TransportAccuracyOrder}(\secparam,\pparam)$
satisfying the exact algebraic, source-independence, and computational
requirements of \cref{def:gaussian-t-aft}. With the endpoint and carry
of \cref{lem:affine-transport-endpoint-carry-bound}, it remains to prove
nonnegative error bounds
\begin{equation*}
  \begin{aligned}
    \StatisticalDistance{\DistributionLaw{\EndpointResidue}}
    {\Unif{\ZeroSumEndpointHyperplane}}&\leq\EndpointMixingError,\\
    \Pr[\EndpointCarry\notin\LatticeGen{\RoundingMatrix}]
    &\leq\CarryFailureError,\\
    \EndpointMixingError+\CarryFailureError
    &\leq\secpar^{-\TransportAccuracyOrder}.
  \end{aligned}
\end{equation*}
These estimates must hold for every
$\secpar\geq\TransportAccuracyThreshold(\TransportAccuracyOrder)$,
with laws averaged jointly over the public sampler and independent
Gaussian source. The error terms may depend on the fixed accuracy order
and security parameter. The endpoint-and-carry lemma then gives the
fiber-transport accuracy required by \cref{def:gaussian-t-aft}.
The next section gives a common finite-dimensional probability
calibration principle. The affine construction and joint endpoint
analysis in \cref{sec:probabilistic-transport,sec:direct-fourier-transport}
apply it twice while establishing these endpoint and carry estimates
for the fixed rounding matrix and source.

%% file: chapters/finite-dimensional-calibration.tex
\section{Finite-Dimensional Probability Calibration}
\label{sec:finite-dimensional-calibration}

We choose weights on public choices so that their outputs satisfy
specified probability constraints. In the local application
(\cref{sec:leg-map-distribution}), the choices are affine maps and the
constraints prescribe color probabilities at every protected input.
In the joint application (\cref{sec:df-matrix-calibration}), the choices
are public controls and the constraints prescribe Fourier coefficients
after averaging the protected source. In both cases the unknown weight
uses only public data. We compare the actual response to a solvable
reference response; a small perturbation then permits stable inversion.

\paragraph{Neumann inversion.}
\label{sec:calibration-neumann}
Let $\ProbCalSpace$ be a finite-dimensional real normed space, with
identity $\ProbCalIdentity$, and let
$\ProbCalOperator:\ProbCalSpace\to\ProbCalSpace$ be linear. If its
defect $\ProbCalDefect:=\ProbCalIdentity-\ProbCalOperator$ satisfies
$\|\ProbCalDefect\|_{\ProbCalSpace\to\ProbCalSpace}
\leq\ProbCalContraction<1$, the Neumann series converges in operator
norm and gives
\[
  \ProbCalOperator^{-1}
  =\sum_{\ProbCalPower\geq0}\ProbCalDefect^{\ProbCalPower},
  \qquad
  \|\ProbCalOperator^{-1}\|_{\ProbCalSpace\to\ProbCalSpace}
  \leq\frac1{1-\ProbCalContraction}.
\]
Once an application constructs such a response and bounds its
perturbation, stable solvability follows.

\paragraph{Finding such a linear response.}
Let $(\FiniteSampleSpace,\ProbCalLaw)$ be the reference probability space
of public choices.
A linear synthesis map
$\ProbCalSynthesis:\ProbCalSpace\to L^1(\ProbCalLaw)$ turns coefficients
into weights. A linear map $\ProbCalResponse$ records their constraints
in a real vector space $\ProbCalOutputSpace$, and a linear map
$\ProbCalEvaluation:\ProbCalSpace\to\ProbCalOutputSpace$ specifies the
reference response. We seek a weight whose response is the prescribed
$\ProbCalEvaluation\ProbCalTarget$.

Factoring the actual response through $\ProbCalEvaluation$ reduces
calibration to a coefficient equation. The theorem bounds its correction
by the residual of any reference coefficient $\ProbCalReference$.

\begin{theorem}[Stable finite-dimensional calibration]
  \label{thm:stable-finite-dimensional-calibration}
  Suppose an explicit linear operator
  $\ProbCalOperator:\ProbCalSpace\to\ProbCalSpace$ satisfies
  \[
    \ProbCalResponse\ProbCalSynthesis=\ProbCalEvaluation\ProbCalOperator,
    \qquad
    \|\ProbCalOperator-\ProbCalIdentity\|_{\ProbCalSpace\to\ProbCalSpace}
    \leq\ProbCalContraction<1.
  \]
  For every specified $\ProbCalTarget\in\ProbCalSpace$, there is a unique
  solution $\ProbCalCoefficient$ of
  $\ProbCalOperator\ProbCalCoefficient=\ProbCalTarget$, and
  $\ProbCalWeight=\ProbCalSynthesis\ProbCalCoefficient$ satisfies
  $\ProbCalResponse(\ProbCalWeight)=\ProbCalEvaluation\ProbCalTarget$.
  For every reference $\ProbCalReference\in\ProbCalSpace$,
  \[
    \ProbCalNorm{\ProbCalCoefficient-\ProbCalReference}
    \leq
    \frac{\ProbCalNorm{\ProbCalTarget-\ProbCalOperator\ProbCalReference}}
    {1-\ProbCalContraction}.
  \]
  If the constraints include total mass and the target mass is one, then
  $\mathbb E_{\ProbCalLaw}\ProbCalWeight=1$.
\end{theorem}

Uniqueness concerns the full coefficient equation; neither
$\ProbCalResponse$ nor $\ProbCalEvaluation$ need be invertible.
Efficient finite computation additionally requires dimension,
precision, and inverse-gap bounds. Proofs are in
\cref{sec:calibration-principle-proofs}.

\paragraph{Probability interpretation.}
Weighted expectations
$\ProbCalResponse(\ProbCalWeight)_{\CoordinateIndex}
=\mathbb E_{\ProbCalLaw}[\ProbCalWeight\ProbCalTest_{\CoordinateIndex}]$
give linear constraints when the products are integrable;
$\ProbCalTest_0=1$ records total mass.
For local calibration these tests record colors at a fixed protected
input. For joint calibration they are conditional expectations of the
real and imaginary parts of endpoint characters, averaging the
protected source and remaining channel randomness given the public
choices. This averaging defines the response; evaluating the
synthesized weight requires no hidden input.

\paragraph{Nonnegative weights.}
\label{sec:calibration-nonnegativity}
Mass one gives a signed law. A pointwise lower bound can establish
nonnegativity directly; high-moment bounds instead support clipping
to an approximate probability law.

\begin{corollary}[One-sided pointwise positivity]
  \label[corollary]{cor:calibration-pointwise-positivity}
  Under the hypotheses of \cref{thm:stable-finite-dimensional-calibration},
  suppose $\ProbCalSpace$ is a function space with the supremum norm.
  The solution coefficient is then itself a function, written
  $\ProbCalFunction=\ProbCalCoefficient
  =\ProbCalOperator^{-1}\ProbCalTarget$; the public weight is
  $\ProbCalSynthesis\ProbCalFunction$. We have
  \[
    \inf\ProbCalFunction\geq\inf\ProbCalTarget
    -\frac{\ProbCalContraction\|\ProbCalTarget\|_\infty}
    {1-\ProbCalContraction}.
  \]
  In particular, if $\ProbCalTarget=1-\ProbCalTest$,
  $\sup\ProbCalTest\leq\ProbCalUpperDefect$ and
  $\|\ProbCalTest\|_\infty\leq\ProbCalAbsoluteDefect$, then
  \[
    \ProbCalFunction\geq1-\ProbCalUpperDefect
    -\frac{\ProbCalContraction(1+\ProbCalAbsoluteDefect)}
    {1-\ProbCalContraction}.
  \]
  For a support set $\ProbCalSupport\subseteq\FiniteSampleSpace$,
  masked synthesis means
  $\ProbCalSynthesis\ProbCalFunction
  =\EventIndicator{\ProbCalSupport}\ProbCalFunction$.
  This weight vanishes outside $\ProbCalSupport$.
  If $\ProbCalFunction\geq0$ on $\ProbCalSupport$ and the masked weight
  has calibrated expectation one under $\ProbCalLaw$, it is a
  probability density relative to $\ProbCalLaw$.
\end{corollary}

\begin{corollary}[High moments and bounded clipping]
  \label[corollary]{cor:calibration-bounded-clipping}
  For a real integrable weight $\ProbCalSignedWeight$, put
  $\ProbCalClippedWeight=\ProbCalClipMap(\ProbCalSignedWeight)
  =\min\{2,\max\{0,\ProbCalSignedWeight\}\}$ and
  $\ProbCalClipError=\mathbb E_{\ProbCalLaw}
  |\ProbCalSignedWeight-\ProbCalClippedWeight|$. For every integer
  $\ProbCalExponent\geq1$,
  \[
    0\leq\ProbCalClippedWeight\leq2,
    \qquad
    |\ProbCalSignedWeight-\ProbCalClippedWeight|
    \leq|\ProbCalSignedWeight-1|^{2\ProbCalExponent}.
  \]
  Consequently,
  \[
    \ProbCalClipError\leq
    \mathbb E_{\ProbCalLaw}|\ProbCalSignedWeight-1|^{2\ProbCalExponent}.
  \]
\end{corollary}

A small coefficient correction need not keep the public weight
pointwise close to one. To control clipping, we instead bound a high
moment of the weight correction under the reference law. Suppose
$\ProbCalSignedWeight=\ProbCalSynthesis\ProbCalCoefficient$ and
$\ProbCalSynthesis\ProbCalReference=1$. For an integer
$\ProbCalExponent\geq1$, assume
\[
  \|\ProbCalSynthesis\ProbCalVector\|_{L^{2\ProbCalExponent}(\ProbCalLaw)}
  \leq\ProbCalBridgeAt{2\ProbCalExponent}\ProbCalNorm{\ProbCalVector}
\]
on a subspace $\ProbCalCorrectionSpace$ containing
$\ProbCalCoefficient-\ProbCalReference$. Since
$\ProbCalSignedWeight-1
=\ProbCalSynthesis(\ProbCalCoefficient-\ProbCalReference)$,
this bound converts the coefficient estimate in
\cref{thm:stable-finite-dimensional-calibration} into a moment estimate
for the public weight. Applying \cref{cor:calibration-bounded-clipping} gives
\[
  \ProbCalClipError\leq
  \left(\frac{\ProbCalBridgeAt{2\ProbCalExponent}
    \ProbCalNorm{\ProbCalTarget-\ProbCalOperator\ProbCalReference}}
  {1-\ProbCalContraction}\right)^{2\ProbCalExponent}.
\]
This bounds the expected change in the public weight caused by clipping,
before conditioning and normalization.

%% file: chapters/probabilistic-transport.tex
\section{Chameleon-Lobster Family with Safe Local Sampling}
\label{sec:probabilistic-transport}

The construction starts from the canonical arrangement in
\cref{fig:lobster-tree-structure}. The horizontal path is the body,
each leg adds two off-body vertices, and the highlighted queue is a
set of body vertices that receive one shared scalar. Solid edges form
a tree; the dashed enclosure denotes the additional queue hyperedge.

\begin{figure}[H]
  \centering
  \begin{tikzpicture}[
      x=1cm,y=1cm,font=\small,
      vertex/.style={circle,draw=black,fill=white,line width=0.7pt,
      minimum size=4.5mm,inner sep=0pt},
      queue vertex/.style={vertex,fill=blue!12,draw=blue!55!black},
      tree edge/.style={line width=0.85pt},
      guide/.style={draw=black!65,line width=0.5pt},
      annotation/.style={font=\small,align=center}
    ]
    \draw[draw=blue!55!black,fill=blue!3,dashed,rounded corners=8pt,
    line width=0.8pt] (7.95,-0.48) rectangle (11.35,0.48);

    \node[vertex] (body1) at (0,0) {};
    \node[vertex] (body2) at (1.2,0) {};
    \node[vertex] (body3) at (2.4,0) {};
    \node[vertex] (body4) at (3.6,0) {};
    \node[vertex] (body5) at (4.8,0) {};
    \node[vertex] (body6) at (6,0) {};
    \node[vertex] (body7) at (7.2,0) {};
    \node[queue vertex] (body8) at (8.4,0) {};
    \node[queue vertex] (body9) at (9.6,0) {};
    \node[queue vertex] (body10) at (10.8,0) {};
    \node[vertex] (middle1) at (0,-0.95) {};
    \node[vertex] (leaf1) at (0,-1.9) {};
    \node[vertex] (middle2) at (1.2,-0.95) {};
    \node[vertex] (leaf2) at (1.2,-1.9) {};
    \node[vertex] (middle3) at (2.4,-0.95) {};
    \node[vertex] (leaf3) at (2.4,-1.9) {};

    \draw[tree edge] (body1) -- (body2);
    \draw[tree edge] (body2) -- (body3);
    \draw[tree edge] (body3) -- (body4);
    \draw[tree edge] (body4) -- (body5);
    \draw[tree edge] (body5) -- (body6);
    \draw[tree edge] (body6) -- (body7);
    \draw[tree edge] (body7) -- (body8);
    \draw[tree edge] (body8) -- (body9);
    \draw[tree edge] (body9) -- (body10);
    \draw[tree edge] (body1) -- (middle1);
    \draw[tree edge] (middle1) -- (leaf1);
    \draw[tree edge] (body2) -- (middle2);
    \draw[tree edge] (middle2) -- (leaf2);
    \draw[tree edge] (body3) -- (middle3);
    \draw[tree edge] (middle3) -- (leaf3);

    \draw[guide] (0,0.72) -- (0,0.9) -- (10.8,0.9) -- (10.8,0.72);
    \node[annotation] at (5.4,1.2) {Body};
    \draw[guide] (2.82,0) -- (3.02,0) -- (3.02,-1.9) -- (2.82,-1.9);
    \node[annotation,anchor=west] at (3.2,-0.95)
    {Leg\\(three vertices)};
    \draw[guide] (8.4,-0.68) -- (8.4,-0.86) -- (10.8,-0.86)
    -- (10.8,-0.68);
    \node[annotation] at (9.6,-1.16) {Queue ($\LARModulus=3$)};
    \node[annotation,text=blue!55!black] at (9.6,-1.95)
    {Dashed enclosure: one hyperedge};
  \end{tikzpicture}
  \caption{The canonical lobster-tree construction with
    $\LARModulus=3$.
    Solid edges form the underlying tree: the body is the horizontal
    path, and each leg adds two vertices to a body vertex.
    The three highlighted vertices at the tail form the queue; the four
    remaining body vertices are available for later calibration.
    The dashed enclosure represents the
  additional hyperedge supported on the queue.}
  \label{fig:lobster-tree-structure}
\end{figure}

We specialize the tree-incidence construction of
\cref{sec:aft-instantiation} to this graph and safe local sampling laws.
Given an admissible public-core sampler,
\cref{lem:joint-discrete-realization} supplies integral,
source-independent affine data, modular invertibility, and a Gaussian
tail bound on carry failure. The joint calibration in
\cref{sec:direct-fourier-transport} chooses the core law to make endpoints
nearly uniform, completing the fixed-order Gaussian
$\RoundingMatrix$-AFT guarantee of \cref{def:gaussian-t-aft}.
The local and joint calibrations both use
\cref{thm:stable-finite-dimensional-calibration}; the construction here
provides their support and source-independence guarantees.

\subsection*{Construction Roadmap and Invariants}

Recall the interface of \cref{sec:probabilistic-fiber-transport-requirements}.
For $\PaddingSource\sample\Gaussian{\ZZ^{\BlockDim}}{\GaussianWidth}$, with
$\LWEError=\PaddingSource\bmod\LARModulus$, sample public affine data
$(\TransportMatrix,\TransportOffset)$ independently of the source.
We require integrality, invertibility of $\TransportMatrix$ modulo
$\LARModulus$, and carry safety on protected inputs. With the endpoint
and carry of \cref{lem:affine-transport-endpoint-carry-bound}, that lemma gives
\[
  \StatisticalDistance{\DistributionLaw{
  \TransportMatrix\LWEError+\TransportOffset\bmod\LARModulus}}
  {\Unif{\LWEZeroFiber}}
  \leq
  \StatisticalDistance{\DistributionLaw{\EndpointResidue}}
  {\Unif{\ZeroSumEndpointHyperplane}}
  +\Pr[\EndpointCarry\notin\LatticeGen{\RoundingMatrix}].
\]
This section bounds the carry term and prepares the endpoint term for
\cref{sec:direct-fourier-transport}. An arbitrary admissible core law
need not make endpoints nearly uniform.

Closing the body path gives all body increments the same marginal law
while keeping their sum zero. Each leg adds control over a primary
color through a safe public affine map; the dependence of the off-body
pair's total load on its second hidden input is canceled by the
attachment's increment. Tree increments
alone cannot change the total load, so the queue supplies the missing
direction by adding one shared scalar to $\LARModulus$ vertices.
The baseline's public charge congruence ensures that the complete
offset has an integer preimage under $\RoundingMatrix$.
The component laws are defined in
\cref{def:body-injection,def:leg-injection,sec:queue-injection,def:admissible-joint-baseline-law}.

The joint placement fixes disjoint attachments and queue vertices.
An independent body permutation moves the cycle, and an independent
final permutation relabels the complete load. These operations retain
the cycle, leg, and queue dependencies; the endpoint analysis must
therefore control their joint law. The affine validity and carry
arguments appear in
\cref{prop:matrix-integer-realization,prop:matrix-carry-safety},
with bounded implementation in \cref{lem:joint-discrete-realization}.

\subsection{Geometry and Placement}
\label{sec:lobster-components}
\label{sec:lobster-combination}

Let $\LARModulus$ be an odd prime. Choose the body size
$\InteriorVertexCount$ and the number $\GadgetCount$ of labeled legs so that
\begin{equation*}
  \BlockDim=\InteriorVertexCount+2\GadgetCount,\qquad
  \InteriorVertexCount>\LARModulus+2\GadgetCount.
\end{equation*}
The graph has the following components, illustrated in
\cref{fig:lobster-tree-structure}.

\begin{itemize}
  \item The \emph{body} is a path on an ordered vertex set
    $\InteriorVertices$ of size $\InteriorVertexCount$.
  \item For $1\leq\GadgetIndex\leq\GadgetCount$, labeled
    \emph{leg} $\GadgetIndex$ is a two-edge path with vertex order
    $({\LegLeftVertex}_{\GadgetIndex},
    {\LegMiddleVertex}_{\GadgetIndex},{\LegRightVertex}_{\GadgetIndex})$.
    The \emph{primary} vertex ${\LegLeftVertex}_{\GadgetIndex}$ and
    \emph{companion} vertex ${\LegMiddleVertex}_{\GadgetIndex}$ lie
    outside the body; the \emph{attachment}
    ${\LegRightVertex}_{\GadgetIndex}$ lies in $\InteriorVertices$.
    The legs have disjoint off-body vertices and distinct attachments.
  \item The \emph{queue} is a set
    $\ChargeCluster\subset\InteriorVertices$ of $\LARModulus$ vertices,
    disjoint from the attachments. It supports one additional hyperedge
    that adds the same scalar to every queue vertex.
\end{itemize}
The body path and the two edges of each leg form a spanning tree with
$\BlockDim-1$ edges. In the canonical arrangement, the attachments are
the first $\GadgetCount$ body vertices in labeled order, and the queue
consists of the last $\LARModulus$ body vertices. The strict inequality
in \cref{sec:lobster-combination} leaves additional body
vertices for the calibration controls of \cref{sec:df-protected-reference}.

The following three public permutations place these components and
relabel the resulting load.
\begin{enumerate}
  \item \emph{Jointly place the legs and queue.}
    A uniform body permutation $\JointPlacementMatrix$ places the queue
    set and labeled attachments on disjoint vertices. The resulting
    \emph{joint placement} $\CalibContext$ consists of the queue set and
    the ordered tuple of labeled attachments, retaining only these
    images of the permutation.
  \item \emph{Independently place the body cycle.}
    The path will be closed by an independent public Gaussian offset
    at its two endpoints, as defined in \cref{def:body-injection}.
    A uniform body permutation $\BodyPlacementMatrix$, independent of
    the placement, initial-load sampling, and Gaussian inputs, moves the
    path and its closure together. It leaves the attachments and queue
    fixed.
  \item \emph{Relabel the complete load.}
    An independent uniform permutation $\GlobalRelabelingMatrix$ acts
    on the entire vertex load, including its initial configuration
    and all increments.
\end{enumerate}
The calibrated sampler postpones the body and final permutations until
the other public sampling steps are complete. Rejection sampling
keeps $\CalibContext$ fixed; the complete sampling order and
independence conditions are in \cref{sec:df-sampling-contract}.
The final permutation preserves the image lattice and carry condition
because both depend on coordinate sums. Conditional on the endpoint
histogram, it makes all arrangements of that histogram equally likely.

\subsection{The Affine Realization Matrix}
\label{sec:affine-realization-matrix}

The affine family uses one shared baseline and three increment blocks.
The public \emph{core} specifies the body centers, leg shifts, and marks,
where each mark is a target primary color and a mixing weight. Given
this core, each leg samples a flag and a complete affine map. The same
map supplies the baseline intercept and the two mixer coefficients;
these quantities must retain their joint law. The deterministic baseline
assembly is defined in \cref{def:shared-baseline-vector}, and
\cref{sec:baseline-shift} specifies the admissible joint sampling laws.
The independent public Gaussian closure contributes to the offset.

To evaluate the affine map, first mix the source coordinates and use
the incidence columns to place the resulting increments on vertices.
Add the sampled baseline and closure offset in these vertex coordinates,
relabel the complete load, and apply $\RoundingMatrix^{-1}$ to return
to the rounding input coordinates. The formulas below separate the linear
part from the offset, retaining their shared public randomness.

The placement $\CalibContext$ of \cref{sec:lobster-combination} determines
the embedding
$\RealMatrixLegEmbedding{\GadgetIndex}:\ZZ^3\to\ZZ^{\BlockDim}$
in the leg vertex order fixed above. Write
$\RealMatrixBodyEmbedding:\ZZ^{\InteriorVertexCount}\to\ZZ^{\BlockDim}$
for zero extension from the body. Order the hidden source blocks as
\begin{equation*}
  \PaddingSource=
  \begin{pmatrix}
    \BodySourceCoordinates\\
    \LegSourceCoordinates{1}\\
    \vdots\\
    \LegSourceCoordinates{\GadgetCount}\\
    \QueueSourceCoordinate
  \end{pmatrix}
  \sample\Gaussian{\ZZ^{\BlockDim}}{\GaussianWidth},
  \qquad
  \begin{gathered}
    \BodySourceCoordinates\in\ZZ^{\InteriorVertexCount-1},\quad
    \LegSourceCoordinates{\GadgetIndex}\in\ZZ^2,\\
    \QueueSourceCoordinate\in\ZZ,\qquad
    \BlockDim=(\InteriorVertexCount-1)+2\GadgetCount+1.
  \end{gathered}
\end{equation*}
The fixed local directions, public leg mixers, and queue column are
\begin{equation*}
  \begin{aligned}
    \BodyIncidenceMatrix
    &=[\StandardBasisVec_1-\StandardBasisVec_2\mid\cdots\mid
      \StandardBasisVec_{\InteriorVertexCount-1}
    -\StandardBasisVec_{\InteriorVertexCount}],\\
    \LegIncidenceMatrix
    &=
    \begin{pmatrix}1&0\\-1&-1\\0&1
    \end{pmatrix},
    &\LegCoordinateBlock{\GadgetIndex}
    &=
    \begin{pmatrix}
      \LegScaleLift_{\GadgetIndex}&\LegShearLift_{\GadgetIndex}\\0&1
    \end{pmatrix},
    &\ChargeColumn
    &=\sum_{\VertexIndex\in\ChargeCluster}\StandardBasisVec_{\VertexIndex}.
  \end{aligned}
\end{equation*}
The mixer entries are the canonical integer lifts of the nonzero scale
and shear in \cref{def:leg-public-map}, sampled by the safe law of
\cref{sec:leg-injection}. The independent body permutation acts only on
the path and closure; the closure Gaussian of \cref{def:body-injection}
is independent of the source and core. Before the independent final
permutation $\GlobalRelabelingMatrix$, define
\begin{equation*}
  \begin{aligned}
    \RealMatrixIncidence(\BodyPlacementMatrix)
    &:=[\RealMatrixBodyEmbedding\BodyPlacementMatrix\BodyIncidenceMatrix
      \mid\RealMatrixLegEmbedding{1}\LegIncidenceMatrix\mid\cdots\mid
      \RealMatrixLegEmbedding{\GadgetCount}\LegIncidenceMatrix
    \mid\ChargeColumn],\\
    \RealMatrixSourceMixer
    &:=\operatorname{diag}(\IdentityMatrix_{\InteriorVertexCount-1},
      \LegCoordinateBlock{1},\ldots,
    \LegCoordinateBlock{\GadgetCount},1),\\
    \PreRelabelOffset
    &:=\PreRelabelBaseline+
    \RealMatrixBodyEmbedding\BodyPlacementMatrix\BodyAffineOffset.
  \end{aligned}
\end{equation*}
Here $\BodyAffineOffset$ is the two-endpoint closure offset defined in
\cref{def:body-injection}, and $\PreRelabelBaseline$ is the single
shared baseline of \cref{def:shared-baseline-vector}, which counts each
attachment center once. The affine realization matrix and offset are
\begin{equation*}
  \begin{aligned}
    \TransportMatrix
    &=\RoundingMatrix^{-1}\GlobalRelabelingMatrix
    \RealMatrixIncidence(\BodyPlacementMatrix)\RealMatrixSourceMixer,
    &\TransportOffset
    &=\RoundingMatrix^{-1}\GlobalRelabelingMatrix\PreRelabelOffset.
  \end{aligned}
\end{equation*}
All inverse products with $\RoundingMatrix$ are computed over $\QQ$
before reduction modulo $\LARModulus$. For an admissible core and safe
leg maps, \cref{prop:matrix-integer-realization} gives integral affine
data: the public charge condition makes the offset integral, and the
nonzero leg scales give modular invertibility.

\paragraph{Public calibration.}
\label{sec:statistic-proof-roadmap}
The Gaussian-convolution viewpoint of~\cite{EC:MicPei12,AC:DevPasSte23}
motivates choosing public randomness to shape the output distribution.
We use \cref{sec:finite-dimensional-calibration} twice.
Locally, restricting affine maps to safe support biases their color
marginals, and marked maps add unwanted bias away from the protected
line. The calibration in \cref{sec:leg-map-distribution} reweights
one public distribution of complete maps to give every protected
input its prescribed marginal simultaneously. Its output is a positive
map law sampled without observing the source.

The joint calibration in \cref{sec:direct-fourier-transport} addresses
the correlations that remain between endpoints. At each fixed placement,
it reweights public controls in the protected experiment before imposing
charge. After independent uniform endpoint relabeling, the resulting
signed law has no nonzero Fourier coefficients up to the chosen support
order. A complete Fourier bound controls the remaining distribution
error. Moment bounds control the error from clipping the public weight
to a bounded nonnegative function; public charge conditioning and
normalization then restore an admissible law. Bounded rejection sampling
keeps placement fixed and uses only public data. Coupling the protected
inputs back to the original Gaussian source gives the AFT bound.

\subsection{Safe Baseline Support}
\label{sec:lobster-injection-targets}

Use vertex labels before the final relabeling. Denote the placed
increments by
\[
  \begin{aligned}
    \PlacedBodyIncrement
    &:=\RealMatrixBodyEmbedding\BodyPlacementMatrix\BodyLoadIncrement,\\
    \PlacedLegIncrement
    &:=\sum_{\GadgetIndex=1}^{\GadgetCount}
    \RealMatrixLegEmbedding{\GadgetIndex}\LegIncidenceMatrix
    \LegCoordinateBlock{\GadgetIndex}
    \LegSourceCoordinates{\GadgetIndex},\\
    \PlacedQueueIncrement&:=\ChargeColumn\QueueSourceCoordinate.
  \end{aligned}
\]
Here $\BodyLoadIncrement$ includes the public closure, as defined in
\cref{def:body-injection}. A superscript $\mathrm{loc}$ denotes a local
increment; subscripts $\mathrm B,\mathrm L,\mathrm Q,\mathrm{BS}$
refer to the body, leg, queue, and shared baseline.

Fix positive integer cutoffs $\PaddingInteriorCutoff$, $\LegCutoff$,
and $\PaddingChargeCutoff$, and an integer margin $\PaddingSafeMargin$
with
\[
  \PaddingInteriorCutoff+\LegCutoff+\PaddingChargeCutoff
  \leq\PaddingSafeMargin,\qquad
  2\PaddingSafeMargin\leq\LARModulus-1.
\]
The body baseline lies in
\[
  \HistogramSafeCenters
  :=\set{\PaddingSafeMargin,\ldots,\LARModulus-1-\PaddingSafeMargin}.
\]
The centers, shifts, and affine maps determine the baseline below;
\cref{sec:baseline-shift} specifies their joint sampling law and charge
constraint. All labels precede the final relabeling, so we suppress
$\mathrm{pre}$ on $\PreRelabelBaseline$.

\begingroup
\mdfsetup{topline=true}
\begin{definition}[Shared baseline vector]
  \label[definition]{def:shared-baseline-vector}
  Given public body centers $\BaselineCenters$, leg shifts
  $\LegBaselineShift$, and one public affine map per leg with canonical
  intercept $\LegBaselineFirst_{\GadgetIndex}$ as in
  \cref{def:leg-public-map}, define $\BaselineShift$ by
  \begin{equation*}
    \begin{aligned}
      (\BaselineShift)_{\VertexIndex}
      &:=\BaselineCenters_{\VertexIndex}
      &&(\VertexIndex\in\InteriorVertices),\\
      (\BaselineShift)_{{\LegLeftVertex}_{\GadgetIndex}}
      &:=\LegBaselineFirst_{\GadgetIndex}
      &&(1\leq\GadgetIndex\leq\GadgetCount),\\
      (\BaselineShift)_{{\LegMiddleVertex}_{\GadgetIndex}}
      &:=\LARModulus-1+\LegBaselineShift_{\GadgetIndex}
      -\LegBaselineFirst_{\GadgetIndex}
      &&(1\leq\GadgetIndex\leq\GadgetCount).
    \end{aligned}
  \end{equation*}
\end{definition}
\endgroup
Each attachment retains its body entry; the leg supplies no second
baseline there. The same sampled map $\LocalChartPoint_{\GadgetIndex}$
determines both the mixer $\LegCoordinateBlock{\GadgetIndex}$ and the
intercept $\LegBaselineFirst_{\GadgetIndex}$, preserving their joint law.

\Cref{sec:baseline-shift} specifies the joint sampling law and public
charge condition. On the protected event of
\cref{def:joint-protected-event}, the margin and safe maps give the
carry guarantee of \cref{prop:matrix-carry-safety}.

\input{chapters/body-injection}
\input{chapters/leg-injection}
\input{chapters/queue-injection}
\input{chapters/baseline-shift}
\input{chapters/discrete-realization}

%% file: chapters/body-injection.tex
\subsection{Cycle Gaussian Body Injection}
\label{sec:body-injection}

The public closure gives the two body endpoints the same Gaussian
difference marginal as the internal vertices. It produces a cycle
of correlated, charge-neutral increments, illustrated in
\cref{fig:body-goal}.

\begin{figure}[H]
  \centering
  \input{figures/body-goal.tex}
  \caption{Cycle Gaussian Body Injection for
    $\InteriorVertexCount=5$. The solid arrows are the four
    hidden path coordinates. The dashed closure denotes the public
    offset $\BodyClosureGaussian(\StandardBasisVec_5-\StandardBasisVec_1)$;
    it is not a transport-matrix column. Its endpoint contributions
    produce the exact cyclic difference law. The body placement moves
  this entire increment together. Baseline, legs, and queue are omitted.}
  \label{fig:body-goal}
\end{figure}

\begingroup
\mdfsetup{topline=true}
\begin{definition}[Cycle Gaussian body injection]
  \label[definition]{def:body-injection}
  For $\GaussianWidth>0$, put
  $\CycleScalarLaw:=\Gaussian{\ZZ}{\GaussianWidth}$ and use
  the unit-weight path matrix $\BodyIncidenceMatrix$ from
  \cref{sec:affine-realization-matrix}, without mixing.
  Draw the hidden source
  $\BodyGaussianInput=(\CycleGaussianScalar_1,\ldots,
  \CycleGaussianScalar_{\InteriorVertexCount-1})^{\trans}
  \sample\CycleScalarLaw^{\otimes(\InteriorVertexCount-1)}$
  and independently the public closure
  $\BodyClosureGaussian\sample\CycleScalarLaw$. Set
  \begin{equation*}
    \begin{aligned}
      \BodyAffineOffset
      &:=\BodyClosureGaussian
      (\StandardBasisVec_{\InteriorVertexCount}-\StandardBasisVec_1)
      =-\BodyClosureGaussian\BodyIncidenceMatrix\OnesVec_{\InteriorVertexCount-1},\\
      \BodyLoadIncrement
      &:=\BodyIncidenceMatrix\BodyGaussianInput+\BodyAffineOffset.
    \end{aligned}
  \end{equation*}
  With $\CycleGaussianScalar_{\InteriorVertexCount}:=\BodyClosureGaussian$
  and
  $\CycleGaussianScalar_0:=\CycleGaussianScalar_{\InteriorVertexCount}$,
  we have
  \begin{equation*}
    (\BodyLoadIncrement)_{\CycleIndex}
    =\CycleGaussianScalar_{\CycleIndex}
    -\CycleGaussianScalar_{\CycleIndex-1}
    \quad(1\leq\CycleIndex\leq\InteriorVertexCount),
    \qquad
    \OnesVec^{\trans}\BodyLoadIncrement=0
    \quad\text{over }\ZZ.
  \end{equation*}
\end{definition}
\endgroup

Both $\BodyAffineOffset$ and $\BodyLoadIncrement$ lie in
$\PaddingInteriorRootLattice$ by \cref{lem:tree-incidence-basis}.
The placement of \cref{sec:affine-realization-matrix} moves the entire
cycle, including both closure contributions.

The correlated cycle law averages over the public closure Gaussian;
fixing its realized value generally changes the endpoint marginals.
Appendix~\ref{sec:cycle-body-law-proof} verifies this law and its
coordinate identities. Appendix~\ref{sec:discrete-body-injection}
implements the public scalar with error $\BodyClosureSamplingError$
in worst-case polynomial time, preserving every integer-lattice
identity. The protected-edge reference and joint error bounds are in
\cref{sec:joint-injection-protection}.

%% file: figures/body-goal.tex
\begin{tikzpicture}[x=1cm,y=1cm,font=\footnotesize,line cap=round]
  \path[use as bounding box] (0,-0.25) rectangle (13.8,3.1);
  \node at (6.9,2.9) {hidden path source coordinates};
  \node[circle,draw,fill=white,minimum size=6mm,inner sep=0pt] (b1) at (1.1,2) {$1$};
  \node[circle,draw,fill=white,minimum size=6mm,inner sep=0pt] (b2) at (4.0,2) {$2$};
  \node[circle,draw,fill=white,minimum size=6mm,inner sep=0pt] (b3) at (6.9,2) {$3$};
  \node[circle,draw,fill=white,minimum size=6mm,inner sep=0pt] (b4) at (9.8,2) {$4$};
  \node[circle,draw,fill=white,minimum size=6mm,inner sep=0pt] (b5) at (12.7,2) {$5$};
  \draw[->,semithick] (b1) -- node[above] {$\CycleGaussianScalar_1$} (b2);
  \draw[->,semithick] (b2) -- node[above] {$\CycleGaussianScalar_2$} (b3);
  \draw[->,semithick] (b3) -- node[above] {$\CycleGaussianScalar_3$} (b4);
  \draw[->,semithick] (b4) -- node[above] {$\CycleGaussianScalar_4$} (b5);
  \node at (1.1,1.35) {$\CycleGaussianScalar_1-\BodyClosureGaussian$};
  \node at (4.0,1.35) {$\CycleGaussianScalar_2-\CycleGaussianScalar_1$};
  \node at (6.9,1.35) {$\CycleGaussianScalar_3-\CycleGaussianScalar_2$};
  \node at (9.8,1.35) {$\CycleGaussianScalar_4-\CycleGaussianScalar_3$};
  \node at (12.7,1.35) {$\BodyClosureGaussian-\CycleGaussianScalar_4$};
  \draw[->,semithick,dashed,rounded corners=4mm] (b5.east)
    -- (13.6,2) -- (13.6,0.12) -- (0.2,0.12) -- (0.2,2) -- (b1.west);
  \node[fill=white,inner sep=3pt] at (6.9,0.12)
    {public closure $\BodyClosureGaussian\sim\Gaussian{\ZZ}{\GaussianWidth}$};
\end{tikzpicture}

%% file: chapters/leg-injection.tex
\subsection{Leg Injection in Tree-Edge Coordinates}
\label{sec:leg-injection}

Use the leg vertex roles from \cref{sec:lobster-components}, suppressing
the leg index. The public mixer shapes the primary color while the
second hidden input transfers bounded load between the companion and
attachment. One public law on affine maps must cancel off-body carries
at every protected input without observing either hidden input.
For a protected input with nonzero first coordinate, it makes the
off-body pair uniform over the allowed interior pairs. When this
coordinate vanishes, it permits a prescribed point-mass component
while retaining the second input's
distribution and hence the attachment marginal.
Endpoint assembly uses the joint law of the off-body pair and
attachment increment.

\subsubsection{Color Pairs and Safe Affine Maps}
\label{sec:leg-safe-affine-maps}

Use the prime field and canonical lifts of
\cref{def:rounding-matrix,def:canonical-lift}.

\begingroup
\mdfsetup{topline=true}
\begin{definition}[Color slice]
  \label[definition]{def:leg-color-slice}
  For an integer offset $\LegSliceShift$ with
  $|\LegSliceShift|<\LARModulus$, define the \emph{color slice} by
  \[
    \LocalSlice_{\LegSliceShift}
    :=\set{\LocalFieldValue\in\set{0,\ldots,\LARModulus-1}:
      \LARModulus-1+\LegSliceShift-\LocalFieldValue
      \in\set{0,\ldots,\LARModulus-1}}.
  \]
\end{definition}
This is the set of allowed primary colors for integer pair sum
$\LARModulus-1+\LegSliceShift$. The companion of
$\LocalFieldValue\in\LocalSlice_{\LegSliceShift}$ is uniquely determined
as $\LARModulus-1+\LegSliceShift-\LocalFieldValue$.

\begin{definition}[Interior colors and trimmed slices]
  \label[definition]{def:leg-trimmed-slice}
  For an integer margin $\LegTrimMargin\geq0$ with
  $2\LegTrimMargin\leq\LARModulus-1$, define the \emph{interior colors}
  and \emph{boundary colors} by
  \[
    \LegInteriorColors:=\set{\LegTrimMargin,\ldots,
      \LARModulus-1-\LegTrimMargin},\qquad
    \LegBoundaryColors:=\set{0,\ldots,\LARModulus-1}
      \setminus\LegInteriorColors.
  \]
  For an integer offset $\LegSliceShift$ with
  $|\LegSliceShift|+2\LegTrimMargin<\LARModulus$, define the
  \emph{trimmed slice} by
  \[
    \LegTrimmedSlice{\LegSliceShift}
    :=\set{\LocalFieldValue\in\LegInteriorColors:
      \LARModulus-1+\LegSliceShift-\LocalFieldValue
      \in\LegInteriorColors}.
  \]
\end{definition}
The trimmed slice keeps both colors at distance at least
$\LegTrimMargin$ from $0$ and $\LARModulus-1$.

Both slices are nonempty under their stated parameter conditions.
Appendix~\ref{sec:leg-safe-slices-proof} gives their interval forms
and cardinalities.
When $\LegTrimMargin=0$, the boundary set is empty and
$\LegTrimmedSlice{\LegSliceShift}=\LocalSlice_{\LegSliceShift}$.

\begin{definition}[Uniform law on trimmed pairs]
  \label[definition]{def:leg-trimmed-pair-law}
  For $\LegSliceShift$ and $\LegTrimMargin$ as in
  \cref{def:leg-trimmed-slice}, define
  \[
    \LegTrimmedLaw{\LegSliceShift}
    :=\Unif{\set{(\LocalFieldValue,
        \LARModulus-1+\LegSliceShift-\LocalFieldValue)}_{
      \LocalFieldValue\in\LegTrimmedSlice{\LegSliceShift}}}.
  \]
\end{definition}
Choosing the primary uniformly from $\LegTrimmedSlice{\LegSliceShift}$
fixes its companion and gives each allowed pair probability
$1/|\LegTrimmedSlice{\LegSliceShift}|$.

\begin{definition}[Protected inputs]
  \label[definition]{def:leg-protected-inputs}
  For a public cutoff $\LegCutoff\in\ZZ_{>0}$, an integer input
  $(\LocalFirstError,\LocalSecondError)$ is \emph{protected} if
  $|\LocalFirstError|<\LegCutoff$ and $|\LocalSecondError|<\LegCutoff$.
  The protected set and its size are
  \[
    \LegProtectedInputs:=\set{-(\LegCutoff-1),\ldots,\LegCutoff-1}^2,
    \qquad \LegInputCount:=|\LegProtectedInputs|=(2\LegCutoff-1)^2.
  \]
  The \emph{protected line} consists of the inputs with
  $\LocalFirstError=0$, namely
  $\set{0}\times\set{-(\LegCutoff-1),\ldots,\LegCutoff-1}$.
  Let $\LegClippedLaw$ denote $\Gaussian{\ZZ}{\GaussianWidth}$
  conditioned on $|\LocalFirstError|<\LegCutoff$.
\end{definition}

In affine evaluations, integer inputs are reduced modulo
$\LARModulus$, while offsets and prescribed pair sums remain integers.
\begin{definition}[Affine evaluations and public leg maps]
  \label[definition]{def:leg-public-map}
  Let
  $\LocalChartPoint=(\LocalChartBase,\GadgetScale,\GadgetShear)
  \in\LARField^3$, with intercept $\LocalChartBase$, scale
  $\GadgetScale$, and shear $\GadgetShear$. At a protected input,
  define the \emph{affine evaluation} by
  \[
    \LegAffineEvaluation_{\LocalFirstError,\LocalSecondError}(\LocalChartPoint)
    :=\CanonicalLift{\LocalChartBase+\GadgetScale\LocalFirstError
      +\GadgetShear\LocalSecondError}\in\set{0,\ldots,\LARModulus-1}.
  \]
  The canonical integer coefficients are
  \[
    \LegBaselineFirst:=\CanonicalLift{\LocalChartBase},\qquad
    \LegScaleLift:=\CanonicalLift{\GadgetScale},\qquad
    \LegShearLift:=\CanonicalLift{\GadgetShear}.
  \]
  A \emph{public leg map} is such a triple with $\GadgetScale\ne0$.
\end{definition}
\endgroup
The evaluation is the primary output color. Since the scale is nonzero,
fixing the second field input makes the public map a bijection from the
first field input onto $\LARField$.

\begingroup
\mdfsetup{topline=true}
\begin{definition}[Safe affine map]
  \label[definition]{def:safe-affine-map}
  Fix a public integer offset $\LegBaselineShift$ satisfying
  $|\LegBaselineShift|+\LegCutoff-1<\LARModulus$.
  A public leg map $\LocalChartPoint$ is \emph{safe} for
  $\LegBaselineShift$ and $\LegCutoff$ if
  \begin{equation*}
    \LegAffineEvaluation_{\LocalFirstError,\LocalSecondError}
    (\LocalChartPoint)
    \in\LocalSlice_{\LegBaselineShift-\LocalSecondError}
    \quad\text{for every }(\LocalFirstError,\LocalSecondError)
    \in\LegProtectedInputs.
  \end{equation*}
\end{definition}
\endgroup

At second input $\LocalSecondError$, the required integer pair sum is
$\LARModulus-1+\LegBaselineShift-\LocalSecondError$.
The offset bound makes this slice nonempty at every protected
input. The distinction between safety and trimming is as follows.
\begin{itemize}
  \item Safety requires the same map to produce a valid primary and
    companion color at every protected input simultaneously. Boundary
    colors are allowed.
  \item To require both colors to lie in $\LegInteriorColors$, impose
    the stronger bound
    \[
      |\LegBaselineShift|+\LegCutoff-1+2\LegTrimMargin<\LARModulus
    \]
    and replace $\LocalSlice_{\LegBaselineShift-\LocalSecondError}$
    in the safety condition by
    $\LegTrimmedSlice{\LegBaselineShift-\LocalSecondError}$.
\end{itemize}
Appendix~\ref{sec:leg-safe-slices-proof} verifies these consequences of
the offset bounds and the bijectivity of public leg maps.

\subsubsection{Tree-Edge Injection and Carry Safety}
\label{sec:leg-edge-injection}
Use $\LegIncidenceMatrix$ from \cref{sec:affine-realization-matrix} in the
vertex order $(\LegLeftVertex,\LegMiddleVertex,\LegRightVertex)$.
Fix an integer shift $\LegBaselineShift$ and an integer bound
$\LegSumBound\in\ZZ_{\geq0}$ with $|\LegBaselineShift|\leq\LegSumBound$.
For a public leg map $\LocalChartPoint$ and the hidden input
$\LegGaussianInput=(\LocalFirstError,\LocalSecondError)^{\trans}$,
write $\LegCoordinateMatrix$ for its block in
\cref{sec:affine-realization-matrix}, suppressing the leg index, and set
$\LegEdgeCoordinates:=\LegCoordinateMatrix\LegGaussianInput$.
The first edge redistributes load within the off-body pair, so its
coefficients can mix colors without changing the pair sum. The second
edge transfers exactly $\LocalSecondError$ from the middle vertex to the
attachment; leaving this coordinate unmixed bounds the attachment
perturbation by $\LegCutoff$ on protected inputs, even when the first
mixed coordinate is large. The mixer is invertible modulo
$\LARModulus$ because
$\det\LegCoordinateMatrix=\LegScaleLift\not\equiv0\pmod{\LARModulus}$.
Written in these tree-edge coordinates, the local increment is
\begin{equation*}
  \LegLoadIncrement:=\LegIncidenceMatrix\LegEdgeCoordinates
  =\LegIncidenceMatrix\LegCoordinateMatrix\LegGaussianInput
  =
  \begin{pmatrix}
    \LegScaleLift\LocalFirstError+\LegShearLift\LocalSecondError\\
    -\LegScaleLift\LocalFirstError-(\LegShearLift+1)\LocalSecondError\\
    \LocalSecondError
  \end{pmatrix}.
\end{equation*}
The vertex-load matrix is therefore the derived product
$\LegInjectionMatrix:=\LegIncidenceMatrix\LegCoordinateMatrix$.
Restricting the shared baseline of \cref{def:shared-baseline-vector} and
adding this increment gives the complete local load:
\begin{equation*}
  \LegBaseLoad:=\bigl(\BaselineShift\bigr)_
    {(\LegLeftVertex,\LegMiddleVertex,\LegRightVertex)},\qquad
  \begin{pmatrix}
    \LocalFirstEndpoint\\\LocalSecondEndpoint\\\LegAttachmentLoad
  \end{pmatrix}
  =\LegBaseLoad+\LegLoadIncrement.
\end{equation*}
The intercept $\LegBaselineFirst=\CanonicalLift{\LocalChartBase}$
and mixer come from the same public map. The attachment center
$\LocalReservoirBase$ is its body entry in $\BaselineShift$.
Thus $\LegBaseLoad$ is a restriction of the shared baseline, and the
leg adds only $\LegLoadIncrement$, with
$\LegPaddingVector=\ZeroVec$ and $\LegAffineOffset=\ZeroVec$.

The off-body pair loses exactly the load transferred to the attachment,
so the complete triple has source-independent integer charge.
For a safe map, the prescribed pair sum fixes the companion residue
from the primary residue, forcing the two off-body carries to cancel.
Appendix~\ref{sec:leg-carry-proof} gives the charge identities and
the precise deterministic statement in \cref{lem:leg-deterministic}.

At the attachment, only the bounded second input is added to the
baseline center. If the body, leg, and queue cutoffs fit within the
shared baseline margin, the assembled attachment stays in
$[0,\LARModulus)$ and has zero carry.
\Cref{lem:leg-attachment-safety} in
Appendix~\ref{sec:leg-attachment-proof} states the exact margin and
contribution bounds and verifies this claim.

\subsubsection{Target Distribution and Parameters}
\label{sec:leg-sampling-parameters}
\paragraph{Choosing the prescribed primary residue.}
Choose the mixing weight and primary color so that
\[
  \LocalAtomWeight\in\QQ\cap[0,1/2],\qquad
  \LegMarkedFirst\in\set{0,\ldots,\LARModulus-1}.
\]
The required pair sum fixes the companion. Write
\[
  \LegMarkedSecond:=\LARModulus-1+\LegBaselineShift-\LegMarkedFirst,
  \qquad
  \LegMarkedSecond-\LocalSecondError\in\LegInteriorColors
  \quad\text{for every }|\LocalSecondError|<\LegCutoff.
\]
For each fixed $\LocalSecondError$, the prescribed pair is therefore
$(\LegMarkedFirst,\LegMarkedSecond-\LocalSecondError)$.
The last condition keeps this companion in the interior for every
protected second input. Thus only the prescribed primary color can
contribute boundary mass. These are the \emph{marked-color
admissibility conditions} used below.

\paragraph{Target pair laws.}
Fix any protected input $(\LocalFirstError,\LocalSecondError)$ and
consider the randomness of the public map. Its primary residue is
$\LocalChartBase+\GadgetScale\LocalFirstError+
\GadgetShear\LocalSecondError$ modulo $\LARModulus$; the fixed pair sum
then determines the companion residue. We want the following two cases.
\begin{description}
  \item[When $\LocalFirstError\ne0$.]
    The residue pair should be uniform over the trimmed pairs: those
    with the required sum and both coordinates in $\LegInteriorColors$.
  \item[When $\LocalFirstError=0$.]
    The residue pair should be a mixture of that uniform law and a
    point mass of weight $\LocalAtomWeight$ at one prescribed pair
    $(\LegMarkedFirst,\LegMarkedSecond-\LocalSecondError)$
    for each fixed $\LocalSecondError$. The primary is chosen in advance;
    its companion varies with $\LocalSecondError$ to preserve the pair
    sum. The baseline calibration controls this extra mass through the
    prescribed primary and mixing weight.
\end{description}
One public sampling law must satisfy both cases at every protected
input without observing the hidden source. The exact pair law is in
\cref{thm:leg-constructive-sampler}. Figure~\ref{fig:leg-goal}
illustrates the resulting profiles after averaging over the protected
Gaussian inputs. The marked contribution occurs only when the first
input vanishes. Its total weight after averaging is
\[
  \LegSignalWeight:=\LocalAtomWeight\LegClippedLaw(0).
\]

\begin{figure}[ht!]
  \centering
  \input{figures/leg-goal.tex}
  \caption{Normalized off-body pair profiles, with the attachment
    increment marginalized, for illustrative parameters
    $\LARModulus=31$, $\GaussianWidth=3$, $\LegCutoff=5$,
    $\LegTrimMargin=3$, and $\LegBaselineShift=-8$.
    These profiles illustrate the target laws; the parameters are not
    required to satisfy the sampler's sufficient feasibility conditions.
    Left: $\LocalAtomWeight=0$.
    Right: $\LocalAtomWeight=1/2$ and $\LegMarkedFirst=0$ give the
    boundary line $(0,22-\LocalSecondError)$ total mass
    $\LegSignalWeight$,
    with $|\LocalSecondError|<\LegCutoff$; the unmarked component has
    weight $1-\LegSignalWeight$.}
  \label{fig:leg-goal}
\end{figure}

\paragraph{Feasibility conditions.}
Use the protected-input count $\LegInputCount$ and trimming margin
$\LegTrimMargin$ from \cref{sec:leg-safe-affine-maps}. We impose three
sufficient conditions.
\begin{enumerate}[label=(\roman*)]
  \item \emph{Distinct protected inputs.}
    Require
    \[
      2(\LegCutoff-1)<\LARModulus.
    \]
    Then different protected inputs remain different after reduction
    modulo $\LARModulus$.
  \item \emph{Room for trimmed pairs.}
    Require
    \[
      \LegSumBound+\LegCutoff-1+2\LegTrimMargin<\LARModulus.
    \]
    Then every protected input has at least one pair with both
    residues in the interior.
  \item \emph{A small enough correction.}
    Filtering unsafe affine maps generally destroys the exact color
    marginals. The correction reweights the surviving safe affine maps;
    it changes neither their coefficients nor the hidden source. The
    following bounds ensure a strictly positive solution to the
    simultaneous marginal correction:
    \[
      \begin{aligned}
        \ChartDeletionBound
        &:=\frac{\LegInputCount(\LegSumBound+\LegCutoff-1+
                    2\LegTrimMargin)+1}{\LARModulus},\\
        \ChartOperatorBound
        &:=(2\LegInputCount+1)\ChartDeletionBound\leq\tfrac14,\\
        \ChartLocalPositivity
        &:=\frac{\LegInputCount}{1-\ChartDeletionBound}
          \left(\ChartDeletionBound+
          \frac{\ChartOperatorBound}{1-\ChartOperatorBound}\right)
          \leq\tfrac13.
      \end{aligned}
    \]
    Here $\ChartDeletionBound$ bounds the deleted fraction of maps,
    including after fixing an allowed evaluation color, and the lost
    fraction of marked scales. The quantity $\ChartOperatorBound$
    bounds the resulting perturbation of the marginal-response operator
    in the function supremum norm. Finally, $\ChartLocalPositivity$
    bounds the possible density decrease from cancelling the marked
    branch's off-line bias after inversion.
    Appendix~\ref{sec:leg-calibration-proof} proves these bounds.
\end{enumerate}
We refer to these three requirements as the \emph{feasibility conditions}.

\subsubsection{First Application: Safe Affine-Map Calibration}
\label{sec:leg-map-distribution}
Filtering uniform affine maps to safe supports biases their color
marginals; the marked branch adds an off-line bias. We compensate by
reweighting one public law on complete maps, enforcing every protected
input's target marginal without observing the hidden input.
The sampler first chooses a regular or marked branch.

The regular branch keeps only maps whose primary and companion
colors are interior at every protected input simultaneously.
The marked branch fixes the intercept to $\LegMarkedFirst$ and the
shear to zero, with nonzero scale. It therefore produces the
prescribed primary exactly on the protected line
$\LocalFirstError=0$; off that line, it is filtered to keep both
colors interior. Marked-color admissibility keeps the companion
interior on the line as well.

Choose the marked branch with probability $\LocalAtomWeight$.
Within that branch, sample uniformly from its safe support; on the
regular branch, reweight the surviving maps to remove both the
filtering bias and the marked branch's off-line bias. The same public
law applies simultaneously at all protected inputs and is sampled
independently of the hidden source. Write $\LegFlagChartLaw$ for
the complete flag and map law and $\ChartLaw$ for its map marginal.
Appendix~\ref{sec:leg-calibration-proof} specifies its finite model,
the supports $\ChartInteriorSupport,\ChartLineSupport$, and the
branch densities $\ChartRegularDensity,\ChartMarkedDensity$ relative
to the uniform law on coefficient triples.

\begingroup
\mdfsetup{topline=true,everyline=true}
\begin{theorem}[Constructive safe affine-map sampler]
  \label{thm:leg-constructive-sampler}
  Fix public core data satisfying the feasibility and marked-color
  admissibility conditions of \cref{sec:leg-sampling-parameters}.
  The source-independent law $\LegFlagChartLaw$ has the following guarantees.
  \begin{enumerate}[label=(\roman*)]
    \item For every protected input
      $(\LocalFirstError,\LocalSecondError)$, sampling from
      $\LegFlagChartLaw$ gives the residue-pair law
      \begin{equation}
        \label{eq:leg-exact-pair-law}
        \begin{aligned}
          &\DistributionLaw{\CanonicalLift{\LocalFirstEndpoint\bmod\LARModulus},
                            \CanonicalLift{\LocalSecondEndpoint\bmod\LARModulus}}\\
          &\quad=
          \begin{cases}
            (1-\LocalAtomWeight)
              \LegTrimmedLaw{\LegBaselineShift-\LocalSecondError}
              +\LocalAtomWeight\LocalPointMass{(\LegMarkedFirst,
                \LegMarkedSecond-\LocalSecondError)},
                &\LocalFirstError=0,\\
            \LegTrimmedLaw{\LegBaselineShift-\LocalSecondError},
                &\LocalFirstError\ne0.
          \end{cases}
        \end{aligned}
      \end{equation}
    \item Every supported flag and map pair has $\GadgetScale\ne0$
      and is safe at every protected input, with cancelling off-body
      carries. At each such input, an off-body residue lies in
      $\LegBoundaryColors$ if and only if $\LegMarkFlag=1$,
      $\LegMarkedFirst\in\LegBoundaryColors$, and
      $\LocalFirstError=0$. The assembled attachment has zero carry
      under the additional hypotheses of \cref{lem:leg-attachment-safety}.
    \item The supports $\ChartInteriorSupport$ and $\ChartLineSupport$ are
      nonempty, and the calibration matrix $\LegGramMatrix$ of
      \cref{sec:leg-branch-calibration} is nonsingular. The branch densities
      $\ChartRegularDensity$ and $\ChartMarkedDensity$ are rational,
      normalized probability densities relative to
      $\Unif{\ChartUnrestrictedSpace}$. The regular density is at least
      $1/3$ on $\ChartInteriorSupport$ and zero outside it; the marked
      density is uniform on $\ChartLineSupport$.
    \item The exact branch tables are computable in time polynomial
      in the numerical values of $\LARModulus$ and $\LegInputCount$
      and the bit lengths of the public rational inputs. For every
      $\LegChartError\in\QQ\cap(0,1)$, a worst-case bounded fair-bit
      sampler approximates $\LegFlagChartLaw$ within statistical distance
      $\LegChartError$, preserves each branch support, and runs in time
      polynomial in the same parameters and $\log(1/\LegChartError)$.
  \end{enumerate}
\end{theorem}
\endgroup
These are marginal laws for the same joint public distribution; outputs
at distinct protected inputs are not asserted to be independent.

\paragraph{Proof sketch.}
Pairwise independence of unrestricted affine evaluations gives an exact
reference response for additive color corrections. Deleting unsafe maps
perturbs this response by at most $\ChartOperatorBound\leq1/4$.
A finite linear solve restores the marginals and cancels the marked
branch's off-line bias. The regular target subtracts the marked
branch's color imbalances, so their positive part directly threatens
nonnegativity. Before excluding zero scale or unsafe colors, each
off-line color has exactly one scale preimage; deleting only a small
fraction of scales bounds this excess. The inverse-response bound
controls the further change caused by filtering. Together,
$\ChartLocalPositivity\leq1/3$ and $\ChartOperatorBound\leq1/4$
keep the regular density at least $1/3$ on its support.
Appendix~\ref{sec:leg-calibration-proof} gives the complete calibration
proof, and \cref{sec:discrete-leg-injection} proves rational table
computation and bounded sampling of the complete flag--map law.

\subsubsection{Bounded Sampling and Leg Injection}
\label{sec:leg-implementation}
At each fixed public core, \cref{alg:leg-tables,alg:leg-parameters} in
Appendix~\ref{sec:discrete-leg-injection} compute exact rational branch
tables and use bounded fair bits to sample within the chosen support.
Rounding and fallbacks preserve branch support, carry safety, and the
theorem's pointwise boundary-activation criterion. Their effect on
activation probabilities is covered by the complete flag--map sampling
error. For the concrete public family,
$\LARModulus\leq\InteriorVertexCount\leq\BlockDim$, so the modulus
contributes only polynomial cost in $\BlockDim$.

\begingroup
\mdfsetup{topline=true}
\begin{definition}[Conditional line-mark leg injection]
  \label[definition]{def:leg-injection}
  Under the conditions of \cref{sec:leg-sampling-parameters}, the
  leg's \emph{conditional public-data kernel} is $\LegFlagChartLaw$
  for $(\LegMarkFlag,\LocalChartPoint)$, given the core data of the
  shared baseline sampler in \cref{sec:baseline-shift}. The public
  draw is independent of
  $\LegGaussianInput\sim\Gaussian{\ZZ^2}{\GaussianWidth}$.
  Given these public data, the leg contributes the increment
  $\LegLoadIncrement$ of \cref{sec:leg-edge-injection}; its complete
  local load is $\LegBaseLoad+\LegLoadIncrement$.
\end{definition}
\endgroup

\subsubsection{Joint Law for Endpoint Assembly}
\label{sec:leg-joint-law}
For analysis, condition both independent Gaussian inputs of
\cref{def:leg-injection} to be protected. Their marginals become
$\LegClippedLaw$, while the independent public kernel remains unchanged.
Keep the two off-body residues together with the integer attachment
increment to retain the dependence needed for charge cancellation.
\Cref{sec:joint-injection-protection} bounds the full-source comparison
and the bounded sampler's error.

\begingroup
\mdfsetup{topline=true}
\begin{definition}[Protected reference experiment]
  \label[definition]{def:leg-reference-experiment}
  Fix public core data satisfying the feasibility and marked-color
  admissibility conditions of \cref{sec:leg-sampling-parameters}.
  Draw
  $\LocalFirstError,\LocalSecondError$ independently from
  $\LegClippedLaw$, independently of
  $(\LegMarkFlag,\LocalChartPoint)\sim\LegFlagChartLaw$. Define the activation event by
  \[
    \LegActivationEvent:=\set{\LegMarkFlag=1,\ \LocalFirstError=0}.
  \]
\end{definition}
Independence of the flag and source inputs gives
$\Pr[\LegActivationEvent]=\LegSignalWeight$, with the weight defined
in \cref{sec:leg-sampling-parameters}.

\begin{definition}[Joint reference law]
  \label[definition]{def:leg-joint-reference-law}
  For the protected reference experiment, define the diffuse and marked
  component laws and their mixture by
  \[
    \begin{aligned}
      \LegUnmarkedReferenceLaw
        &:=\sum_{|\LocalSecondError|<\LegCutoff}
          \LegClippedLaw(\LocalSecondError)
          \LegTrimmedLaw{\LegBaselineShift-\LocalSecondError}
          \otimes\LocalPointMass{\LocalSecondError},\\
      \LegMarkedReferenceLaw
        &:=\sum_{|\LocalSecondError|<\LegCutoff}
          \LegClippedLaw(\LocalSecondError)
          \LocalPointMass{(\LegMarkedFirst,
            \LegMarkedSecond-\LocalSecondError,\LocalSecondError)},\\
      \LegJointReferenceLaw
        &:=(1-\LegSignalWeight)\LegUnmarkedReferenceLaw
          +\LegSignalWeight\LegMarkedReferenceLaw.
    \end{aligned}
  \]
\end{definition}
\endgroup
Averaging \eqref{eq:leg-exact-pair-law} over the two independent
protected inputs gives
\[
  \DistributionLaw{\CanonicalLift{\LocalFirstEndpoint\bmod\LARModulus},
    \CanonicalLift{\LocalSecondEndpoint\bmod\LARModulus},
    \LocalSecondError}
  =\LegJointReferenceLaw.
\]
The third coordinate is the integer attachment increment, before adding
its baseline center. Marked-color admissibility ensures that the
prescribed companion in the marked component is already a canonical
residue for every protected second input.
The two components have the following meaning.
\begin{itemize}
  \item The marked component describes the activated contribution.
    The diffuse component describes the complement after averaging
    both flag branches. In particular, it includes marked-branch
    outcomes with $\LocalFirstError\ne0$.
  \item Each component has attachment-increment marginal
    $\LegClippedLaw$, so their mixture does as well. Activation depends
    only on $\LegMarkFlag$ and $\LocalFirstError$ and therefore leaves
    the second input unrestricted.
  \item The companion's $-\LocalSecondError$ and the attachment's
    $+\LocalSecondError$ always use the same input, including on
    activation. Keeping this input in the joint law preserves their
    cancellation.
\end{itemize}

%% file: figures/leg-goal.tex
\begingroup%
\begin{tikzpicture}[x=1cm,y=1cm,font=\footnotesize,line cap=round]
\path[use as bounding box] (0,0) rectangle (16.200000,8.200000);
\node[inner sep=0pt,font=\small,text=black!85] at (7.000000,7.950000) {Illustrative parameters: $\LARModulus=31,\quad\GaussianWidth=3,\quad\LegTrimMargin=3$};
\path[fill=white,draw=none] (1.150000,1.200000) rectangle (6.400000,6.450000);
\path[fill=black!0.16042578,draw=none]
  (1.658065,3.740323) rectangle (1.827419,3.909677)
  (1.827419,3.570968) rectangle (1.996774,3.740323)
  (1.996774,3.401613) rectangle (2.166129,3.570968)
  (2.166129,3.232258) rectangle (2.335484,3.401613)
  (2.335484,3.062903) rectangle (2.504839,3.232258)
  (2.504839,2.893548) rectangle (2.674194,3.062903)
  (2.674194,2.724194) rectangle (2.843548,2.893548)
  (2.843548,2.554839) rectangle (3.012903,2.724194)
  (3.012903,2.385484) rectangle (3.182258,2.554839)
  (3.182258,2.216129) rectangle (3.351613,2.385484)
  (3.351613,2.046774) rectangle (3.520968,2.216129)
  (3.520968,1.877419) rectangle (3.690323,2.046774)
  (3.690323,1.708065) rectangle (3.859677,1.877419);
\path[fill=black!1.71502751,draw=none]
  (1.658065,3.909677) rectangle (1.827419,4.079032)
  (1.827419,3.740323) rectangle (1.996774,3.909677)
  (1.996774,3.570968) rectangle (2.166129,3.740323)
  (2.166129,3.401613) rectangle (2.335484,3.570968)
  (2.335484,3.232258) rectangle (2.504839,3.401613)
  (2.504839,3.062903) rectangle (2.674194,3.232258)
  (2.674194,2.893548) rectangle (2.843548,3.062903)
  (2.843548,2.724194) rectangle (3.012903,2.893548)
  (3.012903,2.554839) rectangle (3.182258,2.724194)
  (3.182258,2.385484) rectangle (3.351613,2.554839)
  (3.351613,2.216129) rectangle (3.520968,2.385484)
  (3.520968,2.046774) rectangle (3.690323,2.216129)
  (3.690323,1.877419) rectangle (3.859677,2.046774)
  (3.859677,1.708065) rectangle (4.029032,1.877419);
\path[fill=black!9.16842490,draw=none]
  (1.658065,4.079032) rectangle (1.827419,4.248387)
  (1.827419,3.909677) rectangle (1.996774,4.079032)
  (1.996774,3.740323) rectangle (2.166129,3.909677)
  (2.166129,3.570968) rectangle (2.335484,3.740323)
  (2.335484,3.401613) rectangle (2.504839,3.570968)
  (2.504839,3.232258) rectangle (2.674194,3.401613)
  (2.674194,3.062903) rectangle (2.843548,3.232258)
  (2.843548,2.893548) rectangle (3.012903,3.062903)
  (3.012903,2.724194) rectangle (3.182258,2.893548)
  (3.182258,2.554839) rectangle (3.351613,2.724194)
  (3.351613,2.385484) rectangle (3.520968,2.554839)
  (3.520968,2.216129) rectangle (3.690323,2.385484)
  (3.690323,2.046774) rectangle (3.859677,2.216129)
  (3.859677,1.877419) rectangle (4.029032,2.046774)
  (4.029032,1.708065) rectangle (4.198387,1.877419);
\path[fill=black!24.49391047,draw=none]
  (1.658065,4.248387) rectangle (1.827419,4.417742)
  (1.827419,4.079032) rectangle (1.996774,4.248387)
  (1.996774,3.909677) rectangle (2.166129,4.079032)
  (2.166129,3.740323) rectangle (2.335484,3.909677)
  (2.335484,3.570968) rectangle (2.504839,3.740323)
  (2.504839,3.401613) rectangle (2.674194,3.570968)
  (2.674194,3.232258) rectangle (2.843548,3.401613)
  (2.843548,3.062903) rectangle (3.012903,3.232258)
  (3.012903,2.893548) rectangle (3.182258,3.062903)
  (3.182258,2.724194) rectangle (3.351613,2.893548)
  (3.351613,2.554839) rectangle (3.520968,2.724194)
  (3.520968,2.385484) rectangle (3.690323,2.554839)
  (3.690323,2.216129) rectangle (3.859677,2.385484)
  (3.859677,2.046774) rectangle (4.029032,2.216129)
  (4.029032,1.877419) rectangle (4.198387,2.046774)
  (4.198387,1.708065) rectangle (4.367742,1.877419);
\path[fill=black!32.68334964,draw=none]
  (1.658065,4.417742) rectangle (1.827419,4.587097)
  (1.827419,4.248387) rectangle (1.996774,4.417742)
  (1.996774,4.079032) rectangle (2.166129,4.248387)
  (2.166129,3.909677) rectangle (2.335484,4.079032)
  (2.335484,3.740323) rectangle (2.504839,3.909677)
  (2.504839,3.570968) rectangle (2.674194,3.740323)
  (2.674194,3.401613) rectangle (2.843548,3.570968)
  (2.843548,3.232258) rectangle (3.012903,3.401613)
  (3.012903,3.062903) rectangle (3.182258,3.232258)
  (3.182258,2.893548) rectangle (3.351613,3.062903)
  (3.351613,2.724194) rectangle (3.520968,2.893548)
  (3.520968,2.554839) rectangle (3.690323,2.724194)
  (3.690323,2.385484) rectangle (3.859677,2.554839)
  (3.859677,2.216129) rectangle (4.029032,2.385484)
  (4.029032,2.046774) rectangle (4.198387,2.216129)
  (4.198387,1.877419) rectangle (4.367742,2.046774)
  (4.367742,1.708065) rectangle (4.537097,1.877419);
\path[fill=black!21.77236486,draw=none]
  (1.658065,4.587097) rectangle (1.827419,4.756452)
  (1.827419,4.417742) rectangle (1.996774,4.587097)
  (1.996774,4.248387) rectangle (2.166129,4.417742)
  (2.166129,4.079032) rectangle (2.335484,4.248387)
  (2.335484,3.909677) rectangle (2.504839,4.079032)
  (2.504839,3.740323) rectangle (2.674194,3.909677)
  (2.674194,3.570968) rectangle (2.843548,3.740323)
  (2.843548,3.401613) rectangle (3.012903,3.570968)
  (3.012903,3.232258) rectangle (3.182258,3.401613)
  (3.182258,3.062903) rectangle (3.351613,3.232258)
  (3.351613,2.893548) rectangle (3.520968,3.062903)
  (3.520968,2.724194) rectangle (3.690323,2.893548)
  (3.690323,2.554839) rectangle (3.859677,2.724194)
  (3.859677,2.385484) rectangle (4.029032,2.554839)
  (4.029032,2.216129) rectangle (4.198387,2.385484)
  (4.198387,2.046774) rectangle (4.367742,2.216129)
  (4.367742,1.877419) rectangle (4.537097,2.046774)
  (4.537097,1.708065) rectangle (4.706452,1.877419);
\path[fill=black!7.23823018,draw=none]
  (1.658065,4.756452) rectangle (1.827419,4.925806)
  (1.827419,4.587097) rectangle (1.996774,4.756452)
  (1.996774,4.417742) rectangle (2.166129,4.587097)
  (2.166129,4.248387) rectangle (2.335484,4.417742)
  (2.335484,4.079032) rectangle (2.504839,4.248387)
  (2.504839,3.909677) rectangle (2.674194,4.079032)
  (2.674194,3.740323) rectangle (2.843548,3.909677)
  (2.843548,3.570968) rectangle (3.012903,3.740323)
  (3.012903,3.401613) rectangle (3.182258,3.570968)
  (3.182258,3.232258) rectangle (3.351613,3.401613)
  (3.351613,3.062903) rectangle (3.520968,3.232258)
  (3.520968,2.893548) rectangle (3.690323,3.062903)
  (3.690323,2.724194) rectangle (3.859677,2.893548)
  (3.859677,2.554839) rectangle (4.029032,2.724194)
  (4.029032,2.385484) rectangle (4.198387,2.554839)
  (4.198387,2.216129) rectangle (4.367742,2.385484)
  (4.367742,2.046774) rectangle (4.537097,2.216129)
  (4.537097,1.877419) rectangle (4.706452,2.046774)
  (4.706452,1.708065) rectangle (4.875806,1.877419);
\path[fill=black!1.20051926,draw=none]
  (1.658065,4.925806) rectangle (1.827419,5.095161)
  (1.827419,4.756452) rectangle (1.996774,4.925806)
  (1.996774,4.587097) rectangle (2.166129,4.756452)
  (2.166129,4.417742) rectangle (2.335484,4.587097)
  (2.335484,4.248387) rectangle (2.504839,4.417742)
  (2.504839,4.079032) rectangle (2.674194,4.248387)
  (2.674194,3.909677) rectangle (2.843548,4.079032)
  (2.843548,3.740323) rectangle (3.012903,3.909677)
  (3.012903,3.570968) rectangle (3.182258,3.740323)
  (3.182258,3.401613) rectangle (3.351613,3.570968)
  (3.351613,3.232258) rectangle (3.520968,3.401613)
  (3.520968,3.062903) rectangle (3.690323,3.232258)
  (3.690323,2.893548) rectangle (3.859677,3.062903)
  (3.859677,2.724194) rectangle (4.029032,2.893548)
  (4.029032,2.554839) rectangle (4.198387,2.724194)
  (4.198387,2.385484) rectangle (4.367742,2.554839)
  (4.367742,2.216129) rectangle (4.537097,2.385484)
  (4.537097,2.046774) rectangle (4.706452,2.216129)
  (4.706452,1.877419) rectangle (4.875806,2.046774)
  (4.875806,1.708065) rectangle (5.045161,1.877419);
\path[fill=black!0.09931120,draw=none]
  (1.658065,5.095161) rectangle (1.827419,5.264516)
  (1.827419,4.925806) rectangle (1.996774,5.095161)
  (1.996774,4.756452) rectangle (2.166129,4.925806)
  (2.166129,4.587097) rectangle (2.335484,4.756452)
  (2.335484,4.417742) rectangle (2.504839,4.587097)
  (2.504839,4.248387) rectangle (2.674194,4.417742)
  (2.674194,4.079032) rectangle (2.843548,4.248387)
  (2.843548,3.909677) rectangle (3.012903,4.079032)
  (3.012903,3.740323) rectangle (3.182258,3.909677)
  (3.182258,3.570968) rectangle (3.351613,3.740323)
  (3.351613,3.401613) rectangle (3.520968,3.570968)
  (3.520968,3.232258) rectangle (3.690323,3.401613)
  (3.690323,3.062903) rectangle (3.859677,3.232258)
  (3.859677,2.893548) rectangle (4.029032,3.062903)
  (4.029032,2.724194) rectangle (4.198387,2.893548)
  (4.198387,2.554839) rectangle (4.367742,2.724194)
  (4.367742,2.385484) rectangle (4.537097,2.554839)
  (4.537097,2.216129) rectangle (4.706452,2.385484)
  (4.706452,2.046774) rectangle (4.875806,2.216129)
  (4.875806,1.877419) rectangle (5.045161,2.046774)
  (5.045161,1.708065) rectangle (5.214516,1.877419);
\path[draw=black!40,line width=.4pt,fill=none] (1.150000,1.200000) rectangle (6.400000,6.450000);
\path[draw=black!55,line width=.35pt] (1.234677,1.200000) -- (1.234677,1.130000);
\node[inner sep=0pt,text=black!85] at (1.234677,0.990000) {$0$};
\path[draw=black!55,line width=.35pt] (2.081452,1.200000) -- (2.081452,1.130000);
\node[inner sep=0pt,text=black!85] at (2.081452,0.990000) {$5$};
\path[draw=black!55,line width=.35pt] (2.928226,1.200000) -- (2.928226,1.130000);
\node[inner sep=0pt,text=black!85] at (2.928226,0.990000) {$10$};
\path[draw=black!55,line width=.35pt] (3.775000,1.200000) -- (3.775000,1.130000);
\node[inner sep=0pt,text=black!85] at (3.775000,0.990000) {$15$};
\path[draw=black!55,line width=.35pt] (4.621774,1.200000) -- (4.621774,1.130000);
\node[inner sep=0pt,text=black!85] at (4.621774,0.990000) {$20$};
\path[draw=black!55,line width=.35pt] (5.468548,1.200000) -- (5.468548,1.130000);
\node[inner sep=0pt,text=black!85] at (5.468548,0.990000) {$25$};
\path[draw=black!55,line width=.35pt] (6.315323,1.200000) -- (6.315323,1.130000);
\node[inner sep=0pt,text=black!85] at (6.315323,0.990000) {$30$};
\path[draw=black!55,line width=.35pt] (1.150000,1.284677) -- (1.080000,1.284677);
\node[inner sep=0pt,anchor=east,text=black!85] at (1.000000,1.284677) {$0$};
\path[draw=black!55,line width=.35pt] (1.150000,2.131452) -- (1.080000,2.131452);
\node[inner sep=0pt,anchor=east,text=black!85] at (1.000000,2.131452) {$5$};
\path[draw=black!55,line width=.35pt] (1.150000,2.978226) -- (1.080000,2.978226);
\node[inner sep=0pt,anchor=east,text=black!85] at (1.000000,2.978226) {$10$};
\path[draw=black!55,line width=.35pt] (1.150000,3.825000) -- (1.080000,3.825000);
\node[inner sep=0pt,anchor=east,text=black!85] at (1.000000,3.825000) {$15$};
\path[draw=black!55,line width=.35pt] (1.150000,4.671774) -- (1.080000,4.671774);
\node[inner sep=0pt,anchor=east,text=black!85] at (1.000000,4.671774) {$20$};
\path[draw=black!55,line width=.35pt] (1.150000,5.518548) -- (1.080000,5.518548);
\node[inner sep=0pt,anchor=east,text=black!85] at (1.000000,5.518548) {$25$};
\path[draw=black!55,line width=.35pt] (1.150000,6.365323) -- (1.080000,6.365323);
\node[inner sep=0pt,anchor=east,text=black!85] at (1.000000,6.365323) {$30$};
\node[inner sep=0pt,text=black!85] at (3.775000,0.660000) {First off-body residue};
\node[inner sep=0pt,rotate=90,text=black!85] at (0.300000,3.825000) {Second off-body residue};
\node[inner sep=0pt,font=\small\bfseries] at (3.775000,7.400000) {Without boundary line};
\node[inner sep=0pt,text=black!75] at (3.775000,6.970000) {$\LocalAtomWeight=0$};
\node[inner sep=0pt,text=black!85] at (3.775000,0.160000) {Line mass: $0$};
\path[fill=white,draw=none] (4.400000,3.600000) rectangle (6.370000,6.430000);
\path[fill=white,draw=none] (4.860000,3.940000) rectangle (6.110000,6.190000);
\path[fill=black!24.49391047,draw=none]
  (5.610000,3.940000) rectangle (5.860000,4.190000);
\path[fill=black!32.68334964,draw=none]
  (5.610000,4.190000) rectangle (5.860000,4.440000)
  (5.860000,3.940000) rectangle (6.110000,4.190000);
\path[fill=black!21.77236486,draw=none]
  (5.610000,4.440000) rectangle (5.860000,4.690000)
  (5.860000,4.190000) rectangle (6.110000,4.440000);
\path[fill=black!7.23823018,draw=none]
  (5.610000,4.690000) rectangle (5.860000,4.940000)
  (5.860000,4.440000) rectangle (6.110000,4.690000);
\path[fill=black!1.20051926,draw=none]
  (5.610000,4.940000) rectangle (5.860000,5.190000)
  (5.860000,4.690000) rectangle (6.110000,4.940000);
\path[fill=black!0.09931120,draw=none]
  (5.610000,5.190000) rectangle (5.860000,5.440000)
  (5.860000,4.940000) rectangle (6.110000,5.190000);
\path[draw=black!45,line width=.35pt,fill=none] (4.860000,3.940000) rectangle (6.110000,6.190000);
\path[draw=black!10,line width=.15pt] (5.110000,3.940000) -- (5.110000,6.190000);
\path[draw=black!10,line width=.15pt] (5.360000,3.940000) -- (5.360000,6.190000);
\path[draw=black!10,line width=.15pt] (5.610000,3.940000) -- (5.610000,6.190000);
\path[draw=black!10,line width=.15pt] (5.860000,3.940000) -- (5.860000,6.190000);
\path[draw=black!10,line width=.15pt] (4.860000,4.190000) -- (6.110000,4.190000);
\path[draw=black!10,line width=.15pt] (4.860000,4.440000) -- (6.110000,4.440000);
\path[draw=black!10,line width=.15pt] (4.860000,4.690000) -- (6.110000,4.690000);
\path[draw=black!10,line width=.15pt] (4.860000,4.940000) -- (6.110000,4.940000);
\path[draw=black!10,line width=.15pt] (4.860000,5.190000) -- (6.110000,5.190000);
\path[draw=black!10,line width=.15pt] (4.860000,5.440000) -- (6.110000,5.440000);
\path[draw=black!10,line width=.15pt] (4.860000,5.690000) -- (6.110000,5.690000);
\path[draw=black!10,line width=.15pt] (4.860000,5.940000) -- (6.110000,5.940000);
\path[draw=black!55,line width=.35pt] (4.985000,3.940000) -- (4.985000,3.890000);
\node[inner sep=0pt,font=\scriptsize] at (4.985000,3.750000) {$0$};
\path[draw=black!55,line width=.35pt] (5.985000,3.940000) -- (5.985000,3.890000);
\node[inner sep=0pt,font=\scriptsize] at (5.985000,3.750000) {$4$};
\path[draw=black!55,line width=.35pt] (4.860000,4.065000) -- (4.810000,4.065000);
\node[inner sep=0pt,font=\scriptsize,anchor=east] at (4.740000,4.065000) {$18$};
\path[draw=black!55,line width=.35pt] (4.860000,6.065000) -- (4.810000,6.065000);
\node[inner sep=0pt,font=\scriptsize,anchor=east] at (4.740000,6.065000) {$26$};
\node[inner sep=0pt,font=\scriptsize,text=black!75] at (5.485000,6.320000) {Boundary zoom};
\path[draw=black!55,line width=.4pt,densely dashed,fill=none] (1.150000,4.248387) rectangle (1.996774,5.772581);
\path[fill=white,draw=none] (7.550000,1.200000) rectangle (12.800000,6.450000);
\path[fill=black!0.34762760,draw=none]
  (7.550000,4.248387) rectangle (7.719355,4.417742)
  (7.550000,5.603226) rectangle (7.719355,5.772581);
\path[fill=black!4.00217305,draw=none]
  (7.550000,4.417742) rectangle (7.719355,4.587097)
  (7.550000,5.433871) rectangle (7.719355,5.603226);
\path[fill=black!22.92359500,draw=none]
  (7.550000,4.587097) rectangle (7.719355,4.756452)
  (7.550000,5.264516) rectangle (7.719355,5.433871);
\path[fill=black!65.32431207,draw=none]
  (7.550000,4.756452) rectangle (7.719355,4.925806)
  (7.550000,5.095161) rectangle (7.719355,5.264516);
\path[fill=black!92.61305652,draw=none]
  (7.550000,4.925806) rectangle (7.719355,5.095161);
\path[fill=black!0.13368520,draw=none]
  (8.058065,3.740323) rectangle (8.227419,3.909677)
  (8.227419,3.570968) rectangle (8.396774,3.740323)
  (8.396774,3.401613) rectangle (8.566129,3.570968)
  (8.566129,3.232258) rectangle (8.735484,3.401613)
  (8.735484,3.062903) rectangle (8.904839,3.232258)
  (8.904839,2.893548) rectangle (9.074194,3.062903)
  (9.074194,2.724194) rectangle (9.243548,2.893548)
  (9.243548,2.554839) rectangle (9.412903,2.724194)
  (9.412903,2.385484) rectangle (9.582258,2.554839)
  (9.582258,2.216129) rectangle (9.751613,2.385484)
  (9.751613,2.046774) rectangle (9.920968,2.216129)
  (9.920968,1.877419) rectangle (10.090323,2.046774)
  (10.090323,1.708065) rectangle (10.259677,1.877419);
\path[fill=black!1.42915801,draw=none]
  (8.058065,3.909677) rectangle (8.227419,4.079032)
  (8.227419,3.740323) rectangle (8.396774,3.909677)
  (8.396774,3.570968) rectangle (8.566129,3.740323)
  (8.566129,3.401613) rectangle (8.735484,3.570968)
  (8.735484,3.232258) rectangle (8.904839,3.401613)
  (8.904839,3.062903) rectangle (9.074194,3.232258)
  (9.074194,2.893548) rectangle (9.243548,3.062903)
  (9.243548,2.724194) rectangle (9.412903,2.893548)
  (9.412903,2.554839) rectangle (9.582258,2.724194)
  (9.582258,2.385484) rectangle (9.751613,2.554839)
  (9.751613,2.216129) rectangle (9.920968,2.385484)
  (9.920968,2.046774) rectangle (10.090323,2.216129)
  (10.090323,1.877419) rectangle (10.259677,2.046774)
  (10.259677,1.708065) rectangle (10.429032,1.877419);
\path[fill=black!7.64018523,draw=none]
  (8.058065,4.079032) rectangle (8.227419,4.248387)
  (8.227419,3.909677) rectangle (8.396774,4.079032)
  (8.396774,3.740323) rectangle (8.566129,3.909677)
  (8.566129,3.570968) rectangle (8.735484,3.740323)
  (8.735484,3.401613) rectangle (8.904839,3.570968)
  (8.904839,3.232258) rectangle (9.074194,3.401613)
  (9.074194,3.062903) rectangle (9.243548,3.232258)
  (9.243548,2.893548) rectangle (9.412903,3.062903)
  (9.412903,2.724194) rectangle (9.582258,2.893548)
  (9.582258,2.554839) rectangle (9.751613,2.724194)
  (9.751613,2.385484) rectangle (9.920968,2.554839)
  (9.920968,2.216129) rectangle (10.090323,2.385484)
  (10.090323,2.046774) rectangle (10.259677,2.216129)
  (10.259677,1.877419) rectangle (10.429032,2.046774)
  (10.429032,1.708065) rectangle (10.598387,1.877419);
\path[fill=black!20.41114097,draw=none]
  (8.058065,4.248387) rectangle (8.227419,4.417742)
  (8.227419,4.079032) rectangle (8.396774,4.248387)
  (8.396774,3.909677) rectangle (8.566129,4.079032)
  (8.566129,3.740323) rectangle (8.735484,3.909677)
  (8.735484,3.570968) rectangle (8.904839,3.740323)
  (8.904839,3.401613) rectangle (9.074194,3.570968)
  (9.074194,3.232258) rectangle (9.243548,3.401613)
  (9.243548,3.062903) rectangle (9.412903,3.232258)
  (9.412903,2.893548) rectangle (9.582258,3.062903)
  (9.582258,2.724194) rectangle (9.751613,2.893548)
  (9.751613,2.554839) rectangle (9.920968,2.724194)
  (9.920968,2.385484) rectangle (10.090323,2.554839)
  (10.090323,2.216129) rectangle (10.259677,2.385484)
  (10.259677,2.046774) rectangle (10.429032,2.216129)
  (10.429032,1.877419) rectangle (10.598387,2.046774)
  (10.598387,1.708065) rectangle (10.767742,1.877419);
\path[fill=black!27.23552279,draw=none]
  (8.058065,4.417742) rectangle (8.227419,4.587097)
  (8.227419,4.248387) rectangle (8.396774,4.417742)
  (8.396774,4.079032) rectangle (8.566129,4.248387)
  (8.566129,3.909677) rectangle (8.735484,4.079032)
  (8.735484,3.740323) rectangle (8.904839,3.909677)
  (8.904839,3.570968) rectangle (9.074194,3.740323)
  (9.074194,3.401613) rectangle (9.243548,3.570968)
  (9.243548,3.232258) rectangle (9.412903,3.401613)
  (9.412903,3.062903) rectangle (9.582258,3.232258)
  (9.582258,2.893548) rectangle (9.751613,3.062903)
  (9.751613,2.724194) rectangle (9.920968,2.893548)
  (9.920968,2.554839) rectangle (10.090323,2.724194)
  (10.090323,2.385484) rectangle (10.259677,2.554839)
  (10.259677,2.216129) rectangle (10.429032,2.385484)
  (10.429032,2.046774) rectangle (10.598387,2.216129)
  (10.598387,1.877419) rectangle (10.767742,2.046774)
  (10.767742,1.708065) rectangle (10.937097,1.877419);
\path[fill=black!18.14323641,draw=none]
  (8.058065,4.587097) rectangle (8.227419,4.756452)
  (8.227419,4.417742) rectangle (8.396774,4.587097)
  (8.396774,4.248387) rectangle (8.566129,4.417742)
  (8.566129,4.079032) rectangle (8.735484,4.248387)
  (8.735484,3.909677) rectangle (8.904839,4.079032)
  (8.904839,3.740323) rectangle (9.074194,3.909677)
  (9.074194,3.570968) rectangle (9.243548,3.740323)
  (9.243548,3.401613) rectangle (9.412903,3.570968)
  (9.412903,3.232258) rectangle (9.582258,3.401613)
  (9.582258,3.062903) rectangle (9.751613,3.232258)
  (9.751613,2.893548) rectangle (9.920968,3.062903)
  (9.920968,2.724194) rectangle (10.090323,2.893548)
  (10.090323,2.554839) rectangle (10.259677,2.724194)
  (10.259677,2.385484) rectangle (10.429032,2.554839)
  (10.429032,2.216129) rectangle (10.598387,2.385484)
  (10.598387,2.046774) rectangle (10.767742,2.216129)
  (10.767742,1.877419) rectangle (10.937097,2.046774)
  (10.937097,1.708065) rectangle (11.106452,1.877419);
\path[fill=black!6.03172518,draw=none]
  (8.058065,4.756452) rectangle (8.227419,4.925806)
  (8.227419,4.587097) rectangle (8.396774,4.756452)
  (8.396774,4.417742) rectangle (8.566129,4.587097)
  (8.566129,4.248387) rectangle (8.735484,4.417742)
  (8.735484,4.079032) rectangle (8.904839,4.248387)
  (8.904839,3.909677) rectangle (9.074194,4.079032)
  (9.074194,3.740323) rectangle (9.243548,3.909677)
  (9.243548,3.570968) rectangle (9.412903,3.740323)
  (9.412903,3.401613) rectangle (9.582258,3.570968)
  (9.582258,3.232258) rectangle (9.751613,3.401613)
  (9.751613,3.062903) rectangle (9.920968,3.232258)
  (9.920968,2.893548) rectangle (10.090323,3.062903)
  (10.090323,2.724194) rectangle (10.259677,2.893548)
  (10.259677,2.554839) rectangle (10.429032,2.724194)
  (10.429032,2.385484) rectangle (10.598387,2.554839)
  (10.598387,2.216129) rectangle (10.767742,2.385484)
  (10.767742,2.046774) rectangle (10.937097,2.216129)
  (10.937097,1.877419) rectangle (11.106452,2.046774)
  (11.106452,1.708065) rectangle (11.275806,1.877419);
\path[fill=black!1.00041061,draw=none]
  (8.058065,4.925806) rectangle (8.227419,5.095161)
  (8.227419,4.756452) rectangle (8.396774,4.925806)
  (8.396774,4.587097) rectangle (8.566129,4.756452)
  (8.566129,4.417742) rectangle (8.735484,4.587097)
  (8.735484,4.248387) rectangle (8.904839,4.417742)
  (8.904839,4.079032) rectangle (9.074194,4.248387)
  (9.074194,3.909677) rectangle (9.243548,4.079032)
  (9.243548,3.740323) rectangle (9.412903,3.909677)
  (9.412903,3.570968) rectangle (9.582258,3.740323)
  (9.582258,3.401613) rectangle (9.751613,3.570968)
  (9.751613,3.232258) rectangle (9.920968,3.401613)
  (9.920968,3.062903) rectangle (10.090323,3.232258)
  (10.090323,2.893548) rectangle (10.259677,3.062903)
  (10.259677,2.724194) rectangle (10.429032,2.893548)
  (10.429032,2.554839) rectangle (10.598387,2.724194)
  (10.598387,2.385484) rectangle (10.767742,2.554839)
  (10.767742,2.216129) rectangle (10.937097,2.385484)
  (10.937097,2.046774) rectangle (11.106452,2.216129)
  (11.106452,1.877419) rectangle (11.275806,2.046774)
  (11.275806,1.708065) rectangle (11.445161,1.877419);
\path[fill=black!0.08275750,draw=none]
  (8.058065,5.095161) rectangle (8.227419,5.264516)
  (8.227419,4.925806) rectangle (8.396774,5.095161)
  (8.396774,4.756452) rectangle (8.566129,4.925806)
  (8.566129,4.587097) rectangle (8.735484,4.756452)
  (8.735484,4.417742) rectangle (8.904839,4.587097)
  (8.904839,4.248387) rectangle (9.074194,4.417742)
  (9.074194,4.079032) rectangle (9.243548,4.248387)
  (9.243548,3.909677) rectangle (9.412903,4.079032)
  (9.412903,3.740323) rectangle (9.582258,3.909677)
  (9.582258,3.570968) rectangle (9.751613,3.740323)
  (9.751613,3.401613) rectangle (9.920968,3.570968)
  (9.920968,3.232258) rectangle (10.090323,3.401613)
  (10.090323,3.062903) rectangle (10.259677,3.232258)
  (10.259677,2.893548) rectangle (10.429032,3.062903)
  (10.429032,2.724194) rectangle (10.598387,2.893548)
  (10.598387,2.554839) rectangle (10.767742,2.724194)
  (10.767742,2.385484) rectangle (10.937097,2.554839)
  (10.937097,2.216129) rectangle (11.106452,2.385484)
  (11.106452,2.046774) rectangle (11.275806,2.216129)
  (11.275806,1.877419) rectangle (11.445161,2.046774)
  (11.445161,1.708065) rectangle (11.614516,1.877419);
\path[draw=black!40,line width=.4pt,fill=none] (7.550000,1.200000) rectangle (12.800000,6.450000);
\path[draw=black!55,line width=.35pt] (7.634677,1.200000) -- (7.634677,1.130000);
\node[inner sep=0pt,text=black!85] at (7.634677,0.990000) {$0$};
\path[draw=black!55,line width=.35pt] (8.481452,1.200000) -- (8.481452,1.130000);
\node[inner sep=0pt,text=black!85] at (8.481452,0.990000) {$5$};
\path[draw=black!55,line width=.35pt] (9.328226,1.200000) -- (9.328226,1.130000);
\node[inner sep=0pt,text=black!85] at (9.328226,0.990000) {$10$};
\path[draw=black!55,line width=.35pt] (10.175000,1.200000) -- (10.175000,1.130000);
\node[inner sep=0pt,text=black!85] at (10.175000,0.990000) {$15$};
\path[draw=black!55,line width=.35pt] (11.021774,1.200000) -- (11.021774,1.130000);
\node[inner sep=0pt,text=black!85] at (11.021774,0.990000) {$20$};
\path[draw=black!55,line width=.35pt] (11.868548,1.200000) -- (11.868548,1.130000);
\node[inner sep=0pt,text=black!85] at (11.868548,0.990000) {$25$};
\path[draw=black!55,line width=.35pt] (12.715323,1.200000) -- (12.715323,1.130000);
\node[inner sep=0pt,text=black!85] at (12.715323,0.990000) {$30$};
\path[draw=black!55,line width=.35pt] (7.550000,1.284677) -- (7.480000,1.284677);
\node[inner sep=0pt,anchor=east,text=black!85] at (7.400000,1.284677) {$0$};
\path[draw=black!55,line width=.35pt] (7.550000,2.131452) -- (7.480000,2.131452);
\node[inner sep=0pt,anchor=east,text=black!85] at (7.400000,2.131452) {$5$};
\path[draw=black!55,line width=.35pt] (7.550000,2.978226) -- (7.480000,2.978226);
\node[inner sep=0pt,anchor=east,text=black!85] at (7.400000,2.978226) {$10$};
\path[draw=black!55,line width=.35pt] (7.550000,3.825000) -- (7.480000,3.825000);
\node[inner sep=0pt,anchor=east,text=black!85] at (7.400000,3.825000) {$15$};
\path[draw=black!55,line width=.35pt] (7.550000,4.671774) -- (7.480000,4.671774);
\node[inner sep=0pt,anchor=east,text=black!85] at (7.400000,4.671774) {$20$};
\path[draw=black!55,line width=.35pt] (7.550000,5.518548) -- (7.480000,5.518548);
\node[inner sep=0pt,anchor=east,text=black!85] at (7.400000,5.518548) {$25$};
\path[draw=black!55,line width=.35pt] (7.550000,6.365323) -- (7.480000,6.365323);
\node[inner sep=0pt,anchor=east,text=black!85] at (7.400000,6.365323) {$30$};
\node[inner sep=0pt,text=black!85] at (10.175000,0.660000) {First off-body residue};
\node[inner sep=0pt,font=\small\bfseries] at (10.175000,7.400000) {With boundary line};
\node[inner sep=0pt,text=black!75] at (10.175000,6.970000) {$\LocalAtomWeight=1/2$};
\node[inner sep=0pt,text=black!85] at (10.175000,0.160000) {Line mass: $\LegSignalWeight\approx 0.1667$};
\path[fill=white,draw=none] (10.800000,3.600000) rectangle (12.770000,6.430000);
\path[fill=white,draw=none] (11.260000,3.940000) rectangle (12.510000,6.190000);
\path[fill=black!0.34762760,draw=none]
  (11.260000,3.940000) rectangle (11.510000,4.190000)
  (11.260000,5.940000) rectangle (11.510000,6.190000);
\path[fill=black!4.00217305,draw=none]
  (11.260000,4.190000) rectangle (11.510000,4.440000)
  (11.260000,5.690000) rectangle (11.510000,5.940000);
\path[fill=black!22.92359500,draw=none]
  (11.260000,4.440000) rectangle (11.510000,4.690000)
  (11.260000,5.440000) rectangle (11.510000,5.690000);
\path[fill=black!65.32431207,draw=none]
  (11.260000,4.690000) rectangle (11.510000,4.940000)
  (11.260000,5.190000) rectangle (11.510000,5.440000);
\path[fill=black!92.61305652,draw=none]
  (11.260000,4.940000) rectangle (11.510000,5.190000);
\path[fill=black!20.41114097,draw=none]
  (12.010000,3.940000) rectangle (12.260000,4.190000);
\path[fill=black!27.23552279,draw=none]
  (12.010000,4.190000) rectangle (12.260000,4.440000)
  (12.260000,3.940000) rectangle (12.510000,4.190000);
\path[fill=black!18.14323641,draw=none]
  (12.010000,4.440000) rectangle (12.260000,4.690000)
  (12.260000,4.190000) rectangle (12.510000,4.440000);
\path[fill=black!6.03172518,draw=none]
  (12.010000,4.690000) rectangle (12.260000,4.940000)
  (12.260000,4.440000) rectangle (12.510000,4.690000);
\path[fill=black!1.00041061,draw=none]
  (12.010000,4.940000) rectangle (12.260000,5.190000)
  (12.260000,4.690000) rectangle (12.510000,4.940000);
\path[fill=black!0.08275750,draw=none]
  (12.010000,5.190000) rectangle (12.260000,5.440000)
  (12.260000,4.940000) rectangle (12.510000,5.190000);
\path[draw=black!45,line width=.35pt,fill=none] (11.260000,3.940000) rectangle (12.510000,6.190000);
\path[draw=black!10,line width=.15pt] (11.510000,3.940000) -- (11.510000,6.190000);
\path[draw=black!10,line width=.15pt] (11.760000,3.940000) -- (11.760000,6.190000);
\path[draw=black!10,line width=.15pt] (12.010000,3.940000) -- (12.010000,6.190000);
\path[draw=black!10,line width=.15pt] (12.260000,3.940000) -- (12.260000,6.190000);
\path[draw=black!10,line width=.15pt] (11.260000,4.190000) -- (12.510000,4.190000);
\path[draw=black!10,line width=.15pt] (11.260000,4.440000) -- (12.510000,4.440000);
\path[draw=black!10,line width=.15pt] (11.260000,4.690000) -- (12.510000,4.690000);
\path[draw=black!10,line width=.15pt] (11.260000,4.940000) -- (12.510000,4.940000);
\path[draw=black!10,line width=.15pt] (11.260000,5.190000) -- (12.510000,5.190000);
\path[draw=black!10,line width=.15pt] (11.260000,5.440000) -- (12.510000,5.440000);
\path[draw=black!10,line width=.15pt] (11.260000,5.690000) -- (12.510000,5.690000);
\path[draw=black!10,line width=.15pt] (11.260000,5.940000) -- (12.510000,5.940000);
\path[draw=black!55,line width=.35pt] (11.385000,3.940000) -- (11.385000,3.890000);
\node[inner sep=0pt,font=\scriptsize] at (11.385000,3.750000) {$0$};
\path[draw=black!55,line width=.35pt] (12.385000,3.940000) -- (12.385000,3.890000);
\node[inner sep=0pt,font=\scriptsize] at (12.385000,3.750000) {$4$};
\path[draw=black!55,line width=.35pt] (11.260000,4.065000) -- (11.210000,4.065000);
\node[inner sep=0pt,font=\scriptsize,anchor=east] at (11.140000,4.065000) {$18$};
\path[draw=black!55,line width=.35pt] (11.260000,6.065000) -- (11.210000,6.065000);
\node[inner sep=0pt,font=\scriptsize,anchor=east] at (11.140000,6.065000) {$26$};
\node[inner sep=0pt,font=\scriptsize,text=black!75] at (11.885000,6.320000) {Boundary zoom};
\path[draw=black!55,line width=.4pt,densely dashed,fill=none] (7.550000,4.248387) rectangle (8.396774,5.772581);
\shade[bottom color=white,top color=black] (14.200000,1.200000) rectangle (14.500000,6.450000);
\path[draw=black!40,line width=.35pt,fill=none] (14.200000,1.200000) rectangle (14.500000,6.450000);
\path[draw=black!55,line width=.35pt] (14.500000,1.200000) -- (14.570000,1.200000);
\node[inner sep=0pt,anchor=west,text=black!85] at (14.650000,1.200000) {$0.000$};
\path[draw=black!55,line width=.35pt] (14.500000,2.250000) -- (14.570000,2.250000);
\node[inner sep=0pt,anchor=west,text=black!85] at (14.650000,2.250000) {$0.012$};
\path[draw=black!55,line width=.35pt] (14.500000,3.300000) -- (14.570000,3.300000);
\node[inner sep=0pt,anchor=west,text=black!85] at (14.650000,3.300000) {$0.024$};
\path[draw=black!55,line width=.35pt] (14.500000,4.350000) -- (14.570000,4.350000);
\node[inner sep=0pt,anchor=west,text=black!85] at (14.650000,4.350000) {$0.036$};
\path[draw=black!55,line width=.35pt] (14.500000,5.400000) -- (14.570000,5.400000);
\node[inner sep=0pt,anchor=west,text=black!85] at (14.650000,5.400000) {$0.048$};
\path[draw=black!55,line width=.35pt] (14.500000,6.450000) -- (14.570000,6.450000);
\node[inner sep=0pt,anchor=west,text=black!85] at (14.650000,6.450000) {$0.060$};
\node[inner sep=0pt,align=center,text=black!85] at (14.350000,6.970000) {Joint probability\\per residue pair};
\node[inner sep=0pt,text=black!85] at (14.350000,0.730000) {Shared scale};
\end{tikzpicture}
\endgroup%

%% file: chapters/queue-injection.tex
\subsection{Queue Injection}
\label{sec:queue-injection}

The column $\ChargeColumn$ from \cref{sec:affine-realization-matrix} adds
one scalar to every queue vertex. The queue set is marginally uniform
and is independent of the source and body permutation; its dependence
on the attachments is fixed by \cref{sec:lobster-combination}.
For a public cutoff
$\PaddingChargeCutoff\in\ZZ_{>0}$, the analytical reference draws
$\PaddingChargeError\sample\QueueClippedLaw
:=\Gaussian{\ZZ}{\GaussianWidth}
\mid\set{|\PaddingChargeError|<\PaddingChargeCutoff}$ and sets
$\DiscreteReferenceQueueIncrement:=\PaddingChargeError\ChargeColumn$.
All queue vertices share this bounded increment, whose total charge is
$\LARModulus\PaddingChargeError\equiv0\pmod{\LARModulus}$.
The actual injection $\PlacedQueueIncrement$ of
\cref{sec:lobster-injection-targets} uses the untruncated source without
padding. Its truncation error is exactly
the single-scalar tail probability in
\cref{sec:joint-injection-protection}, independently of the queue size.

Conditioning on a placement-dependent core can bias the queue's
posterior law, so the joint averaging in \cref{sec:discrete-realization}
retains that dependence.

%% file: chapters/baseline-shift.tex
\subsection{Shared Baseline Sampler: Admissible Joint Distributions}
\label{sec:baseline-shift}

The shared sampler first chooses body centers, leg shifts, and marks.
It then draws a complete flag--map pair for each leg and assembles one
baseline vector.
Assume the geometry and safety margins of
\cref{sec:lobster-injection-targets,sec:lobster-combination} and the leg
feasibility conditions of \cref{sec:leg-sampling-parameters}.

\subsubsection{Admissible Core Laws}
\label{sec:baseline-core-laws}
Use the safe-center interval $\HistogramSafeCenters$ from
\cref{sec:lobster-injection-targets}.

\begingroup
\mdfsetup{topline=true}
\begin{definition}[Admissible joint baseline law]
  \label[definition]{def:admissible-joint-baseline-law}
  A public law $\BaselineCoreLaw$ on body centers $\BaselineCenters$,
  leg shifts $\LegBaselineShift$, and marks $\BaselineMarks$ is
  \emph{admissible} if it uses only permitted public parameters, is
  independent of the hidden Gaussian source and secret-dependent
  information, and has support satisfying
  \begin{equation*}
    \begin{aligned}
      \BaselineCenters&=(\BaselineCenters_{\VertexIndex})_{
      \VertexIndex\in\InteriorVertices}
      \in\HistogramSafeCenters^{\InteriorVertexCount},\\
      \LegBaselineShift&=(\LegBaselineShift_1,\ldots,
      \LegBaselineShift_{\GadgetCount})
      \in([-\LegSumBound,\LegSumBound]\cap\ZZ)^{\GadgetCount},\\
      \sum_{\VertexIndex\in\InteriorVertices}\BaselineCenters_{\VertexIndex}
      +\sum_{\GadgetIndex=1}^{\GadgetCount}\LegBaselineShift_{\GadgetIndex}
      &\equiv\GadgetCount\pmod{\LARModulus}.
    \end{aligned}
  \end{equation*}
  For each leg, $\BaselineMarks$ records a target and mixing weight
  satisfying the marked-color admissibility conditions of
  \cref{sec:leg-sampling-parameters}.
  The common public truncation margin and all leg parameters satisfy
  the feasibility conditions of \cref{sec:leg-sampling-parameters}.
\end{definition}
\endgroup
The primary target may be a boundary color. Only its moving companion
is required to stay in the interior at every protected second input.

\subsubsection{Dependence and Sampling Requirements}
\label{sec:baseline-dependence}
Admissibility permits the following dependencies.
\begin{itemize}
  \item The centers, shifts, and marks may be arbitrarily correlated.
    In particular, the shifts may be randomized conditionally on the
    centers.
  \item The core may depend on preceding public template or local
    placement data. In that case, $\BaselineCoreLaw$ denotes the
    prescribed conditional law given those data. This sampling retains
    the joint disjoint placement law of \cref{sec:lobster-combination}.
    Conditioning on the resulting core may nevertheless bias the
    posterior law of this initial joint placement, even though the
    sampling procedure preserves its marginal.
  \item The public cycle-closing Gaussian and the final common
    relabeling remain independent of the baseline choices.
\end{itemize}

We separately require nonempty admissible support and a bounded,
efficient public sampler preserving that support, as specified in
Appendix~\ref{sec:discrete-baseline-shift}. Admissibility alone does not
provide such a sampler.

The direct-Fourier construction in \cref{sec:direct-fourier-transport}
uses a more restricted subclass:
\begin{itemize}
  \item The joint placement $\CalibContext$ records only the labeled
    attachments and queue set. The core and rejection tests do not read
    the separate body-cycle permutation $\BodyPlacementMatrix$ or the
    closure Gaussian. These are sampled independently after core
    acceptance, as specified in \cref{sec:df-sampling-contract}.
  \item The sampler first constructs an uncharged auxiliary proposal,
    which is not yet admissible, then imposes positive public charge
    for each fixed placement. All rejection trials retain the same
    placement, so acceptance does not reweight its marginal law.
\end{itemize}
These additional restrictions preserve the independence needed for the
cycle density estimates. They are not consequences of admissibility.

\subsubsection{Joint Public Sampling}
\label{sec:baseline-public-sampling}

\begingroup
\mdfsetup{topline=true}
\begin{definition}[Complete baseline data]
  \label[definition]{def:complete-baseline-data}
  Draw $(\BaselineCenters,\LegBaselineShift,\BaselineMarks)
  \sample\BaselineCoreLaw$. Given this core, draw
  $({\LegMarkFlag}_{\GadgetIndex},\LocalChartPoint_{\GadgetIndex})$
  conditionally independently for $1\leq\GadgetIndex\leq\GadgetCount$
  from the corresponding kernels $\LegFlagChartLaw$ of
  \cref{def:leg-injection}, using fresh public randomness independent
  of the hidden source. Define $\BaselineDataLaw$ as the joint law of
  \[
    (\BaselineCenters,\LegBaselineShift,\BaselineMarks,
      (({\LegMarkFlag}_{\GadgetIndex},\LocalChartPoint_{\GadgetIndex}))
    _{\GadgetIndex=1}^{\GadgetCount}),
  \]
  conditional on the same preceding public data when applicable.
\end{definition}
\endgroup
The complete law retains every flag and map with its dependence on the
core and uses the full two-branch kernel of
\cref{thm:leg-constructive-sampler}; the regular branch alone generally
has nonuniform off-line marginals.

\subsubsection{Baseline Vector and Carry Safety}
\label{sec:baseline-vector-safety}

Assemble the complete data of \cref{def:complete-baseline-data} using
\cref{def:shared-baseline-vector}, retaining the same maps for the
baseline intercepts and leg mixers.

\begin{proposition}[Joint baseline realization]
  \label[proposition]{prop:joint-baseline-realization}
  Every admissible core sample and its safe leg maps yield
  $\BaselineShift\in\LatticeGen{\RoundingMatrix}$, with body entries in
  $\HistogramSafeCenters$ and off-body entries in
  $\set{0,\ldots,\LARModulus-1}$. On the joint protected event, the body
  carries vanish and the off-body carries cancel pairwise. The sampling
  is source-independent.
\end{proposition}

Appendix~\ref{sec:baseline-realization-proof} verifies this assembly.
The support and sampling-error guarantees are in
\cref{sec:joint-injection-protection}.

%% file: chapters/discrete-realization.tex
\subsection{Joint Protection and Approximation}
\label{sec:joint-injection-protection}

Assume the geometry, safe maps, and baseline support of
\cref{sec:lobster-combination,def:leg-injection,def:admissible-joint-baseline-law}.
The public core and conditional maps are independent of the hidden
source; the closure Gaussian is independent of both. The actual sampler
uses untruncated Gaussian inputs. We introduce protected inputs only
for the analysis.

The protected reference in \cref{sec:joint-protected-references} clips
the latent body edge scalars including the public closure, both inputs
of every leg, and the single shared queue scalar. It retains the cycle,
same-source leg triples, and shared queue correlations. This is the
reference used for public calibration in
\cref{sec:direct-fourier-transport}. For an admissible positive core
law, a successful source coupling also certifies carry safety.

The source comparisons below also apply to a positive, uncharged core
law with the same local safety support. Integer-lattice validity and the
zero-fiber conclusion require the admissible charge constraint.

\subsubsection{Protected Sources and Tail Bounds}
\label{sec:joint-protected-event}

\begingroup
\mdfsetup{topline=true}
\begin{definition}[Joint protected event]
  \label[definition]{def:joint-protected-event}
  Set
  \[
    \DFEdgeCutoff:=\left\lfloor
      \frac{\PaddingInteriorCutoff-1}{2}\right\rfloor.
  \]
  The event $\PaddingProtectedEvent$ requires all of the following:
  \begin{enumerate}[label=(\roman*)]
    \item Every body edge scalar, including the public closure, has
      magnitude at most $\DFEdgeCutoff$.
    \item Both inputs of every leg have magnitude less than
      $\LegCutoff$.
    \item The queue scalar has magnitude less than
      $\PaddingChargeCutoff$.
  \end{enumerate}
\end{definition}
\endgroup
The latent-scalar bounds protect every cyclic difference and retain
the cycle's dependence and zero sum. Each leg contributes two scalar
tails; the shared queue contributes one, with no factor $\LARModulus$.

The scalar budgets $\BodyTailProbability$, $\LegTailProbability$,
and $\QueueTailProbability$ are defined and bounded in
\cref{lem:joint-source-tails} in
Appendix~\ref{sec:protected-comparison-proof}. Their sum, with one
body term, two leg terms per leg, and one queue term, bounds the
failure of $\PaddingProtectedEvent$. The proof controls latent scalars
before taking differences, so no independence of endpoint increments
is needed.

\subsubsection{Baseline Support and Carry Safety}
\label{sec:joint-carry-safety}

Under both the ideal baseline-data law $\BaselineDataLaw$ and its
support-preserving implementation $\BaselineImplementedDataLaw$,
\[
  \Pr\!\left[
    \BaselineCenters\notin
      \HistogramSafeCenters^{\InteriorVertexCount}
    \ \text{or}\
    \BaselineShift\notin
      \LatticeGen{\RoundingMatrix}\cap
      \set{0,\ldots,\LARModulus-1}^{\BlockDim}
  \right]=0.
\]
Thus sampling approximation adds no baseline support failure, including
on fallback branches. The joint protected event then gives the following
carry guarantee.

\begingroup
\mdfsetup{topline=true}
\begin{proposition}[Carry safety of the matrix realization]
  \label[proposition]{prop:matrix-carry-safety}
  Assume an admissible core, safe leg maps, and
  $\PaddingInteriorCutoff+\LegCutoff+\PaddingChargeCutoff
  \leq\PaddingSafeMargin$.
  For the natural-source load in \cref{sec:affine-realization-matrix},
  $\PaddingProtectedEvent$ implies
  \[
    \OnesVec^{\trans}\EndpointCarry=0,\qquad
    \TransportOutput=\CanonicalEndpointPullback(\EndpointResidue)
    \in\LWEZeroFiber.
  \]
  For any positive source-independent sampling law with this support,
  \[
    \StatisticalDistance{\DistributionLaw{\TransportOutput}}
      {\Unif{\LWEZeroFiber}}
    \leq
    \StatisticalDistance{\DistributionLaw{\EndpointResidue}}
      {\Unif{\ZeroSumEndpointHyperplane}}
      +\Pr[\PaddingProtectedEvent^c].
  \]
\end{proposition}
\endgroup
The baseline margin keeps body loads canonical, and each off-body pair
has cancelling carries. The common relabeling preserves their total.
Appendix~\ref{sec:matrix-carry-proof} proves the endpoint reduction and
its invariance under the choice of source lift.

\subsubsection{Protected References and Coupling}
\label{sec:joint-protected-references}

The fully protected reference uses the leg and queue laws of
\cref{def:leg-reference-experiment,sec:queue-injection} and protects the
latent body scalars, including the public closure, as follows.

\begingroup
\mdfsetup{topline=true}
\begin{definition}[Protected body cycle]
  \label[definition]{def:protected-body-cycle}
  Define the protected edge law and its one-vertex residue marginal by
  \[
    \DFProtectedEdgeLaw
      :=\Gaussian{\ZZ}{\GaussianWidth}\mid
        \set{|\CycleGaussianScalar|\leq\DFEdgeCutoff},\qquad
    \DFProtectedRootMarginal
      :=(\DFProtectedEdgeLaw*\DFProtectedEdgeLaw)\bmod\LARModulus.
  \]
  The \emph{protected body cycle} consists of the cyclic differences
  of $\InteriorVertexCount$ independent samples from
  $\DFProtectedEdgeLaw$.
\end{definition}
\endgroup
These differences sum to zero and satisfy the body cutoff pointwise.
Clipping the latent scalars gives a different law from conditioning
the full cycle on safe differences.

Appendix~\ref{sec:protected-comparison-proof} bounds the whole-cycle
coupling budget $\DFBodyClippingError$ by $\BodyTailProbability$ and
proves the component comparisons. The following joint bound is the
interface used by the endpoint analysis.

\emph{Joint coupling.}
For an exact positive, source-independent public-core law and exact
conditional public sampling, the coupling in
Appendix~\ref{sec:protected-comparison-proof} gives
\[
  \Pr[\text{coupling failure}]
  \leq\DFBodyClippingError+2\GadgetCount\LegTailProbability
    +\QueueTailProbability.
\]
Coupling latent sources before assembly and sharing the complete public
data preserves the body cycle, same-source leg triples, and shared queue
scalar, with all prescribed public dependencies. On success the complete
endpoint vectors agree. For an admissible core, success also implies
$\PaddingProtectedEvent$ and hence carry safety by
\cref{prop:matrix-carry-safety}. After charge conditioning and positive
repair, \cref{thm:df-constructive-aft} therefore pays this source cost
once.

\subsubsection{Public Sampling Errors}
\label{sec:joint-public-sampling-errors}

Couple the implemented core first, then each complete conditional
flag--map pair. Uniform conditional error bounds give total error at
most $\BaselineJointSamplingError+\GadgetCount\LegChartError$ in the
complete baseline data; these include the Bernoulli flags and are
counted once. The formal statement and coupling proof are in
\cref{lem:joint-baseline-sampling-error} in
Appendix~\ref{sec:protected-comparison-proof}.

\subsection{Gadget Assembly and Realization}
\label{sec:discrete-realization}

Use the source order of \cref{sec:affine-realization-matrix}, with
$\LegSourceCoordinates{\GadgetIndex}
=(\LocalFirstError_{\GadgetIndex},\LocalSecondError_{\GadgetIndex})^{\trans}$,
and the joint baseline sampler of \cref{sec:baseline-shift}.
We establish affine validity and bounded implementation under these
sampling assumptions. Endpoint approximation uses the protected
comparison of \cref{sec:joint-protected-references}.

\subsubsection{Integer Affine Validity}
\label{sec:assembly-affine-validity}
The body path and legs form one spanning tree. The queue column supplies
the remaining direction, so the incidence construction applies to the
matrix in \cref{sec:affine-realization-matrix}.

\begin{proposition}[Integer affine realization]
  \label[proposition]{prop:matrix-integer-realization}
  For an admissible public core and its complete safe leg maps, the
  affine data of \cref{sec:affine-realization-matrix} satisfy
  \[
    \TransportMatrix\in\ZZ^{\BlockDim\times\BlockDim},\qquad
    \TransportOffset\in\ZZ^{\BlockDim},\qquad
    \det\TransportMatrix
      =\pm\prod_{\GadgetIndex=1}^{\GadgetCount}
        \LegScaleLift_{\GadgetIndex}
      \not\equiv0\pmod{\LARModulus}.
  \]
  These conclusions hold on every support-preserving fallback.
\end{proposition}
\paragraph{Proof sketch.}
The tree columns span the zero-sum integer lattice, and the queue
column has total charge $\LARModulus$. The general incidence result
therefore makes the inverse product integral. The admissible baseline
makes the offset integral, and the triangular leg mixers contribute
exactly their nonzero scales to the determinant.
Appendix~\ref{sec:matrix-integer-proof} gives the complete verification
using \cref{cor:general-tree-incidence-transport}.

\paragraph{Sampling assumptions and numerical budgets.}
Assume the geometry, safety margins, and local feasibility conditions
above, with rational $\GaussianWidth^2\geq2$ and nonempty admissible
core support. The public template and core samplers must be
source-independent, run in worst-case polynomial time in the dimension
and public input and precision lengths, and preserve admissibility
and polynomial output length on every branch. Their conditional errors
are uniform over the permitted preceding public data and histories.
Appendix~\ref{sec:assembly-sampling-assumptions} specifies the complete
contract and separate core, map, closure, and permutation budgets.

The carry-failure budget is
\[
  \CarryFailureError
  :=\BodyTailProbability+\BodyClosureSamplingError
    +2\GadgetCount\LegTailProbability+\QueueTailProbability.
\]
It bounds failure of the protected event in the actual experiment;
support-preserving core, map, and permutation draws add no carry error.
The final AFT analysis uses the single source coupling of
\cref{sec:joint-protected-references}, which covers endpoint disagreement
and unsafe carries together, so it does not add this carry budget again.

\subsubsection{Bounded Realization}
\label{sec:assembly-bounded-realization}
All probabilities below average over the public sampler randomness and
the independent Gaussian source.

\begingroup
\mdfsetup{everyline=true}
\begin{lemma}[Gadget assembly and realization]
  \label[lemma]{lem:joint-discrete-realization}
  Under the assumptions of \cref{sec:assembly-sampling-assumptions},
  there is a bounded public sampler of affine data
  $(\TransportMatrix,\TransportOffset)$ with the following guarantees.
  \begin{enumerate}[label=(\roman*)]
    \item It runs in worst-case polynomial time in $\BlockDim$, the
      binary public-input lengths, and the requested precision lengths.
      Every output is integral and source-independent, with
      \[
        \TransportMatrix\bmod\LARModulus\in
        \GeneralLinearGroup_{\BlockDim}(\LARField),
      \]
      and determinant as in \cref{prop:matrix-integer-realization}.
    \item For
      $\TransportOutput:=\TransportMatrix\LWEError+\TransportOffset
      \bmod\LARModulus$, where
      $\LWEError=\PaddingSource\bmod\LARModulus$,
      \[
        \Pr[\TransportOutput\notin\LWEZeroFiber]
          \leq\CarryFailureError.
      \]
    \item Each conditional per-leg flag--map draw is within statistical
      distance $\LegChartError$ of $\LegFlagChartLaw$, uniformly over
      admissible public histories. Every such output, including rounding
      and fallback, preserves its branch support and is safe with
      nonzero scale.
  \end{enumerate}
\end{lemma}
\endgroup
\paragraph{Proof sketch.}
Each bounded sampler preserves its exact support, so finite-bit errors
cannot invalidate the affine data or local safety. Couple its closure
to an ideal Gaussian and apply the protected-event carry guarantee;
the remaining sampling errors are charged to the public-data coupling.
Exact rational assembly then gives the stated matrix and running time.
Appendix~\ref{sec:gadget-assembly-proof} supplies the full implementation
and error accounting.

The public calibration in
\cref{sec:direct-fourier-transport} controls the fully protected joint
law before positive repair, source coupling, and numerical implementation.

%% file: chapters/direct-fourier-transport.tex
\section{Calibrating Public Randomness for Affine Fiber Transport}
\label{sec:direct-fourier-transport}

\Cref{sec:probabilistic-transport} supplies the affine construction and
bounds carry failure. We must still choose a positive, source-independent
law for the public randomness so that the joint endpoint law, averaged
over both the public choices and the source, is close to uniform on
$\ZeroSumEndpointHyperplane$. Balancing individual endpoints does not
suffice: the protected body cycle, same-source leg triples, and shared
queue scalar create dependencies between endpoints.

We begin with a positive proposal that balances the expected endpoint
histogram, temporarily leaving the public charge constraint aside.
At each fixed placement, the proposal gives independent groups of public
controls that we can reweight to correct correlations.
The response bounds below place the calibration
operator near the identity, so
\cref{thm:stable-finite-dimensional-calibration} yields a stable finite
correction. After independent uniform final relabeling, this correction
cancels all nonzero Fourier coefficients with support size at most the
chosen fixed calibration order, but it may assign negative weights.
We bound the total Fourier energy of the
remaining modes to control the endpoint error, and use high moments to
clip the weight to a bounded nonnegative function with negligible loss.
Conditioning on public charge then returns the endpoints to
$\ZeroSumEndpointHyperplane$, amplifying the error by only a factor of
order $\LARModulus$. Rejection sampling keeps the placement fixed, and
a single source coupling gives endpoint agreement and carry safety,
completing the transport guarantee of \cref{thm:df-constructive-aft}.

\subsection{The Joint Distribution Target}
\label{sec:df-target}

Once carries are safe, the endpoint-to-fiber correspondence of
\cref{sec:zero-fiber-geometric-structure} reduces the transport problem
to uniformity on $\ZeroSumEndpointHyperplane$. Imposing charge now would
couple the public controls, so we first work on the full ambient space
$\LARField^{\BlockDim}$ under the uncharged proposal. The linear
calibration may produce a signed endpoint law. We bound its ambient
error before restoring nonnegativity and charge in
\cref{sec:df-positive-sampling}. The parameters remain symbolic as in
\cref{sec:lobster-combination} until we choose their sizes in
\cref{sec:df-parameter-choice}.

For a mass-one real signed endpoint law $\SignedEndpointLaw$ on
$\LARField^{\BlockDim}$ and frequency
$\DFFrequency\in\LARField^{\BlockDim}$, we define the positive-sign
characters and Fourier transform by
\[
 \DFCharacter{\DFFrequency}(\HoeffdingEndpointVector)
   =e^{2\pi\mathrm i\langle\DFFrequency,\HoeffdingEndpointVector\rangle/
       \LARModulus},\qquad
 \DFTransform(\DFFrequency)=
     \mathbb E_{\SignedEndpointLaw}
               \DFCharacter{\DFFrequency}(\HoeffdingEndpointVector).
\]

Independent uniform final relabeling makes the uncharged protected
endpoint law exchangeable. To control correlations at every support
size, we collect the Fourier coefficients in one polynomial. Choosing
a nonconstant character from exactly $\HoeffdingDegree$ endpoint
factors records a frequency with support size $\HoeffdingDegree$.
The product form will let us bound all degrees together using the
local endpoint blocks.

\begin{definition}[Weighted character-polynomial norm]
\label[definition]{def:df-weighted-character-norm}
For a mass-one real signed law $\SignedEndpointLaw$ on
$\LARField^{\BlockDim}$, define its character polynomial in the formal
variables
$\DFPolynomialVariables=(\DFPolynomialVariable_{\DFFrequencyEntry})_{
\DFFrequencyEntry\in\DFNonzeroField}$ by
\[
 \DFCharacterPolynomial(\DFPolynomialVariables)
 :=\mathbb E_{\SignedEndpointLaw}
 \prod_{\VertexIndex=1}^{\BlockDim}
 \left(1+\sum_{\DFFrequencyEntry\in\DFNonzeroField}
 \DFPolynomialVariable_{\DFFrequencyEntry}
 \DFCharacter{\DFFrequencyEntry}(\HoeffdingEndpointVector_{\VertexIndex})
 \right),
\]
using the same character convention for scalar arguments.
Its ordered symmetric coefficient tensors
$\DFFourierTensor_{\HoeffdingDegree}$ are specified by
\[
 \DFCharacterPolynomial(\DFPolynomialVariables)
 =\sum_{\HoeffdingDegree=0}^{\BlockDim}
 \sum_{\DFFrequencyEntry_1,\ldots,\DFFrequencyEntry_{\HoeffdingDegree}
                    \in\DFNonzeroField}
 (\DFFourierTensor_{\HoeffdingDegree})_{
       \DFFrequencyEntry_1,\ldots,\DFFrequencyEntry_{\HoeffdingDegree}}
 \prod_{\CoordinateIndex=1}^{\HoeffdingDegree}
       \DFPolynomialVariable_{\DFFrequencyEntry_{\CoordinateIndex}}.
\]
The tensor norm is Euclidean on ordered frequency tuples; repeated
and distinct entries use this same convention, with no additional
factorial in a tensor. In particular, $\DFFourierTensor_0=1$.
For an analysis radius $\FockRadius>0$, define
\[
 \DFFourierNorm_{\FockRadius}(\SignedEndpointLaw)^2
 :=\sum_{\HoeffdingDegree=0}^{\BlockDim}
 \frac{\HoeffdingDegree!\FockRadius^{2\HoeffdingDegree}}
      {\BlockDim^{\HoeffdingDegree}}
 \|\DFFourierTensor_{\HoeffdingDegree}\|_2^2.
\]
\end{definition}

For distinct
$\DFFrequencyEntry_1,\DFFrequencyEntry_2\in\DFNonzeroField$,
a coefficient of one on $\DFPolynomialVariable_{\DFFrequencyEntry_1}^2$
is stored once, as
$(\DFFourierTensor_2)_{\DFFrequencyEntry_1,\DFFrequencyEntry_1}=1$.
A coefficient of one on
$\DFPolynomialVariable_{\DFFrequencyEntry_1}
 \DFPolynomialVariable_{\DFFrequencyEntry_2}$
is split equally between the two ordered entries:
$(\DFFourierTensor_2)_{\DFFrequencyEntry_1,\DFFrequencyEntry_2}
=(\DFFourierTensor_2)_{\DFFrequencyEntry_2,\DFFrequencyEntry_1}=1/2$.
These contribute $1$ and $1/2$, respectively, to
$\|\DFFourierTensor_2\|_2^2$; the degree-two norm weight applies
the same outer factor $2!$ to either contribution.

Exchangeability ensures that this polynomial retains the full Fourier
information: all choices of the support vertices give the same
marginal coefficient tensor. The polynomial sums over
$\binom{\BlockDim}{\HoeffdingDegree}$ such choices, whereas Fourier
energy sums their squared magnitudes. The normalization
$\HoeffdingDegree!/\BlockDim^{\HoeffdingDegree}$ compensates for this
counting difference with at most a factor $e^{\HoeffdingDegree}$
loss, uniformly over all degrees, as proved in
\cref{lem:df-fourier-isometry}.
Thus the norm at radius $\FockRadius$ controls the total Fourier
energy with support weight
$(\FockRadius/\sqrt e)^{2\HoeffdingDegree}$.
When $\FockRadius>\sqrt e$, higher support sizes receive larger weights.
After low-support cancellation, a bounded norm therefore forces the
entire remaining tail to be small. The constant term contributes
exactly one, which explains the subtraction in the following criterion.

\begin{lemma}[Fourier criterion for the joint distribution target]
\label[lemma]{lem:df-joint-target}
Let $\SignedEndpointLaw$ be an exchangeable real signed law of mass one
on $\LARField^{\BlockDim}$. Let $\HoeffdingOrder$ be an integer with
$0\leq\HoeffdingOrder<\BlockDim$, and let
$\FockRadius>\sqrt e$, with $\DFChargeRadius=\FockRadius/\sqrt e$.
If
\[
 \DFTransform(\DFFrequency)=0
 \qquad\text{whenever}\qquad
 1\leq|\operatorname{supp}\DFFrequency|\leq\HoeffdingOrder,
\]
then
\[
 \begin{aligned}
 \tfrac12\|\SignedEndpointLaw-\Unif{\LARField^{\BlockDim}}\|_1
 &\leq\tfrac12
 \left(\sum_{\DFFrequency\ne0}
              |\DFTransform(\DFFrequency)|^2\right)^{1/2}\\
 &\leq\frac{
       \bigl(\DFFourierNorm_{\FockRadius}(\SignedEndpointLaw)^2-1\bigr)^{1/2}}
       {2\DFChargeRadius^{\HoeffdingOrder+1}}.
 \end{aligned}
\]
For a probability law, the left-hand side is its statistical distance
from the ambient uniform law.
\end{lemma}

\paragraph{Proof sketch.}
Exchangeability relates the character-polynomial norm to the complete
weighted Fourier energy. After cancellation, every surviving frequency
has support size at least $\HoeffdingOrder+1$, so its weight is at least
$\DFChargeRadius^{2(\HoeffdingOrder+1)}$.
Parseval and Cauchy--Schwarz then give the distributional bound.
The full proof, using
\cref{lem:df-ambient-target,lem:df-fourier-isometry}, appears in
Appendix~\ref{sec:df-energy-proof}.

\input{chapters/direct-fourier-baseline}

\input{chapters/direct-fourier-matrices}

\subsection{Fourier Tails and Weight Moments}
\label{sec:df-complete-control}

The finite correction does not by itself bound the total energy of the
remaining Fourier modes, and its weight may be negative. We therefore
bound both that energy and the loss from clipping the weight.
Both estimates concern the same uncharged weight
at a fixed placement. The Fourier estimate averages the protected
endpoint experiment, retaining the cycle, same-source leg inputs,
shared queue scalar, and independent final relabeling. The moment
estimate averages over the independent public controls under the
uncharged proposal and will bound the clipping loss.

\begin{proposition}[Complete Fourier tails and weight moments]
 \label[proposition]{prop:df-corrected-weight-bounds}
 Use the public family of \cref{sec:df-parameter-choice} and the
 analysis radius $\FockRadius=\HoeffdingParameterBase^{1/4}$,
 independent of the calibration and moment orders.
 For every fixed pair of positive integers
 $\HoeffdingOrder,\FockMomentOrder$, there are positive constants
 $\DFProofConstant_{\HoeffdingOrder}$ and
 $\DFProofConstant_{\HoeffdingOrder,\FockMomentOrder}$ such that the
 following bounds hold for all sufficiently large
 $\HoeffdingParameterBase$, depending only on these fixed orders,
 uniformly over disjoint placements.
 Let $\FockWeight_{\HoeffdingOrder}$ and
 $\SignedEndpointLaw_{\FockWeight_{\HoeffdingOrder}}$ be the
 uncharged weight and protected endpoint law of
 \cref{thm:df-actual-table-inverse}, including independent uniform final
 relabeling, and write $\DFTransform$ for this law's Fourier transform.
 Then
 \begin{align*}
 \DFFourierNorm_{\FockRadius}
   (\SignedEndpointLaw_{\FockWeight_{\HoeffdingOrder}})
   &\leq\DFProofConstant_{\HoeffdingOrder},\\
 \sum_{\DFFrequency\ne0}|\DFTransform(\DFFrequency)|^2
   &\leq\DFProofConstant_{\HoeffdingOrder}
              \HoeffdingParameterBase^{-(\HoeffdingOrder+1)/2},\\
 \tfrac12\|\SignedEndpointLaw_{\FockWeight_{\HoeffdingOrder}}
       -\Unif{\LARField^{\BlockDim}}\|_1
   &\leq\DFProofConstant_{\HoeffdingOrder}
              \HoeffdingParameterBase^{-(\HoeffdingOrder+1)/4},\\
 \|\FockWeight_{\HoeffdingOrder}-1\|_{2\FockMomentOrder}
   &\leq\DFProofConstant_{\HoeffdingOrder,\FockMomentOrder}
              \HoeffdingParameterBase^{-3}.
 \end{align*}
 The moment norm uses the uncharged public proposal. For fixed
 $\HoeffdingOrder$, the weight is the same for all moment orders.
\end{proposition}
\paragraph{Proof sketch.}
After conditioning on the protected cycle and shared queue scalar,
we bound the product of the local character polynomials across all
degrees. Exact low-support cancellation then converts this bound into
control of the remaining Fourier energy. For the weight moments,
canonical centering uses the large pools of independent public controls
to turn coefficient bounds into a bound on every fixed moment.
The full proof, including the local estimates, is in
\cref{sec:df-corrected-bounds-proof,sec:df-tail-appendix}.

\subsection{Public Charge and Bounded Positive Sampling}
\label{sec:df-positive-sampling}

We recover a probability law by clipping the signed weight and then
imposing public charge. For a fixed placement $\CalibContext$, the
charge event is
\[
 \DFContextChargeEvent=
 \set{\sum_{\VertexIndex\in\InteriorVertices}
              \BaselineCenters_{\VertexIndex}
       +\sum_{\GadgetIndex=1}^{\GadgetCount}
              \LegBaselineShift_{\GadgetIndex}
           \equiv\GadgetCount\pmod{\LARModulus}}.
\]
This test reads only the permitted core centers and leg shifts.
For every choice of the remaining randomness, it is equivalent to the
protected endpoint lying in $\ZeroSumEndpointHyperplane$, as verified
in \cref{sec:df-public-charge-verification}. With
$\DFTableClippedWeight=\DFClipMap(\DFTableWeight)$ from
\cref{cor:calibration-bounded-clipping}, the desired public-core law is
\[
 d\DFContextPositiveMeasure(\CalibCore)
 =\frac{\DFTableClippedWeight(\CalibCore)
                          \EventIndicator{\DFContextChargeEvent}(\CalibCore)}
      {\mathbb E_{\DFContextIdealProposal}
                  \DFTableClippedWeight\EventIndicator{\DFContextChargeEvent}}
       \,d\DFContextIdealProposal(\CalibCore).
\]
We normalize separately at each placement and keep that placement
through every rejection trial. This prevents differences in acceptance
rates from biasing the placement marginal.
Passing the accepted core through the protected endpoint channel,
including final relabeling, gives the probability law
$\DFContextPositiveEndpointLaw$.
In the estimates below, we suppress placement subscripts and write
$\DFIdealProposal$ for the uncharged proposal and $\CalibChargeEvent$
for its charge event.

\begingroup
\mdfsetup{topline=true}
\begin{theorem}[Bounded clipping and public conditioning]
 \label[theorem]{thm:df-positive-conditioning}
 Fix a disjoint placement and the calibrated real weight
 $\FockWeight=\DFTableWeight$ of \cref{thm:df-actual-table-inverse},
 with $\mathbb E_{\DFIdealProposal}\FockWeight=1$.
 Let $\SignedEndpointLaw$ be its protected endpoint image,
 including independent uniform final relabeling, and put
 $\DFClippedWeight=\DFClipMap(\FockWeight)$. Define
 \begin{gather*}
 \DFAmbientError
   =\tfrac12\|\SignedEndpointLaw-\Unif{\LARField^{\BlockDim}}\|_1,
 \qquad
 \DFClippingError=\mathbb E_{\DFIdealProposal}
                      |\FockWeight-\DFClippedWeight|,\\
 \DFConditioningError=2\DFAmbientError+\DFClippingError,
 \qquad
 \DFPositiveMass=\mathbb E_{\DFIdealProposal}
                 \DFClippedWeight\EventIndicator{\CalibChargeEvent}.
 \end{gather*}
 If $\DFConditioningError\leq1/(2\LARModulus)$, then
 \[
 \DFPositiveMass\geq1/(4\LARModulus),\qquad
 \StatisticalDistance{\DFContextPositiveEndpointLaw}
                         {\Unif{\ZeroSumEndpointHyperplane}}
       \leq2\LARModulus\DFConditioningError.
 \]
\end{theorem}
\endgroup
\paragraph{Proof sketch.}
Passing through the endpoint channel cannot increase the weight's
$L^1$ clipping loss. The ambient uniform target assigns mass
$1/\LARModulus$ to the charge hyperplane, so restricting to that
hyperplane and renormalizing amplifies the error by a factor
proportional to $\LARModulus$.
The full proof is in \cref{sec:df-clipping-conditioning-proof}.

\begingroup
\mdfsetup{topline=true}
\begin{corollary}[Quantitative clipping loss]
 \label[corollary]{lem:df-positive-repair}
 Use the public family of \cref{sec:df-parameter-choice}.
 For every fixed pair of positive integers
 $\HoeffdingOrder,\FockMomentOrder$, there is a positive constant
 $\DFProofConstant_{\HoeffdingOrder,\FockMomentOrder}$ such that,
 for all sufficiently large $\HoeffdingParameterBase$, the clipping
 loss of \cref{thm:df-positive-conditioning} satisfies
 \[
  \DFClippingError\leq
  \DFProofConstant_{\HoeffdingOrder,\FockMomentOrder}
              \HoeffdingParameterBase^{-6\FockMomentOrder},
 \]
 uniformly over disjoint placements. For each fixed calibration order
 $\HoeffdingOrder$, both $\DFClippingError$ and
 $\LARModulus\DFClippingError$ are negligible in
 $\HoeffdingParameterBase$.
\end{corollary}
\endgroup
\paragraph{Proof sketch.}
Clipping changes the weight only where its deviation from one is at
least one, so any fixed even moment bounds the clipping loss.
Every such moment estimate applies to the same calibrated weight;
increasing the moment order strengthens the analysis without changing
the sampler. The full proof is in
\cref{sec:df-clipping-conditioning-proof}.

\paragraph{Realizing the bounded positive core.}
To sample this law, draw a fresh core from the uncharged proposal at
the retained placement and accept with probability
\[
 \Pr[\text{accept}\mid\CalibCore]
 =\EventIndicator{\DFContextChargeEvent}(\CalibCore)
                  \frac{\DFTableClippedWeight(\CalibCore)}{2}.
\]
The clipped weight lies in $[0,2]$, giving a fixed rejection envelope.
The high-moment bound controls the statistical cost of clipping, while
finite-atom estimates control the numerical error in evaluating the
acceptance probability.

The ideal acceptance rate is at least $1/(8\LARModulus)$. For a
requested timeout bound $0<\DFTrialError<1$, a cap of
$\lceil8\LARModulus\log(1/\DFTrialError)\rceil$ trials gives failure
probability at most $\DFTrialError$. The same bounded envelope lets
us account for proposal and weight errors before charge normalization,
which amplifies their numerical contribution by only
$O(\LARModulus)$. The geometric timeout estimate and normalization
bounds are made precise in \cref{lem:df-bounded-rejection},
Appendix~\ref{sec:df-numerical-stability}, with the corresponding retry
caps and numerical budgets. After acceptance, we complete the remaining
public draws according to \cref{sec:df-sampling-contract}.

\subsection{Choosing the Public Parameters}
\label{sec:df-parameter-choice}

The estimates above require large control pools and small response
defects. We choose one polynomial-size public family that meets these
requirements for every fixed accuracy order. Its parameters are verified
in \cref{sec:df-parameter-verification}, and the resulting transport
guarantee is stated in \cref{thm:df-constructive-aft}.

\paragraph{One public choice.}
\begin{equation*}
  \begin{gathered}
    \HoeffdingParameterBase=\secpar+2,\quad
    \DFPublicScale=\HoeffdingParameterBase^8,\quad
    \SecretDim=\secpar,\quad \GaussianWidth=\HoeffdingParameterBase,\\
    \LARModulus=\text{the least prime at least }
    2^{16}\DFPublicScale^8,\\
    \InteriorVertexCount=\LARModulus\DFPublicScale^{12},\quad
    \GadgetCount=\DFPublicScale^{16},\quad
    \BlockDim=\InteriorVertexCount+2\GadgetCount,\quad \OutputDim=1,\\
    \PaddingInteriorCutoff=\LegCutoff=\PaddingChargeCutoff
    =\HoeffdingParameterBase^2,\quad
    \PaddingSafeMargin=3\HoeffdingParameterBase^2,\quad
    \CalibAnchorWidth=\DFPublicScale^6,\quad
    \LegSumBound=3\CalibAnchorWidth,\quad \LegTrimMargin=1 .
  \end{gathered}
\end{equation*}
In particular, $\LARModulus=\Theta(\HoeffdingParameterBase^{64})$
and $\BlockDim=\Theta(\HoeffdingParameterBase^{160})$.
For the Fourier analysis, we take
$\FockRadius=\HoeffdingParameterBase^{1/4}$.
This ensures $\DFChargeRadius>1$ for sufficiently large
$\HoeffdingParameterBase$, as needed to control the uncanceled modes.

The complete Fourier estimates remain valid up to radius
$2\FockRadius$. These radii serve only the analysis and are independent
of the calibration and moment orders.
All public sizes are polynomial in $\secpar$. The family, including $\LARModulus$, $\BlockDim$,
$\SecretDim$, $\GaussianWidth$, and the rounding map, is independent
of accuracy order, sample budget, and adversary. Outside the certified
regime, the finite-prefix fallback of
\cref{sec:df-numerical-verification} preserves affine validity;
the stated accuracy holds for all sufficiently large $\secpar$.

\subsection{Bounded Implementation and Fixed-Order Accuracy}
\label{sec:df-implementation}

With the public family fixed, we obtain the requested fixed positive
accuracy order $\TransportAccuracyOrder$ by choosing a sufficiently
large calibration order and a sufficiently small numerical budget:
\[
 \HoeffdingOrder\geq4\TransportAccuracyOrder+267,\qquad
 \DFNumericalBudget\leq\tfrac14
                   \HoeffdingParameterBase^{-(\TransportAccuracyOrder+2)}.
\]
Write
$\DFNegligibleError_{\HoeffdingOrder}(\HoeffdingParameterBase)$
for the nonnegative negligible bound on charge-amplified clipping and
source-protection errors established in \cref{sec:df-final-aft-proof},
uniformly over placements.
The calibration order changes the weight but stays fixed as
$\secpar$ grows. Increasing the moment order only strengthens the
clipping estimate for that same weight. The numerical budget, in turn,
determines implementation precision and may shrink with $\secpar$.
Thus every fixed accuracy order has its own polynomial-time sampler;
the degree of its running-time polynomial may depend on that order,
while the public family remains the same.

\input{chapters/direct-fourier-implementation}

\begingroup
\mdfsetup{topline=true}
\begin{theorem}[Constructive fixed-order Gaussian AFT]
 \label{thm:df-constructive-aft}
 For the family in \cref{sec:df-parameter-choice}, each fixed
 positive integer $\TransportAccuracyOrder$ admits a uniform PPT
 sampler $\TransportSampler_{\TransportAccuracyOrder}$ satisfying
 \cref{def:gaussian-t-aft}. Every execution, including fallback,
 returns integral source-independent affine data with
 $\TransportMatrix\bmod\LARModulus\in
       \GeneralLinearGroup_{\BlockDim}(\LARField)$.
 For the fixed calibration order $\HoeffdingOrder$ chosen above
 and all sufficiently large $\secpar$, its accuracy is bounded by
 \begin{equation*}
 \begin{aligned}
 &\StatisticalDistance{
  \DistributionLaw{\TransportMatrix\LWEError+\TransportOffset
                                  \bmod\LARModulus}}
               {\Unif{\LWEZeroFiber}}\\
 &\quad\leq
 \DFProofConstant_{\HoeffdingOrder}\LARModulus
                    \HoeffdingParameterBase^{-(\HoeffdingOrder+1)/4}\\
 &\qquad\quad
 +\DFNegligibleError_{\HoeffdingOrder}(\HoeffdingParameterBase)
 +\DFNumericalBudget ,
 \qquad
 \LWEError\sample\Gaussian{\ZZ^{\BlockDim}}{\GaussianWidth}.
 \end{aligned}
 \end{equation*}
\end{theorem}
\endgroup
The budget $\DFNumericalBudget$ accounts for errors in drawing the
joint placement, numerically sampling the charged core, capping
rejection, completing the flag--affine-map pairs, sampling the closure,
and drawing the two subsequent independent permutations. The charged-core
contribution includes the amplification from normalization.
\paragraph{Proof sketch.}
The Fourier and clipping bounds first give a positive protected
endpoint law close to uniform after charge conditioning. We then apply
one source coupling to the underlying positive, source-independent
public-core law. On a match,
the protected and actual endpoints agree and carries are safe, so the
endpoint-to-fiber correspondence gives the affine transport bound with
a single source-protection cost. The numerical implementation preserves
valid affine data on every branch, and its errors fit the stated budget.
The full proof, including the accuracy-order calculation, is in
\cref{sec:df-final-aft-proof}.

%% file: chapters/direct-fourier-baseline.tex
\subsection{Protected Cycle Reference and Independent Public Controls}
\label{sec:df-protected-reference}
\label{sec:df-sampling-contract}

To correct the joint endpoint law, we reweight public controls and
measure their effect through the protected channel of
\cref{sec:joint-protected-references}. The root marginal
$\DFProtectedRootMarginal$ determines the averaged profiles used
below. The channel itself retains the zero-sum cycle, the same-source
leg triples of \cref{def:leg-joint-reference-law}, and one shared queue
scalar.

\paragraph{Placements and pools.}
\label{sec:df-control-context}
We work at a fixed placement $\CalibContext$ in the geometry of
\cref{sec:lobster-combination}, with labeled attachments
$\HoeffdingAttachment_{\GadgetIndex}={\LegRightVertex}_{\GadgetIndex}$.
The attachment set and the remaining free body vertices are
\[
 \begin{gathered}
 \DFAttachmentSet=\set{\HoeffdingAttachment_{\GadgetIndex}}_{
                     \GadgetIndex=1}^{\GadgetCount},\qquad
 \DFFreeSet=\InteriorVertices\setminus
                 (\DFAttachmentSet\cup\DFQueueSet),\\
 \DFFreeCount=\InteriorVertexCount-\GadgetCount-\LARModulus.
 \end{gathered}
\]
Each leg supplies one primary control and will need its own
compensator to offset the companion response. We order the free
vertices by their public body labels and assign the first
$\GadgetCount$, in that order, to the labeled legs as compensators.
The remaining free vertices form the ordinary-body pool, so
\begin{equation*}
 \ActivityPrimaryPool=\GadgetCount,\qquad
 \ActivityBodyPool=\DFFreeCount-\GadgetCount
        =\InteriorVertexCount-\LARModulus-2\GadgetCount.
\end{equation*}

Disjointness couples the queue and attachment tuple. Their joint law
comes from a uniform initial permutation, which gives equal numbers
of preimages to every permitted queue set and ordered attachment
tuple. Calibration at a positive integer order
$\BoundaryInverseOrder$ requires selections of up to that many
distinct controls. Such selections are available whenever both pool
sizes are at least $\BoundaryInverseOrder$.
\Cref{lem:df-disjoint-context} in
Appendix~\ref{sec:bounded-permutation-sampling} gives the precise
placement law, pool conditions, and bounded implementation.

\paragraph{Public sampling contract.}
To keep calibration independent of the source, we restrict the
information used by the weight and the charge test. The public core
$\CalibCore$ contains the centers, primary colors, anchors, and shifts
prescribed below. The weight uses only the primary colors, their
assigned compensator centers, the ordinary-body centers, and
$\CalibContext$; the charge test may also use all body centers and leg
shifts. Neither uses hidden source coordinates, actual roots, the
shared queue scalar, closure randomness, body or final permutations,
or completed flag--affine-map draws.

Normalization and every rejection trial keep $\CalibContext$ fixed,
preserving its marginal. Once a core is accepted, we sample complete
flag--affine-map pairs conditional on that core, and independently
draw the closure Gaussian, body-cycle permutation, and final global
permutation. These latter draws and the hidden sources are independent
of $\CalibContext$. The body and final permutations are separate from
the initial permutation that chooses the joint placement.

At each fixed placement, the uncharged proposal has independent
primary and body controls. Pairing each primary with a distinct
compensator therefore gives the independent groups used to form
canonical coefficients. Signed correction and charge conditioning
can subsequently correlate these controls.

We estimate the finite response matrices using
\cref{thm:df-cycle-density}. Finite support bounds the likelihood along
each selected path, while centering eliminates configurations with an
isolated selected site. Under the independent body permutation, the
remaining path configurations are rare. This gives the fixed-order
bound proved in Appendix~\ref{sec:df-cycle-proofs}. The later estimate
of the complete Fourier energy still uses the entire protected cycle.

\subsection{An Explicit Positive Uncharged Proposal}
\label{sec:df-positive-baseline}

Our first goal is to balance the expected endpoint histogram,
averaging over the protected sources. Since a primary affects both
its own endpoint and its companion, we choose the primary and
free-center laws together at each fixed disjoint placement.
Throughout this construction, convolutions are cyclic on $\LARField$,
and support assertions use canonical integer lifts. We impose the
structural relations
\[
 \PaddingInteriorCutoff=\LegCutoff=\PaddingChargeCutoff,\qquad
 \PaddingSafeMargin=3\LegCutoff.
\]
The common cutoff and all other parameter sizes remain free until
\cref{sec:df-parameter-choice}. We use the reference laws and activation
factor
\[
 \CalibSafeUniform=\Unif{\HistogramSafeCenters},\qquad
 \CalibUniform=\Unif{\LARField},\qquad
 \CalibActivation=\tfrac12\LegClippedLaw(0).
\]
The queue and leg clipped scalar laws therefore coincide.

\paragraph{Convolution profiles.}
To express the averaged endpoint profiles, let
$\BaseMatrixBodyConvolution$ and $\BaseMatrixLegConvolution$ denote
convolution with $\DFProtectedRootMarginal$ and $\LegClippedLaw$,
respectively, acting on mass profiles. The two body profiles we need
are
\[
 \DFBodyProfile_{\CycleIndex}
   =\BaseMatrixBodyConvolution\BaseMatrixLegConvolution^{\CycleIndex}
          \CalibSafeUniform\quad(\CycleIndex=0,1).
\]
Body inputs are extended by zero outside their safe support.
\Cref{sec:df-baseline-proof} gives the matrix-entry conventions.

\paragraph{Inward anchors and the independent blur.}
The companion must remain interior for every protected second input.
We obtain its anchor by reflecting the primary color and shifting the
result inward. We then add an independent public blur with law
$\DFProtectedRootMarginal$ and choose the leg shift to preserve the
required pair sum, with mixing weight $\LocalAtomWeight=1/2$.
The blur is a fresh draw from this law and uses no actual root or
closure value.
We impose the anchor and general leg-feasibility conditions specified
in Appendix~\ref{sec:df-baseline-proof}, which also gives the exact
anchor kernel, blur assignments, and diffuse profiles.
We write $\DFSmoothedAnchor$ for the anchor kernel followed by
convolution with $\LegClippedLaw$, as defined there.

\paragraph{Balancing the primary and companion together.}
The companion contribution and the free-center contribution pass
through the same body convolution. We can therefore adjust the
free-center law to cancel the companion mean, then choose the primary
law to balance the remaining profiles. The resulting target
$\BaseMatrixTarget$ is obtained from $\BlockDim\CalibUniform$ by
subtracting the body and diffuse off-body means after this
cancellation. Appendix~\ref{sec:df-baseline-proof} gives its full
profile expansion. The compensation factor is
\[
 \CalibCompensation=\frac{\GadgetCount\CalibActivation}{\DFFreeCount}.
\]
The two proposal laws are specified by
\begin{equation}
\label{eq:df-baseline-block-system}
 \begin{aligned}
 \GadgetCount\CalibActivation\DFBaselinePrimary&=\BaseMatrixTarget,\\
 \DFBaselineCenter&=(1+\CalibCompensation)\CalibSafeUniform
       -\CalibCompensation\DFSmoothedAnchor\DFBaselinePrimary.
 \end{aligned}
\end{equation}
The proposal samples free centers independently from
$\DFBaselineCenter$, all other body centers independently from
$\CalibSafeUniform$, and all leg primaries independently from
$\DFBaselinePrimary$. It then samples the anchors and public blurs
using the kernel specified in
Appendix~\ref{sec:df-baseline-proof}.

\begin{theorem}[Positive proposal and exact averaged mean]
\label{thm:df-positive-baseline}
For the public family of \cref{sec:df-parameter-choice} and all
sufficiently large $\HoeffdingParameterBase$, uniformly over
disjoint placements, the proposal laws above are
reflection-symmetric probability laws and
\begin{align*}
 &\begin{gathered}
 \DFBaselinePrimary\geq(2\LARModulus)^{-1},\\
 \tfrac12\CalibSafeUniform\leq\DFBaselineCenter
       \leq(1+\CalibCompensation)\CalibSafeUniform
       \leq4/\LARModulus
             \quad\hbox{on }\HistogramSafeCenters,
 \end{gathered}
\end{align*}
The center law vanishes off $\HistogramSafeCenters$.
In the auxiliary protected experiment, the expected histogram is
exactly $\BlockDim\CalibUniform$.
\end{theorem}
\paragraph{Proof sketch.}
The inward anchors keep every sampled core on the safe local support.
The numerator defining the primary law has mass
$\GadgetCount\CalibActivation$, and the profile losses are small
enough to leave the stated positive margin. The free-center
adjustment cancels the companion mean, so the primary equation
balances the remaining histogram.
\Cref{lem:df-symbolic-baseline} gives the general loss conditions and
quantitative bounds; its proof also verifies this public family.

This proposal has not yet imposed public charge. Once a positive
charge-correct core law has been obtained, conditional completion by
\cref{def:leg-injection} gives affine validity through
\cref{prop:joint-baseline-realization}.

\subsection{The First Response}
\label{sec:df-first-response}

To correct higher correlations, we need to understand how reweighting
one primary changes the expected endpoint histogram. Its companion
also responds, so we pair the centered primary score with a score on
its distinct free compensator. Their common body convolution allows
the compensator score to cancel the marked companion response.
For the fixed proposal, let
$\BaseMatrixPrimaryDiagonal=\operatorname{diag}(\DFBaselinePrimary)$.
For a centered primary density score $\DFDensityInput$, write
$\CalibProfile=\BaseMatrixPrimaryDiagonal\DFDensityInput$ for its
mass-zero profile.

The resulting paired lift is
\begin{equation*}
 (\ActivityCoreLift\DFDensityInput)(\CalibColor,\CalibOtherColor)
 =\DFDensityInput(\CalibColor)
 -\CalibActivation
  \frac{(\DFSmoothedAnchor\BaseMatrixPrimaryDiagonal
                         \DFDensityInput)(\CalibOtherColor)}
       {\DFBaselineCenter(\CalibOtherColor)}.
\end{equation*}
The primary and compensator inputs have independent laws
$\DFBaselinePrimary$ and $\DFBaselineCenter$, respectively. The ratio
is used only on the safe support.

After companion cancellation, the remaining perturbation comes from
the diffuse target's dependence on the marked color. The reflected
anchor cancels that color from the pair sum, leaving only the choice
of side of the midpoint; at the midpoint the two choices are averaged.
Thus there are only two diffuse profiles, as specified in
Appendix~\ref{sec:df-baseline-proof}. Using these profiles and the anchor sign,
let $\BaseMatrixSignVector$ be the
mass-coordinate column of
$\CalibSignResponse=(1-\CalibActivation)
(\CalibSliceMean_{+1}-\CalibSliceMean_{-1})$, and let
$\BaseMatrixSignFunctional$ be the column with entries
$\DFAnchorSign(\CalibColor)$. For the mass-zero adjustment
$\CalibProfile$, the scalar
$\BaseMatrixSignFunctional^{\mathsf T}\CalibProfile$ is the change in
mass above the midpoint minus the change below it. Multiplying this
scalar by the fixed profile $\BaseMatrixSignVector$ gives the diffuse
residual. After division by the source-averaged activation probability,
the response is
\begin{equation*}
 \DFFirstResponse=\IdentityMatrix
       +\CalibActivation^{-1}
                   \BaseMatrixSignVector\BaseMatrixSignFunctional^{\mathsf T}.
\end{equation*}

\begin{lemma}[Exact and stable first response]
\label[lemma]{lem:df-first-inverse}
For the public family of \cref{sec:df-parameter-choice} and all
sufficiently large $\HoeffdingParameterBase$, uniformly over disjoint
placements, the paired lift of every centered primary score is
centered under its independent proposal inputs. Its complete first
histogram response, averaging the protected roots and the leg inputs, is
$\CalibActivation\DFFirstResponse\CalibProfile$.
The operator $\DFFirstResponse$ preserves mass-zero profiles.
In the norm
\begin{equation*}
 \|\CalibProfile\|_{\DFBaselinePrimary^{-1}}^2
 =\sum_{\CalibColor}\CalibProfile(\CalibColor)^2/
                        \DFBaselinePrimary(\CalibColor),
\end{equation*}
it satisfies
\begin{equation*}
 \|\DFFirstResponse-\mathrm{Id}\|
       =O(\HoeffdingParameterBase^{-5}),\qquad
 \|\DFFirstResponse^{-1}\|\leq2.
\end{equation*}
\end{lemma}
\paragraph{Proof sketch.}
The compensator score cancels the companion response, leaving the
primary point-mass contribution and the diffuse sign term. Dividing
by the activation factor makes the former the identity. The sign term
has rank one and is small in the primary-reference norm, which gives
the inverse bound. The attachment has zero first response, although
its higher joint responses remain in the leg triple.
\Cref{lem:df-symbolic-first-response} gives the general bound and
exact inverse; \cref{sec:df-first-response-proof} proves the claims.

The joint calibration applies this inverse separately in each primary
slot, using the graded coefficient norm of
\cref{sec:df-actual-inversion}. \Cref{lem:df-norm-comparison} relates
this norm control to the point evaluations needed for numerical
implementation.

%% file: chapters/direct-fourier-matrices.tex
\subsection{Hoeffding-Type Coordinates and Response Matrices}
\label{sec:df-actual-inversion}

To correct joint correlations, we reweight groups of independent public
controls at a fixed placement. Fix a positive integer calibration order
$\BoundaryInverseOrder$, a labeled disjoint placement, and its positive
uncharged proposal $\DFIdealProposal$ from
\cref{thm:df-positive-baseline}. Each primary is paired with its assigned
compensator, while ordinary-body centers form a separate pool; a
compensator consumes no ordinary-body slot. As in a Hoeffding
decomposition, the degree counts selected groups and each kernel is
centered in every selected slot. The paired lift $\ActivityCoreLift$
of \cref{sec:df-first-response} adds the compensator score to a primary
correction, and $\CoreUStatistic$ averages over distinct indices in the
two pools.

The coefficient space $\DFMatrixSpace$ records these corrections as
a scalar part and centered, separately symmetric density kernels on
the independent primary and body controls.
\Cref{def:df-control-coordinates} in
Appendix~\ref{sec:df-table-algebra} gives the full definition and the
coordinate map $\DFMatrixCoord$, whose constant-one vector is
$\DFMatrixZero$. The bound
$\DFMatrixDimension\leq(2\LARModulus+\BoundaryInverseOrder)^{\BoundaryInverseOrder}$
makes the system polynomial in size at every fixed order.
All estimates below use the parameters of
\cref{sec:df-parameter-choice} and hold uniformly over placements at
fixed $\BoundaryInverseOrder$.

\paragraph{Three maps and the calibration equation.}
Calibration must connect a coefficient vector to both a public weight
and its endpoint response. The physical row
$\DFMatrixPhysical(\CalibCore)$ maps input coefficients
to a public weight: it sends $\DFMatrixZero$ to one and sends
$\DFMatrixCoord(\DFTable)$ of positive degree to
$(\CoreUStatistic\ActivityCoreLift\DFTable)(\CalibCore)$.
On the output side, leading evaluation $\DFMatrixLeading$ converts
coefficients to ordered Fourier tensors by
\cref{lem:df-table-algebra}. The full response $\DFMatrixLifted$ acts
on the input space and records the protected channel's effect before
leading evaluation. For every coefficient vector $\DFMatrixSolution$,
\cref{sec:df-common-queue-inverse} gives
\[
 \DFTableWeight=\DFMatrixPhysical(\CalibCore)\DFMatrixSolution,
 \qquad
 [\DFMatrixWeightedPolynomial]_{\leq\BoundaryInverseOrder}
 =\DFMatrixLeading\DFMatrixLifted\DFMatrixSolution.
\]
Here $\DFMatrixWeightedPolynomial$ is the unnormalized character
polynomial of the protected endpoint experiment weighted by
$\DFTableWeight$ relative to the uncharged proposal $\DFIdealProposal$.
The response averages the hidden source conditional on public controls,
without normalizing the weighted law, and treats the Fourier output as
a real vector space. Since leading evaluation may be rectangular and
noninjective, the Fourier equations need not determine the input
coefficients. Calibration instead solves the full square system.

The background coefficient separates the contribution of unselected
controls. Products in the coefficient algebra use disjoint slots;
physical multiplication also includes overlapping control indices.
Appendix~\ref{sec:df-table-algebra} defines the product precisely.
Let $\DFTableTotalBackground$ be the unit-constant background
coefficient constructed in \cref{sec:df-common-queue-inverse}, put
$\DFMatrixBackgroundVector=\DFMatrixCoord(\DFTableTotalBackground)$,
and let $\DFMatrixBackground$ represent multiplication by it. Then
\[
 \DFMatrixLifted=\DFMatrixBackground\DFMatrixNormalized
 =\DFMatrixBackground(\IdentityMatrix+\DFMatrixRaising+\DFMatrixRoot).
\]
After removing the background, $\DFMatrixRaising$ strictly raises
degree, while $\DFMatrixRoot$ records protected-root effects and can
mix types without lowering degree. Both annihilate $\DFMatrixZero$
after the raw root constant column has been absorbed into the
background. Each complete expression uses one common queue scalar;
we average these expressions before normalization and inversion.

Centering explains the triangular structure. Consider one summand in
the distinct-index average, with protected roots fixed. If a selected
group emits no output, integrating its centered score makes that
summand vanish. Every surviving summand therefore has output degree
at least its input degree. Averaging over selections and multiplying
by the unit-constant background preserve this grading. The normalized
and full responses are consequently block lower triangular, with common
diagonal blocks
$\DFMatrixDiagonalBlock_{\HoeffdingDegree\HoeffdingDegree}
=\IdentityMatrix+(\DFMatrixRoot)_{\HoeffdingDegree,\HoeffdingDegree}$.
These blocks retain the degree-preserving protected-root effects,
as verified in \cref{sec:df-canonical-first-degree}.

\paragraph{A norm adapted to the controls.}
The graded weighted Hilbert norm $\DFMatrixNorm{\cdot}$\newline of
Appendix~\ref{sec:df-graded-norm} controls both response inversion and
public weight moments. Its population weights match averaging over distinct
control groups, while its body-radius weight matches the Fourier
estimate. The induced operator norm measures the response perturbation,
and canonical centering transfers coefficient control to physical
moment bounds. Clipping will use these moments, since coefficient
control alone gives no pointwise closeness of the weight to one.

\subsection{Second Application: Joint Endpoint Calibration}
\label{sec:df-matrix-calibration}

We enforce low-order cancellation by choosing coefficients whose full
response is the constant-one vector. We apply
\cref{thm:stable-finite-dimensional-calibration} on $\DFMatrixSpace$,
with the graded norm above, synthesis $\DFMatrixPhysical$, the
unnormalized truncated character response, and target evaluation
$\DFMatrixLeading$. The reference coefficient operator is the identity
and the actual operator is $\DFMatrixLifted$; both the target and the
reference coefficient vector are $\DFMatrixZero$.

\begin{theorem}[Finite-order joint endpoint calibration]
\label{thm:df-actual-table-inverse}
For the family and radius of \cref{sec:df-parameter-choice} and every
fixed positive integer $\BoundaryInverseOrder$, the following holds
for all sufficiently large $\HoeffdingParameterBase$, with a threshold
depending only on the fixed order and uniformly over placements.
Each labeled disjoint placement admits the
unique full coefficient solution
\begin{equation*}
\DFMatrixLifted\DFMatrixSolution=\DFMatrixZero,
\qquad
\DFTableWeight(\CalibCore)
 =\DFMatrixPhysical(\CalibCore)\DFMatrixSolution,
\qquad
\DFMatrixSolution=\DFMatrixZero+\DFMatrixCorrection.
\end{equation*}
Its scalar coordinate is one, the correction has scalar coordinate
zero, and
\begin{equation*}
\DFMatrixNorm{\DFMatrixCorrection}
 =O_{\BoundaryInverseOrder}(\HoeffdingParameterBase^{-3}),
\qquad
\mathbb E_{\DFIdealProposal}\DFTableWeight=1.
\end{equation*}
For the signed, uncharged protected-cycle endpoint law weighted by
$\DFTableWeight$,
\begin{equation*}
[\DFMatrixWeightedPolynomial]_{\leq\BoundaryInverseOrder}
 =\DFMatrixLeading\DFMatrixLifted\DFMatrixSolution=1.
\end{equation*}
After independent uniform final relabeling, every nonzero Fourier
coefficient of support at most $\BoundaryInverseOrder$ vanishes.
\end{theorem}

The vector $\DFMatrixCorrection$ represents the canonical kernel
$\DFTableCorrection$. The calibrated weight may be signed; bounded
clipping and public conditioning recover a positive sampling law in
\cref{sec:df-positive-sampling}.

\paragraph{Response estimates.}
Stability requires a quantitative bound on the full response. The
background, raised-response, and protected-root estimates of
\cref{prop:df-actual-background,lem:df-independent-response,lem:df-actual-cycle-defect}
give
\[
 \DFMatrixNorm{\DFMatrixBackgroundVector-\DFMatrixZero}
 +\DFMatrixNorm{\DFMatrixRaising}
 +\DFMatrixNorm{\DFMatrixRoot}
 \leq\DFProofConstant_{\BoundaryInverseOrder}
                  \HoeffdingParameterBase^{-3}.
\]
Combining these estimates after the common-queue assembly of
\cref{sec:df-common-queue-inverse}, submultiplicativity gives
$\DFMatrixNorm{\DFMatrixLifted-\IdentityMatrix}
=O_{\BoundaryInverseOrder}(\HoeffdingParameterBase^{-3})$.
The full proof is in \cref{sec:df-graded-application}.

\paragraph{Finite block solution.}
Lower triangularity lets us solve one degree at a time: subtract the
response of previously determined coefficients, then invert the current
diagonal block. The full perturbation bound makes each diagonal block
stably invertible, so the recurrence remains controlled.
Appendix~\ref{sec:df-finite-coefficient-computation} gives the explicit
recurrence and certified finite computation.
The proposal already balances degree one before leading evaluation, so
$(\DFMatrixBackgroundVector)_1=(\DFMatrixCorrection)_1=0$ by
\cref{sec:df-canonical-first-degree,sec:df-graded-application}.
For $\BoundaryInverseOrder\geq2$, the first possible correction therefore
lies in the degree-two block, where primary--primary, primary--body,
and body--body kernels interact.

%% file: chapters/direct-fourier-implementation.tex
The sampler draws one disjoint placement and constructs the finite
calibration tables for it. It then draws fresh proposal cores at that
placement until the clipped public-charge rule of
\cref{sec:df-positive-sampling} accepts or the retry cap is reached.
After acceptance, it completes the conditional leg maps, independently
samples the closure and the two subsequent permutations, and assembles
the affine data. If certification fails or trials time out, the prescribed
fallbacks still return integral affine data with modular invertibility.
Every branch uses public data alone, preserving source independence.

\paragraph{Implementation sketch.}
Table construction and weight evaluation use rational intervals with
certified error bounds. Deterministic work limits and retry caps make
the worst-case running time polynomial for each fixed calibration order.
The precision is chosen so that charge amplification and all subsequent
public draws fit the numerical budget. Algorithm~\ref{alg:df-core-sampler} in
Appendix~\ref{sec:df-numerical-verification} gives the complete procedure,
precision allocation, and support-preserving fallbacks.

%% file: chapters/applications.tex
\section{Applications}
\label{sec:applications}

We give three consequences for the fixed parameter family of
\cref{sec:df-parameter-choice}, with the rounding matrix and quotient
encoding of \cref{def:explicit-rounding-matrix} and the transport of
\cref{thm:df-constructive-aft}.
All dimensions and moduli are polynomially bounded in $\secpar$.
The transport is used in the security reduction; the functions evaluated
in the applications are deterministic rounding functions.

\subsection{Basing LAR on Worst-Case Lattice Problems}
\label{sec:applications-worst-case}

The direction of the computational reductions is
\begin{equation*}
 \set{\GapSVP_{\LatticeApproxFactor},\SIVP_{\LatticeApproxFactor}}
 \ \leq_{\mathrm{quantum}}\ \text{Gaussian }\LWE
 \ \leq_{\mathrm{classical}}\ \text{uniform-zero-fiber }\LWE
 \ \leq_{\mathrm{classical}}\ \DecisionalLAR.
\end{equation*}
Thus worst-case lattice hardness implies decisional $\LAR$ hardness.
The classical arrows describe the sample transformations: with a QPT
downstream distinguisher, their compositions are QPT on classical
sample tables, as in \cref{sec:lar-assumptions}.
Here $\GapSVP_{\LatticeApproxFactor}$ is the decision shortest-vector
problem with approximation gap $\LatticeApproxFactor$, and
$\SIVP_{\LatticeApproxFactor}$ asks for $\SecretDim$ independent lattice
vectors of length at most $\LatticeApproxFactor$ times the last
successive minimum.

The worst-case reduction uses continuous Gaussian errors, whereas
our affine transport takes a discrete Gaussian source. Gaussian
randomized rounding connects these interfaces while preserving the
uniform experiment up to negligible implementation error, as verified in
\cref{sec:applications-worst-case-proof}.

For this public family, the approximation factor is
\[
 \LatticeApproxFactor
 =\widetilde{O}\!\left(\frac{\SecretDim\LARModulus}{\GaussianWidth}\right)
 =\widetilde{O}(\LARModulus).
\]

\begin{corollary}[Worst-case foundation for decisional $\LAR$]
\label[corollary]{cor:applications-worst-case}
For the fixed public family and the approximation factor above,
a uniform QPT decisional $\LAR$ distinguisher on polynomially many
classical random examples with nonnegligible advantage yields uniform
quantum polynomial-time algorithms
for worst-case $\GapSVP_{\LatticeApproxFactor}$ and
$\SIVP_{\LatticeApproxFactor}$ in dimension $\SecretDim$.
\end{corollary}
If the distinguishing advantage is guaranteed only on infinitely many
security parameters, the resulting algorithms succeed on those parameters.
Consequently, quantum worst-case hardness that excludes such solvers
succeeding on infinitely many dimensions implies decisional $\LAR$ hardness
against uniform QPT distinguishers for every polynomial classical sample count.
The proof in \cref{sec:applications-worst-case-proof} chooses the
transport accuracy after the adversary's sample and advantage bounds,
then composes with the Gaussian rounding interface. This order of
choices supports every polynomial sample count within one fixed public
family.

\subsection{Direct Field-Valued Weak PRFs in Constant-Depth Threshold Circuits}
\label{sec:applications-weak-prf}

Deterministic rounding as a source of weak pseudorandom functions,
including $\TCZero$ evaluation, has a substantial
history~\cite{EC:BanPeiRos12,C:AKPW13}. Here we obtain a direct
field-valued construction whose fixed polynomial parameters support
every polynomial random-example budget. Its input law is exactly
uniform on the full field-matrix space, and its evaluation is the
explicit quotient-rounding map; the affine transport is needed only
to prove security.

Let $\PRFKeySpace:=\LARField^{\SecretDim}\setminus\set{\ZeroVec}$.
Sample a key $\LARSecret\sample\Unif{\PRFKeySpace}$ and define
\begin{equation*}
 \WeakPRFFunction_{\LARSecret}\colon
 \LARField^{\BlockDim\times\SecretDim}\longrightarrow\LARField,
 \qquad
 \WeakPRFFunction_{\LARSecret}(\LARPublicMatrix)
 :=\LARLabelMap(\LARPublicMatrix\LARSecret).
\end{equation*}
Excluding the zero key makes each single output exactly uniform
for every fixed key under uniform inputs. Security concerns the joint
law of examples sharing that key. This marginal property differs from
the key-uniformity condition of
Ding, Jain, and Komargodski~\cite[Definition~15]{ARXIV:DinJaiKom26},
which fixes an input and randomizes the key.

For implementation, use a bounded polynomial-time key sampler supported
on $\PRFKeySpace$ whose law is within $2^{-\secpar}$ in SD of
$\Unif{\PRFKeySpace}$. Its sampling error and the effect of repeated
inputs are negligible for every polynomial sample count. The
joint key-and-input evaluation circuits are polynomial-time
constructible. Security is against uniform QPT adversaries
receiving any polynomial number of independent classical random examples
from one fixed key, in the access model of \cref{sec:lar-assumptions}.

\begin{proposition}[A field-valued weak PRF in $\TCZero$]
\label[proposition]{prop:applications-tc0}
Under the worst-case hardness condition of
\cref{cor:applications-worst-case}, the bounded implementation above has
post-quantum weak pseudorandomness against uniform QPT random-example
distinguishers receiving polynomially many classical examples.
Its joint key-and-input evaluation is in $\TCZero$, and its single-output
marginals are exactly uniform for every fixed key under uniform
field-matrix inputs.
\end{proposition}
The evaluation circuit uses a fixed sequence of arithmetic stages:
inner products, coordinatewise floors, and the quotient sum. Their
polynomially bounded ranges permit parallel threshold comparisons.
\Cref{sec:applications-weak-prf-proof} verifies key sampling, circuit
construction, and security against a uniformly random function.

\subsection{Noise-Free Learning on a Uniform Finite Grid}
\label{sec:applications-learning}

The weak PRF of \cref{prop:applications-tc0} supplies a family of
noise-free learning targets: each label is a deterministic function
of a uniform input and a fixed unknown key. A learner that predicts
fresh field labels with an inverse-polynomial advantage over the
uniform-guess baseline $1/\LARModulus$ yields a decisional $\LAR$
distinguisher, also when training is QPT. Thus the same worst-case lattice
hardness condition rules out such efficient prediction for these polynomial-size
$\TCZero$ functions under uniform inputs in
$\LARField^{\BlockDim\times\SecretDim}$. The learner may be improper,
provided it outputs a classical hypothesis with efficient deterministic
classical evaluation, as specified below. This is the standard
learning-to-distinguishing argument used for rounding-based targets;
see \cite[Theorem~4.1]{NeuroIPS:CGKM22} and the verification in
\cref{sec:applications-learning-proof}.

For the ReLU representation, set
$\LearningInputDim:=\BlockDim\SecretDim$ and
$\LearningGrid:=\set{0,\ldots,\LARModulus-1}^{\LearningInputDim}$,
identify each input matrix with its canonical entries in
$\LearningGrid$, and define
\begin{equation*}
 \LearningFunction_{\LARSecret}(\LARPublicMatrix)
 :=\frac{\CanonicalLift{
               \WeakPRFFunction_{\LARSecret}(\LARPublicMatrix)}}{
               \LARModulus}.
\end{equation*}
Grid coordinates are encoded as binary integers and labels as exact
rational numerator and denominator pairs. Network depth counts ReLU
hidden layers, excluding the final affine scalar output. For our
public family,
$\LARModulus\leq\BlockDim\leq\LearningInputDim
=\operatorname{poly}(\secpar)$, so time polynomial in the learning
dimension is also polynomial in the security parameter.

\begingroup
\mdfsetup{topline=true}
\begin{lemma}[Two-hidden-layer representation on a finite grid]
\label[lemma]{lem:applications-relu-representation}
For every $\LARSecret\in\PRFKeySpace$, the function
$\LearningFunction_{\LARSecret}$ on $\LearningGrid$ is representable
by a two-hidden-layer ReLU network of size polynomial in
$\LearningInputDim$, with rational weights and biases of polynomially
bounded magnitude and bit length.
\end{lemma}
\endgroup
The first layer implements the coordinatewise floors, whose sum gives
an integer score. The second implements reduction modulo
$\LARModulus$ and normalization on the score's bounded integer range.
The construction is polynomial-time in $\LearningInputDim$ given the
key and yields a $[0,1]$-valued extension to all real inputs.
\Cref{sec:applications-relu-proof} gives the complete construction,
including the $O(\LearningInputDim\LARModulus^2)$ unit bound and
$O(\log\LearningInputDim)$ parameter bit length.

\paragraph{Learning model.}
A learner is a uniform QPT algorithm whose running time
is polynomial in $\LearningInputDim$ on every input transcript.
It receives independent classical noiseless examples under
$\Unif{\LearningGrid}$ and outputs a classical description of a possibly
improper hypothesis $\LearningHypothesis$.
A uniform deterministic classical evaluator must, for
every output hypothesis and every grid point, approximate its
prediction clipped to $[0,1]$ by a rational number to additive error
$1/1000$ in time polynomial in $\LearningInputDim$.
No quantum state is part of the hypothesis or supplied to its evaluator.
For a target $\LearningFunction_{\LARSecret}$, squared loss means
\[
 \mathbb E_{\LARPublicMatrix\sample\Unif{\LearningGrid}}
 \bigl(\LearningHypothesis(\LARPublicMatrix)
       -\LearningFunction_{\LARSecret}(\LARPublicMatrix)\bigr)^2,
\]
where the input is fresh and independent of training. Success
probability is over the training examples and the learner's internal
randomness and measurement outcomes.

\begin{corollary}[Learning hardness on a uniform finite grid]
\label[corollary]{cor:applications-learning}
Under the preceding learning model, a learner attaining squared loss
at most $1/100$ with success probability at least $2/3$ for every
target of \cref{lem:applications-relu-representation} yields a uniform
QPT decisional $\LAR$ distinguisher on classical samples with advantage at least $1/2$
for all sufficiently large $\secpar$.
\end{corollary}
Independent validation separates the learner's small squared loss
from the constant variance of an independent uniform label.
\Cref{sec:applications-learning-proof} gives the quantitative argument,
including the evaluation error and the zero-secret event.

Together with \cref{cor:applications-worst-case}, the corollary rules
out such a QPT learner under the stated quantum worst-case lattice
hardness condition. In particular, it rules out learning all
polynomial-size two-hidden-layer ReLU networks with polynomially
bounded rational weights and biases of polynomial bit length under
the same classical sampling and hypothesis-evaluation conventions.
In contrast, \cref{lem:lar-information-theoretic-identifiability} gives
exact identification from $O(\SecretDim)$ classical examples at fixed
confidence by computationally inefficient exhaustive search.
The hardness concerns efficient learning in the stated classical-example
model; it does not cover coherent quantum examples or quantum membership
queries.

%% file: chapters/appendix-foundations-proofs.tex
\section{Foundations and the Main Reduction}
\label{sec:foundations-proofs}

We complete the main reduction, then verify the lattice, probability,
rounding, and calibration facts used in the main text.

\subsection{The main reduction}
\label{sec:main-reduction-proof}

\begin{proof}[Proof of \cref{thm:introduction-main}]
Fix the public family of \cref{sec:df-parameter-choice} and
the rounding matrix of \cref{def:explicit-rounding-matrix} before
choosing an accuracy order, sample count, or distinguisher.
Given a uniform QPT distinguisher with nonnegligible advantage on
polynomially many classical examples, choose a fixed accuracy order
as in \cref{cor:gaussian-lwe-to-lar-composition}.
\Cref{thm:df-constructive-aft} supplies its source-independent
uniform PPT sampler. It preserves
integral affine validity and modular invertibility at every security
parameter, with the approximation guarantee holding eventually.
The composition corollary therefore gives a uniform QPT Gaussian
$\LWE$ distinguisher with nonnegligible advantage. This sample-table
composition preserves the block count, uses
$\SampleCount\BlockDim$ scalar samples, and invokes the original
distinguisher once after classical preprocessing.

Applying \cref{cor:applications-worst-case} gives the claimed uniform
QPT algorithms for worst-case $\GapSVP_{\LatticeApproxFactor}$ and
$\SIVP_{\LatticeApproxFactor}$, with
$\LatticeApproxFactor=\widetilde O(\LARModulus)$.
Its proof in \cref{sec:applications-worst-case-proof} supplies the
Gaussian interface and the quantum subroutine implementation.
If the distinguishing advantage is guaranteed only infinitely often,
the lattice guarantees hold on the corresponding dimensions.
The running-time polynomial and accuracy threshold may depend on the
fixed order and adversary bounds; the public family and source law
remain unchanged.
\end{proof}

\subsection{Smith coordinates and the image lattice}
\label{sec:foundations-smith}

\begin{proof}[Proof of \cref{lem:rounding-smith-properties}]
Fix a rounding matrix $\RoundingMatrix$ together with matrices
$\SmithLeft,\SmithRight$ as in \cref{def:rounding-matrix}.
The reductions of $\SmithLeft$ and $\SmithRight$ modulo $\LARModulus$
are invertible over $\LARField$. Hence the same identity modulo
$\LARModulus$ transforms $\RoundingMatrix\bmod\LARModulus$ into
$\operatorname{diag}(\IdentityMatrix_{\RoundingRank},
\ZeroMatrix_{\OutputDim\times\OutputDim})$.
Consequently,
\[
  \operatorname{rank}_{\LARField}(\RoundingMatrix\bmod\LARModulus)
  =\RoundingRank,\qquad
  \dim_{\LARField}\ker(\RoundingMatrix\bmod\LARModulus)
  =\OutputDim.
\]
Taking determinants gives
$|\det\RoundingMatrix|=\LARModulus^{\OutputDim}>0$, so
$\RoundingMatrix$ has full rank $\BlockDim$ over $\QQ$.

Since $\SmithRight\ZZ^{\BlockDim}=\ZZ^{\BlockDim}$, the lattice
generated by the columns of $\RoundingMatrix$ satisfies
\[
  \SmithLeft\LatticeGen{\RoundingMatrix}
  =\ZZ^{\RoundingRank}\times\LARModulus\ZZ^{\OutputDim}.
\]
The unimodular map $\SmithLeft$ preserves both $\ZZ^{\BlockDim}$
and $\LARModulus\ZZ^{\BlockDim}$, so
$\LARModulus\ZZ^{\BlockDim}\subseteq
\LatticeGen{\RoundingMatrix}\subseteq\ZZ^{\BlockDim}$.
Its quotient is
\[
  \ZZ^{\BlockDim}/\LatticeGen{\RoundingMatrix}
  \cong
  \ZZ^{\BlockDim}/
    (\ZZ^{\RoundingRank}\times\LARModulus\ZZ^{\OutputDim})
  \cong\LARField^{\OutputDim}.
\]
For a fixed choice of $\SmithLeft$, this isomorphism sends the coset
of $\IntegerVec\in\ZZ^{\BlockDim}$ to the last $\OutputDim$ coordinates
of $\SmithLeft\IntegerVec$, reduced modulo $\LARModulus$.
Thus the quotient has $\LARModulus^{\OutputDim}$ cosets.

If $\RoundingRank=0$, then
$\RoundingMatrix\bmod\LARModulus=\ZeroMatrix$,
$\LatticeGen{\RoundingMatrix}=\LARModulus\ZZ^{\BlockDim}$, and the
quotient is $\LARField^{\BlockDim}$.
If $\RoundingRank=\BlockDim$, then $\RoundingMatrix$ is unimodular,
$\LatticeGen{\RoundingMatrix}=\ZZ^{\BlockDim}$, and the quotient has
one element.
For $0<\RoundingRank<\BlockDim$, the modular image and the quotient
both have positive dimension strictly less than $\BlockDim$.
The condition $\OutputDim\geq1$ is exactly the requirement that
the quotient contain more than one coset.

\end{proof}

\subsection{Conditioning and statistical distance}
\label{sec:foundations-conditioning}

\begin{proof}[Proof of the conditioning identities in \cref{sec:probability-notation}]
On $\ProbabilityEvent$, conditioning rescales each probability by
$1/\FiniteProbabilityLaw(\ProbabilityEvent)$, and off the event it
sets that probability to zero. Thus
\[
\begin{split}
2\StatisticalDistance{\FiniteProbabilityLaw}
 {\FiniteProbabilityLaw(\cdot\mid\ProbabilityEvent)}
 &=\sum_{\FiniteSamplePoint\in\ProbabilityEvent}
 \FiniteProbabilityLaw(\FiniteSamplePoint)
 \left(\frac1{\FiniteProbabilityLaw(\ProbabilityEvent)}-1\right)
 +\FiniteProbabilityLaw(\ProbabilityEvent^c)\\
 &=2\FiniteProbabilityLaw(\ProbabilityEvent^c).
\end{split}
\]
For independent variables restricted separately to coordinate events,
both the joint point probability and the probability of the restriction
factor. Dividing the former by the latter proves the product conditional
law and hence preservation of independence.
\end{proof}

The contraction, coupling, and product bounds in
\cref{sec:probability-notation} are standard finite-state
inequalities~\cite[Chapter~4]{LevinPeres2017}. Their proofs extend
to countable spaces by absolute summability.

\subsection{Gaussian exponential moments}
\label{sec:foundations-gaussian}

\begin{proof}[Proof of \cref{lem:gaussian-linear-mgf}]
  Completing the square expresses the expectation as the right-hand
  exponential times
  \[
    \frac{\GaussianWeight_{\GaussianWidth,\ZeroVec}
      (\Lattice-\GaussianWidth^2\GaussianLaplaceParameter
    \GaussianTestDirection/(2\pi))}
    {\GaussianWeight_{\GaussianWidth,\ZeroVec}(\Lattice)}.
  \]
  For rank zero, both sides of the claimed inequality equal one.
  For positive rank, regard $\Lattice$ as a full-rank lattice in its
  Euclidean span. Poisson summation within $\Span_{\RR}(\Lattice)$
  expresses the translated Gaussian mass as a sum of positive dual
  Gaussian coefficients multiplied by cosines. Each cosine is at
  most one, and all are one at zero shift, so the ratio is at most one.
\end{proof}

\subsection{Quotient labels and translation equivariance}
\label{sec:foundations-lar}

\begin{proof}[Proof of \cref{lem:lar-properties}]
  \begin{description}
    \item[Well-defined quotient encoding.]
      For integer vectors $\RoundingInput,\IntegerVec$, the identity
      \[
        \floor{\RoundingMatrix\RoundingInput/\LARModulus
        +\RoundingMatrix\IntegerVec}
        =\floor{\RoundingMatrix\RoundingInput/\LARModulus}
        +\RoundingMatrix\IntegerVec
      \]
      proves the stated translation formula.
      By \cref{lem:rounding-smith-properties}, the quotient projection
      has kernel $\LatticeGen{\RoundingMatrix}$ and is surjective.
      The change $\RoundingMatrix\IntegerVec$ lies in this kernel,
      proving independence of the integer lift.

    \item[Translation equivariance.]
      Since $\SmithRight$ is unimodular, its reduction modulo $\LARModulus$
      is invertible, so $\RoundingDirection$ has full column rank. Moreover,
      \[
        \RoundingMatrix\RoundingDirectionInt
        =\LARModulus\SmithLeft^{-1}\QuotientSelector.
      \]
      Thus $(\RoundingMatrix\bmod\LARModulus)\RoundingDirection
      =\ZeroMatrix$ over $\LARField$. Since the modular kernel has dimension
      $\OutputDim$ by \cref{sec:rounding-matrices}, the columns of
      $\RoundingDirection$ form a basis of that kernel.
      Applying coordinatewise floor to integer lifts therefore gives
      \[
        \RoundingMap\bigl(\CanonicalLift{\RoundingInput}
        +\RoundingDirectionInt\CanonicalLift{\RoundingOffset}\bigr)
        =\RoundingMap(\CanonicalLift{\RoundingInput})
        +\SmithLeft^{-1}\QuotientSelector\CanonicalLift{\RoundingOffset}.
      \]
      By the first property, reducing the shifted input to its canonical lift
      does not change the encoded label. Applying $\QuotientProjection$
      modulo $\LARModulus$ proves the translation-equivariance claim of
      \cref{lem:lar-properties}.

    \item[Balanced fibers.]
      The image of $\RoundingDirection$ is an $\OutputDim$-dimensional
      subspace of $\LARField^{\BlockDim}$.
      For each fixed $\RoundingInput\in\LARField^{\BlockDim}$, the map
      $\RoundingOffset\mapsto
      \RoundingInput+\RoundingDirection\RoundingOffset$ is a bijection
      from $\LARField^{\OutputDim}$ onto the coset
      $\RoundingInput+\operatorname{im}(\RoundingDirection)$.
      By translation equivariance in \cref{lem:lar-properties}, its label is
      $\LARLabelMap(\RoundingInput)+\RoundingOffset$, so each label
      occurs exactly once on that coset. There are
      $\LARModulus^{\BlockDim-\OutputDim}=\LARModulus^{\RoundingRank}$ such
      cosets, so each label has exactly that many preimages.\qedhere
  \end{description}
\end{proof}

\subsection{Information-theoretic identifiability}
\label{sec:foundations-lar-identifiability}

\begin{proof}[Proof of \cref{lem:lar-information-theoretic-identifiability}]
  Let $\LARPublicMatrix$ be uniform on
  $\LARField^{\BlockDim\times\SecretDim}$, and fix distinct secrets
  $\LARSecret,\LARSecret'\in\LARField^{\SecretDim}$.
  Choose $\LARSecretSeparator\in\LARField^{\SecretDim}$ with
  $\LARSecretSeparator^{\trans}(\LARSecret-\LARSecret')=1$;
  such a vector exists because $\LARSecret-\LARSecret'\ne\ZeroVec$.
  For any fixed $\LARLabelShift\in\LARField^{\OutputDim}$, the matrix
  $\LARPublicMatrix':=\LARPublicMatrix+
  (\RoundingDirection\LARLabelShift)\LARSecretSeparator^{\trans}$
  is also uniform. Translation equivariance from
  \cref{lem:lar-properties} gives
  \[
    \begin{aligned}
      \LARLabelMap(\LARPublicMatrix'\LARSecret)
        -\LARLabelMap(\LARPublicMatrix'\LARSecret')
      &=\LARLabelMap(\LARPublicMatrix\LARSecret)
        +\LARLabelShift(\LARSecretSeparator^{\trans}\LARSecret)\\
      &\quad-\LARLabelMap(\LARPublicMatrix\LARSecret')
        -\LARLabelShift(\LARSecretSeparator^{\trans}\LARSecret')\\
      &=\LARLabelMap(\LARPublicMatrix\LARSecret)
        -\LARLabelMap(\LARPublicMatrix\LARSecret')+\LARLabelShift.
    \end{aligned}
  \]
  The difference law is translation-invariant on the finite label space,
  hence uniform. The two labels therefore coincide with probability
  $\LARModulus^{-\OutputDim}$.

  Fix the true secret and one incorrect candidate.
  By this uniform difference law and independence of the matrices,
  the candidate matches all
  labels with probability $\LARModulus^{-\OutputDim\SampleCount}$.
  A union bound over the $\LARModulus^{\SecretDim}-1$ incorrect candidates
  proves the stated recovery bound.

  Exhaustive search recovers the secret whenever it is the unique
  candidate explaining every label. For $\OutputDim=1$, take
  $\SampleCount=\SecretDim+\LARRecoverySlack$ with an integer
  $\LARRecoverySlack\geq0$. The recovery failure probability is at most
  \[
    (1-\LARModulus^{-\SecretDim})\LARModulus^{-\LARRecoverySlack}
    <\LARModulus^{-\LARRecoverySlack}.
  \]
  For fixed output dimension and confidence, $O(\SecretDim)$ examples
  therefore suffice for exact recovery, which implies information-theoretic
  probably approximately correct (PAC) learnability under uniform inputs.

  The consistency test accepts exactly when some secret explains
  every label. The sampled common secret always explains a real table,
  so the test accepts $\LARReal_{\SampleCount}$ with probability one.
  In the ideal experiment, condition on any public matrix table and fix
  a candidate secret. The independent uniform labels match its outputs
  with probability $\LARModulus^{-\OutputDim\SampleCount}$.
  A union bound over all $\LARModulus^{\SecretDim}$ candidates bounds
  the acceptance probability on $\LARIdeal_{\SampleCount}$ by
  $\LARModulus^{\SecretDim-\OutputDim\SampleCount}$.
  Thus the distinguishing advantage is at least
  $1-\LARModulus^{\SecretDim-\OutputDim\SampleCount}$ when
  $\OutputDim\SampleCount>\SecretDim$.
  Both procedures enumerate $\LARModulus^{\SecretDim}$ secrets and generally
  take exponential time, so neither contradicts
  \cref{def:decisional-lar-assumption} or the later conditional hardness
  results for efficient learning.
\end{proof}

\subsection{From uniform-zero-fiber $\LWE$ to decisional $\LAR$}
\label{sec:zero-fiber-lar-reduction-proof}

\begin{proof}[Proof of \cref{thm:uniform-zero-fiber-lwe-to-dlar}]
  Fix $\LARDistinguisher$ and $\SampleCount$. We use the following
  sequence of games, using the standard indistinguishability and
  failure-event transitions described in~\cite[Section~1.1]{EPRINT:Shoup04}.
  Each game constructs the indexed family
  $\set{(\LARPublicMatrix_{\SampleIndex},\RoundingLabel_{\SampleIndex})}
  _{\SampleIndex=1}^{\SampleCount}$, retaining repeated entries,
  gives it to $\LARDistinguisher$ together with
  $(\secparam,\pparam)$, and outputs its decision bit.
  The displays below specify how this family is generated.
  The \emph{view} of $\LARDistinguisher$ means this input family;
  its public parameters are fixed throughout the games. We suppress these
  common public inputs in the acceptance probabilities below. Auxiliary
  random choices are kept internal to the game. Every sampling instruction uses
  fresh independent randomness; variables sampled before a loop are
  shared across its iterations. Vector and matrix arithmetic is over
  $\LARField$. The games specify probability experiments with exact
  uniform draws; a worst-case polynomial-time implementation is
  constructed at the end of the proof.
  All views are classical, so their distributional identities and
  mixture formulas are independent of the implementation of
  $\LARDistinguisher$. The contraction bound in
  \cref{sec:probability-notation} applies throughout.
  The reduction generates the complete table classically, invokes
  $\LARDistinguisher$ once, and returns its measured output bit without
  inspecting or rewinding its quantum state.
  \gamedes{0}
  This is the real $\LAR$ experiment:
  \[
    \set{
      \set{(\LARPublicMatrix_{\SampleIndex},\RoundingLabel_{\SampleIndex})}
      _{\SampleIndex=1}^{\SampleCount}
      \;:\;
      \begin{array}{@{}l@{}}
        \LWESecret\sample\Unif{\LARField^{\SecretDim}}\\
        \cfor\ \SampleIndex=1,\ldots,\SampleCount\ \text{\bfseries do}\\
        \quad\LARPublicMatrix_{\SampleIndex}\sample
        \Unif{\LARField^{\BlockDim\times\SecretDim}};\;
        \RoundingLabel_{\SampleIndex}:=
        \LARLabelMap(\LARPublicMatrix_{\SampleIndex}\LWESecret)
      \end{array}
    }
  \]

  \emph{Difference.}
  This is the starting game, with no modification to the real experiment:
  the view of $\LARDistinguisher$ has law $\LARReal_{\SampleCount}$. Thus
  \[
    \bigl|\Pr[\GAME_0=1]
    -\Pr[\LARDistinguisher(\LARReal_{\SampleCount})=1]\bigr|=0.
  \]
  \gamedes{1}
  We modify $\GAME_0$ by defining the event
  $\ZeroSecretBadEvent:=\set{\LWESecret=\ZeroVec}$.
  On this event, replace the labels by independent uniform vectors
  in $\LARField^{\OutputDim}$:
  \[
    \set{
      \set{(\LARPublicMatrix_{\SampleIndex},\RoundingLabel_{\SampleIndex})}
      _{\SampleIndex=1}^{\SampleCount}
      \;:\;
      \begin{array}{@{}l@{}}
        \LWESecret\sample\Unif{\LARField^{\SecretDim}};\;
        \ZeroSecretBadEvent:=\set{\LWESecret=\ZeroVec}\\
        \cfor\ \SampleIndex=1,\ldots,\SampleCount\ \text{\bfseries do}\\
        \quad\LARPublicMatrix_{\SampleIndex}\sample
        \Unif{\LARField^{\BlockDim\times\SecretDim}}\\
        \quad\cif\ \neg\ZeroSecretBadEvent\ \text{\bfseries then}\ \RoundingLabel_{\SampleIndex}:=
        \LARLabelMap(\LARPublicMatrix_{\SampleIndex}\LWESecret)\\
        \quad\text{\bfseries else}\ \RoundingLabel_{\SampleIndex}\sample\Unif{\LARField^{\OutputDim}}
      \end{array}
    }
  \]

  \emph{Difference.}
  Couple the two games to use the same classical secret and matrices.
  Their input views agree whenever $\ZeroSecretBadEvent$
  does not occur. Let $\LARNonzeroReal_{\SampleCount}$ denote the law
  of the output family in the real $\LAR$ experiment conditioned on
  its sampled secret being nonzero.
  The conditional view in $\GAME_1$ has this law on
  $\overline{\ZeroSecretBadEvent}$ and has law
  $\LARIdeal_{\SampleCount}$ on $\ZeroSecretBadEvent$.
  Its unconditional law is therefore
  \[
    (1-\LARModulus^{-\SecretDim})\LARNonzeroReal_{\SampleCount}
    +\LARModulus^{-\SecretDim}\LARIdeal_{\SampleCount}.
  \]
  On $\ZeroSecretBadEvent$, the labels in $\GAME_0$ are
  all zero because $\LARLabelMap(\ZeroVec)=\ZeroVec$; in $\GAME_1$
  they are independent and uniform. The SD between the conditional
  laws of the two views on this event is exactly
  $1-\LARModulus^{-\OutputDim\SampleCount}$. Since the secret is uniform,
  the difference in acceptance probabilities satisfies
  \begin{equation}
    \label{eq:uzf-zero-secret-reserve}
    \bigl|\Pr[\GAME_0=1]-\Pr[\GAME_1=1]\bigr|
    \leq\LARModulus^{-\SecretDim}
       \bigl(1-\LARModulus^{-\OutputDim\SampleCount}\bigr).
  \end{equation}
  \gamedes{2}
  Independently sample
  $\ReductionVector\sample\Unif{\LARField^{\SecretDim}}$ and define
  \[
    \ReductionBadEvent:=
    \set{\ReductionVector^{\trans}\LWESecret\ne1}.
  \]
  Modify $\GAME_1$ by using independent uniform labels whenever
  $\ReductionBadEvent$ occurs, and real labels otherwise.
  Since $\ReductionVector^{\trans}\ZeroVec=0\ne1$, we have
  $\ZeroSecretBadEvent\subseteq\ReductionBadEvent$.
  Thus the zero-secret case still uses uniform labels:
  \[
    \set{
      \set{(\LARPublicMatrix_{\SampleIndex},\RoundingLabel_{\SampleIndex})}
      _{\SampleIndex=1}^{\SampleCount}
      \;:\;
      \begin{array}{@{}l@{}}
        \LWESecret\sample\Unif{\LARField^{\SecretDim}};\;
        \ReductionVector\sample\Unif{\LARField^{\SecretDim}};\;
        \ReductionBadEvent:=\set{\ReductionVector^{\trans}\LWESecret\ne1}\\
        \cfor\ \SampleIndex=1,\ldots,\SampleCount\ \text{\bfseries do}\\
        \quad\LARPublicMatrix_{\SampleIndex}\sample
        \Unif{\LARField^{\BlockDim\times\SecretDim}}\\
        \quad\cif\ \neg\ReductionBadEvent\ \text{\bfseries then}\ \RoundingLabel_{\SampleIndex}:=
        \LARLabelMap(\LARPublicMatrix_{\SampleIndex}\LWESecret)\\
        \quad\text{\bfseries else}\ \RoundingLabel_{\SampleIndex}\sample\Unif{\LARField^{\OutputDim}}
      \end{array}
    }
  \]

  \emph{Difference.}
  For each nonzero $\LWESecret$, exactly a
  $1/\LARModulus$ fraction of the choices of $\ReductionVector$ satisfy
  $\ReductionVector^{\trans}\LWESecret=1$: the nonzero linear
  functional $\ReductionVector\mapsto
  \ReductionVector^{\trans}\LWESecret$ has exactly
  $\LARModulus^{\SecretDim-1}$ preimages of each field element.
  Thus
  \[
    \Pr[\overline{\ReductionBadEvent}]
    =\frac{\Pr[\overline{\ZeroSecretBadEvent}]}{\LARModulus}
    =\frac{1-\LARModulus^{-\SecretDim}}{\LARModulus},
  \]
  and $\LWESecret$ conditioned on $\overline{\ReductionBadEvent}$ is
  uniform over $\LARField^{\SecretDim}\setminus\set{\ZeroVec}$.
  The conditional view in $\GAME_2$ has law
  $\LARNonzeroReal_{\SampleCount}$ on $\overline{\ReductionBadEvent}$
  and law $\LARIdeal_{\SampleCount}$ on $\ReductionBadEvent$.
  Its unconditional law is
  \[
    \frac{1-\LARModulus^{-\SecretDim}}{\LARModulus}
      \LARNonzeroReal_{\SampleCount}
    +\left(1-\frac{1-\LARModulus^{-\SecretDim}}{\LARModulus}\right)
      \LARIdeal_{\SampleCount}.
  \]
  \gamedes{3}
  We replace the direct sampling and event-dependent label rule in
  $\GAME_2$ by a transformation of real uniform-zero-fiber $\LWE$ blocks:
  \begin{equation*}
    \set{
      \set{(\LARPublicMatrix_{\SampleIndex},\RoundingLabel_{\SampleIndex})}
      _{\SampleIndex=1}^{\SampleCount}
      \;:\;
      \begin{array}{@{}l@{}}
        \LWESecret\sample\Unif{\LARField^{\SecretDim}};\;
        \ReductionVector\sample\Unif{\LARField^{\SecretDim}}\\
        \cfor\ \SampleIndex=1,\ldots,\SampleCount\ \text{\bfseries do}\\
        \quad\LWEPublicMatrix_{\SampleIndex}\sample
        \Unif{\LARField^{\BlockDim\times\SecretDim}};\;
        \LWEError_{\SampleIndex}\sample\Unif{\LWEZeroFiber};\;
        \RoundingLabel_{\SampleIndex}\sample\Unif{\LARField^{\OutputDim}}\\
        \quad\LWEResponse_{\SampleIndex}:=
        \LWEPublicMatrix_{\SampleIndex}\LWESecret+\LWEError_{\SampleIndex};\;
        \LARPublicMatrix_{\SampleIndex}:=
        -\LWEPublicMatrix_{\SampleIndex}
        +(\LWEResponse_{\SampleIndex}
        +\RoundingDirection\RoundingLabel_{\SampleIndex})\ReductionVector^{\trans}
      \end{array}
    }
  \end{equation*}
  Here the secret, input matrices, and errors are drawn by the real
  $\LWE$ experiment. A procedure given only the blocks
  $\set{(\LWEPublicMatrix_{\SampleIndex},\LWEResponse_{\SampleIndex})}
  _{\SampleIndex=1}^{\SampleCount}$ can produce the displayed output
  matrices by sampling $\ReductionVector$ and the labels and applying
  the last assignment. It does not need the secret or errors.

  \emph{Difference.}
  This modification preserves the output distribution. To prove this, define
  $\RoundingInput_{\SampleIndex}:=\LWEError_{\SampleIndex}
  +\RoundingDirection\RoundingLabel_{\SampleIndex}$.
  By \cref{lem:zero-fiber-decomposition}, these vectors are independent
  and uniform over $\LARField^{\BlockDim}$, independent of all input
  matrices $\LWEPublicMatrix_{\SampleIndex}$, the secret, and
  $\ReductionVector$, with
  $\RoundingLabel_{\SampleIndex}=\LARLabelMap(\RoundingInput_{\SampleIndex})$.
  The matrices in the rank-one transformation above satisfy
  \begin{equation*}
    \LARPublicMatrix_{\SampleIndex}
    =-\LWEPublicMatrix_{\SampleIndex}
    \bigl(\IdentityMatrix_{\SecretDim}
    -\LWESecret\ReductionVector^{\trans}\bigr)
    +\RoundingInput_{\SampleIndex}\ReductionVector^{\trans}.
  \end{equation*}
  Fix $\LWESecret$ and $\ReductionVector$ such that
  $\ReductionBadEvent$ does not occur. Multiplying this identity by
  $\LWESecret$ gives
  \[
    \LARPublicMatrix_{\SampleIndex}\LWESecret
    =\RoundingInput_{\SampleIndex},\qquad
    \LARLabelMap(\LARPublicMatrix_{\SampleIndex}\LWESecret)
    =\RoundingLabel_{\SampleIndex}.
  \]
  The linear map from
  $(\LWEPublicMatrix_{\SampleIndex},\RoundingInput_{\SampleIndex})$
  to $\LARPublicMatrix_{\SampleIndex}$ is surjective: any prescribed
  $\LARPublicMatrix_{\SampleIndex}$ is obtained by taking
  $\LWEPublicMatrix_{\SampleIndex}=-\LARPublicMatrix_{\SampleIndex}$ and
  $\RoundingInput_{\SampleIndex}=\LARPublicMatrix_{\SampleIndex}\LWESecret$.
  Each fiber of this surjective linear map is a translate of its
  kernel and hence has the same cardinality. The map therefore sends
  the uniform input pair to a uniform matrix.
  The blocks remain independent for these fixed choices of
  $\LWESecret$ and $\ReductionVector$, exactly as in $\GAME_2$.

  Now fix $\LWESecret$ and $\ReductionVector$ such that
  $\ReductionBadEvent$ occurs. Then
  \[
    \det\bigl(\IdentityMatrix_{\SecretDim}
    -\LWESecret\ReductionVector^{\trans}\bigr)
    =1-\ReductionVector^{\trans}\LWESecret\ne0.
  \]
  For these fixed values, conditional on all $\RoundingInput_{\SampleIndex}$,
  the transformed-matrix expression above defines an affine bijection in
  each input matrix $\LWEPublicMatrix_{\SampleIndex}$.
  Hence the output matrices are independent and
  uniform, independently of the vectors $\RoundingInput_{\SampleIndex}$
  and their labels. These labels are themselves independent and uniform.
  This conditional output law is $\LARIdeal_{\SampleCount}$ for
  every such fixed secret and auxiliary vector, so it remains the
  same after averaging over their conditional law.
  On $\overline{\ReductionBadEvent}$ the previous argument gives the
  real branch of $\GAME_2$. The secret and auxiliary vector have the
  same joint law in both games, proving equality of the output laws.
  Therefore,
  \[
    \bigl|\Pr[\GAME_2=1]-\Pr[\GAME_3=1]\bigr|=0.
  \]
  \gamedes{4}
  We replace the real uniform-zero-fiber $\LWE$ blocks in $\GAME_3$
  by ideal $\LWE$ blocks, keeping the transformation unchanged:
  \[
    \set{
      \set{(\LARPublicMatrix_{\SampleIndex},\RoundingLabel_{\SampleIndex})}
      _{\SampleIndex=1}^{\SampleCount}
      \;:\;
      \begin{array}{@{}l@{}}
        \ReductionVector\sample\Unif{\LARField^{\SecretDim}}\\
        \cfor\ \SampleIndex=1,\ldots,\SampleCount\ \text{\bfseries do}\\
        \quad\LWEPublicMatrix_{\SampleIndex}\sample
        \Unif{\LARField^{\BlockDim\times\SecretDim}};\;
        \LWEResponse_{\SampleIndex}\sample\Unif{\LARField^{\BlockDim}};\;
        \RoundingLabel_{\SampleIndex}\sample\Unif{\LARField^{\OutputDim}}\\
        \quad\LARPublicMatrix_{\SampleIndex}:=
        -\LWEPublicMatrix_{\SampleIndex}
        +(\LWEResponse_{\SampleIndex}
        +\RoundingDirection\RoundingLabel_{\SampleIndex})\ReductionVector^{\trans}
      \end{array}
    }
  \]

  \emph{Difference.}
  The gap between $\GAME_3$ and $\GAME_4$ is exactly the advantage of
  the following procedure $\IdealLWEReduction$, defined using exact
  uniform auxiliary choices. On input
  $\set{(\LWEPublicMatrix_{\SampleIndex},\LWEResponse_{\SampleIndex})}
  _{\SampleIndex=1}^{\SampleCount}$, it samples
  $\ReductionVector\sample\Unif{\LARField^{\SecretDim}}$ and
  $\RoundingLabel_{\SampleIndex}\sample\Unif{\LARField^{\OutputDim}}$
  independently of the input and one another, computes
  \[
    \LARPublicMatrix_{\SampleIndex}:=
      -\LWEPublicMatrix_{\SampleIndex}
      +(\LWEResponse_{\SampleIndex}
        +\RoundingDirection\RoundingLabel_{\SampleIndex})
        \ReductionVector^{\trans},
  \]
  and gives the resulting indexed family of matrices and labels,
  together with $(\secparam,\pparam)$, to $\LARDistinguisher$.
  It returns the distinguisher's output bit.
  It simulates $\GAME_3$ on real input and $\GAME_4$ on ideal
  input. This procedure performs polynomially many field operations and one
  call to $\LARDistinguisher$. It does not test either bad event or
  reveal $\ReductionVector$. Hence
  \begin{equation}
    \label{eq:uzf-lwe-hybrid-hop}
    \bigl|\Pr[\GAME_3=1]-\Pr[\GAME_4=1]\bigr|
    =\LWEAdvantage{\IdealLWEReduction}(\secpar,\SampleCount).
  \end{equation}
  \gamedes{5}
  We replace the affine construction in $\GAME_4$ by direct sampling
  of independent uniform matrices and labels:
  \[
    \set{
      \set{(\LARPublicMatrix_{\SampleIndex},\RoundingLabel_{\SampleIndex})}
      _{\SampleIndex=1}^{\SampleCount}
      \;:\;
      \begin{array}{@{}l@{}}
        \cfor\ \SampleIndex=1,\ldots,\SampleCount\ \text{\bfseries do}\\
        \quad\LARPublicMatrix_{\SampleIndex}\sample
        \Unif{\LARField^{\BlockDim\times\SecretDim}};\;
        \RoundingLabel_{\SampleIndex}\sample\Unif{\LARField^{\OutputDim}}
      \end{array}
    }
  \]

  \emph{Difference.}
  This modification preserves the output distribution.
  To see this, condition in $\GAME_4$ on $\ReductionVector$,
  all $\LWEResponse_{\SampleIndex}$, and all $\RoundingLabel_{\SampleIndex}$.
  The rank-one matrix transformation defined above is an affine bijection
  in each independent uniform $\LWEPublicMatrix_{\SampleIndex}$.
  Thus the output matrices are independent and uniform, independently
  of the labels, as in $\GAME_5$. By its sampling rule, $\GAME_5$ is
  exactly the ideal $\LAR$ experiment. Consequently,
  \[
    \bigl|\Pr[\GAME_4=1]-\Pr[\GAME_5=1]\bigr|=0.
  \]

  \paragraph{Combining the games.}
  The initial and final games satisfy
  \[
    \LARDecisionalAdv{\LARDistinguisher}(\secpar,\SampleCount)
    =\bigl|\Pr[\GAME_0=1]-\Pr[\GAME_5=1]\bigr|.
  \]
  Since $\GAME_2$ and $\GAME_3$ have the same output law, as do
  $\GAME_4$ and $\GAME_5$, \cref{eq:uzf-lwe-hybrid-hop} gives
  \[
    \bigl|\Pr[\GAME_2=1]-\Pr[\GAME_5=1]\bigr|
    =\LWEAdvantage{\IdealLWEReduction}(\secpar,\SampleCount).
  \]
  The laws of the views in $\GAME_1$ and $\GAME_2$ are mixtures of the same two
  distributions, $\LARNonzeroReal_{\SampleCount}$ and
  $\LARIdeal_{\SampleCount}$, with respective real-branch weights
  $\Pr[\overline{\ZeroSecretBadEvent}]$ and
  $\Pr[\overline{\ZeroSecretBadEvent}]/\LARModulus$.
  Subtracting the ideal acceptance probability and taking absolute values gives
  \begin{equation*}
    \begin{aligned}
      \bigl|\Pr[\GAME_1=1]-\Pr[\GAME_5=1]\bigr|
      &=\LARModulus\,
      \bigl|\Pr[\GAME_2=1]-\Pr[\GAME_5=1]\bigr|\\
      &=\LARModulus\,
      \LWEAdvantage{\IdealLWEReduction}(\secpar,\SampleCount).
    \end{aligned}
  \end{equation*}
  The triangle inequality and \eqref{eq:uzf-zero-secret-reserve} yield
  \begin{equation}
    \label{eq:uzf-ideal-advantage}
    \begin{aligned}
      \LARDecisionalAdv{\LARDistinguisher}(\secpar,\SampleCount)
      &\leq\bigl|\Pr[\GAME_0=1]-\Pr[\GAME_1=1]\bigr|
      +\bigl|\Pr[\GAME_1=1]-\Pr[\GAME_5=1]\bigr|\\
      &\leq\LARModulus\,
        \LWEAdvantage{\IdealLWEReduction}(\secpar,\SampleCount)
        +\LARModulus^{-\SecretDim}
          \bigl(1-\LARModulus^{-\OutputDim\SampleCount}\bigr).
    \end{aligned}
  \end{equation}

  Implement the $\SecretDim+\OutputDim\SampleCount$ auxiliary field
  choices of $\IdealLWEReduction$ by standard capped rejection sampling
  \cite[Section~III.5.1]{Devroye1986}, independently of the input blocks,
  with joint statistical error at most
  \[
    \ReductionFieldSamplingError
    :=\tfrac12\LARModulus^{-(\SecretDim+\OutputDim\SampleCount+1)}.
  \]
  Both the number of choices and
  $\log(1/\ReductionFieldSamplingError)$ are polynomially bounded in
  $\secpar$, so the resulting $\LWEReduction$ is uniform QPT with a
  worst-case polynomial running-time bound, using the same input blocks
  and distinguisher call.
  The standard failure-event bound \cite[Section~1.1]{EPRINT:Shoup04},
  applied in each $\LWE$ experiment, changes the advantage by at most
  $2\ReductionFieldSamplingError$.
  Thus the additional loss
  $2\LARModulus\ReductionFieldSamplingError
  =\LARModulus^{-\SecretDim-\OutputDim\SampleCount}$
  is absorbed by the unused zero-secret allowance in
  \eqref{eq:uzf-ideal-advantage}, proving the bound in
  \cref{thm:uniform-zero-fiber-lwe-to-dlar}.
\end{proof}

\subsection{From Gaussian $\LWE$ to uniform-zero-fiber $\LWE$}
\label{sec:gaussian-zero-fiber-reduction-proof}

\begin{proof}[Proof of \cref{thm:gaussian-lwe-to-uniform-zero-fiber-lwe}]
  Fix a positive integer $\TransportAccuracyOrder$, a distinguisher
  $\LWEDistinguisher$, and a block count $\SampleCount$. Use the sampler
  $\TransportSampler_{\TransportAccuracyOrder}$ from
  \cref{def:gaussian-t-aft}. The reduction below is defined at every
  security parameter; we prove its advantage bound for
  $\secpar\geq\TransportAccuracyThreshold(\TransportAccuracyOrder)$.
  We use the sequence-of-games method~\cite[Section~1.1]{EPRINT:Shoup04}.
  Each game gives
  $\set{(\LWEPublicMatrix'_{\SampleIndex},\LWEResponse'_{\SampleIndex})}
  _{\SampleIndex=1}^{\SampleCount}$ to $\LWEDistinguisher$, together with
  $(\secparam,\pparam)$, and outputs its decision bit. We suppress these
  common public inputs in the acceptance probabilities below. Every
  sampling instruction uses fresh independent randomness; variables
  sampled before a loop are shared across its iterations. The transport
  pairs remain internal to the games. All arithmetic is over $\LARField$.
  The sampler and all table transformations are classical. Equality of
  table laws is preserved by the quantum distinguisher, and the
  contraction bound in \cref{sec:probability-notation} controls the
  approximate transition.
  \gamedes{0}
  This is the real uniform-zero-fiber $\LWE$ experiment:
  \[
    \set{
      \set{(\LWEPublicMatrix'_{\SampleIndex},\LWEResponse'_{\SampleIndex})}
      _{\SampleIndex=1}^{\SampleCount}
      \;:\;
      \begin{array}{@{}l@{}}
        \LWESecret\sample\Unif{\LARField^{\SecretDim}}\\
        \cfor\ \SampleIndex=1,\ldots,\SampleCount\ \text{\bfseries do}\\
        \quad\LWEPublicMatrix'_{\SampleIndex}\sample
        \Unif{\LARField^{\BlockDim\times\SecretDim}};\;
        \LWEError'_{\SampleIndex}\sample\Unif{\LWEZeroFiber}\\
        \quad\LWEResponse'_{\SampleIndex}:=
        \LWEPublicMatrix'_{\SampleIndex}\LWESecret+\LWEError'_{\SampleIndex}
      \end{array}
    }
  \]

  \emph{Difference.}
  This is the starting game, whose sample distribution is exactly
  $\LWEReal_{\Unif{\LWEZeroFiber},\SampleCount}$. Hence
  \[
    \bigl|\Pr[\GAME_0=1]
    -\Pr[\LWEDistinguisher(\LWEReal_{\Unif{\LWEZeroFiber},\SampleCount})=1]
    \bigr|=0.
  \]
  \gamedes{1}
  We modify $\GAME_0$ by replacing each uniform zero-fiber error with
  an independently transported Gaussian error:
  \[
    \set{
      \set{(\LWEPublicMatrix'_{\SampleIndex},\LWEResponse'_{\SampleIndex})}
      _{\SampleIndex=1}^{\SampleCount}
      \;:\;
      \begin{array}{@{}l@{}}
        \LWESecret\sample\Unif{\LARField^{\SecretDim}}\\
        \cfor\ \SampleIndex=1,\ldots,\SampleCount\ \text{\bfseries do}\\
        \quad\LWEPublicMatrix'_{\SampleIndex}\sample
        \Unif{\LARField^{\BlockDim\times\SecretDim}};\;
        \LWEError_{\SampleIndex}\sample\GaussianErrorLaw\\
        \quad(\TransportMatrix_{\SampleIndex},\TransportOffset_{\SampleIndex})
        \sample\TransportSampler_{\TransportAccuracyOrder}(\secparam,\pparam)\\
        \quad\LWEError'_{\SampleIndex}:=
        \TransportEvaluator(\TransportMatrix_{\SampleIndex},
        \TransportOffset_{\SampleIndex},\LWEError_{\SampleIndex})\\
        \quad\LWEResponse'_{\SampleIndex}:=
        \LWEPublicMatrix'_{\SampleIndex}\LWESecret+\LWEError'_{\SampleIndex}
      \end{array}
    }
  \]

  \emph{Difference.}
  By \cref{def:gaussian-t-aft}, the law of each
  $\LWEError'_{\SampleIndex}$, averaged over both its Gaussian source
  error and its transport pair, is within
  $\secpar^{-\TransportAccuracyOrder}$ of
  $\Unif{\LWEZeroFiber}$. The transported errors are independent across
  blocks, so the product bound for statistical distance
  \cite[Exercise~4.4]{LevinPeres2017} gives
  \[
    \StatisticalDistance{
      \DistributionLaw{\set{\LWEError'_{\SampleIndex}}_{
    \SampleIndex=1}^{\SampleCount}}}{
    \Unif{\LWEZeroFiber}^{\otimes\SampleCount}}
    \leq\SampleCount\secpar^{-\TransportAccuracyOrder}.
  \]
  The entire public-matrix table and the shared secret are independent
  of these errors. Appending them and forming the responses cannot
  increase statistical distance, nor can quantum postprocessing
  distinguish the resulting classical tables with a larger gap. Therefore,
  \[
    \bigl|\Pr[\GAME_0=1]-\Pr[\GAME_1=1]\bigr|
    \leq\SampleCount\secpar^{-\TransportAccuracyOrder}.
  \]
  \gamedes{2}
  We replace the direct sampling of the public matrices in $\GAME_1$
  by a transformation of real Gaussian $\LWE$ blocks:
  \begin{equation*}
    \set{
      \set{(\LWEPublicMatrix'_{\SampleIndex},\LWEResponse'_{\SampleIndex})}
      _{\SampleIndex=1}^{\SampleCount}
      \;:\;
      \begin{array}{@{}l@{}}
        \LWESecret\sample\Unif{\LARField^{\SecretDim}}\\
        \cfor\ \SampleIndex=1,\ldots,\SampleCount\ \text{\bfseries do}\\
        \quad\LWEPublicMatrix_{\SampleIndex}\sample
        \Unif{\LARField^{\BlockDim\times\SecretDim}};\;
        \LWEError_{\SampleIndex}\sample\GaussianErrorLaw\\
        \quad(\TransportMatrix_{\SampleIndex},\TransportOffset_{\SampleIndex})
        \sample\TransportSampler_{\TransportAccuracyOrder}(\secparam,\pparam)\\
        \quad\LWEResponse_{\SampleIndex}:=
        \LWEPublicMatrix_{\SampleIndex}\LWESecret+\LWEError_{\SampleIndex}\\
        \quad\LWEPublicMatrix'_{\SampleIndex}:=
        \TransportMatrix_{\SampleIndex}\LWEPublicMatrix_{\SampleIndex};\;
        \LWEResponse'_{\SampleIndex}:=
        \TransportEvaluator(\TransportMatrix_{\SampleIndex},
        \TransportOffset_{\SampleIndex},\LWEResponse_{\SampleIndex})
      \end{array}
    }
  \end{equation*}

  \emph{Difference.}
  The affine evaluation rule gives
  \[
    \LWEResponse'_{\SampleIndex}
    =\LWEPublicMatrix'_{\SampleIndex}\LWESecret
    +\LWEError'_{\SampleIndex},\qquad
    \LWEError'_{\SampleIndex}:=
    \TransportMatrix_{\SampleIndex}\LWEError_{\SampleIndex}
    +\TransportOffset_{\SampleIndex}.
  \]
  Fix the shared secret and apply \cref{lem:t-aft-properties}
  independently in each block. The entire transformed matrix table is
  exactly uniform and independent of the table of transported errors;
  these joint laws do not depend on the fixed secret. The errors have
  exactly the joint law used in $\GAME_1$, with the same shared secret
  preserved. Thus
  \[
    \bigl|\Pr[\GAME_1=1]-\Pr[\GAME_2=1]\bigr|=0.
  \]
  \gamedes{3}
  We replace the real Gaussian $\LWE$ blocks in $\GAME_2$ by ideal
  $\LWE$ blocks, keeping the affine transformation unchanged:
  \[
    \set{
      \set{(\LWEPublicMatrix'_{\SampleIndex},\LWEResponse'_{\SampleIndex})}
      _{\SampleIndex=1}^{\SampleCount}
      \;:\;
      \begin{array}{@{}l@{}}
        \cfor\ \SampleIndex=1,\ldots,\SampleCount\ \text{\bfseries do}\\
        \quad\LWEPublicMatrix_{\SampleIndex}\sample
        \Unif{\LARField^{\BlockDim\times\SecretDim}};\;
        \LWEResponse_{\SampleIndex}\sample\Unif{\LARField^{\BlockDim}}\\
        \quad(\TransportMatrix_{\SampleIndex},\TransportOffset_{\SampleIndex})
        \sample\TransportSampler_{\TransportAccuracyOrder}(\secparam,\pparam)\\
        \quad\LWEPublicMatrix'_{\SampleIndex}:=
        \TransportMatrix_{\SampleIndex}\LWEPublicMatrix_{\SampleIndex};\;
        \LWEResponse'_{\SampleIndex}:=
        \TransportEvaluator(\TransportMatrix_{\SampleIndex},
        \TransportOffset_{\SampleIndex},\LWEResponse_{\SampleIndex})
      \end{array}
    }
  \]

  \emph{Difference.}
  Define the Gaussian $\LWE$ distinguisher
  $\LWEReduction_{\TransportAccuracyOrder}$ as follows.
  On receiving its input blocks, it samples a fresh transport pair for
  each block using
  $\TransportSampler_{\TransportAccuracyOrder}(\secparam,\pparam)$,
  independently of the entire input experiment, and applies
  the matrix and response transformation in $\GAME_2$. It gives only the transformed
  classical table to $\LWEDistinguisher$ and returns its measured output
  bit, without inspecting or rewinding its quantum state. For fixed
  $\TransportAccuracyOrder$, the sampler runs in worst-case polynomial
  time at every security parameter, so
  the classical preprocessing followed by the QPT call makes
  $\LWEReduction_{\TransportAccuracyOrder}$ uniform QPT. It uses
  $\SampleCount$ input blocks and one call to $\LWEDistinguisher$,
  without needing to compute
  $\TransportAccuracyThreshold(\TransportAccuracyOrder)$.
  It simulates $\GAME_2$ on real Gaussian
  input and $\GAME_3$ on ideal input. Consequently,
  \begin{equation*}
    \bigl|\Pr[\GAME_2=1]-\Pr[\GAME_3=1]\bigr|
    =\GaussianLWEAdvantage{\LWEReduction_{\TransportAccuracyOrder}}
    (\secpar,\SampleCount).
  \end{equation*}
  \gamedes{4}
  We replace the affine construction in $\GAME_3$ by direct sampling
  of independent uniform matrices and responses:
  \[
    \set{
      \set{(\LWEPublicMatrix'_{\SampleIndex},\LWEResponse'_{\SampleIndex})}
      _{\SampleIndex=1}^{\SampleCount}
      \;:\;
      \begin{array}{@{}l@{}}
        \cfor\ \SampleIndex=1,\ldots,\SampleCount\ \text{\bfseries do}\\
        \quad\LWEPublicMatrix'_{\SampleIndex}\sample
        \Unif{\LARField^{\BlockDim\times\SecretDim}};\;
        \LWEResponse'_{\SampleIndex}\sample\Unif{\LARField^{\BlockDim}}
      \end{array}
    }
  \]

  \emph{Difference.}
  Condition in $\GAME_3$ on all transport pairs. By
  \cref{lem:t-aft-properties}, each transformed pair consists of an
  independent uniform matrix and response, and these pairs remain
  independent across blocks. This product law does not depend on the
  conditioned values, so averaging gives exactly
  $\LWEIdeal_{\SampleCount}$, the distribution in $\GAME_4$. Hence
  \[
    \bigl|\Pr[\GAME_3=1]-\Pr[\GAME_4=1]\bigr|=0.
  \]

  \paragraph{Combining the games.}
  The initial and final games are the real and ideal
  uniform-zero-fiber $\LWE$ experiments, respectively. The triangle
  inequality and the bounds above give
  \[
    \begin{aligned}
      \ZeroFiberLWEAdvantage{\LWEDistinguisher}(\secpar,\SampleCount)
      &=\bigl|\Pr[\GAME_0=1]-\Pr[\GAME_4=1]\bigr|\\
      &\leq\bigl|\Pr[\GAME_0=1]-\Pr[\GAME_1=1]\bigr|
      +\bigl|\Pr[\GAME_2=1]-\Pr[\GAME_3=1]\bigr|\\
      &\leq\SampleCount\secpar^{-\TransportAccuracyOrder}
      +\GaussianLWEAdvantage{\LWEReduction_{\TransportAccuracyOrder}}
      (\secpar,\SampleCount).
    \end{aligned}
  \]

  For the hardness implication, suppose that some uniform QPT
  distinguisher $\LWEDistinguisher$, polynomially bounded $\SampleCount$,
  and positive polynomial $\AdvantagePolynomial$ satisfy
  $\ZeroFiberLWEAdvantage{\LWEDistinguisher}(\secpar,\SampleCount)
  \geq1/\AdvantagePolynomial(\secpar)$ for infinitely many $\secpar$.
  Choose fixed positive integers $\SampleCountExponent,\AdvantageExponent$
  such that, for all sufficiently large $\secpar$,
  \[
    \SampleCount(\secpar)\leq\secpar^{\SampleCountExponent},
    \qquad
    \AdvantagePolynomial(\secpar)\leq\secpar^{\AdvantageExponent}.
  \]
  Fix one integer
  $\TransportAccuracyOrder>\SampleCountExponent+\AdvantageExponent+2$.
  For all sufficiently large $\secpar$,
  \[
    \SampleCount(\secpar)\secpar^{-\TransportAccuracyOrder}
    \leq\secpar^{\SampleCountExponent-\TransportAccuracyOrder}
    \leq\frac{1}{2\secpar^{\AdvantageExponent}}
    \leq\frac{1}{2\AdvantagePolynomial(\secpar)}.
  \]
  Apply the quantitative bound just proved with the fixed sampler
  $\TransportSampler_{\TransportAccuracyOrder}$.
  For infinitely many sufficiently large $\secpar$, it gives
  \[
    \GaussianLWEAdvantage{\LWEReduction_{\TransportAccuracyOrder}}
    (\secpar,\SampleCount)
    \geq\frac{1}{2\AdvantagePolynomial(\secpar)}.
  \]
  Since $\TransportAccuracyOrder$ is fixed,
  $\LWEReduction_{\TransportAccuracyOrder}$ is uniform QPT, contradicting
  Gaussian decisional $\LWE$ hardness. This order depends on the fixed
  distinguisher and its polynomial bounds; the public parameter family
  and Gaussian source remain unchanged.
\end{proof}

\subsection{Finite-dimensional calibration}
\label{sec:calibration-principle-proofs}

\begin{proof}[Proof of \cref{thm:stable-finite-dimensional-calibration}]
  The defect of \cref{sec:calibration-neumann} satisfies
  $\|\ProbCalDefect^{\ProbCalPower}\|\leq\ProbCalContraction^{\ProbCalPower}$,
  so its Neumann series converges in operator norm. Multiplying the partial
  sum through power $\ProbCalPower$ by
  $\ProbCalOperator=\ProbCalIdentity-\ProbCalDefect$ on either side
  gives $\ProbCalIdentity-\ProbCalDefect^{\ProbCalPower+1}$, which tends
  to the identity. Thus the series is the inverse, with norm at most
  $\sum_{\ProbCalPower\geq0}\ProbCalContraction^{\ProbCalPower}
  =(1-\ProbCalContraction)^{-1}$.
  Take $\ProbCalCoefficient=\ProbCalOperator^{-1}\ProbCalTarget$,
  use the response identity, and apply the inverse bound to
  $\ProbCalCoefficient-\ProbCalReference
  =\ProbCalOperator^{-1}(\ProbCalTarget-\ProbCalOperator\ProbCalReference)$.
  The mass constraint gives the final claim.
\end{proof}

\begin{proof}[Proof of \cref{cor:calibration-pointwise-positivity}]
  Apply \cref{thm:stable-finite-dimensional-calibration} with reference
  $\ProbCalReference=\ProbCalTarget$; its residual has norm at most
  $\ProbCalContraction\|\ProbCalTarget\|_\infty$. For the second claim use
  $\inf\ProbCalTarget\geq1-\ProbCalUpperDefect$ and
  $\|\ProbCalTarget\|_\infty\leq1+\ProbCalAbsoluteDefect$.
\end{proof}

\begin{proof}[Proof of \cref{cor:calibration-bounded-clipping}]
  Inside $[0,2]$ the clipping error is zero; outside it,
  $|\ProbCalSignedWeight-1|\geq1$ and the clipping error is at most
  $|\ProbCalSignedWeight-1|$. Integration gives the stated loss bound.
\end{proof}

%% file: chapters/appendix-incidence-background.tex
\section{Incidence Geometry and Affine Transport Proofs}
\label{sec:incidence-background}

This appendix fixes the incidence conventions and verifies the geometric
and algebraic statements of \cref{sec:aft-instantiation}.
In the construction, $\TreeIncidenceMatrix$ maps tree-edge coefficients
to vertex loads; the product
$\ExtendedTreeIncidenceMatrix\ExtendedEdgeCoordinateMatrix$ includes
the source-coordinate mixing and the separate charge-injection column
of \cref{cor:general-tree-incidence-transport}.

\begingroup
\mdfsetup{everyline=true,topline=true}

\begin{definition}[Hypergraph]
  \label[definition]{def:hypergraph}
  Let $\BlockDim\geq1$ and use the vertex set $\TransportVertices$
  fixed in \cref{sec:graph-interpretation}. A finite \emph{hypergraph} $\TransportGraph=(\TransportVertices,\GraphEdges)$
  consists of the vertex set $\TransportVertices$, a finite set
  $\GraphEdges$ of edge labels, and a nonempty support
  $\HyperedgeSupport{\GraphEdge}\subseteq\TransportVertices$ for each
  $\GraphEdge\in\GraphEdges$. Distinct labels may have the same support.
  A vertex $\VertexIndex$ is \emph{incident} to $\GraphEdge$ if
  $\VertexIndex\in\HyperedgeSupport{\GraphEdge}$; its \emph{degree}
  is the number of incident hyperedges. An \emph{ordinary loopless graph}
  is a hypergraph with $|\HyperedgeSupport{\GraphEdge}|=2$ for every
  $\GraphEdge\in\GraphEdges$; parallel edges are allowed.
\end{definition}

Orienting individual vertex--edge incidences is standard in bidirected
graph theory~\cite[Section~2]{Zaslavsky1991}. For hyperedges, we use the
head--tail representation and signed incidence matrix of Gallo et
al.~\cite[Section~2]{GalloLongoPallottinoNguyen1993}, with the opposite
global sign convention: $+1$ at tails and $-1$ at heads.

\begin{definition}[Oriented hypergraph and signed incidence matrix]
  \label[definition]{def:oriented-hypergraph}
  An \emph{orientation} of a hypergraph $\TransportGraph$ assigns to each
  $\GraphEdge\in\GraphEdges$ a partition
  $\EdgeTail{\GraphEdge}\mathbin{\dot\cup}\EdgeHead{\GraphEdge}
  =\HyperedgeSupport{\GraphEdge}$ into disjoint tail and head sets.
  Either set may be empty. The hypergraph together with these choices
  is an \emph{oriented hypergraph}. Its \emph{signed incidence matrix} is
  \[
    \GraphIncidenceMatrix\in\set{-1,0,1}^{\BlockDim\times|\GraphEdges|},
    \qquad
    (\GraphIncidenceMatrix)_{\VertexIndex,\GraphEdge}
    :=
    \begin{cases}
      +1 & \text{if }\VertexIndex\in\EdgeTail{\GraphEdge},\\
      -1 & \text{if }\VertexIndex\in\EdgeHead{\GraphEdge},\\
      0 & \text{otherwise.}
    \end{cases}
  \]
  An \emph{ordinary directed graph} is the special case with
  $|\EdgeTail{\GraphEdge}|=|\EdgeHead{\GraphEdge}|=1$ for every edge.
\end{definition}

An ordinary edge column is the tail basis vector minus the head basis
vector; reversing orientation negates it. Mixing tree-edge coefficients
changes the source-to-load matrix while retaining this incidence convention.

\begin{definition}[Integer-weighted incidence matrix]
  \label[definition]{def:integer-weighted-incidence-matrix}
  Let $\TransportGraph$ be an oriented hypergraph. An
  \emph{integer-weighted incidence matrix} of $\TransportGraph$ is a
  matrix $\WeightedIncidenceMatrix\in\ZZ^{\BlockDim\times|\GraphEdges|}$
  satisfying, for every vertex $\VertexIndex$ and hyperedge $\GraphEdge$,
  \[
    (\WeightedIncidenceMatrix)_{\VertexIndex,\GraphEdge}\in
    \begin{cases}
      \ZZ_{>0} & \text{if }\VertexIndex\in\EdgeTail{\GraphEdge},\\
      \ZZ_{<0} & \text{if }\VertexIndex\in\EdgeHead{\GraphEdge},\\
      \set{0} & \text{otherwise.}
    \end{cases}
  \]
  The absolute value of a nonzero entry is its incidence weight;
  weights may differ within a hyperedge.
\end{definition}

\begin{definition}[Paths, trees, and spanning trees]
  \label[definition]{def:graph-tree}
  Let $\TransportGraph$ be an ordinary loopless graph.
  \begin{description}
    \item[Path.]
      A \emph{path} is an alternating sequence of vertices and edges
      in which each edge joins its two neighboring vertices and no
      vertex is repeated.
    \item[Cycle.]
      A \emph{cycle} is a closed alternating sequence of vertices and
      edges satisfying the same incidence condition, with at least
      two edges, no repeated edge, and no repeated vertex other than
      the coincident endpoints.
    \item[Connected graph.]
      A graph is \emph{connected} if every two vertices are joined by a path.
    \item[Tree.]
      A graph is a \emph{tree} if it is connected and contains no cycle.
    \item[Spanning tree.]
      A \emph{spanning tree} of $\TransportGraph$ is a subgraph that is
      a tree and contains all of $\TransportVertices$.
  \end{description}
  For directed graphs these notions refer to the underlying graph
  obtained by forgetting edge orientations.
\end{definition}

\endgroup

\subsection{Proof of the tree incidence basis}
\label{sec:tree-incidence-basis-proof}

\begin{proof}[Proof of \cref{lem:tree-incidence-basis}]
  For $\BlockDim=1$, the empty incidence matrix generates the trivial
  lattice. For $\BlockDim>1$, a tree has a leaf, and deleting the leaf
  and its unique incident edge leaves a tree. Induction therefore
  gives $\BlockDim-1$ edges. Every incidence column has zero sum, so
  its integer span is contained in $\RootLattice$.
  Conversely, let $\IntegerVec\in\RootLattice$. At a leaf, the
  prescribed vertex value determines the unique incident edge
  coefficient up to its orientation sign. Subtract that multiple of
  the incidence column and delete the leaf. The remaining vector
  still has zero sum, so induction determines all remaining edge
  coefficients uniquely. The final vertex equation is automatic.
  Thus every vector in $\RootLattice$ has a unique integral expansion
  in the tree columns, proving the basis and rank assertions.
\end{proof}

\subsection{Image lattice and quotient coordinates}
\label{sec:incidence-image-lattice-proof}

\begin{proof}[Proof of \cref{lem:rounding-image-lattice}]
  Every path column has coordinate sum zero, and $\RoundingColumn$ has
  coordinate sum $\LARModulus$. Hence every vector in
  $\LatticeGen{\RoundingMatrix}$ has coordinate sum divisible by
  $\LARModulus$.
  Conversely, let $\IntegerVec\in\ZZ^{\BlockDim}$ satisfy this
  divisibility condition and set
  $\IntegerCharge:=\OnesVec^{\trans}\IntegerVec/\LARModulus\in\ZZ$.
  Then
  \[
    \IntegerVec-\IntegerCharge\RoundingColumn
    \in\RootLattice
    =\PathIncidenceMatrix\ZZ^{\BlockDim-1},
  \]
  where the equality follows from \cref{lem:tree-incidence-basis}.
  Thus $\IntegerVec\in\LatticeGen{\RoundingMatrix}$, proving the characterization.

  The coordinate-sum homomorphism
  $\IntegerVec\mapsto\OnesVec^{\trans}\IntegerVec\bmod\LARModulus$
  from $\ZZ^{\BlockDim}$ onto $\LARField$ has kernel
  $\LatticeGen{\RoundingMatrix}$. Its quotient is therefore cyclic of
  order $\LARModulus$, which gives the index and the claimed Smith
  normal form. The induced quotient coordinate map is the one specified in
  \cref{def:explicit-rounding-matrix}. The rounding-matrix parameters follow from
  \cref{def:rounding-matrix}.
\end{proof}

\subsection{Canonical endpoint parametrization}
\label{sec:incidence-pullback-proof}

\begin{proof}[Proof of \cref{lem:canonical-endpoint-pullback}]
  For $\EndpointResidue\in\ZeroSumEndpointHyperplane$, the integer vector
  $\RoundingMatrix^{-1}\CanonicalLift{\EndpointResidue}$ is a lift of
  $\CanonicalEndpointPullback(\EndpointResidue)$. Using lift independence
  from \cref{lem:lar-properties}, we obtain
  \[
    \LARLabelMap\bigl(\CanonicalEndpointPullback(\EndpointResidue)\bigr)
    =\QuotientProjection
    \floor{\frac{\CanonicalLift{\EndpointResidue}}{\LARModulus}}
    \bmod\LARModulus
    =\ZeroVec.
  \]
  Thus the pullback indeed takes values in $\LWEZeroFiber$.
  Multiplying its defining congruence by the integer matrix
  $\RoundingMatrix$ gives
  \[
    \RoundingMatrix
    \CanonicalLift{\CanonicalEndpointPullback(\EndpointResidue)}
    \equiv\CanonicalLift{\EndpointResidue}\pmod{\LARModulus}.
  \]
  Hence the proposed inverse recovers $\EndpointResidue$, proving
  injectivity.

  The hyperplane $\ZeroSumEndpointHyperplane$ has
  $\LARModulus^{\BlockDim-1}$ elements. By
  \cref{lem:rounding-image-lattice}, $\RoundingRank=\BlockDim-1$;
  the balanced fibers of \cref{lem:lar-properties} therefore give
  $|\LWEZeroFiber|=\LARModulus^{\BlockDim-1}$.
  The injection is consequently a bijection between these finite sets,
  and the preceding congruence gives the stated inverse on all of
  $\LWEZeroFiber$.
\end{proof}

\subsection{Endpoint and carry criterion}
\label{sec:incidence-carry-proof}

\begin{proof}[Proof of \cref{lem:affine-transport-endpoint-carry-bound}]
  The endpoint $\EndpointResidue$ lies in $\ZeroSumEndpointHyperplane$
  because $\OnesVec^{\trans}\RoundingMatrix\equiv\ZeroVec^{\trans}
  \pmod{\LARModulus}$. Substituting the definitions of the affine output
  and canonical pullback gives the following difference in
  $\LARField^{\BlockDim}$, with
  all products involving $\RoundingMatrix^{-1}$ evaluated over $\QQ$
  before reduction modulo $\LARModulus$:
  \[
    \begin{aligned}
      \TransportOutput-\CanonicalEndpointPullback(\EndpointResidue)
      &=\bigl(\TransportMatrix\CanonicalLift{\TransportInput}+\TransportOffset
      -\RoundingMatrix^{-1}\CanonicalLift{\EndpointResidue}\bigr)
      \bmod\LARModulus\\
      &=\RoundingMatrix^{-1}\bigl(
        \RoundingMatrix(\TransportMatrix\CanonicalLift{\TransportInput}
      +\TransportOffset)-\CanonicalLift{\EndpointResidue}\bigr)
      \bmod\LARModulus\\
      &=\RoundingMatrix^{-1}
      \bigl(\EndpointLoad-\CanonicalLift{\EndpointResidue}\bigr)
      \bmod\LARModulus\\
      &=\LARModulus\RoundingMatrix^{-1}\EndpointCarry\bmod\LARModulus
    \end{aligned}
  \]

  In particular, $\LARModulus\RoundingMatrix^{-1}\EndpointCarry$ is an
  integer vector. Its reduction modulo $\LARModulus$ vanishes exactly when
  $\RoundingMatrix^{-1}\EndpointCarry\in\ZZ^{\BlockDim}$, or equivalently
  $\EndpointCarry\in\LatticeGen{\RoundingMatrix}$.
  Combining this with the lattice characterization of
  \cref{lem:rounding-image-lattice} proves the carry criterion of
  \cref{lem:affine-transport-endpoint-carry-bound}.

  Couple $\TransportOutput$ and
  $\CanonicalEndpointPullback(\EndpointResidue)$ using the same transport
  data and source input. The carry criterion just proved and the
  coupling inequality give
  \[
    \StatisticalDistance{\DistributionLaw{\TransportOutput}}
    {\DistributionLaw{\CanonicalEndpointPullback(\EndpointResidue)}}
    \leq\Pr[\EndpointCarry\notin\LatticeGen{\RoundingMatrix}].
  \]
  By \cref{lem:canonical-endpoint-pullback}, the map
  $\CanonicalEndpointPullback$ is a bijection from
  $\ZeroSumEndpointHyperplane$ onto $\LWEZeroFiber$, so
  \[
    \StatisticalDistance
    {\DistributionLaw{\CanonicalEndpointPullback(\EndpointResidue)}}
    {\Unif{\LWEZeroFiber}}
    =\StatisticalDistance{\DistributionLaw{\EndpointResidue}}
    {\Unif{\ZeroSumEndpointHyperplane}}.
  \]
  The claim follows from the triangle inequality.
\end{proof}

\subsection{Extended incidence determinants}
\label{sec:incidence-extended-determinant-proof}

\begin{proof}[Proof of \cref{lem:tree-incidence-determinant}]
  The tree columns are an integer basis of $\RootLattice$ by
  \cref{lem:tree-incidence-basis}. Summing coordinates identifies
  $\ZZ^{\BlockDim}/\RootLattice$ with $\ZZ$. If
  $\OnesVec^{\trans}\TreeExtensionColumn=0$, the appended column lies in
  $\RootLattice$, so the determinant vanishes. Otherwise, the image of
  the lattice generated by the appended matrix in this quotient is
  $(\OnesVec^{\trans}\TreeExtensionColumn)\ZZ$. Hence its index in
  $\ZZ^{\BlockDim}$ is $|\OnesVec^{\trans}\TreeExtensionColumn|$.
  This finite index equals the absolute determinant of the appended
  square matrix, proving the claim.
\end{proof}

\subsection{Integral transport from an incidence basis}
\label{sec:incidence-general-transport-proof}

\begin{proof}[Proof of \cref{cor:general-tree-incidence-transport}]
  Every column of $\TreeIncidenceMatrix\EdgeCoordinateMatrix$ has
  integer sum zero, and the appended column has sum $\LARModulus$.
  By \cref{lem:rounding-image-lattice}, these columns and the admissible
  initial load lie in $\LatticeGen{\RoundingMatrix}$. Their
  $\RoundingMatrix$-preimages are therefore integral, and every integer
  input yields a load in that lattice and an endpoint in
  $\ZeroSumEndpointHyperplane$.
  Lemma~\ref{lem:tree-incidence-determinant} gives
  $|\det\ExtendedTreeIncidenceMatrix|=\LARModulus$.
  Taking determinants in the defining product and dividing by
  $\det\RoundingMatrix$, whose absolute value is $\LARModulus$ by
  \cref{lem:rounding-image-lattice}, gives
  the determinant identities of \cref{cor:general-tree-incidence-transport}.
  A nonzero determinant modulo the prime $\LARModulus$ gives the
  asserted modular invertibility.
\end{proof}

%% file: chapters/appendix-affine-proofs.tex
\section{Local Kernels and Bounded Sampling}
\label{sec:affine-realization-proofs}

This appendix verifies the body and leg laws of
\cref{sec:probabilistic-transport} and their bounded public samplers.
It defines the color-fiber operators used for local invertibility and
positivity, followed by the rational calibration system and its
implementation.

\subsection{The cycle body law}
\label{sec:cycle-body-law-proof}

The coordinate identity in \cref{def:body-injection} follows by multiplying
the path block of \cref{sec:affine-realization-matrix} and inserting the
two closure entries specified in that definition. Summing its coordinates
telescopes to zero.

With an ideal draw of $\BodyClosureGaussian$, the
$\InteriorVertexCount-1$ independent body source coordinates and this
one independent public Gaussian are $\InteriorVertexCount$ independent
samples from $\Gaussian{\ZZ}{\GaussianWidth}$. Their consecutive cyclic
differences are exactly the body increment $\BodyLoadIncrement$ of
\cref{def:body-injection}, without convolution or smoothing loss.
This joint-law identity remains valid after the prescribed body
placement and conditional on public data sampled independently of the
closure and hidden source. It retains all shared-coordinate dependencies.
The identity is not conditioned on the realized closure or affine offset.

\input{chapters/discrete-body-injection}

\subsection{Safe color slices}
\label{sec:leg-safe-slices-proof}

Under their respective parameter conditions, the slices have the
following interval forms and positive sizes:
\begin{equation*}
  \begin{aligned}
    \LocalSlice_{\LegSliceShift}
    &=\set{\max(0,\LegSliceShift),\ldots,
      \min(\LARModulus-1,\LARModulus-1+\LegSliceShift)},\\
    |\LocalSlice_{\LegSliceShift}|
    &=\LARModulus-|\LegSliceShift|>0,\\
    \LegTrimmedSlice{\LegSliceShift}
    &=\set{\LegTrimMargin+\max(0,\LegSliceShift),\ldots,
      \LARModulus-1-\LegTrimMargin+\min(0,\LegSliceShift)},\\
    |\LegTrimmedSlice{\LegSliceShift}|
    &=\LARModulus-2\LegTrimMargin-|\LegSliceShift|>0.
  \end{aligned}
\end{equation*}

\begin{proof}[Proof of the slice identities following \cref{def:leg-color-slice,def:leg-trimmed-slice}]
For an ordered pair with sum
$\LARModulus-1+\LegSliceShift$, requiring both coordinates to lie
between zero and $\LARModulus-1$ imposes
\[
 0\leq\LocalFieldValue\leq\LARModulus-1,
 \qquad
 \LegSliceShift\leq\LocalFieldValue
             \leq\LARModulus-1+\LegSliceShift.
\]
Intersecting the intervals gives the safe slice of
\cref{def:leg-color-slice}; counting
its integer points gives $\LARModulus-|\LegSliceShift|$.
Requiring both coordinates instead to lie in $\LegInteriorColors$
adds $\LegTrimMargin$ to both lower bounds and subtracts it from
both upper bounds. Their intersection gives the trimmed slice of
\cref{def:leg-trimmed-slice}, with
$\LARModulus-2\LegTrimMargin-|\LegSliceShift|$ integer points.
The stated hypotheses make these counts positive.

For a protected second input,
$|\LegBaselineShift-\LocalSecondError|
\leq|\LegBaselineShift|+\LegCutoff-1<\LARModulus$.
Thus every slice in \cref{def:safe-affine-map} is defined and nonempty.
The stronger trimmed bound gives the same conclusion after deleting
$\LegTrimMargin$ colors at each endpoint. Membership ensures that the
evaluated primary and its prescribed integer companion are both
colors. Finally $\GadgetScale\ne0$ makes the affine function of the
first field input a bijection for each fixed second input.
\end{proof}

\subsection{Off-body carry cancellation}
\label{sec:leg-carry-proof}

For every integer input $(\LocalFirstError,\LocalSecondError)\in\ZZ^2$,
the load formulas give the charge identities
\begin{equation*}
  \begin{aligned}
    \LocalFirstEndpoint+\LocalSecondEndpoint
    &=\LARModulus-1+\LegBaselineShift-\LocalSecondError,\\
    \LocalFirstEndpoint+\LocalSecondEndpoint+\LegAttachmentLoad
    &=\LARModulus-1+\LegBaselineShift+\LocalReservoirBase.
  \end{aligned}
\end{equation*}

We restrict the sampling support to maps $\LocalChartPoint$ for which
the pair sum satisfies
\begin{equation*}
  \CanonicalLift{\LocalFirstEndpoint\bmod\LARModulus}
  \leq\LocalFirstEndpoint+\LocalSecondEndpoint
  =\LARModulus-1+\LegBaselineShift-\LocalSecondError
  <\CanonicalLift{\LocalFirstEndpoint\bmod\LARModulus}+\LARModulus
\end{equation*}
simultaneously for every protected input
$(\LocalFirstError,\LocalSecondError)$ with
$|\LocalFirstError|,|\LocalSecondError|<\LegCutoff$.
\begin{lemma}[Off-body carry cancellation]
  \label[lemma]{lem:leg-deterministic}
  Fix a public map $\LocalChartPoint$ with $\GadgetScale\ne0$.
  If the off-body carry condition above holds for every protected
  input, then for every such input,
  \[
    \floor{\LocalFirstEndpoint/\LARModulus}
    +\floor{\LocalSecondEndpoint/\LARModulus}=0.
  \]
\end{lemma}

\begin{proof}[Proof of \cref{lem:leg-deterministic}]
  Multiplying the local incidence and mixer blocks reproduces the increment
  of \cref{def:leg-injection}. Adding the restriction of the shared baseline
  gives its complete local load. Summing the first two, or all three,
  coordinates gives the pair and triple charge identities above.
  By the stated off-body support condition,
  the integer
  $\LocalFirstEndpoint+\LocalSecondEndpoint
  -\CanonicalLift{\LocalFirstEndpoint\bmod\LARModulus}$
  lies in $\set{0,\ldots,\LARModulus-1}$ and is congruent to
  $\LocalSecondEndpoint$ modulo $\LARModulus$. It therefore equals
  $\CanonicalLift{\LocalSecondEndpoint\bmod\LARModulus}$.
  Subtracting the two canonical residues from the pair sum and dividing
  by $\LARModulus$ gives the claimed cancellation.
\end{proof}

\subsection{Attachment safety}
\label{sec:leg-attachment-proof}

Attachment safety is controlled by the shared baseline margin. We require
\begin{equation*}
  \begin{gathered}
    \PaddingSafeMargin\leq\LocalReservoirBase
    \leq\LARModulus-1-\PaddingSafeMargin,\qquad
    \PaddingInteriorCutoff+\LegCutoff+\PaddingChargeCutoff
    \leq\PaddingSafeMargin,\\
    |(\PlacedBodyIncrement)_{\LegRightVertex}|<\PaddingInteriorCutoff,
    \qquad
    |(\PlacedQueueIncrement)_{\LegRightVertex}|<\PaddingChargeCutoff,
  \end{gathered}
\end{equation*}
\begin{lemma}[Attachment safety]
  \label[lemma]{lem:leg-attachment-safety}
  For every protected leg input and any body and queue contributions
  satisfying the attachment margin and contribution bounds above,
  \[
    0\leq\LegAttachmentLoad
    +(\PlacedBodyIncrement)_{\LegRightVertex}
    +(\PlacedQueueIncrement)_{\LegRightVertex}<\LARModulus.
  \]
  In particular, the assembled attachment has zero carry:
  \[
    \floor{\frac{\LegAttachmentLoad
      +(\PlacedBodyIncrement)_{\LegRightVertex}
      +(\PlacedQueueIncrement)_{\LegRightVertex}}{\LARModulus}}=0.
  \]
\end{lemma}

\begin{proof}[Proof of \cref{lem:leg-attachment-safety}]
  By the local-load construction of \cref{def:leg-injection},
  $\LegAttachmentLoad=\LocalReservoirBase+\LocalSecondError$.
  Since $|\LocalSecondError|<\LegCutoff$, the triangle inequality gives
  \[
    \left|\LocalSecondError
      +(\PlacedBodyIncrement)_{\LegRightVertex}
      +(\PlacedQueueIncrement)_{\LegRightVertex}\right|
    <\LegCutoff+\PaddingInteriorCutoff+\PaddingChargeCutoff
    \leq\PaddingSafeMargin.
  \]
  Combining this bound with
  $\PaddingSafeMargin\leq\LocalReservoirBase
  \leq\LARModulus-1-\PaddingSafeMargin$ places the assembled attachment
  load in $[0,\LARModulus)$, so its carry is zero.
\end{proof}

\subsection{Finite calibration of the line-mark law}
\label{sec:leg-calibration-proof}

We first specify the finite coefficient model and the complete public
law used in \cref{thm:leg-constructive-sampler}.

\begingroup
\mdfsetup{topline=true}
\begin{definition}[Finite affine-map model]
  \label[definition]{def:leg-affine-map-model}
  Equip $\ChartUnrestrictedSpace:=\LARField^3$ with its uniform law.
  Enumerate the protected inputs as
  $\set{(\LocalFirstError_{\LegInputIndex},\LocalSecondError_{\LegInputIndex})}_{
  \LegInputIndex=1}^{\LegInputCount}$, and set
  \[
    \LegAffineEvaluation_{\LegInputIndex}
      :=\LegAffineEvaluation_{\LocalFirstError_{\LegInputIndex},
                             \LocalSecondError_{\LegInputIndex}},
    \qquad
    \LegAllowedColors_{\LegInputIndex}
      :=\LegTrimmedSlice{\LegBaselineShift-\LocalSecondError_{\LegInputIndex}}.
  \]
  Let $\ChartLineIndices$ index the inputs with
  $\LocalFirstError_{\LegInputIndex}=0$, and let
  $\ChartOffLineIndices$ index the remaining inputs.
\end{definition}

\begin{definition}[Regular and marked supports]
  \label[definition]{def:leg-branch-supports}
  Define the \emph{regular support} and \emph{marked support} by
  \[
    \ChartInteriorSupport
      :=\set{\LocalChartPoint\in\ChartUnrestrictedSpace:\GadgetScale\ne0,\
        \LegAffineEvaluation_{\LegInputIndex}(\LocalChartPoint)
        \in\LegAllowedColors_{\LegInputIndex}
        \text{ for every }1\leq\LegInputIndex\leq\LegInputCount},
  \]
  \[
    \ChartLineSupport
      :=\set{\begin{aligned}
        (\LegMarkedFirst,\GadgetScale,0)\in\ChartUnrestrictedSpace:\ &
          \GadgetScale\ne0,\\
          &\LegAffineEvaluation_{\LegInputIndex}
            (\LegMarkedFirst,\GadgetScale,0)
            \in\LegAllowedColors_{\LegInputIndex}
            \text{ for every }\LegInputIndex\in\ChartOffLineIndices
      \end{aligned}}.
  \]
\end{definition}
\endgroup
Regular maps keep both colors interior. Marked maps do so off the line;
on the line their primary is $\LegMarkedFirst$ and their companion is
interior by marked-color admissibility. Every protected second input
remains available, as required by the joint law in \cref{sec:leg-joint-law}.

\paragraph{The complete public law.}
\label{sec:leg-complete-public-law}
Fix public data satisfying \cref{sec:leg-sampling-parameters}.
All densities below are relative to $\Unif{\ChartUnrestrictedSpace}$.
Let $\ChartRegularDensity$ be the regular-branch density constructed
below, with exact rational coordinates in
\cref{sec:leg-branch-calibration}.
Define the uniform marked-branch density by
\begin{equation*}
  \ChartMarkedDensity(\LocalChartPoint)
  :=\frac{\LARModulus^3\EventIndicator{\ChartLineSupport}(\LocalChartPoint)}
  {|\ChartLineSupport|}.
\end{equation*}
The flag $\LegMarkFlag$ equals $0$ on the regular branch and $1$ on
the marked branch.
The joint flag and map law is
\[
  \begin{aligned}
    \LegFlagChartLaw(0,\LocalChartPoint)
      &:=\frac{(1-\LocalAtomWeight)
                 \ChartRegularDensity(\LocalChartPoint)}
                {\LARModulus^3},\\
    \LegFlagChartLaw(1,\LocalChartPoint)
      &:=\frac{\LocalAtomWeight
                 \ChartMarkedDensity(\LocalChartPoint)}
                {\LARModulus^3}.
  \end{aligned}
\]
Its flag marginal is $\BernoulliLaw{\LocalAtomWeight}$; write
$\ChartLaw$ for its map marginal.

We construct the regular-branch density and prove the pair law in
\cref{thm:leg-constructive-sampler}, using the finite model, supports,
and marked density specified above. Fix public data satisfying
\cref{sec:leg-sampling-parameters}; \cref{sec:leg-branch-calibration}
gives the rational coordinates.
Expectations over coefficient triples use the unrestricted uniform law
unless otherwise specified.

\paragraph{Affine two-point counts.}
Under the unrestricted uniform law on coefficient triples, for distinct
indices $\LegInputIndex,\LegSecondInputIndex$ and any colors
$\LocalFieldValue,\LegSecondColor$,
\[
  \Pr[\LegAffineEvaluation_{\LegInputIndex}=\LocalFieldValue,
  \LegAffineEvaluation_{\LegSecondInputIndex}=\LegSecondColor]
  =\LARModulus^{-2}.
\]
Indeed, the two equations over $\LARField$ are
\[
  \begin{aligned}
    \LocalChartBase+\GadgetScale\LocalFirstError_{\LegInputIndex}
    +\GadgetShear\LocalSecondError_{\LegInputIndex}&=\LocalFieldValue,\\
    \LocalChartBase+\GadgetScale\LocalFirstError_{\LegSecondInputIndex}
    +\GadgetShear\LocalSecondError_{\LegSecondInputIndex}&=\LegSecondColor.
  \end{aligned}
\]
Subtracting them gives one nonzero linear equation in
$(\GadgetScale,\GadgetShear)$: distinct protected inputs remain distinct
modulo $\LARModulus$ by the input-separation condition of
\cref{sec:leg-injection}. It has exactly
$\LARModulus$ solutions, each determining $\LocalChartBase$ uniquely.
A single evaluation equation has $\LARModulus^2$ solutions. If the scale
$\GadgetScale$ is also fixed, each of the $\LARModulus$ choices of shear
again determines one intercept. Hence each evaluation is uniform and
independent of the scale, and distinct evaluations are pairwise
independent. These assertions apply only before excluding zero scale
or imposing either safe support; the deletion bounds below account for
all such restrictions.

\begin{lemma}[Safe supports and deletion bounds]
  \label[lemma]{lem:leg-safe-supports}
  Under the feasibility and marked-color admissibility conditions of
  \cref{sec:leg-injection},
  both supports are nonempty and consist of safe maps. For
  $\LocalFieldValue\in\LegAllowedColors_{\LegInputIndex}$,
  \begin{equation*}
    \begin{gathered}
      \Pr(\ChartInteriorSupport^c),\quad
      \Pr(\ChartInteriorSupport^c\mid
      \LegAffineEvaluation_{\LegInputIndex}=\LocalFieldValue)
      \leq\ChartDeletionBound,\\
      \ChartMarkedSurvival:=\Pr(\ChartLineSupport\mid
      \LocalChartBase=\LegMarkedFirst,\GadgetShear=0)
      =\frac{|\ChartLineSupport|}{\LARModulus}
      \geq1-\ChartDeletionBound.
    \end{gathered}
  \end{equation*}
\end{lemma}
\begin{proof}
  Each input excludes at most
  $\LegSumBound+\LegCutoff-1+2\LegTrimMargin$ colors. The unrestricted
  counts above and the zero-scale probability $1/\LARModulus$ give
  the unconditional union bound. Conditional on an allowed color at
  input $\LegInputIndex$, its own color constraint cannot fail;
  each other evaluation and the scale remain uniform. The same union
  bound therefore proves the conditional estimate.
  On the unfiltered marked fiber
  $\LocalChartBase=\LegMarkedFirst,\GadgetShear=0$, each off-line
  evaluation is uniform as $\GadgetScale$ ranges over all of
  $\LARField$. A union bound over its color exclusions and the event
  $\GadgetScale=0$ proves the survival estimate; no joint independence
  on this fiber is needed. Since $\ChartDeletionBound<1$, both supports
  are nonempty. Regular maps and off-line marked evaluations use
  trimmed slices. On the marked line, the primary is a canonical
  color and the companion is interior by the marked-color admissibility
  condition of \cref{sec:leg-injection}.
  Thus every supported map is safe.
\end{proof}

\begin{definition}[Additive color-function space]
  \label[definition]{def:leg-calibration-space}
  Equip the real function space
  \[
    \LegFunctionSpace:=\RR 1\mathbin{\oplus}
    \bigoplus_{\LegInputIndex=1}^{\LegInputCount}
    \left\{\LocalComponentFunction_{\LegInputIndex}\circ
      \LegAffineEvaluation_{\LegInputIndex}:
      \begin{array}{l}
        \operatorname{supp}(\LocalComponentFunction_{\LegInputIndex})
          \subseteq\LegAllowedColors_{\LegInputIndex},\\
        \displaystyle\sum_{\LocalFieldValue\in\LegAllowedColors_{\LegInputIndex}}
          \LocalComponentFunction_{\LegInputIndex}(\LocalFieldValue)=0
      \end{array}\right\}
  \]
  with the function supremum norm on $\ChartUnrestrictedSpace$.
\end{definition}

\begin{proof}[Proof of \cref{thm:leg-constructive-sampler}]
  For a real function $\LegAveragedFunction$ on coefficient triples,
  define its color-fiber averages and centered color imbalances by
  \begin{equation}
    \label{eq:leg-color-averages}
    \begin{aligned}
      \LegColorMean_{\LegInputIndex}^{\LegAveragedFunction}(\LocalFieldValue)
      &:=\LARModulus^{-2}
      \sum_{\substack{\LocalChartPoint\in\ChartUnrestrictedSpace\\
      \LegAffineEvaluation_{\LegInputIndex}(\LocalChartPoint)=\LocalFieldValue}}
      \LegAveragedFunction(\LocalChartPoint),\\
      (\LegColorImbalance_{\LegInputIndex}\LegAveragedFunction)(\LocalChartPoint)
      &:=\EventIndicator{\set{\LegAffineEvaluation_{\LegInputIndex}(\LocalChartPoint)
      \in\LegAllowedColors_{\LegInputIndex}}}
      \left(\LegColorMean_{\LegInputIndex}^{\LegAveragedFunction}
      (\LegAffineEvaluation_{\LegInputIndex}(\LocalChartPoint))
      -\frac1{|\LegAllowedColors_{\LegInputIndex}|}
      \sum_{\LocalFieldValue\in\LegAllowedColors_{\LegInputIndex}}
      \LegColorMean_{\LegInputIndex}^{\LegAveragedFunction}(\LocalFieldValue)\right),\\
      \LegCountReconstruction\LegAveragedFunction
      &:=\LARModulus^{-3}\sum_{\LocalChartPoint\in\ChartUnrestrictedSpace}
      \LegAveragedFunction(\LocalChartPoint)
      +\sum_{\LegInputIndex=1}^{\LegInputCount}
      \LegColorImbalance_{\LegInputIndex}\LegAveragedFunction.
    \end{aligned}
  \end{equation}
  In the unrestricted uniform $L^2$ space,
  $\LegColorImbalance_{\LegInputIndex}$ is the orthogonal projection
  onto the corresponding centered summand of $\LegFunctionSpace$:
  first condition on the evaluation, then subtract the average over
  allowed colors and set the remaining colors to zero. The two-point counts make these
  summands mutually orthogonal and orthogonal to constants.
  Thus $\LegCountReconstruction$ is the orthogonal projection onto
  $\LegFunctionSpace$. It preserves mass and centered color moments
  and restricts to the identity there. For
  $\LocalTestFunction\in\LegFunctionSpace$, define the masked synthesis
  and deletion operator by
  \[
    \LegWeightSynthesis\LocalTestFunction
      :=\EventIndicator{\ChartInteriorSupport}\LocalTestFunction,
    \qquad
    \ChartDeletionOperator\LocalTestFunction
      :=\LegCountReconstruction
        (\EventIndicator{\ChartInteriorSupport^c}\LocalTestFunction).
  \]
  The actual coefficient response on $\LegFunctionSpace$ is therefore
  \[
    \LegCalibrationResponse:=\LegCountReconstruction\LegWeightSynthesis
      =\LocalIdentityOperator-\ChartDeletionOperator.
  \]
  For $\LegAveragedFunction=\EventIndicator{\ChartInteriorSupport^c}
  \LocalTestFunction$, \cref{lem:leg-safe-supports} bounds
  $|\mathbb E\LegAveragedFunction|$ and every
  $|\LegColorMean_{\LegInputIndex}^{\LegAveragedFunction}(\LocalFieldValue)|$
  at an allowed color by
  $\ChartDeletionBound\|\LocalTestFunction\|_\infty$.
  Bounding the mean and the two terms per centered component gives
  \begin{equation*}
    \|\ChartDeletionOperator\|_{\infty\to\infty}
    \leq(2\LegInputCount+1)\ChartDeletionBound
    =\ChartOperatorBound\leq\tfrac14.
  \end{equation*}

  Define the branch-selection odds and marked off-line defect by
  \[
    \ChartMarkOdds:=\frac{\LocalAtomWeight}{1-\LocalAtomWeight},
    \qquad
    \ChartMarkDefect:=\sum_{\LegInputIndex\in\ChartOffLineIndices}
      \LegColorImbalance_{\LegInputIndex}\ChartMarkedDensity.
  \]
  The defect is the sum of projections of $\ChartMarkedDensity$ onto
  the off-line summands.
  Its mass and its color-difference moments at line inputs vanish;
  at each off-line input those moments equal those of $\ChartMarkedDensity$.
  To cancel these moments, set
  $\LegCalibrationTarget:=1-\ChartMarkOdds\ChartMarkDefect$ and solve
  \[
    \LegCalibrationResponse\LocalTestFunction=\LegCalibrationTarget.
  \]
  By \cref{thm:stable-finite-dimensional-calibration}, this equation has
  a unique solution $\LocalTestFunction\in\LegFunctionSpace$. Define
  the regular-branch density by
  \[
    \ChartRegularDensity:=\LegWeightSynthesis\LocalTestFunction
      =\EventIndicator{\ChartInteriorSupport}\LocalTestFunction.
  \]

  For positivity, fix an off-line input. Before excluding
  $\GadgetScale=0$ or filtering unsafe colors on the marked fiber,
  each color has exactly one scale preimage. Thus
  \[
    0\leq\LegColorMean_{\LegInputIndex}^{\ChartMarkedDensity}(\LocalFieldValue)
    =\LARModulus\Pr_{\mathrm{mark}}
    [\LegAffineEvaluation_{\LegInputIndex}=\LocalFieldValue]
    \leq\ChartMarkedSurvival^{-1}\leq(1-\ChartDeletionBound)^{-1}.
  \]
  This mean vanishes outside $\LegAllowedColors_{\LegInputIndex}$,
  and its average on that set is
  $\LARModulus/|\LegAllowedColors_{\LegInputIndex}|
  \in[1,\ChartMarkedSurvival^{-1}]$; the upper bound also follows from
  $|\ChartLineSupport|\leq|\LegAllowedColors_{\LegInputIndex}|$.
  Subtracting the average in
  \eqref{eq:leg-color-averages} and summing gives
  \begin{equation*}
    \sup\ChartMarkDefect\leq
    \frac{\LegInputCount\ChartDeletionBound}{1-\ChartDeletionBound},
    \qquad
    \|\ChartMarkDefect\|_\infty\leq
    \frac{\LegInputCount}{1-\ChartDeletionBound}.
  \end{equation*}
  Apply \cref{cor:calibration-pointwise-positivity} to
  $\LegCalibrationTarget=1-\ChartMarkOdds\ChartMarkDefect$.
  Using the respective one-sided and absolute bounds gives
  \begin{equation*}
    \LocalTestFunction
    \geq1-\frac{\ChartOperatorBound}{1-\ChartOperatorBound}
      -\ChartMarkOdds\ChartLocalPositivity
    \geq\tfrac13,
  \end{equation*}
  by $\ChartMarkOdds\leq1$ and the feasibility bounds in
  \cref{sec:leg-sampling-parameters}.
  Reconstruction preserves mass and $\mathbb E\ChartMarkDefect=0$, so
  $\mathbb E[\EventIndicator{\ChartInteriorSupport}\LocalTestFunction]=1$.
  This normalizes the masked density; $\LocalTestFunction$ itself
  need not have mean one. The marked density is normalized by definition.
  The equivalent rational coordinate system in
  \cref{sec:leg-branch-calibration} gives rational solution and branch
  tables, together with the stated Gram nonsingularity. Its bounded
  implementation is proved in \cref{sec:discrete-leg-injection}.

  The color-difference equations, normalization, and support give
  the regular color probabilities:
  \begin{equation*}
    \Pr_{\mathrm{reg}}[\LegAffineEvaluation_{\LegInputIndex}=\LocalFieldValue]
    =\begin{cases}
      |\LegAllowedColors_{\LegInputIndex}|^{-1},
      &\LegInputIndex\in\ChartLineIndices,\\
      |\LegAllowedColors_{\LegInputIndex}|^{-1}
      -\ChartMarkOdds\bigl(\Pr_{\mathrm{mark}}
      [\LegAffineEvaluation_{\LegInputIndex}=\LocalFieldValue]
      -|\LegAllowedColors_{\LegInputIndex}|^{-1}\bigr),
      &\LegInputIndex\in\ChartOffLineIndices,
    \end{cases}
  \end{equation*}
  for $\LocalFieldValue\in\LegAllowedColors_{\LegInputIndex}$.
  Since $(1-\LocalAtomWeight)\ChartMarkOdds=\LocalAtomWeight$, the
  off-line mixture is uniform on this set. On the line, the regular
  primary is uniform there, whereas the marked branch gives the
  integer loads $\LocalFirstEndpoint=\LegMarkedFirst$ and
  $\LocalSecondEndpoint=\LegMarkedSecond-\LocalSecondError$.
  For every supported map, safety determines the second residue
  uniquely from the first and the prescribed integer pair sum.
  This proves the exact pair law. All regular colors and
  all off-line marked colors are interior, as is every marked
  companion, proving the boundary-activation claim.
  Off-body carry cancellation follows from \cref{lem:leg-deterministic},
  and attachment safety follows from \cref{lem:leg-attachment-safety}
  under its stated margin and contribution bounds.
  Normalization of the branches gives the Bernoulli flag law.
  All table entries depend only on public data; sampling uses no
  rejection or correction conditioned on the hidden inputs.
  If $\LegCutoff=1$, then $\ChartOffLineIndices=\varnothing$ and
  $\ChartMarkDefect=0$, and the same argument applies.
\end{proof}

\input{chapters/discrete-leg-injection}

%% file: chapters/discrete-body-injection.tex
\subsubsection{Bounded Sampling of the Public Cycle-Closing Gaussian}
\label{sec:discrete-body-injection}

The body sampler draws only the public closure
$\BodyClosureGaussian\sample\Gaussian{\ZZ}{\GaussianWidth}$ of
\cref{def:body-injection}, retaining the hidden body coordinates.
We implement the scalar Gaussian rejection rule
of~\cite[full version, Section~4.1]{STOC:GenPeiVai08} with bounded running
time, accounting for truncation and finite precision. Assume
$\GaussianWidth^2\in\QQ$ and $\GaussianWidth^2\geq2$, as in
\cref{lem:joint-discrete-realization}, and fix a rational
$0<\BodyClosureSamplingError<1$.

\paragraph{Truncation and trial cap.}
Use the integer parameters
\begin{equation*}
  \begin{aligned}
    \ClosureSamplingRadius
      &:=\left\lceil\sqrt{\left\lceil
        \log_2(8/\BodyClosureSamplingError)\right\rceil}\right\rceil,
      &\ClosureSamplingCutoff
      &:=\ClosureSamplingRadius\lceil\GaussianWidth\rceil,\\
    \ClosureProposalSize&:=2\ClosureSamplingCutoff+1,
      &\ClosureTrialCap&:=\left\lceil120\ClosureSamplingRadius
        \left\lceil\log_2(4/\BodyClosureSamplingError)\right\rceil
        \right\rceil.
  \end{aligned}
\end{equation*}
The ideal trial proposes an integer $\ClosureProposal$ uniformly from
$[-\ClosureSamplingCutoff,\ClosureSamplingCutoff]\cap\ZZ$ and accepts
it with probability $e^{-\pi\ClosureProposal^2/\GaussianWidth^2}$.
The first accepted proposal has the Gaussian law conditioned on that
interval. By \cref{lem:gaussian-linear-mgf} and Chernoff's bound,
\[
  \Pr_{\GaussianPoint\sample\Gaussian{\ZZ}{\GaussianWidth}}
    [|\GaussianPoint|>\ClosureSamplingCutoff]
  \leq2e^{-\pi(\ClosureSamplingCutoff+1)^2/\GaussianWidth^2}
  \leq2e^{-\pi\ClosureSamplingRadius^2}
  \leq\BodyClosureSamplingError/4.
\]
This uses the integer threshold $\ClosureSamplingCutoff+1$; no
smoothing assumption is required.

The interval $[-\GaussianWidth,\GaussianWidth]$ contains at least
$\GaussianWidth$ integers, all with acceptance weight at least
$e^{-\pi}$. Since $\GaussianWidth\geq1$ and
$\ClosureSamplingRadius\geq1$,
$\ClosureProposalSize\leq5\ClosureSamplingRadius\GaussianWidth$.
Consequently every ideal trial succeeds with probability at least
$e^{-\pi}/(5\ClosureSamplingRadius)>1/(120\ClosureSamplingRadius)$.
Stop after at most $\ClosureTrialCap$ trials and return
$\BodyClosureGaussian=0$ if none succeeds. The failure probability is
at most
\[
  \exp\bigl(-\ClosureTrialCap/(120\ClosureSamplingRadius)\bigr)
  \leq\BodyClosureSamplingError/4.
\]
The zero fallback preserves exact integer affine validity.

\paragraph{Finite precision.}
Choose the fixed bit counts
\begin{equation*}
  \begin{aligned}
    \ClosureProposalBits
      &:=\left\lceil\log_2
        (8\ClosureProposalSize\ClosureTrialCap/
        \BodyClosureSamplingError)\right\rceil,\\
    \ClosureAcceptanceBits
      &:=\left\lceil\log_2
        (16\ClosureTrialCap/\BodyClosureSamplingError)\right\rceil.
  \end{aligned}
\end{equation*}
Reduce a uniform $\ClosureProposalBits$-bit integer modulo
$\ClosureProposalSize$ and subtract $\ClosureSamplingCutoff$ to obtain
the proposal. Its statistical distance from uniform is at most
$\ClosureProposalSize\,2^{-\ClosureProposalBits}
\leq\BodyClosureSamplingError/(8\ClosureTrialCap)$.
Evaluate the Gaussian acceptance weight with absolute error at most
$2^{-\ClosureAcceptanceBits}$, clamp the approximation to $[0,1]$, and
round down to a multiple of $2^{-\ClosureAcceptanceBits}$.
A fresh uniform $\ClosureAcceptanceBits$-bit integer implements this
coin exactly. The acceptance probability differs from its ideal value
by at most $2^{1-\ClosureAcceptanceBits}
\leq\BodyClosureSamplingError/(8\ClosureTrialCap)$.
Coupling all proposals and coins adds at most
$\BodyClosureSamplingError/4$ over the capped execution. Together with
truncation and the failure branch, this proves
\begin{equation*}
  \StatisticalDistance{\DistributionLaw{\BodyClosureGaussian}}
    {\Gaussian{\ZZ}{\GaussianWidth}}
  \leq\BodyClosureSamplingError.
\end{equation*}
All random bits are fresh and independent of the hidden source and
all other sampling blocks.

\paragraph{Worst-case complexity.}
Exact integer and rational comparisons on the binary inputs determine
$\lceil\GaussianWidth\rceil$ and all parameter and precision ceilings.
No proposal-interval enumeration is needed. Each candidate
and output has $O(\log\lceil\GaussianWidth\rceil+
\log(\ClosureSamplingRadius+1))$ bits. On this interval the absolute
Gaussian exponent is at most $4\pi\ClosureSamplingRadius^2$, so
standard rational approximations to $\pi$ and the exponential evaluate
the acceptance weight to the stated absolute precision in polynomial
time. The cap $\ClosureTrialCap$ and both bit counts are polynomial in
the binary input lengths and $\log(1/\BodyClosureSamplingError)$.
Thus every execution has polynomial running time and output length.
The sampler never observes, conditions on, or rejects a hidden source
coordinate.

%% file: chapters/discrete-leg-injection.tex
\subsubsection{Sampling the Sharp Line-Mark Flag and Affine Map}
\label{sec:discrete-leg-injection}

We prove the table-computation and bounded-sampling claims of
\cref{thm:leg-constructive-sampler} for $\LegFlagChartLaw$.
All parameters are public and satisfy the feasibility and marked-color
admissibility conditions of \cref{sec:leg-injection}.
Use the supports of \cref{def:leg-branch-supports}, enumerating the
marked support separately on the fiber
$\LocalChartBase=\LegMarkedFirst$, $\GadgetShear=0$.  Only the
$\LocalFirstError\ne0$ inputs impose trimmed color constraints there.
The companion safety test holds on every point of the marked line.

\paragraph{Public support enumeration.}
Enumerate the $\LARModulus^3$ coefficient triples in
$\ChartUnrestrictedSpace=\LARField^3$ and compute
$\LegAffineEvaluation_{\LegInputIndex}(\LocalChartPoint)$ at every
protected input to construct both supports.
This takes $O(\LARModulus^3\LegInputCount)$ field operations and
membership tests. These are evaluations at publicly listed protected
inputs, not at the hidden source. By
\cref{lem:leg-safe-supports}, both supports are nonempty, so
one affine map in each may be stored as a certified representative.

\paragraph{The finite rational calibration system.}
\label{sec:leg-branch-calibration}
For each protected input, choose the least allowed canonical color
$\LegReferenceColor_{\LegInputIndex}:=\min\LegAllowedColors_{\LegInputIndex}$
as reference, and define
\begin{equation*}
  \LegBasisFunction_0(\LocalChartPoint):=1,\qquad
  \LegBasisFunction_{\LegInputIndex,\LocalFieldValue}(\LocalChartPoint)
  :=\EventIndicator{\set{\LegAffineEvaluation_{\LegInputIndex}(\LocalChartPoint)
  =\LocalFieldValue}}
  -\EventIndicator{\set{\LegAffineEvaluation_{\LegInputIndex}(\LocalChartPoint)
  =\LegReferenceColor_{\LegInputIndex}}},
\end{equation*}
for $\LocalFieldValue\in\LegAllowedColors_{\LegInputIndex}
\setminus\set{\LegReferenceColor_{\LegInputIndex}}$.
Let $\LegBasisIndices$ index the constant and these real-valued
indicator differences. Their number is
\[
  |\LegBasisIndices|
  =1+\sum_{\LegInputIndex=1}^{\LegInputCount}
    (|\LegAllowedColors_{\LegInputIndex}|-1)
  \leq1+\LegInputCount(\LARModulus-1).
\]
For $\LegBasisIndex,\LegSecondBasisIndex\in\LegBasisIndices$, define
the rational matrix and right-hand side by
\begin{equation*}
  \begin{gathered}
    (\LegGramMatrix)_{\LegBasisIndex,\LegSecondBasisIndex}
    :=\LARModulus^{-3}\sum_{\LocalChartPoint\in\ChartInteriorSupport}
    \LegBasisFunction_{\LegBasisIndex}(\LocalChartPoint)
    \LegBasisFunction_{\LegSecondBasisIndex}(\LocalChartPoint),
    \qquad (\LegSolveRHS)_0:=1,\\
    (\LegSolveRHS)_{\LegInputIndex,\LocalFieldValue}
    :=-\frac{\ChartMarkOdds\EventIndicator{\set{\LocalFirstError_{\LegInputIndex}\ne0}}}
    {|\ChartLineSupport|}
    \sum_{\LocalChartPoint\in\ChartLineSupport}
    \LegBasisFunction_{\LegInputIndex,\LocalFieldValue}(\LocalChartPoint).
  \end{gathered}
\end{equation*}
The coefficient vector $\LegSolveCoefficients$ is determined by
\begin{equation}
  \label{eq:leg-gram-system}
  \LegGramMatrix\LegSolveCoefficients=\LegSolveRHS.
\end{equation}

The constant and indicator differences form a basis of
$\LegFunctionSpace$. Write
\[
  \LocalTestFunction=\sum_{\LegBasisIndex\in\LegBasisIndices}
    (\LegSolveCoefficients)_{\LegBasisIndex}\LegBasisFunction_{\LegBasisIndex}.
\]
In the unrestricted uniform $L^2$ inner product, the orthogonal
projection property of \cref{sec:leg-calibration-proof} gives
\[
 \langle\LegBasisFunction_{\LegBasisIndex},
          \LegCalibrationResponse\LocalTestFunction\rangle
 =\langle\LegBasisFunction_{\LegBasisIndex},
          \EventIndicator{\ChartInteriorSupport}\LocalTestFunction\rangle
 =(\LegGramMatrix\LegSolveCoefficients)_{\LegBasisIndex}.
\]
The marked color-moment identities there give
$\langle\LegBasisFunction_{\LegBasisIndex},\LegCalibrationTarget\rangle
 =(\LegSolveRHS)_{\LegBasisIndex}$.
Both sides of the function response equation lie in $\LegFunctionSpace$,
so testing against its basis makes that equation equivalent to
\eqref{eq:leg-gram-system}. The proved invertibility of the function
response therefore makes $\LegGramMatrix$ nonsingular.
The regular density is then evaluated using its masked-function
definition in \cref{sec:leg-calibration-proof}.
The contraction uses the function supremum norm, not coordinate
$\ell_\infty$ or closeness of $\LegGramMatrix$ to the identity.

Direct assembly uses at most
$O(\LARModulus^3|\LegBasisIndices|^2)$ arithmetic operations.
Every Gram entry is an integer of absolute value at most
$\LARModulus^3$, divided by $\LARModulus^3$.
Every nonconstant right-hand-side entry is
$-\ChartMarkOdds$ times an integer average over $\ChartLineSupport$,
or zero. Here $\ChartMarkOdds=\LocalAtomWeight/(1-\LocalAtomWeight)$
has polynomial bit length in the rational input, and
$1\leq|\ChartLineSupport|\leq\LARModulus$.
Thus all entries are rational with polynomial bit length.

Fraction-free elimination with row interchanges to select nonzero
pivots computes the unique solution~\cite[Section~V]{Bareiss1968};
the original leading principal minors need not be nonzero.
Clearing denominators in the polynomial-size matrix and right-hand side
gives integer entries of polynomial bit length. Minors of the augmented
matrix, including after row permutations, control the intermediate
integers. The determinant bound therefore gives polynomial bit lengths
for these integers and every solution numerator and denominator.
Polynomially many arithmetic operations and exact divisions give
polynomial bit time, and the branch-density formulas of
\cref{sec:leg-map-distribution,sec:leg-calibration-proof} have polynomial
bit length at every map.
Their nonnegativity and normalization were established in
\cref{sec:leg-calibration-proof}; no numerical positivity test or
clipping is used.

The normalized branch tables have probabilities
$\ChartRegularDensity(\LocalChartPoint)/\LARModulus^3$ and
$\ChartMarkedDensity(\LocalChartPoint)/\LARModulus^3$ on their
respective supports. Multiplying these by $1-\LocalAtomWeight$ and
$\LocalAtomWeight$ gives a finite rational table for the complete
flag--affine-map law, for any admissible rational
$0\leq\LocalAtomWeight\leq1/2$.
These probabilities need not be dyadic; bounded fair-bit sampling
therefore approximates the exact rational table.

\paragraph{Fixed random bits and branch-preserving rounding.}
For $0<\LegChartError<1$, choose an integer $\LocalSamplingBits$ such that
\begin{equation*}
 (1+|\ChartInteriorSupport|+|\ChartLineSupport|)
          2^{-\LocalSamplingBits}\leq\LegChartError.
\end{equation*}
For the flag draw, replace $\LocalAtomWeight$ by
\[
 \LegRoundedAtomWeight
 =2^{-\LocalSamplingBits}
   \lfloor2^{\LocalSamplingBits}\LocalAtomWeight\rfloor.
\]
Both branch tables remain those computed at the original rational
$\LocalAtomWeight$.
Round each probability of each normalized branch table down to a
multiple of $2^{-\LocalSamplingBits}$, and assign all remaining mass
to the stored representative of that same branch support.  Each
rounded table remains normalized and nonnegative.  Its SD from the
exact table is at most its support size times
$2^{-\LocalSamplingBits}$: the total mass rounded down is at most
that amount, and only this mass is moved to the representative.
The construction never clips a negative exact table entry.

Use $\LocalSamplingBits$ unbiased bits to draw the flag with probability
$\LegRoundedAtomWeight$, and draw the selected table with exactly
$\LocalSamplingBits$ further unbiased bits, using its integer cumulative
probabilities. Both bit blocks are drawn even for a deterministic branch.
Coupling the flag and then the selected table bounds the complete
flag--map error by the prescribed rounding budget of
\cref{sec:discrete-leg-injection}, and hence by $\LegChartError$.
Every output in either branch remains in its certified support,
including all rounding-remainder outcomes.  Nonzero scale,
protected carry cancellation, and the boundary-activation identity
therefore hold exactly for the bounded implementation.  The exact
marginal identities hold for the ideal rational law; their numerical
perturbation is covered by this complete-pair SD allowance.

\paragraph{Sharp-strength implementation.}
For $\LocalAtomWeight=1/2$, one unbiased bit samples the flag exactly.
The complete-pair error is then at most
$\tfrac12(|\ChartInteriorSupport|+|\ChartLineSupport|)
2^{-\LocalSamplingBits}$, so the extra flag-rounding allowance can be omitted.

The bounds are uniform over admissible public histories. Enumeration,
exact elimination, table construction, and sampling take time polynomial
in the numerical values of $\LARModulus$ and $\LegInputCount$,
the public rational input bit lengths, and $\log(1/\LegChartError)$.
The enumeration depends polynomially on $\LARModulus$, rather than
$\log\LARModulus$. In the assembly,
$\LARModulus\leq\InteriorVertexCount\leq\BlockDim$ and
$2(\LegCutoff-1)<\LARModulus$ imply
$\LegInputCount=O(\LARModulus^2)$; hence these bounds are
polynomial in the overall public dimension and precision lengths.
All computations and random bits are independent of the hidden source.

\paragraph{Computing the public leg parameters.}
Given the core, including the attachment center $\LocalReservoirBase$,
\cref{alg:leg-parameters} computes and samples one leg's public affine
data. It calls \cref{alg:leg-tables} for the exact branch laws and keeps
all finite-bit outputs in their supports, preserving safety on protected
inputs. If the specification instead gives a
Gaussian width $\GaussianWidth>0$ and a desired local error
$\LegApproximationError\in\QQ\cap(0,1)$, a sufficient allocation is
\begin{equation*}
  \LegChartError:=\frac{\LegApproximationError}{2},\qquad
  \LegCutoff:=\max\!\left\{1,
    \left\lceil\GaussianWidth
      \sqrt{\frac{\log(8/\LegApproximationError)}{\pi}}
    \right\rceil\right\}.
\end{equation*}
Any integer at least this cutoff gives the same error bound, provided
the feasibility tests of \cref{alg:leg-parameters} pass; a preassigned
cutoff and sampling budget may instead be supplied directly. The shift
bound $\LegSumBound$ can be set to $|\LegBaselineShift|$ for one fixed
shift, or kept as a common bound for a family of legs. Within the two
algorithms, $\LegUniformBits(\LocalSamplingBits)$ is the integer encoded
by $\LocalSamplingBits$ fresh independent fair bits, and $\LegExactSolve$
solves a rational linear system exactly. Lexicographic order on affine
maps is taken in their canonical coordinates.

\begin{algorithm}[H]
 \caption{Public leg parameters}
 \label{alg:leg-parameters}
 \begin{algorithmic}[1]
 \Require $\LARModulus\in\ZZ_{\geq3}$, $\LegPrimeTest(\LARModulus)$;
 $\LegCutoff\in\ZZ_{>0}$, $\LegTrimMargin,\LegSumBound\in\ZZ_{\geq0}$,
 $\LegBaselineShift,\LocalReservoirBase\in\ZZ$,
 $|\LegBaselineShift|\leq\LegSumBound$;
 $\LegMarkedFirst\in\set{0,\ldots,\LARModulus-1}$;
 $\LocalAtomWeight\in\QQ\cap[0,1/2]$, $\LegChartError\in\QQ\cap(0,1)$.
 \State $\LegInputCount\gets(2\LegCutoff-1)^2$;
 $\LegMarkedSecond\gets\LARModulus-1+\LegBaselineShift-\LegMarkedFirst$
 \State $\ChartDeletionBound\gets
 [\LegInputCount(\LegSumBound+\LegCutoff-1+2\LegTrimMargin)+1]/\LARModulus$;
 $\ChartOperatorBound\gets(2\LegInputCount+1)\ChartDeletionBound$
 \If{$2(\LegCutoff-1)\geq\LARModulus\ \lor\
 \LegSumBound+\LegCutoff-1+2\LegTrimMargin\geq\LARModulus\ \lor\
 \ChartOperatorBound>1/4$}
   \State \Return $\bot$
 \EndIf
 \State $\ChartLocalPositivity\gets
 \dfrac{\LegInputCount}{1-\ChartDeletionBound}
 \left(\ChartDeletionBound+
 \dfrac{\ChartOperatorBound}{1-\ChartOperatorBound}\right)$
 \If{$\ChartLocalPositivity>1/3\ \lor\
 \LegMarkedSecond\notin
 [\LegTrimMargin+\LegCutoff-1,\LARModulus-\LegTrimMargin-\LegCutoff]$}
   \State \Return $\bot$
 \EndIf
 \State $(\ChartInteriorSupport,\ChartRegularDensity,
 \ChartLineSupport,\ChartMarkedDensity)\gets
 \LegTableRoutine(\LARModulus,\LegCutoff,\LegTrimMargin,
 \LegBaselineShift,\LegMarkedFirst,\LocalAtomWeight)$
 \State $\LocalSamplingBits\gets
 \left\lceil\log_2\bigl((1+|\ChartInteriorSupport|
 +|\ChartLineSupport|)/\LegChartError\bigr)\right\rceil$;
 $\LocalSamplingInteger\gets\LegUniformBits(\LocalSamplingBits)$
 \State $\LegMarkFlag\gets
 \EventIndicator{\set{\LocalSamplingInteger<
 \lfloor2^{\LocalSamplingBits}\LocalAtomWeight\rfloor}}$
 \If{$\LegMarkFlag=0$}
   \State $\LocalChartPoint\gets\LegDyadicRoutine
   (\ChartInteriorSupport,\ChartRegularDensity/\LARModulus^3,\LocalSamplingBits)$
 \Else
   \State $\LocalChartPoint\gets\LegDyadicRoutine
   (\ChartLineSupport,\ChartMarkedDensity/\LARModulus^3,\LocalSamplingBits)$
 \EndIf
 \State $(\LocalChartBase,\GadgetScale,\GadgetShear)\gets\LocalChartPoint$;
 $(\LegBaselineFirst,\LegScaleLift,\LegShearLift)\gets
 (\CanonicalLift{\LocalChartBase},\CanonicalLift{\GadgetScale},
 \CanonicalLift{\GadgetShear})$
 \State Form $\LegCoordinateMatrix$, $\LegInjectionMatrix$, and
 $\LegBaseLoad$ using
 \cref{sec:affine-realization-matrix,sec:leg-injection},
 with attachment entry $\LocalReservoirBase$ in \cref{sec:baseline-shift}.
 \State \Return $(\LegMarkFlag,\LocalChartPoint,\LegCoordinateMatrix,
 \LegInjectionMatrix,\LegBaseLoad)$
 \Statex
 \Function{\LegDyadicRoutine}{$\LegSamplingSupport,\LegSamplingLaw,\LocalSamplingBits$}
   \State $\LocalSamplingDenominator\gets2^{\LocalSamplingBits}$;
   $\LegStoredMap\gets\min_{\mathrm{lex}}\LegSamplingSupport$
   \ForAll{$\LocalChartPoint\in\LegSamplingSupport$}
     \State $\LegSamplingCounts(\LocalChartPoint)\gets
     \lfloor\LocalSamplingDenominator\LegSamplingLaw(\LocalChartPoint)\rfloor$
   \EndFor
   \State $\LegSamplingCounts(\LegStoredMap)\gets
   \LegSamplingCounts(\LegStoredMap)+\LocalSamplingDenominator
   -\sum_{\LocalChartPoint\in\LegSamplingSupport}\LegSamplingCounts(\LocalChartPoint)$
   \State $\LocalSamplingInteger\gets\LegUniformBits(\LocalSamplingBits)$
   \State \Return $\min_{\mathrm{lex}}\set{\LocalChartPoint\in\LegSamplingSupport:
   \LocalSamplingInteger<\sum_{\substack{\LegSecondChartPoint\in\LegSamplingSupport\\
   \LegSecondChartPoint\leq_{\mathrm{lex}}\LocalChartPoint}}
   \LegSamplingCounts(\LegSecondChartPoint)}$
 \EndFunction
 \end{algorithmic}
\end{algorithm}

\begin{algorithm}[H]
 \caption{Exact leg tables}
 \label{alg:leg-tables}
 \begin{algorithmic}[1]
 \Function{\LegTableRoutine}{$\LARModulus,\LegCutoff,\LegTrimMargin,
 \LegBaselineShift,\LegMarkedFirst,\LocalAtomWeight$}
   \State Enumerate $\ChartUnrestrictedSpace$ and the protected inputs
   lexicographically; form both supports of \cref{def:leg-branch-supports}.
   \State Choose the least allowed color as each reference color and form
   the indicator differences in \cref{sec:leg-branch-calibration}.
   \State Assemble $\LegGramMatrix$ and $\LegSolveRHS$ by the finite sums
   in \cref{sec:leg-branch-calibration}.
   \State $\LegSolveCoefficients\gets\LegExactSolve(\LegGramMatrix,\LegSolveRHS)$.
   \State Recover $\LocalTestFunction$ in the chosen basis and evaluate
   the branch densities of
   \cref{sec:leg-map-distribution,sec:leg-calibration-proof}.
   \State \Return $(\ChartInteriorSupport,\ChartRegularDensity,
   \ChartLineSupport,\ChartMarkedDensity)$
 \EndFunction
 \end{algorithmic}
\end{algorithm}

The interval test on $\LegMarkedSecond$ is equivalent to the companion
condition in \cref{sec:leg-injection}, because the second
protected input ranges from $-(\LegCutoff-1)$ to $\LegCutoff-1$. On
success, \cref{thm:leg-constructive-sampler} gives complete flag--map error
at most $\LegChartError$, while preserving exact carry cancellation and
the pointwise boundary-activation criterion on protected inputs. Together,
the error-budget choice above and the comparison of
\cref{sec:joint-injection-protection} bound the local output SD from $\LegJointReferenceLaw$ by
$\LegApproximationError$. A return value of $\bot$ records only that one
of the sufficient tests has failed. The algorithm performs one exact
linear solve and has the polynomial bit complexity established above.
All computations and random bits are source-independent. The baseline
and mixer use the same sampled map, and assembly counts the shared
attachment center once. Its zero-carry guarantee follows from
\cref{lem:leg-attachment-safety} under its stated margin and contribution bounds.

%% file: chapters/appendix-bounded-sampling.tex
\section{Shared Baseline and Joint Realization}
\label{sec:bounded-sampling-appendix}

We verify the shared baseline, source comparisons, carry safety, and
integer affine realization using the local samplers of
\cref{sec:discrete-body-injection,sec:discrete-leg-injection} and the
placement sampler of \cref{sec:bounded-permutation-sampling}.

\subsection{Shared baseline realization}
\label{sec:baseline-realization-proof}

\begin{proof}[Proof of \cref{prop:joint-baseline-realization}]
  By admissibility and the off-body pair sums,
  \begin{equation*}
    \begin{aligned}
      \OnesVec^{\trans}\BaselineShift
      &=\sum_{\VertexIndex\in\InteriorVertices}\BaselineCenters_{\VertexIndex}
        +\sum_{\GadgetIndex=1}^{\GadgetCount}
        (\LARModulus-1+\LegBaselineShift_{\GadgetIndex})\\
      &=\sum_{\VertexIndex\in\InteriorVertices}\BaselineCenters_{\VertexIndex}
        +\GadgetCount(\LARModulus-1)
        +\sum_{\GadgetIndex=1}^{\GadgetCount}\LegBaselineShift_{\GadgetIndex}
        \equiv0\pmod{\LARModulus}.
    \end{aligned}
  \end{equation*}
  Hence \cref{lem:rounding-image-lattice} gives lattice membership.
  Body support is part of admissibility; off-body support follows from
  \cref{def:safe-affine-map} at zero input. The carry statement is
  \cref{prop:matrix-carry-safety}, whose proof below uses exactly this
  safe-center support and local carry lemmas. All draws use public
  data and fresh randomness without inspecting, conditioning on, or
  rejecting the hidden source.
\end{proof}

\input{chapters/discrete-baseline-shift}

\subsection{Protected sources and comparison bounds}
\label{sec:protected-comparison-proof}

For the source comparisons, fix only preceding public template and
placement data permitted by \cref{sec:baseline-dependence}. Share the
exact positive, source-independent core law and all complete conditional
flag--map draws, retaining their prescribed dependencies. The closure
Gaussian and final relabeling remain independent of the core. The
initial placement and conditional core retain their prescribed
dependence, so conditioning on the core may bias that placement.
In the direct-Fourier sampler, the separate body-cycle permutation
$\BodyPlacementMatrix$ is independent of the placement, core, and
latent cycle, as required by \cref{sec:df-sampling-contract}.
The cycle comparison averages over the closure.

\begingroup
\mdfsetup{topline=true}
\begin{lemma}[Source tail bounds]
  \label[lemma]{lem:joint-source-tails}
  For the ideal Gaussian inputs and public closure,
  \begin{equation*}
    \begin{aligned}
      \Pr[\norm{\BodyLoadIncrement}_\infty\geq\PaddingInteriorCutoff]
      &\leq\BodyTailProbability
      :=\InteriorVertexCount\Pr_{\CycleGaussianScalar\sim\CycleScalarLaw}
        [|\CycleGaussianScalar|>\DFEdgeCutoff]\\
      &\leq2\InteriorVertexCount
        \exp\!\left(-\frac{\pi
        \lceil\PaddingInteriorCutoff/2\rceil^2}{\GaussianWidth^2}\right).
    \end{aligned}
  \end{equation*}
  The leg and queue scalar tails satisfy
  \begin{align*}
    \LegTailProbability
    &:=\Pr_{\LocalFirstError\sim\Gaussian{\ZZ}{\GaussianWidth}}
        [|\LocalFirstError|\geq\LegCutoff]
      \leq2\exp\!\left(-\frac{\pi\LegCutoff^2}{\GaussianWidth^2}\right),
      \\
    \QueueTailProbability
    &:=\Pr_{\PaddingChargeError\sim\Gaussian{\ZZ}{\GaussianWidth}}
        [|\PaddingChargeError|\geq\PaddingChargeCutoff]
      \leq\min\!\left\{1,\,
        2\exp\!\left(-\frac{\pi\PaddingChargeCutoff^2}{\GaussianWidth^2}\right)
        \right\}.
  \end{align*}
  Consequently,
  \[
    \Pr[\PaddingProtectedEvent^c]
    \leq\BodyTailProbability+2\GadgetCount\LegTailProbability
      +\QueueTailProbability.
  \]
\end{lemma}
\endgroup
Here $\BodyTailProbability$ is a union-bound budget, rather than the
body's exact failure probability. All bounds hold under every permitted
placement and final relabeling. If the implemented closure law has
statistical error at most $\BodyClosureSamplingError$, add that amount
to the body bound and the joint event bound.

\paragraph{Proof of the scalar tail bounds.}
The two-sided Chernoff estimate from \cref{lem:gaussian-linear-mgf},
applied at an integer threshold, gives the leg and queue bounds in
\cref{lem:joint-source-tails}; the queue estimate
is also bounded by one. For the body, there are
$\InteriorVertexCount$ independent Gaussian edge scalars, including the
public closure. If every scalar has magnitude at most $\DFEdgeCutoff$,
each cyclic difference has magnitude at most
$2\DFEdgeCutoff<\PaddingInteriorCutoff$. The union bound and
$\DFEdgeCutoff+1=\lceil\PaddingInteriorCutoff/2\rceil$ prove
the body tail bound in \cref{lem:joint-source-tails}. Taking the union over these edge
scalars, the two coordinates of every leg, and the single queue scalar
proves the joint protected-event bound. These events concern source
coordinates, so supported placements and relabelings preserve the
bounds. A maximal coupling of the public closure to its ideal Gaussian
adds at most $\BodyClosureSamplingError$ to either body-containing bound.

The three source blocks admit the following comparisons.
\begin{description}
  \item[Body.]
    Coupling the edge scalars gives a coupling of the entire body
    cycle, with failure bound $\DFBodyClippingError$ satisfying
    \begin{equation}
      \label{eq:df-edge-coupling}
      \DFBodyClippingError
      \leq\InteriorVertexCount\Pr_{\CycleGaussianScalar\sim\CycleScalarLaw}
        [|\CycleGaussianScalar|>\DFEdgeCutoff]
      =\BodyTailProbability.
    \end{equation}
    This comparison retains all correlations among the cyclic
    differences.
  \item[Legs.]
    For each fixed admissible core, keep the complete public map and
    the same second input for the companion and attachment. The
    implemented local law satisfies
    \[
      \begin{aligned}
        &\StatisticalDistance{\DistributionLaw{(
          \CanonicalLift{\LocalFirstEndpoint\bmod\LARModulus},
          \CanonicalLift{\LocalSecondEndpoint\bmod\LARModulus},
          \LocalSecondError)}}{\LegJointReferenceLaw}\\
        &\qquad\leq\LegChartError+1-(1-\LegTailProbability)^2
          \leq\LegChartError+2\LegTailProbability.
      \end{aligned}
    \]
    The bound remains valid after translating and reducing the
    attachment coordinate, or taking a histogram. Map-sampling error
    adds no carry failure because each branch support is preserved.
  \item[Queue.]
    For a fixed queue subset, and after averaging any common
    source-independent placement law,
    \[
      \StatisticalDistance{\DistributionLaw{\PlacedQueueIncrement}}
        {\DistributionLaw{\DiscreteReferenceQueueIncrement}}
      =\QueueTailProbability.
    \]
    This equality concerns the integer increment. Reduction modulo
    $\LARModulus$ and passage to a histogram can only decrease the
    distance.
\end{description}

\paragraph{Proof of the component comparisons.}
Independently couple each latent edge scalar, including the public
closure, maximally to its protected law:
\[
  \CycleGaussianScalar\sim\Gaussian{\ZZ}{\GaussianWidth},\qquad
  \CycleReferenceScalar\sim\DFProtectedEdgeLaw.
\]
The couplings fail with probability at most the sum of their tail
probabilities. On a match every cyclic difference agrees, proving
the stated whole-cycle comparison.
The protected cycle retains its telescoping sum zero and the dependence
among its vertex differences.

\paragraph{Legs and the shared queue.}
The two independent inputs of a leg both lie in their protected interval
with probability $(1-\LegTailProbability)^2$. Their joint conditional
law is $\LegClippedLaw^{\otimes2}$, so conditioning has exact statistical
distance at the input level
\[
  1-(1-\LegTailProbability)^2\leq2\LegTailProbability.
\]
Keep the same complete flag--map pair on a match, and use
the same second input for the companion and attachment. The exact law
of \cref{thm:leg-constructive-sampler} and a coupling of the implemented
pair at cost at most $\LegChartError$ give
the stated leg comparison. Data processing
preserves it under translation, reduction, or taking a histogram.
Exact map support preserves carry safety even if this coupling fails.

Conditioning the single queue scalar costs exactly
$\QueueTailProbability$. For each queue subset, the increment determines
its scalar by dividing its total integer charge by $\LARModulus$.
This recovery also works after averaging any common source-independent
placement law. Data processing in both directions therefore gives
the exact queue clipping distance stated above,
retaining the shared scalar and the dependence between queue and attachments.

\paragraph{Joint comparison and public sampling.}
Under the exact positive public-core law fixed above, the union bound
over the source couplings proves the joint comparison bound of
\cref{sec:joint-protected-references}. On a match, the entire integer
load agrees. Assembly, modular reduction, and the shared final
permutation therefore give agreement of the complete endpoint vectors,
with all cycle, leg, and queue dependencies retained. A match also
places the actual source in $\PaddingProtectedEvent$; for an admissible
core, \cref{prop:matrix-carry-safety} then supplies carry safety. This
single source cost covers both conclusions. Positivity is required for
the probability coupling, and admissibility for the carry and zero-fiber
conclusions. Clipping is an analytical comparison and never tests or
rejects a hidden source value.

\begin{lemma}[Complete baseline-data sampling error]
  \label[lemma]{lem:joint-baseline-sampling-error}
  For the same preceding public data, suppose the implemented core
  law has statistical error at most
  $\BaselineJointSamplingError$, and each conditional complete
  flag--map law has error at most $\LegChartError$, uniformly over
  admissible preceding public template and placement data and public
  histories, respectively. Then
  \[
    \StatisticalDistance{\BaselineImplementedDataLaw}{\BaselineDataLaw}
    \leq\BaselineJointSamplingError+\GadgetCount\LegChartError.
  \]
\end{lemma}

\paragraph{Proof of the public sampling bound.}
Couple the implemented core to its target, conditional on the
same preceding public data. On a match couple each complete conditional
flag--map pair using its uniform error guarantee. The union bound proves
the bound of \cref{lem:joint-baseline-sampling-error}.
The leg comparison uses these same map errors, charged once in assembly.
Support preservation leaves
the zero baseline failure probability exact, including every fallback.

\subsection{Carry safety and source representatives}
\label{sec:matrix-carry-proof}

\begin{proof}[Proof of \cref{prop:matrix-carry-safety}]
  Every body vertex satisfies
  \[
    \left|\bigl(\PlacedBodyIncrement+\PlacedLegIncrement
       +\PlacedQueueIncrement\bigr)_{\VertexIndex}\right|
    \leq\PaddingInteriorCutoff+\LegCutoff+\PaddingChargeCutoff
    \leq\PaddingSafeMargin.
  \]
  Its baseline is in $\HistogramSafeCenters$, so its total load stays
  in $[0,\LARModulus)$ and its carry is zero.
  For each off-body pair, \cref{lem:leg-deterministic} gives cancellation
  of the two carries, using the same complete safe map and its shared
  second source coordinate. Body and queue increments vanish off the
  body. Summing these contributions proves the integer equality for
  the total carry, which survives the final common permutation.
  By \cref{lem:rounding-image-lattice}, the carry lies in
  $\LatticeGen{\RoundingMatrix}$.
  For the natural and canonical source lifts, write
\[
  \CanonicalLift{\PaddingSource\bmod\LARModulus}
  =\PaddingSource+\LARModulus\AssemblyLiftDifference,
  \qquad \AssemblyLiftDifference\in\ZZ^{\BlockDim}.
\]
Their loads differ by
$\LARModulus\HypergraphIncidenceMatrix\AssemblyLiftDifference$, so
their canonical carries differ by
$\HypergraphIncidenceMatrix\AssemblyLiftDifference
\in\LatticeGen{\RoundingMatrix}$. Hence changing the lift preserves
the carry-membership criterion of \cref{lem:affine-transport-endpoint-carry-bound}.

  The zero-fiber assertion and statistical bound therefore follow from
  \cref{lem:affine-transport-endpoint-carry-bound}.
\end{proof}

\subsection{Integer affine realization}
\label{sec:matrix-integer-proof}

The following table records the dimensions and public dependencies of
the realization objects defined in \cref{sec:affine-realization-matrix}.

\begin{table}[ht]
  \centering
  \small
  \renewcommand{\arraystretch}{1.15}
  \begin{tabular}{p{0.24\linewidth}p{0.29\linewidth}p{0.37\linewidth}}
    Object & Domain and codomain & Public dependence and construction\\
    \hline
    $\RealMatrixBodyEmbedding$,
    $\RealMatrixLegEmbedding{\GadgetIndex}$
    & $\ZZ^{\InteriorVertexCount}\to\ZZ^{\BlockDim}$,
    $\ZZ^3\to\ZZ^{\BlockDim}$
    & Body labels and $\CalibContext$; explicit sparse embeddings.\\
    $\RealMatrixIncidence(\BodyPlacementMatrix)$
    & $\ZZ^{\BlockDim}\to\ZZ^{\BlockDim}$
    & Placement and body permutation; explicit incidence columns.\\
    $\RealMatrixSourceMixer$
    & $\ZZ^{\BlockDim}\to\ZZ^{\BlockDim}$
    & Complete public leg maps; explicit block diagonal matrix.\\
    $\PreRelabelOffset$
    & An element of $\ZZ^{\BlockDim}$
    & Baseline, closure, and body permutation; explicit vector.\\
    $\TransportMatrix$, $\TransportOffset$
    & $\ZZ^{\BlockDim}\to\ZZ^{\BlockDim}$,
    an element of $\ZZ^{\BlockDim}$
    & All preceding data and final permutation; explicitly formed
    for an admissible core.\\
  \end{tabular}
  \caption{Integer realization objects. Matrix dimensions are codomain
    dimension by domain dimension. The probability-response matrices in
    \cref{sec:direct-fourier-transport} act on different, real coefficient
  spaces.}
  \label{tab:affine-realization-objects}
\end{table}

\begin{proof}[Proof of \cref{prop:matrix-integer-realization}]
The first $\BlockDim-1$ columns of
$\RealMatrixIncidence(\BodyPlacementMatrix)$ form a spanning-tree
incidence matrix: the placed body path spans the body, and each leg
joins two new vertices to its specified attachment. Hence the general
objects of \cref{cor:general-tree-incidence-transport} specialize to
\begin{equation*}
  \begin{aligned}
    \ExtendedTreeIncidenceMatrix
      &=[\TreeIncidenceMatrix\mid\ChargeColumn]
       =\RealMatrixIncidence(\BodyPlacementMatrix),\\
    \EdgeCoordinateMatrix
      &=\operatorname{diag}(\IdentityMatrix_{\InteriorVertexCount-1},
          \LegCoordinateBlock{1},\ldots,
          \LegCoordinateBlock{\GadgetCount}),\\
    \ExtendedEdgeCoordinateMatrix
      &=\operatorname{diag}(\EdgeCoordinateMatrix,1)
       =\RealMatrixSourceMixer,\\
    \PreRelabelIncidenceMatrix
      &=\RealMatrixIncidence(\BodyPlacementMatrix)\RealMatrixSourceMixer,\\
    \HypergraphIncidenceMatrix
      &=\GlobalRelabelingMatrix\PreRelabelIncidenceMatrix.
  \end{aligned}
\end{equation*}
Multiplying the source blocks gives, respectively, the zero extension of
$\BodyPlacementMatrix\BodyIncidenceMatrix\BodySourceCoordinates$,
the increments $\RealMatrixLegEmbedding{\GadgetIndex}
\LegIncidenceMatrix\LegCoordinateBlock{\GadgetIndex}
\LegSourceCoordinates{\GadgetIndex}$ from
the leg construction of \cref{def:leg-injection}, and
$\ChargeColumn\QueueSourceCoordinate$.
The closure term in the matrix assembly of
\cref{sec:affine-realization-matrix} is exactly the
zero extension of $\BodyPlacementMatrix\BodyAffineOffset$ from
\cref{def:body-injection}. The assembly definition in
\cref{sec:affine-realization-matrix} therefore gives
\[
  \InitialVertexConfiguration
  =\GlobalRelabelingMatrix\bigl(\PreRelabelBaseline+
    \RealMatrixBodyEmbedding\BodyPlacementMatrix\BodyAffineOffset\bigr).
\]
The assembled load adds the three increments to the shared baseline
before the common relabeling.

  The queue column has total charge $\LARModulus$. The shared baseline
  satisfies the charge condition of \cref{def:admissible-joint-baseline-law}, and the closure offset
  has zero charge. These are precisely the lattice hypotheses of
  \cref{cor:general-tree-incidence-transport}; the final permutation
  preserves them because it preserves coordinate sums. That corollary
  therefore gives integral matrix and offset with the dimensions in
  \cref{tab:affine-realization-objects}, and determinant
  $\pm\det\EdgeCoordinateMatrix$. The block diagonal form above and
  the triangular local mixers give the scale product in
  the determinant formula of \cref{prop:matrix-integer-realization}. Each scale is nonzero
  modulo $\LARModulus$; the final permutation contributes only a sign.
  All inverse products are evaluated over $\QQ$, as required by the
  general incidence construction.
\end{proof}

\input{chapters/appendix-gadget-assembly}

%% file: chapters/discrete-baseline-shift.tex
\subsubsection{Baseline Data}
\label{sec:discrete-baseline-shift}

Let $\BaselineImplementedCoreLaw$ be the law returned by the public-core
implementation assumed in \cref{sec:assembly-sampling-assumptions},
with ideal target $\BaselineCoreLaw$ from \cref{sec:baseline-shift}.
That contract requires uniform conditional error
$\BaselineJointSamplingError$, bounded running time, and exact
admissibility on every execution. Marking probabilities are rational.
Any timeout fallback must run within the same bound, return an
admissible tuple for the current public context, and charge its
probability to $\BaselineJointSamplingError$; nonempty support alone
does not provide a fallback.

After the core draw, use fresh independent public randomness for each
complete flag--affine-map pair
$(\LegMarkFlag_{\GadgetIndex},\LocalChartPoint_{\GadgetIndex})$,
using Appendix~\ref{sec:discrete-leg-injection}. Its conditional error
is at most $\LegChartError$ relative to the full joint law
$\LegFlagChartLaw$ of \cref{def:leg-injection}, uniformly
over admissible public histories, including the sampled shift and mark.
This error includes the finite-bit Bernoulli draw. Branch zero always
returns an affine map in $\ChartInteriorSupport$, and branch one always
returns an affine map in $\ChartLineSupport$; in particular every output
satisfies \cref{def:safe-affine-map}. The complete implemented law
$\BaselineImplementedDataLaw$ is on the same tuple of core data and
complete flag--map pairs as its ideal target $\BaselineDataLaw$ in
\cref{def:complete-baseline-data}. The resulting baseline is the
deterministic image specified in \cref{def:shared-baseline-vector}.
The $\GadgetCount$ conditional pair errors are separate from the core
error and counted once each in \cref{sec:joint-injection-protection}.
All steps are independent of hidden source and secret-dependent
information; none inspects, conditions on, or rejects that information.

%% file: chapters/appendix-gadget-assembly.tex
\subsection{Gadget Assembly and Realization}
\label{sec:gadget-assembly-proof}

We prove \cref{lem:joint-discrete-realization} for the affine data and
sampling assumptions of \cref{sec:discrete-realization}, using the
bounded implementations in
\cref{sec:discrete-body-injection,sec:discrete-leg-injection,sec:discrete-baseline-shift,sec:bounded-permutation-sampling}.

\paragraph{Sampling assumptions and numerical budgets.}
\label{sec:assembly-sampling-assumptions}
Assume the geometry and safety margins of
\cref{sec:lobster-injection-targets,sec:lobster-combination}, the
feasibility and marked-color admissibility conditions of
\cref{sec:leg-sampling-parameters} for every leg, and an admissible core
law $\BaselineCoreLaw$ with nonempty support as in
\cref{def:admissible-joint-baseline-law}. Require
\[
  \LARModulus\text{ odd prime},\qquad
  \InteriorVertexCount\geq\LARModulus,\qquad
  \GaussianWidth^2\in\QQ,\qquad \GaussianWidth^2\geq2.
\]
The public template and core samplers must satisfy two conditions.
\begin{enumerate}[label=(\roman*)]
  \item They are source-independent and run in worst-case polynomial
    time in $\BlockDim$, the binary public-input lengths, and the
    requested precision lengths.
  \item The core sampler has statistical error at most
    $\BaselineJointSamplingError$ relative to $\BaselineCoreLaw$,
    uniformly over admissible preceding public template and placement
    data. Every output, including any fallback, has polynomial encoding
    length and satisfies all admissibility constraints exactly.
\end{enumerate}
Fix $\NumericalSamplingError\in\QQ\cap(0,1)$, rational
$\BodyClosureSamplingError,\BodyPermutationSamplingError\in(0,1)$, and
$\LegChartError\in\QQ\cap(0,1)$, with
\[
  \BodyClosureSamplingError+\BodyPermutationSamplingError
  \leq\NumericalSamplingError.
\]
Here $\BodyClosureSamplingError$ bounds the total variation error of the
public closure Gaussian sampler. The quantity
$\BodyPermutationSamplingError$ bounds the joint error of the two local
permutations and the common relabeling.

The carry-failure budget is defined in
\cref{sec:discrete-realization}; the core and map errors are charged
separately from the source comparison.

\paragraph{Finite-sampling interface.}
\label{sec:discrete-public-sampler}
Use the prescribed public template and admissible core baseline laws of
\cref{sec:lobster-combination,sec:baseline-shift}, retaining all their
permitted dependence on preceding public data. The bounded samplers of
\cref{sec:discrete-body-injection,sec:bounded-permutation-sampling}
return an integer closure within $\BodyClosureSamplingError$ of
$\Gaussian{\ZZ}{\GaussianWidth}$ and valid permutations with joint error
at most $\BodyPermutationSamplingError$ from the three independent
uniform laws. The closure's zero fallback is integral, and placement preserves
the disjointness of \cref{lem:df-disjoint-context}. These draws use fresh
independent randomness, independent of the hidden source, with the
independence prescribed in
\cref{sec:lobster-combination,sec:baseline-dependence}.
The core retains its prescribed conditional law given preceding
template and placement data, and its tests never read the body-cycle
or final permutation.
\Cref{sec:discrete-baseline-shift} supplies the complete conditional
flag--map extension with per-leg error at most $\LegChartError$,
uniformly over admissible public histories, and exact branch support.
The core error remains $\BaselineJointSamplingError$ under
\cref{sec:assembly-sampling-assumptions}.
\begin{proof}[Proof of \cref{lem:joint-discrete-realization}]
\emph{Exact affine validity.}
Every supported placement is valid, every returned leg map has nonzero
scale, and every returned core is admissible. The closure sampler returns
an integer, including on its zero fallback. Hence
\cref{prop:matrix-integer-realization} gives integral affine data with
the stated determinant on every execution.
The public sampling contract above makes the complete matrix and offset
source-independent; their assembly is exactly that of
\cref{sec:affine-realization-matrix}.

\emph{Protected carry.}
By \cref{prop:matrix-carry-safety}, failure of the zero-fiber conclusion
is contained in the complement of the protected event. Coupling the
implemented public closure to its ideal law adds
$\BodyClosureSamplingError$ to the joint protected-event bound of
\cref{sec:joint-injection-protection}, which
is exactly the carry budget in \cref{sec:discrete-realization}.
Baseline and flag--map sampling preserve safe support, and valid
permutations preserve the protected bounds even for biased finite-bit
draws. Their errors add no carry-failure term, proving the stated
zero-fiber failure bound.

\emph{Bounded implementation and numerical errors.}
The cited local samplers have deterministic work and bit caps and
polynomial output length. Their exact leg-table counting and linear
algebra are polynomial in the numerical modulus, protected-input count,
and public rational bit lengths; since
$\LARModulus\leq\InteriorVertexCount\leq\BlockDim$ and
the leg-feasibility conditions of \cref{sec:leg-injection} bound $\LegCutoff$ by
$O(\LARModulus)$, this is polynomial in the overall dimension.
The core contract and permutation sampler give the same bounded-time
guarantees for the remaining draws. Exact integer and rational linear
algebra compute the realization of
\cref{sec:affine-realization-matrix} in polynomial time and bit length.
Every branch returns valid affine data without reading the hidden source.

Couple the closure and the three permutations to their ideal laws at
cost at most $\BodyClosureSamplingError+\BodyPermutationSamplingError$.
Conditional on matching placements and the same preceding public data,
\cref{lem:joint-baseline-sampling-error} couples the core and complete
flag--map pairs at cost at most
$\BaselineJointSamplingError+\GadgetCount\LegChartError$.
Thus the numerical error in the complete public data is at most
$\NumericalSamplingError+\BaselineJointSamplingError
+\GadgetCount\LegChartError$. This coupling retains the prescribed
dependencies and does not require placements to remain uniform
conditional on the core. Applying the same independent source and
affine assembly cannot increase its error. Source truncation is
accounted for separately by the protected comparison of
\cref{sec:joint-protected-references}.
For every fixed accuracy order $\TransportAccuracyOrder$, all these
numerical tolerances can be chosen inverse-polynomially while retaining
a uniform worst-case polynomial implementation. The public
$\LARModulus,\BlockDim,\SecretDim,\GaussianWidth,\RoundingMatrix$
and the $\BlockDim$ hidden source coordinates remain fixed.
\end{proof}

\paragraph{The selected parameters and protected cutoffs.}
For the fixed family of \cref{sec:df-parameter-choice},
\cref{sec:df-parameter-verification} verifies feasibility and
negligible source tails for sufficiently large $\secpar$.
The calibrated sampler of \cref{alg:df-core-sampler} meets the
admissible-sampling hypotheses in its certified regime and preserves
exact affine validity on the remaining finite prefix, without computing
an asymptotic accuracy threshold. Its fiber bound in
\cref{thm:df-constructive-aft} uses the fully protected reference:
one source-coupling cost covers both endpoint disagreement and unsafe
carries, so the separate carry budget above is not added again.

%% file: chapters/appendix-direct-fourier-reference.tex
\section{Fourier Targets and Protected Reference Tools}
\label{sec:df-reference-tools}

\input{chapters/appendix-direct-fourier-energy.tex}
\input{chapters/appendix-public-placement.tex}
\input{chapters/appendix-direct-fourier-cycle.tex}

%% file: chapters/appendix-direct-fourier-energy.tex
\subsection{Direct Fourier Targets and Energy Identities}
\label{sec:df-energy-proofs}

Use the characters and mass-one real signed laws of \cref{sec:df-target}.

\subsubsection{The ambient uniform target}
\label{sec:df-ambient-proof}

\begin{lemma}[Ambient Fourier energy and distributional error]
 \label[lemma]{lem:df-ambient-target}
 Let $\SignedEndpointLaw$ be a mass-one real signed law on
 $\LARField^{\BlockDim}$, and let
 $\FiniteDensity=\SignedEndpointLaw/\Unif{\LARField^{\BlockDim}}$.
 Then
 \begin{align*}
  \|\FiniteDensity-1\|_{L^2(\Unif{\LARField^{\BlockDim}})}^2
    &=\sum_{\DFFrequency\ne0}|\DFTransform(\DFFrequency)|^2,
       \\
  \frac12\|\SignedEndpointLaw-\Unif{\LARField^{\BlockDim}}\|_1
    &\leq\frac12
       \left(\sum_{\DFFrequency\ne0}|\DFTransform(\DFFrequency)|^2\right)^{1/2}.
 \end{align*}
\end{lemma}
For a probability law, the second inequality bounds statistical distance.

\begin{proof}[Proof of \cref{lem:df-ambient-target}]
The ambient characters form an orthonormal basis under
$\Unif{\LARField^{\BlockDim}}$. Mass one gives zero-frequency
coefficient one. Parseval with this coefficient removed proves the
energy identity, and Cauchy--Schwarz gives the distance bound.
\end{proof}

\subsubsection{The exact support-energy identity}
\label{sec:df-energy-proof}

Use the character polynomial, ordered coefficient tensors, and norm
of \cref{def:df-weighted-character-norm}. For $\DFChargeRadius>0$,
define the weighted support energy by
\[
 \DFEnergy_{\DFChargeRadius}(\SignedEndpointLaw)
 :=\sum_{\DFFrequency\ne0}
 \DFChargeRadius^{2|\operatorname{supp}\DFFrequency|}
 |\DFTransform(\DFFrequency)|^2.
\]

\begin{lemma}[Character polynomial and support energy]
 \label[lemma]{lem:df-fourier-isometry}
 Let $\SignedEndpointLaw$ be an exchangeable real signed law of mass one
 on $\LARField^{\BlockDim}$. For every
 $0\leq\HoeffdingDegree\leq\BlockDim$,
 \begin{equation}
 \label{eq:df-exact-isometry}
 \sum_{|\operatorname{supp}\DFFrequency|=\HoeffdingDegree}
       |\DFTransform(\DFFrequency)|^2
 =\frac{\HoeffdingDegree!}{(\BlockDim)_{\HoeffdingDegree}}
       \|\DFFourierTensor_{\HoeffdingDegree}\|_2^2.
 \end{equation}
 For every $\FockRadius>0$, with $\DFChargeRadius=\FockRadius/\sqrt e$,
 \[
 \DFEnergy_{\DFChargeRadius}(\SignedEndpointLaw)
 \leq\DFFourierNorm_{\FockRadius}(\SignedEndpointLaw)^2-1.
 \]
\end{lemma}

\begin{proof}[Proof of \cref{lem:df-fourier-isometry}]
 For a fixed ordered nonzero frequency tuple, exchangeability makes the
 corresponding marginal Fourier coefficient independent of the chosen
 ordered distinct vertices. The coefficient tensor is therefore
 $\binom{\BlockDim}{\HoeffdingDegree}$ times that marginal tensor.
 The support energy in \cref{lem:df-fourier-isometry} is
 $\binom{\BlockDim}{\HoeffdingDegree}$ times the squared norm of the
 marginal tensor. Dividing by the square of the binomial coefficient
 proves the identity, including degrees zero and $\BlockDim$.

 For the all-degree comparison, the geometric mean of the largest
 $\HoeffdingDegree$ elements of $\set{1,\ldots,\BlockDim}$ is at least
 that of the entire set. Also
 \[
  \log(\BlockDim!)
   =\sum_{\CoordinateIndex=1}^{\BlockDim}\log\CoordinateIndex
   \geq\int_1^{\BlockDim}\log\HoeffdingScalarValue\,\mathrm d\HoeffdingScalarValue
   \geq\BlockDim\log\BlockDim-\BlockDim .
 \]
 Consequently $(\BlockDim)_{\HoeffdingDegree}\geq
 (\BlockDim/e)^{\HoeffdingDegree}$. Substitute this inequality into
 \eqref{eq:df-exact-isometry}, multiply by
 $\DFChargeRadius^{2\HoeffdingDegree}$, and sum the positive degrees.
 This bound holds at every degree.
\end{proof}

\begin{proof}[Proof of \cref{lem:df-joint-target}]
Low-support cancellation removes all nonzero frequencies of support
size at most $\HoeffdingOrder$. Since $\DFChargeRadius>1$,
\cref{lem:df-fourier-isometry} gives
\[
 \begin{aligned}
 \sum_{\DFFrequency\ne0}|\DFTransform(\DFFrequency)|^2
 &\leq\DFChargeRadius^{-2(\HoeffdingOrder+1)}
       \DFEnergy_{\DFChargeRadius}(\SignedEndpointLaw)\\
 &\leq\DFChargeRadius^{-2(\HoeffdingOrder+1)}
       \bigl(\DFFourierNorm_{\FockRadius}(\SignedEndpointLaw)^2-1\bigr).
 \end{aligned}
\]
\Cref{lem:df-ambient-target} supplies the first inequality in the
claimed conclusion. Substituting this energy bound and taking square
roots gives the second. For a probability law, half the $\ell_1$
distance is statistical distance.
\end{proof}

\paragraph{Complex coefficient convention.}
\label{sec:df-complex-coefficients}
Our complex Hilbert inner product is conjugate linear in the first
argument and linear in the second. The character-polynomial expansion
in \cref{def:df-weighted-character-norm} uses a bilinear coefficient
contraction,
equivalently the Hilbert pairing of
$\overline{\DFFourierTensor_{\HoeffdingDegree}}$ with
$\DFPolynomialVariables^{\otimes\HoeffdingDegree}$. Real channels obey
$\DFTransform(-\DFFrequency)=\overline{\DFTransform(\DFFrequency)}$.
All subsequent inverses act on real probability or signed profile
tables, so their physical weights are real.

%% file: chapters/appendix-public-placement.tex
\subsection{Placements and Relabeling}
\label{sec:bounded-permutation-sampling}

Use the placement, free vertices, and control pools of
\cref{sec:df-control-context}.

\begin{lemma}[Disjoint placement and control pools]
\label[lemma]{lem:df-disjoint-context}
Every sampled placement satisfies
$\DFAttachmentSet\cap\DFQueueSet=\varnothing$ and the pool counts of \cref{sec:df-control-context}.
Under the ideal joint placement law, the queue is a uniform
$\LARModulus$-subset, and the ordered attachments are conditionally
uniform without replacement in its complement.
The attachment tuple also has a uniform-without-replacement
unconditional marginal.
For every positive integer $\BoundaryInverseOrder$, any selection
requiring at most $\BoundaryInverseOrder$ distinct controls from either
pool is available when
$\GadgetCount\geq\BoundaryInverseOrder$ and
$\InteriorVertexCount-\LARModulus-2\GadgetCount
\geq\BoundaryInverseOrder$.
The placement and compensators admit bounded implementations whose
running time is polynomial in the public dimension and the requested
precision length.
\end{lemma}

\paragraph{Proof of the placement law and pool claims.}
We verify \cref{lem:df-disjoint-context} for the joint permutation of
\cref{sec:lobster-combination}. Fix the public order on the body.
For rank decoding, write $\ReferenceQueueSet$ for the last
$\LARModulus$ reference vertices and
$\ReferenceAttachmentSet=\set{\ReferenceAttachmentVertex_{\GadgetIndex}}_{
\GadgetIndex=1}^{\GadgetCount}$ for the first
$\GadgetCount$, in labeled order. Their images are
\begin{equation*}
  \DFQueueSet=\JointPlacementMatrix(\ReferenceQueueSet),\qquad
  {\LegRightVertex}_{\GadgetIndex}
  =\JointPlacementMatrix(\ReferenceAttachmentVertex_{\GadgetIndex}).
\end{equation*}
The reference sets are disjoint by the geometric requirements, and
permutations preserve their disjointness.

For a queue set $\PlacementQueueSet$ of size $\LARModulus$ and an
ordered distinct attachment tuple
$(\PlacementAttachmentVertex_1,\ldots,
\PlacementAttachmentVertex_{\GadgetCount})$ in its complement, the
exact placement probability is
\begin{equation*}
 \Pr\!\left[\begin{gathered}\DFQueueSet=\PlacementQueueSet,\\
       (\HoeffdingAttachment_{\GadgetIndex})_{\GadgetIndex=1}^{\GadgetCount}
       =(\PlacementAttachmentVertex_{\GadgetIndex})_{\GadgetIndex=1}^{\GadgetCount}
       \end{gathered}\right]
 =\frac{1}{\binom{\InteriorVertexCount}{\LARModulus}
                 (\InteriorVertexCount-\LARModulus)_{\GadgetCount}}.
\end{equation*}
Indeed, there are
$\LARModulus!(\InteriorVertexCount-\LARModulus-\GadgetCount)!$
permutations realizing those images, out of
$\InteriorVertexCount!$ permutations. Summing over the
$(\InteriorVertexCount-\LARModulus)_{\GadgetCount}$ permitted attachment
tuples gives a uniform queue subset. Conversely, a fixed attachment
tuple permits $\binom{\InteriorVertexCount-\GadgetCount}{\LARModulus}$
queue sets, so its unconditional probability is
$1/(\InteriorVertexCount)_{\GadgetCount}$. The conditional attachment
law is uniform in the queue complement; the queue and attachments are
dependent because they are disjoint.

Removing the queue and attachments leaves the free vertices defined in
\cref{sec:df-protected-reference}. Assigning the first $\GadgetCount$
of them to the labeled legs leaves the ordinary-body pool in
\cref{sec:df-control-context}. The geometry inequality in
\cref{sec:lobster-combination} provides distinct compensators
and a nonempty ordinary-body pool. Each pool supports every selection
of at most $\BoundaryInverseOrder$ distinct controls under the two
size inequalities in \cref{lem:df-disjoint-context}.

The independent body-cycle permutation moves the path columns and
closure together, and the independent final permutation relabels the
whole load. Both obey \cref{sec:df-sampling-contract}. Distributional
averages use \cref{lem:df-disjoint-context} with the prescribed conditional
core law; conditioning on the core need not preserve queue uniformity.

\paragraph{Bounded implementation.}
Approximate the three independent uniform permutation laws using
independent fixed-length random bit strings. Choose
\[
  \AssemblyPermutationBits
  :=\left\lceil\log_2
    (3\BlockDim!/\BodyPermutationSamplingError)\right\rceil.
\]
For each of the two local permutations and the final common relabeling,
reduce a uniform $\AssemblyPermutationBits$-bit integer modulo the number
of permutations of the relevant vertex set, and decode this rank into a
permutation. Each rank probability differs from uniform by less than
$2^{-\AssemblyPermutationBits}$, so each permutation has statistical
error at most $\BodyPermutationSamplingError/3$ and their joint error is
at most $\BodyPermutationSamplingError$.
Factorials and permutation ranks have $O(\BlockDim\log\BlockDim)$ bits;
ranking inversion and the fixed bit count are polynomial in
$\BlockDim$ and $\log(1/\BodyPermutationSamplingError)$. Every output is a
valid permutation preserving disjointness of the reference sets.
Scanning the free
vertices in public order assigns all compensators in
$O(\InteriorVertexCount+\GadgetCount)$ operations. All exact affine and
carry properties are preserved. Core sampling and rejection retain the
resulting placement; the body-cycle and final permutations may be drawn
after acceptance. Rank decoding enumerates neither permutations nor
placements.

%% file: chapters/appendix-direct-fourier-cycle.tex
\subsection{Finite-Support Path Likelihoods and Protected Cycle Correlations}
\label{sec:df-cycle-proofs}

The protected latent edges of \cref{def:protected-body-cycle} are
independent and supported on
$[-\DFEdgeCutoff,\DFEdgeCutoff]\cap\ZZ$. These two properties suffice
for the fixed-order likelihood bounds below. Constants may depend on
the fixed order. Likelihoods are taken over integer differences until
reduction modulo $\LARModulus$ and are set to zero outside their
reference support.

\subsubsection{Fixed-order root kernels}
\label{sec:df-root-kernel-definition}

Put $\DFClippedIntegerMarginal=\DFProtectedEdgeLaw*\DFProtectedEdgeLaw$,
the integer difference marginal by symmetry, and define
\[
 \DFRootTagSize:=
 \sqrt{\frac{4\DFEdgeCutoff+1}{\InteriorVertexCount}}.
\]
Fix $\DFRootOrder$ different labeled body positions after an independent
uniform body permutation. Let $\DFJointRootLaw$ be their protected
integer joint law, define its likelihood
$\DFJointRootDensity=\DFJointRootLaw/
 \DFClippedIntegerMarginal^{\otimes\DFRootOrder}$, and define
\[
 \DFCenteredRootDensity_{\DFRootOrder}
 =\prod_{\CycleIndex=1}^{\DFRootOrder}
       (\mathrm{Id}-\mathbb E_{\DFClippedIntegerMarginal,\CycleIndex})
          \DFJointRootDensity.
\]
Here the coordinate expectations integrate only the indicated
coordinate under the product reference.

\begin{theorem}[Centered finite-support cycle density]
\label{thm:df-cycle-density}
For each fixed positive integer $\DFRootOrder$, assume
$\InteriorVertexCount\geq2\DFRootOrder$. Then
\begin{align*}
 \DFCenteredRootDensity_1&=0,\\
 \|\DFCenteredRootDensity_{\DFRootOrder}\|_
 {L^2(\DFClippedIntegerMarginal^{\otimes\DFRootOrder})}
 &\leq \DFProofConstant_{\DFRootOrder}
      (4\DFEdgeCutoff+1)^{\DFRootOrder/2}
      \InteriorVertexCount^{-\lceil\DFRootOrder/2\rceil},\\
 &\leq\DFProofConstant_{\DFRootOrder}
       \DFRootTagSize^{\DFRootOrder}.
\end{align*}
The same bounds hold after coordinatewise reduction modulo
$\LARModulus$, relative to
$(\DFProtectedRootMarginal)^{\otimes\DFRootOrder}$.
They hold conditionally on any data independent of the body permutation
and all latent edge variables.
\end{theorem}
The proof combines a finite-support path bound with the probability
that the selected sites have no isolated vertex.

\subsubsection{A finite-support path bound}

\begin{lemma}[Finite-support path likelihood]
\label[lemma]{lem:df-finite-support-path}
For every positive integer $\DFPathLength$, let
$\DFClippedPathLaw$ be the law of $\DFPathLength$ successive
differences of $\DFPathLength+1$
independent samples from $\DFProtectedEdgeLaw$. Then
\[
 \left\|\frac{\DFClippedPathLaw}
 {\DFClippedIntegerMarginal^{\otimes\DFPathLength}}\right\|_
 {L^2(\DFClippedIntegerMarginal^{\otimes\DFPathLength})}
 \leq(4\DFEdgeCutoff+1)^{\DFPathLength/2}.
\]
\end{lemma}
\begin{proof}
Index a vector and a family of matrices by integer latent values
$\CycleGaussianScalar,\CycleReferenceScalar
\in[-\DFEdgeCutoff,\DFEdgeCutoff]$. Set
\[
\begin{aligned}
 (\DFPathBoundaryVector)_{\CycleGaussianScalar}
   &:=\sqrt{\DFProtectedEdgeLaw(\CycleGaussianScalar)},\\
 (\DFPathTransferMatrix(\DFPathDifferenceEntry))_{
          \CycleGaussianScalar,\CycleReferenceScalar}
   &:=\sqrt{\DFProtectedEdgeLaw(\CycleGaussianScalar)
             \DFProtectedEdgeLaw(\CycleReferenceScalar)}
       \EventIndicator{\{\CycleReferenceScalar-\CycleGaussianScalar
                         =\DFPathDifferenceEntry\}}.
\end{aligned}
\]
The vector has Euclidean norm one, and
$\|\DFPathTransferMatrix(\DFPathDifferenceEntry)\|_{\mathrm F}^2
=\DFClippedIntegerMarginal(\DFPathDifferenceEntry)$, where
$\|\cdot\|_{\mathrm F}$ is the Frobenius norm. Expanding the matrix
product, ordered by increasing $\CycleIndex$, counts each latent
probability exactly once. Thus, for every
$\DFPathDifference\in\ZZ^{\DFPathLength}$,
\[
\begin{aligned}
 \DFClippedPathLaw(\DFPathDifference)
 &=\DFPathBoundaryVector^{\trans}
     \left(\prod_{\CycleIndex=1}^{\DFPathLength}
       \DFPathTransferMatrix(\DFPathDifferenceEntry_{\CycleIndex})\right)
        \DFPathBoundaryVector\\
 &\leq\prod_{\CycleIndex=1}^{\DFPathLength}
      \|\DFPathTransferMatrix(\DFPathDifferenceEntry_{\CycleIndex})\|_{\mathrm F}
 =\prod_{\CycleIndex=1}^{\DFPathLength}
      \DFClippedIntegerMarginal(\DFPathDifferenceEntry_{\CycleIndex})^{1/2}.
\end{aligned}
\]
On the product reference support, the squared path probability divided
by the product marginal is therefore at most one. Each difference
has at most $4\DFEdgeCutoff+1$ possible values. Summing over all such
tuples bounds the squared likelihood norm by
$(4\DFEdgeCutoff+1)^{\DFPathLength}$.
\end{proof}

\subsubsection{Random placement and centering}

\begin{proof}[Proof of \cref{thm:df-cycle-density}]
Condition first on the selected positions in the underlying cycle.
Their induced graph is a disjoint union of paths, since fewer than all
cycle vertices were selected. A path of $\DFPathLength$ vertices
uses $\DFPathLength+1$ consecutive latent variables. Distinct components
use disjoint latent variables, so the conditional likelihood is the
tensor product of their path likelihoods. By
\cref{lem:df-finite-support-path}, its norm is at most
$(4\DFEdgeCutoff+1)^{\DFRootOrder/2}$,
uniformly over the selected positions.

If a selected vertex is isolated, its difference is independent of all
other selected differences and has marginal
$\DFClippedIntegerMarginal$. Centering in that slot makes the whole
conditional likelihood zero. A forest on $\DFRootOrder$ vertices with
no isolated vertices has at most $\lfloor\DFRootOrder/2\rfloor$
components and hence at least $\lceil\DFRootOrder/2\rceil$ edges.
For any specified labeled path forest with
$\DFPlacementForestEdges$ edges, expose one vertex per component
freely and then expose each adjacent vertex. There are at most two
possible cycle neighbors at each such exposure, among at least
$\InteriorVertexCount-\DFRootOrder$ remaining positions. Thus its
embedding probability is at most
$(4/\InteriorVertexCount)^{\DFPlacementForestEdges}$.
There are only a number depending on $\DFRootOrder$ of possible
labeled induced forests. A union bound gives probability at most
$\DFProofConstant_{\DFRootOrder}
 \InteriorVertexCount^{-\lceil\DFRootOrder/2\rceil}$
of having no isolated vertex.

As an orthogonal projection, coordinate centering contracts product
$L^2$. Minkowski's inequality multiplies the path likelihood bound by
the probability of having no isolated vertex, without a square root.
This gives the first bound and also
$\DFCenteredRootDensity_1=0$. The second bound uses
$\lceil\DFRootOrder/2\rceil\geq\DFRootOrder/2$ and the root-tag definition.

For a deterministic coordinatewise map $\DFRootPushforward$, the
pushforward likelihood is the conditional expectation of the original
likelihood given the mapped coordinates, under the product reference.
This conditional expectation contracts $L^2$ and commutes with
coordinate centering. Applying it to reduction modulo
$\LARModulus$ proves the residue bound. Conditioning on the independent
data in the statement preserves the argument.
\end{proof}

An unchanged placement marginal does not suffice for the conditional
clause. A core depending on the body permutation or an acceptance event
reading the actual closure Gaussian need not satisfy its independence
condition.

%% file: chapters/appendix-direct-fourier-calibration.tex
\section{Positive Proposal and Response Calibration}
\label{sec:df-response-calibration}

\input{chapters/appendix-calibration-proofs.tex}
\input{chapters/appendix-direct-fourier-tables.tex}

%% file: chapters/appendix-calibration-proofs.tex
\subsection{Positive Proposal and First-Response Proofs}
\label{sec:df-calibration-proofs}

We specify the proposal data and verify the positive proposal and
first-response estimates of \cref{sec:direct-fourier-transport}.

\subsubsection{Anchor support and positive proposal laws}
\label{sec:df-baseline-proof}

\paragraph{Inward anchors and the independent blur.}
Let $\DFMidpoint=(\LARModulus-1)/2$ and let
$\DFAnchorSign(\CalibColor)$ equal $-1,0,1$ below, at, and above
$\DFMidpoint$. Draw $\DFAnchorIndex$ uniformly from
$\set{\CalibAnchorWidth,\ldots,2\CalibAnchorWidth-1}$.
Given $\CalibColor$, set $\DFSignChoice=\DFAnchorSign(\CalibColor)$
off the midpoint, and choose $\DFSignChoice$ fairly from $\set{-1,1}$
at the midpoint. Define the Markov kernel $\CalibAnchorLaw$ by
\begin{equation*}
 \DFAnchorPoint=\LARModulus-1-\CalibColor
                       +\DFSignChoice\DFAnchorIndex.
\end{equation*}
Writing $\BaseMatrixAnchor$ for this kernel in mass coordinates, its
smoothed version is
\begin{equation*}
 \DFSmoothedAnchor
   =\BaseMatrixLegConvolution\BaseMatrixAnchor.
\end{equation*}
Draw $\CalibPublicBlur\sim\DFProtectedRootMarginal$ independently,
using its supported integer lift, and set
\begin{equation*}
 \LegMarkedFirst=\CalibColor,\qquad
 \LegMarkedSecond=\DFAnchorPoint+\CalibPublicBlur,\qquad
 \LegBaselineShift=\DFSignChoice\DFAnchorIndex+\CalibPublicBlur,\qquad
 \LocalAtomWeight=\tfrac12.
\end{equation*}
The blur copies a distribution, not any actual root or closure value.
Require
\begin{equation*}
 \begin{gathered}
 \LARModulus\geq6\CalibAnchorWidth+1,\quad
 \CalibAnchorWidth\geq4\LegCutoff,\quad
 \CalibAnchorWidth\geq\LegTrimMargin+2\LegCutoff,\quad
 \LegSumBound\geq2\CalibAnchorWidth+\LegCutoff.
 \end{gathered}
\end{equation*}
Here $\CalibAnchorWidth$ is a positive integer, and the general leg
feasibility conditions of \cref{sec:leg-injection} remain in force.

\paragraph{A diffuse profile defined before the primary law.}
For each fixed $\DFSignChoice\in\set{-1,1}$, let
$\CalibSliceMean_{\DFSignChoice}$ be either endpoint marginal of
\[
 \mathbb E_{\DFAnchorIndex,\CalibPublicBlur,\LocalSecondError}
 \LegTrimmedLaw{\DFSignChoice\DFAnchorIndex+
                         \CalibPublicBlur-\LocalSecondError},
 \qquad
 \LocalSecondError\sim\LegClippedLaw,
 \qquad
 \CalibSliceMean=\tfrac12
       (\CalibSliceMean_{+1}+\CalibSliceMean_{-1}).
\]
These profiles use only the anchor and scalar laws. Write
$\DFReflection(\CalibColor)=\LARModulus-1-\CalibColor$
for reflection.

\paragraph{The balancing target.}
The profile used in \cref{sec:df-positive-baseline} is
\begin{equation*}
\begin{split}
 \BaseMatrixTarget={}&\BlockDim\CalibUniform
    -(\DFFreeCount+\GadgetCount\CalibActivation)\DFBodyProfile_0\\
 &-(\GadgetCount+\LARModulus)\DFBodyProfile_1
    -2\GadgetCount(1-\CalibActivation)\CalibSliceMean.
\end{split}
\end{equation*}

\paragraph{Matrix coordinates.}
Mass profiles are columns; matrix rows index outputs and columns index
inputs. The convolution and anchor operators have entries
\[
 \begin{aligned}
 (\BaseMatrixBodyConvolution)_{\CalibOtherColor,\CalibColor}
   &=\DFProtectedRootMarginal(\CalibOtherColor-\CalibColor),&
 (\BaseMatrixLegConvolution)_{\CalibOtherColor,\CalibColor}
   &=\LegClippedLaw(\CalibOtherColor-\CalibColor),\\
 (\BaseMatrixAnchor)_{\CalibOtherColor,\CalibColor}
   &=\Pr[\DFAnchorPoint=\CalibOtherColor\mid\CalibColor].
 \end{aligned}
\]
These matrices and $\DFSmoothedAnchor$ act on
$\mathbb R^{\LARModulus}$ and are column-stochastic.

Before choosing parameter sizes, define the losses
\begin{align*}
 \DFSliceDeletion&:=2\LegTrimMargin+2\CalibAnchorWidth+2\LegCutoff,
 \\
 \DFBaselineLoss&:=
 \frac{(\InteriorVertexCount+\GadgetCount\CalibActivation)
             2\PaddingSafeMargin}
 {\GadgetCount\CalibActivation
                   (\LARModulus-2\PaddingSafeMargin)}
 +\frac{2(1-\CalibActivation)\DFSliceDeletion}
 {\CalibActivation(\LARModulus-\DFSliceDeletion)},
 \\
 \DFCenterLoss&:=
       \frac{\CalibCompensation
                    (\LARModulus-2\PaddingSafeMargin)}
            {\CalibAnchorWidth},
 \\
 \DFPrimaryDeviation&:=
 \frac{4(\InteriorVertexCount+\GadgetCount\CalibActivation)
                    \PaddingSafeMargin}
      {\GadgetCount\CalibActivation\LARModulus}
 +\frac{4(1-\CalibActivation)\DFSliceDeletion}
       {\CalibActivation\LARModulus}.
\end{align*}
The denominators are positive under the anchor-feasibility conditions
above.

\begingroup
\mdfsetup{topline=true}
\begin{lemma}[Symbolic positive-proposal bounds]
\label[lemma]{lem:df-symbolic-baseline}
Assume the structural relations, anchor-feasibility conditions, and
general leg-feasibility conditions of \cref{sec:df-positive-baseline}.
If
\begin{equation*}
 \DFBaselineLoss,\DFCenterLoss\leq\tfrac12,
 \qquad \CalibCompensation\leq1,
\end{equation*}
then the proposal has the symmetry, atom, support, and averaged-mean
guarantees of \cref{thm:df-positive-baseline}, uniformly over disjoint
placements. Every sampled core satisfies the leg-feasibility conditions.
Moreover,
\begin{align*}
&\begin{gathered}
 \|\DFBaselinePrimary\|_\infty
       \leq\frac{\BlockDim}
                    {\GadgetCount\CalibActivation\LARModulus},
 \qquad
 \|\DFBaselinePrimary-\CalibUniform\|_1\leq\DFPrimaryDeviation,
 \end{gathered}
 \\
 &\begin{gathered}
 \CalibActivation
       \|\DFSmoothedAnchor\DFBaselinePrimary\|_\infty
 \leq\CalibActivation
       \left(\frac4{\LARModulus}
                  +\frac{\DFPrimaryDeviation}{\CalibAnchorWidth}\right),
 \\
 \|\DFBaselineCenter-\CalibUniform\|_{\CalibUniform^{-1}}
 +\|\CalibSafeUniform-\CalibUniform\|_{\CalibUniform^{-1}}\\
 \leq(2+\CalibCompensation)
       \sqrt{\frac{2\PaddingSafeMargin}
                       {\LARModulus-2\PaddingSafeMargin}}
      +\CalibCompensation
         \sqrt{3+\frac{\LARModulus\DFPrimaryDeviation}
                              {\CalibAnchorWidth}}.
 \end{gathered}
\end{align*}
\end{lemma}
\endgroup

\begin{proof}[Proof of \cref{lem:df-symbolic-baseline,thm:df-positive-baseline}]
Every anchor lies in
$[\CalibAnchorWidth,\LARModulus-1-\CalibAnchorWidth]$:
below the midpoint its lower endpoint is at least
$\DFMidpoint-2\CalibAnchorWidth+2\geq\CalibAnchorWidth$,
and its upper endpoint is at most
$\LARModulus-1-\CalibAnchorWidth$; reflection handles the other side.
At the midpoint, either sign gives an anchor in
$[\DFMidpoint-2\CalibAnchorWidth+1,
  \DFMidpoint+2\CalibAnchorWidth-1]$, which lies in the same safe
interval because $\DFMidpoint\geq3\CalibAnchorWidth$.
Consequently $\DFSmoothedAnchor$ is supported on
$\HistogramSafeCenters$, and
$\LegMarkedSecond-\LocalSecondError\in\LegInteriorColors$
for every protected second input. Also
$|\LegBaselineShift|<2\CalibAnchorWidth+\LegCutoff
\leq\LegSumBound$.
The primary and moving companion have the prescribed ordinary pair
sum, so the primary belongs to the untrimmed safe slice required by
\cref{def:leg-injection}. Thus every chosen core satisfies the leg-feasibility conditions of
\cref{sec:leg-injection}.
The anchor and smoothed anchor kernels have height at most
$1/\CalibAnchorWidth$.

The exact populations satisfy
$\DFFreeCount+\GadgetCount+\LARModulus=\InteriorVertexCount$.
Consequently the body coefficients in the primary numerator sum to
$\DFFreeCount+\GadgetCount\CalibActivation+\GadgetCount+\LARModulus
=\InteriorVertexCount+\GadgetCount\CalibActivation$.
Adding the diffuse coefficient gives
$\BlockDim-\GadgetCount\CalibActivation$, so the primary has mass
one. The center has mass one because the anchor is Markov.
Exchange symmetry makes the two endpoint marginals of each diffuse
slice equal. Reflection exchanges the two sign profiles and preserves
the body profiles and $\CalibSliceMean$, hence
the primary. The anchor intertwines reflection and symmetric scalar
convolution preserves it, so the center is symmetric. The symmetric
primary law and fair midpoint convention give equal sign probabilities,
so the diffuse profile is the prescribed mixture $\CalibSliceMean$.

Convolution bounds each body height by
$(\LARModulus-2\PaddingSafeMargin)^{-1}$.
Each diffuse slice deletes at most $\DFSliceDeletion$ colors,
so its height is at most
$(\LARModulus-\DFSliceDeletion)^{-1}$.
Dividing the first equation of \eqref{eq:df-baseline-block-system} by
$\GadgetCount\CalibActivation$ and subtracting the expression with
every profile replaced by uniform gives
$\LARModulus\DFBaselinePrimary\geq1-\DFBaselineLoss$.
Anchor height and support imply
$(1-\DFCenterLoss)\CalibSafeUniform
\leq\DFBaselineCenter
\leq(1+\CalibCompensation)\CalibSafeUniform$.
The assumed loss bounds prove the asserted positivity;
its upper height bound also uses
$\LARModulus-2\PaddingSafeMargin\geq\LARModulus/2$ and
$\CalibCompensation\leq1$.

The total variation distance of the safe uniform from full uniform
is $2\PaddingSafeMargin/\LARModulus$; for each diffuse slice it is
at most $\DFSliceDeletion/\LARModulus$.
Convolution contracts this distance. Inserting the corresponding
$\ell_1$ bounds into the first equation of
\eqref{eq:df-baseline-block-system} proves
$\|\DFBaselinePrimary-\CalibUniform\|_1\leq\DFPrimaryDeviation$.
Positivity of the subtracted profiles gives height at most
$\BlockDim/(\GadgetCount\CalibActivation\LARModulus)$.
At most $2\CalibAnchorWidth+2$ primary colors can contribute to one
anchor output, and each conditional probability is at most
$1/\CalibAnchorWidth$. Thus
$\|\DFSmoothedAnchor\CalibUniform\|_\infty\leq4/\LARModulus$.
Adding $\DFPrimaryDeviation/\CalibAnchorWidth$ yields
\[
 \|\DFSmoothedAnchor\DFBaselinePrimary\|_\infty
 \leq4/\LARModulus+\DFPrimaryDeviation/\CalibAnchorWidth,
\]
Also,
$\|\CalibSafeUniform-\CalibUniform\|_{\CalibUniform^{-1}}^2
=2\PaddingSafeMargin/(\LARModulus-2\PaddingSafeMargin)$.
For any probability vector, its squared distance from uniform in the
uniform-reference norm is at most $\LARModulus$ times its height
minus one. Applying this to
$\DFSmoothedAnchor\DFBaselinePrimary$ and using
the second equation of \eqref{eq:df-baseline-block-system} with the
triangle inequality proves
the remaining center bound.

The $\DFFreeCount$ free body vertices have marginal
$\BaseMatrixBodyConvolution\DFBaselineCenter$; the $\GadgetCount$
attachment-only and $\LARModulus$ queue-only vertices have marginal
$\DFBodyProfile_1$. The latter equality uses the common widths and
cutoffs of the clipped second leg input and the shared queue scalar.
By the exact leg kernel, the two off-body coordinates per leg have
total mean
\[
 \CalibActivation\DFBaselinePrimary
 +\CalibActivation\BaseMatrixBodyConvolution
                 \DFSmoothedAnchor\DFBaselinePrimary
 +2(1-\CalibActivation)\CalibSliceMean.
\]
Multiplying the second equation of \eqref{eq:df-baseline-block-system} by
$\DFFreeCount\BaseMatrixBodyConvolution$ gives the companion
cancellation described in \cref{sec:df-positive-baseline}, since
$\DFFreeCount\CalibCompensation=\GadgetCount\CalibActivation$.
Summing the remaining profiles gives the primary-law numerator,
which proves the mean claim after averaging the protected source.
This proves the symbolic lemma. The direct substitutions in
\cref{sec:df-parameter-verification} verify its structural and
feasibility conditions for the public family, and show that
$\DFBaselineLoss,\DFCenterLoss,\DFPrimaryDeviation$ and
$\CalibCompensation$ tend to zero. Its hypotheses therefore
hold for all sufficiently large $\HoeffdingParameterBase$, proving
\cref{thm:df-positive-baseline}.
\end{proof}

\subsubsection{The first response}
\label{sec:df-first-response-proof}

\begin{lemma}[Symbolic first-response bounds]
\label[lemma]{lem:df-symbolic-first-response}
Under the hypotheses of \cref{lem:df-symbolic-baseline}, the paired
lift of every centered primary score is centered and its complete
first histogram response is
$\CalibActivation\DFFirstResponse\CalibProfile$.
The operator $\DFFirstResponse$ preserves mass-zero profiles and,
in the primary-reference norm of \cref{lem:df-first-inverse},
\begin{equation*}
 \|\DFFirstResponse-\mathrm{Id}\|
       =O(\CalibActivation^{-1}
                 \sqrt{\CalibAnchorWidth/\LARModulus}).
\end{equation*}
For any such parameter family satisfying
$\CalibActivation^{-1}\sqrt{\CalibAnchorWidth/\LARModulus}=o(1)$,
eventually
\begin{equation*}
 \DFFirstResponse^{-1}
 =\mathrm{Id}
    -\frac{\BaseMatrixSignVector\BaseMatrixSignFunctional^{\mathsf T}}
            {\CalibActivation+
                \BaseMatrixSignFunctional^{\mathsf T}\BaseMatrixSignVector},
 \qquad \|\DFFirstResponse^{-1}\|\leq2.
\end{equation*}
\end{lemma}

\begin{proof}[Proof of \cref{lem:df-symbolic-first-response,lem:df-first-inverse}]
The primary score is centered by assumption. The compensator score
is centered because $\DFSmoothedAnchor$ is column-stochastic and
$\CalibProfile$ has mass zero. Independence of the two proposal
inputs then gives a centered paired lift. The sign profile also has
mass zero, so $\DFFirstResponse$ preserves centering.

At fixed primary color the two off-body mean profiles sum to
\[
 \CalibActivation\LocalPointMass{\CalibColor}
 +\CalibActivation\BaseMatrixBodyConvolution
                         (\DFSmoothedAnchor)_{\cdot,\CalibColor}
 +2(1-\CalibActivation)\CalibSliceMean
 +\CalibSignResponse\DFAnchorSign(\CalibColor).
\]
The primary score gives the corresponding signed combination. The
attachment marginal is independent of the primary color, so its first
response is zero. The compensator contributes
$-\CalibActivation\BaseMatrixBodyConvolution
                                 \DFSmoothedAnchor\CalibProfile$.
The centered public score integrates all other coordinate responses
to zero even with correlated roots, leaving exact companion cancellation.

By the anchor-feasibility conditions of \cref{sec:df-positive-baseline},
$\DFSliceDeletion=O(\CalibAnchorWidth)$ and
$\LARModulus-\DFSliceDeletion\geq\LARModulus/3$.
Each sign slice differs from uniform in $\ell_1$ by
$O(\CalibAnchorWidth/\LARModulus)$ and has height
$O(\LARModulus^{-1})$. Hence
\[
 \|\CalibSignResponse\|_1=O(\CalibAnchorWidth/\LARModulus),\quad
 \|\CalibSignResponse\|_\infty=O(\LARModulus^{-1}),\quad
 \|\CalibSignResponse\|_{\DFBaselinePrimary^{-1}}
        =O(\sqrt{\CalibAnchorWidth/\LARModulus}).
\]
The dual norm of the sign functional is at most one by
Cauchy--Schwarz. The perturbation norm is therefore
$O(\CalibActivation^{-1}
                 \sqrt{\CalibAnchorWidth/\LARModulus})$.
Under the lemma's smallness hypothesis, the denominator correction has
absolute value at most
$\|\CalibSignResponse\|_{\DFBaselinePrimary^{-1}}
=o(\CalibActivation)$.
Multiplying the two rank-one formulas proves the inverse formula of
\cref{lem:df-symbolic-first-response}; the perturbation bound gives
inverse norm at most two eventually.
For the public family, \cref{sec:df-parameter-verification} gives
$\CalibActivation^{-1}\sqrt{\CalibAnchorWidth/\LARModulus}
=O(\HoeffdingParameterBase/\DFPublicScale)
=O(\HoeffdingParameterBase^{-5})$.
Together with \cref{thm:df-positive-baseline}, this proves
\cref{lem:df-first-inverse}.
\end{proof}

%% file: chapters/appendix-direct-fourier-tables.tex
\subsection{Construction and Bounds for the Graded Response Matrices}
\label{sec:df-actual-tables}

Use the parameters and radius of \cref{sec:df-parameter-choice},
including $\DFPublicScale$, and fix a labeled disjoint placement from
\cref{thm:df-positive-baseline}. We construct coefficients in
$\DFMatrixSpace$ as product-reference density kernels, following the
coordinate convention of \cref{def:df-control-coordinates}.
All expectations are uncharged.
The independent-root comparison retains the protected marginal
$\DFProtectedRootMarginal$, the complete same-source leg channel,
and one common queue scalar. The protected-cycle response retains the dependent cycle law.
Use the sampling contract of \cref{sec:df-sampling-contract} and the
populations and pairings of \cref{sec:df-control-context}. Constants
may depend on fixed $\BoundaryInverseOrder$, but not on placement.

\subsubsection{Canonical coefficient algebra and its Hilbert norm}
\label{sec:df-table-algebra}

\begin{definition}[Canonical public-control coordinates]
\label[definition]{def:df-control-coordinates}
Let
\[
\begin{split}
\DFMatrixPrimarySpace
 &=\set{\DFTable:\LARField\to\mathbb R:
                    \mathbb E_{\DFBaselinePrimary}\DFTable=0},\\
\DFMatrixBodySpace
 &=\set{\DFTable:\HistogramSafeCenters\to\mathbb R:
                    \mathbb E_{\DFBaselineCenter}\DFTable=0},\\
\DFMatrixDegreeSpace_{\HoeffdingDegree}
 &=\bigoplus_{\ActivityPrimaryDegree+\ActivityBodyDegree=\HoeffdingDegree}
   \operatorname{Sym}^{\ActivityPrimaryDegree}(\DFMatrixPrimarySpace)
   \otimes\operatorname{Sym}^{\ActivityBodyDegree}(\DFMatrixBodySpace),\\
\DFMatrixSpace
 &=\mathbb R\oplus\bigoplus_{\HoeffdingDegree=1}^{\BoundaryInverseOrder}
                         \DFMatrixDegreeSpace_{\HoeffdingDegree}.
\end{split}
\]
Let $\DFMatrixCoord:\DFMatrixSpace\to\mathbb R^{\DFMatrixDimension}$
be the coordinate isomorphism constructed in \cref{sec:df-table-algebra},
and put $\DFMatrixZero=\DFMatrixCoord(1)$.
\end{definition}

Positive-degree coefficients are separately symmetric density kernels relative to
$\DFBaselinePrimary^{\otimes\ActivityPrimaryDegree}
 \DFBaselineCenter^{\otimes\ActivityBodyDegree}$, centered in every
input slot. The coordinate dimension satisfies
$\DFMatrixDimension\leq(2\LARModulus+\BoundaryInverseOrder)^{\BoundaryInverseOrder}$;
\cref{sec:df-table-algebra} supplies the basis and conversion bounds.
All estimates use \cref{sec:df-parameter-choice}, at fixed
$\BoundaryInverseOrder$, uniformly over placements.
Throughout, $\DFMatrixCoord$ identifies coefficient kernels with their
vectors and coefficient-space endomorphisms with their matrices.
Vector and induced operator norms are transported through this same
isomorphism.

\paragraph{The disjoint-slot product.}
The formal product $\CoreDisjointProduct$ tensors disjoint slots,
symmetrizes each type and truncates above degree $\BoundaryInverseOrder$.
It differs from physical multiplication, which includes overlaps.

\paragraph{A graded weighted Hilbert norm.}
\label{sec:df-graded-norm}
For a coefficient vector $\DFMatrixVector=\DFMatrixCoord(\DFTable)$,
put $\DFMatrixNorm{\DFMatrixVector}:=\DFTableNorm(\DFTable)$, where
\[
 \DFTableNorm(\DFTable)
 :=\sum_{\ActivityPrimaryDegree+\ActivityBodyDegree\leq\BoundaryInverseOrder}
 \FockRadius^{\ActivityBodyDegree}
 \ActivityPrimaryPool^{-\ActivityPrimaryDegree/2}
 \ActivityBodyPool^{-\ActivityBodyDegree/2}
 \|\DFTable_{\ActivityPrimaryDegree,\ActivityBodyDegree}\|_
 {L^2(\DFBaselinePrimary^{\otimes\ActivityPrimaryDegree}
       \otimes\DFBaselineCenter^{\otimes\ActivityBodyDegree})}.
\]
Add contributions of the same arity before taking their Hilbert norm.
The population factors match distinct-index averaging; the body-radius
factor matches the Fourier estimate.
\Cref{lem:df-table-algebra,lem:df-norm-comparison} establish
submultiplicativity and the retained-input fiber estimates used below.
Operator norms below are induced by this norm. It controls physical
moments but does not imply pointwise closeness of the weight to one.

\begin{lemma}[Finite-atom comparison]
\label[lemma]{lem:df-norm-comparison}
For a kernel of type $(\ActivityPrimaryDegree,\ActivityBodyDegree)$,
possibly taking values in a Hilbert space, fixing input subsets
$\CorePrimaryFixed,\CoreBodyFixed$ gives
\[
 \sup_{\mathrm{fixed}\ \CorePrimaryFixed,\CoreBodyFixed}
 \|\DFTable_{\ActivityPrimaryDegree,\ActivityBodyDegree}\|_
                         {L^2(\mathrm{remaining})}
 \leq(2\LARModulus)^{(|\CorePrimaryFixed|+|\CoreBodyFixed|)/2}
       \|\DFTable_{\ActivityPrimaryDegree,\ActivityBodyDegree}\|_2.
\]
\end{lemma}
\begin{proof}
In the squared product-reference norm, the summand for any fixed
input tuple has probability at least
$(2\LARModulus)^{-|\CorePrimaryFixed|-|\CoreBodyFixed|}$ by
\cref{thm:df-positive-baseline}. Dropping the other nonnegative
summands proves the fiber bound, also for a Hilbert-valued kernel.
\end{proof}

\paragraph{Coordinate construction.}
Coordinates are ordered by total degree and then by hybrid type, with
columns indexing inputs and rows outputs.
Choose reference colors $\DFMatrixReferenceColor\in\LARField$ and
$\DFMatrixReferenceCenter\in\HistogramSafeCenters$. For each
nonreference color use the centered point-mass density bases
\[
\begin{split}
\DFMatrixPrimaryBasis_{\CalibColor}
 &=\frac{\LocalPointMass{\CalibColor}}{\DFBaselinePrimary(\CalibColor)}
   -\frac{\LocalPointMass{\DFMatrixReferenceColor}}
          {\DFBaselinePrimary(\DFMatrixReferenceColor)},\\
\DFMatrixBodyBasis_{\CalibOtherColor}
 &=\frac{\LocalPointMass{\CalibOtherColor}}{\DFBaselineCenter(\CalibOtherColor)}
   -\frac{\LocalPointMass{\DFMatrixReferenceCenter}}
          {\DFBaselineCenter(\DFMatrixReferenceCenter)}.
\end{split}
\]
For each nondecreasing tuple of basis indices, average the tensor
product over all permutations within the primary slots and,
separately, all permutations within the body slots, including repeated
permutations. These averaged tensors, together with the scalar unit,
form the chosen basis. This defines the coordinate isomorphism $\DFMatrixCoord$ declared in
\cref{def:df-control-coordinates}.

In degree one, the coefficient of $\DFMatrixPrimaryBasis_{\CalibColor}$ in
$\DFTable$ is $\DFBaselinePrimary(\CalibColor)\DFTable(\CalibColor)$;
centering recovers its value at the reference color. The body formula
uses $\DFBaselineCenter$. In several slots, evaluate at nonreference
colors and multiply by their product reference probability. These
are the ordered tensor coefficients. Multiply by the orbit size
$\DFMatrixOrbitSize$ of distinct separate permutations to obtain the
averaged symmetric coefficients; divide by it to recover the ordered
expansion. Iterated centering recovers tuples containing reference colors.
A basis tensor is specified by a multiset of at most
$\BoundaryInverseOrder$ labels from the two basis types, which together
have at most $2\LARModulus$ labels. Order these labels and pad with a
distinct symbol to obtain a tuple of length $\BoundaryInverseOrder$.
This injective encoding gives
$\DFMatrixDimension\leq(2\LARModulus+1)^{\BoundaryInverseOrder}
\leq(2\LARModulus+\BoundaryInverseOrder)^{\BoundaryInverseOrder}$
for $\BoundaryInverseOrder\geq1$, proving the stated dimension bound.
The lower atom bounds in \cref{thm:df-positive-baseline} make both
basis transformations polynomially bounded at fixed order: each
uses at most $\BoundaryInverseOrder$ reference probabilities or their
inverses and permutation factors at most $\BoundaryInverseOrder!$.
The Fourier output of $\DFMatrixLeading$ stores the ordered symmetric
coefficient tensors of \cref{sec:df-target}, including degree
zero; \cref{lem:df-table-algebra} specifies the evaluation maps.

\paragraph{Multiplication entries.}
For $\DFMatrixVector=\DFMatrixCoord(\DFTable)$ and
$\DFMatrixOtherVector=\DFMatrixCoord(\DFOtherTable)$, define
\[
 \DFMatrixMultiply(\DFMatrixVector)\DFMatrixOtherVector
 =\DFMatrixCoord(\DFTable\CoreDisjointProduct\DFOtherTable),
 \qquad
 \DFMatrixBackground=\DFMatrixMultiply(\DFMatrixBackgroundVector).
\]
Let $\DFMatrixBasis_{\DFMatrixBasisIndex}$
range over the chosen basis, including the unit. Its structure
constants are defined by
\[
\DFMatrixBasis_{\DFMatrixBasisIndex}\CoreDisjointProduct
\DFMatrixBasis_{\DFMatrixSecondIndex}
 =\sum_{\DFMatrixThirdIndex}
 \DFMatrixStructure_{\DFMatrixThirdIndex,\DFMatrixBasisIndex,\DFMatrixSecondIndex}
 \DFMatrixBasis_{\DFMatrixThirdIndex}.
\]
The resulting matrix entries are
\begin{equation*}
 (\DFMatrixMultiply(\DFMatrixVector))_{
                 \DFMatrixThirdIndex,\DFMatrixSecondIndex}
 =\sum_{\DFMatrixBasisIndex}
 \DFMatrixStructure_{\DFMatrixThirdIndex,\DFMatrixBasisIndex,\DFMatrixSecondIndex}
 (\DFMatrixVector)_{\DFMatrixBasisIndex}.
\end{equation*}

\begin{lemma}[Algebra and leading evaluation]
\label[lemma]{lem:df-table-algebra}
The finite coefficient algebra is commutative and associative, and
\[
\DFTableNorm(\DFTable\CoreDisjointProduct\DFOtherTable)
 \leq\DFTableNorm(\DFTable)\DFTableNorm(\DFOtherTable).
\]
Define its leading evaluation by first multiplying a density by its
product reference measure, then applying
\begin{equation*}
\DFTableEvaluation_{\ActivityPrimaryDegree,\ActivityBodyDegree}
 =\DFTableSym\left[
 (\CalibActivation\DFFirstResponse)^{\otimes\ActivityPrimaryDegree}
 \otimes(\DFProtectedRootMarginal*)^{\otimes\ActivityBodyDegree}
 \right].
\end{equation*}
Here $\DFTableSym$ averages over all permutations of the output slots.
Evaluation gives symmetric tensors of mass-zero profiles in positive
degree and is a unital algebra homomorphism, not necessarily injective.
Its Fourier transform is $\DFMatrixLeading$. In coordinates,
\[
\begin{split}
\DFMatrixMultiply(\DFMatrixVector)\DFMatrixMultiply(\DFMatrixOtherVector)
 &=\DFMatrixMultiply\bigl(\DFMatrixCoord(
                      \DFTable\CoreDisjointProduct\DFOtherTable)\bigr),\\
\DFMatrixMultiply(\DFMatrixZero)&=\IdentityMatrix.
\end{split}
\]
For a positive-degree coefficient define the finite formal operations
\[
 \exp_{\CoreDisjointProduct}(\DFTable)
 =\sum_{\DFTablePower=0}^{\BoundaryInverseOrder}
       \frac{\DFTable^{\DFTablePower}}{\DFTablePower!},
 \qquad
 \log_{\CoreDisjointProduct}(1+\DFTable)
 =\sum_{\DFTablePower=1}^{\BoundaryInverseOrder}
       \frac{(-1)^{\DFTablePower+1}}{\DFTablePower}\DFTable^{\DFTablePower}.
\]
All powers use $\CoreDisjointProduct$. These operations are mutually
inverse, and the logarithm takes products of unit-constant
coefficients to sums. Multiplication by a unit-constant coefficient
is unit block lower triangular. Its reciprocal and inverse bound are in
\cref{prop:df-finite-coefficient-computation}.
These formal identities do not assert pointwise identities for physical weights:
\[
 \CoreUStatistic\ActivityCoreLift(\DFTable\CoreDisjointProduct\DFOtherTable)
 \ne(\CoreUStatistic\ActivityCoreLift\DFTable)
    (\CoreUStatistic\ActivityCoreLift\DFOtherTable)
 \quad\hbox{in general},
\]
because the product on the right includes overlapping indices.
\end{lemma}
\begin{proof}
Before symmetrization, the product-reference Hilbert norm of a tensor
product is the product of its two norms. The population and radius
weights multiply because their exponents add. Separate symmetrization
averages isometries and therefore contracts the Hilbert norm. Summing
over arities proves submultiplicativity; truncation only drops
nonnegative terms from the bound.
Associativity follows either by averaging all permutations once or
by viewing these tensors as symmetric polynomial coefficients.
The density-to-profile multiplication also factors over disjoint
slots. The slotwise linear maps commute with averaged symmetrization,
proving multiplicativity on the full coefficient algebra.
In coordinates, associativity gives the multiplication-matrix
homomorphism and multiplication by the scalar unit gives the identity.
Every product of more than $\BoundaryInverseOrder$ positive-degree
coefficients vanishes. The logarithm and exponential identities thus
follow from the corresponding formal power-series identities modulo
degree $\BoundaryInverseOrder+1$, with only finitely many terms per
coefficient. Multiplication by a
positive-degree coefficient strictly raises degree, giving the asserted
triangular form.
\end{proof}

Equalities with character polynomials below implicitly Fourier-transform
the evaluated profiles. Evaluation followed by this transform is
$\DFMatrixLeading\DFMatrixCoord$, with the coefficient convention of
\cref{def:df-control-coordinates}.

\subsubsection{Finite computation in the coefficient algebra}
\label{sec:df-finite-coefficient-computation}

Finite algebra implements the coefficient solve; stability uses
\cref{thm:stable-finite-dimensional-calibration}. The background
reciprocal is formed before solving the response system.

\begin{proposition}[Finite coefficient computation]
\label[proposition]{prop:df-finite-coefficient-computation}
Use the algebra and graded Hilbert norm of \cref{lem:df-table-algebra},
with coordinates respecting degree.
For any unit-constant coefficient $\DFTableTotalBackground$, let
$\DFMatrixBackground$ denote multiplication by it. The formal
reciprocal and multiplication inverse are
\[
 \DFTableTotalBackground^{-1}
 =\sum_{\DFTablePower=0}^{\BoundaryInverseOrder}
        (1-\DFTableTotalBackground)^{\DFTablePower},
 \qquad
 \DFMatrixBackground^{-1}\DFMatrixZero
       =\DFMatrixCoord(\DFTableTotalBackground^{-1}),
\]
where every power uses $\CoreDisjointProduct$. No smallness assumption
is needed for these finite identities. If
$\DFTableNorm(\DFTableTotalBackground-1)
 \leq\DFCalibrationBackgroundSize<1$, then
\[
 \DFTableNorm(\DFTableTotalBackground^{-1}-1)
 \leq\frac{\DFCalibrationBackgroundSize}{1-\DFCalibrationBackgroundSize},
 \qquad
 \DFMatrixNorm{\DFMatrixBackground^{-1}}
 \leq\frac1{1-\DFCalibrationBackgroundSize}.
\]

Suppose $\DFMatrixNormalized=\IdentityMatrix+\DFMatrixRaising+\DFMatrixRoot$,
where $\DFMatrixRaising$ strictly raises degree,
$\DFMatrixRoot$ does not decrease degree, both annihilate
$\DFMatrixZero$, and
\[
 \DFMatrixNorm{\DFMatrixRaising}+\DFMatrixNorm{\DFMatrixRoot}
       \leq\DFCalibrationResponseSize<1.
\]
The scalar block is the identity, with no coupling to positive degrees.
On positive degrees, solve in increasing degree: subtract the response
of the already determined lower-degree coefficients, then solve with
the current diagonal block. The same small-perturbation bound makes
every diagonal block invertible.
Under both displayed smallness hypotheses, the solution to
$\DFMatrixBackground\DFMatrixNormalized\DFMatrixSolution=\DFMatrixZero$
is unique and is computed using the background reciprocal followed by
these degree-block solves. The correction bound and its scalar and
degree-one structure for the constructed response are proved in
\cref{sec:df-graded-application}.

At fixed $\BoundaryInverseOrder$, computation uses polynomially many
arithmetic operations in the matrix dimension.
\Cref{sec:df-numerical-verification} certifies the data approximations,
coordinate conditioning, precision and work bounds for the public
family, preserving structural zero blocks and the scalar identity.
Nonrational data are approximated rather than computed exactly in
finite bits.
\end{proposition}
\begin{proof}
Multiplication by $\DFTableTotalBackground-1$ is nilpotent, giving
the finite reciprocal. Submultiplicativity bounds
$\|\DFMatrixBackground-\IdentityMatrix\|$ by
$\DFCalibrationBackgroundSize$; applying
\cref{thm:stable-finite-dimensional-calibration} to
$\DFMatrixBackground$ gives the stated
inverse bounds. The reference-residual estimate with reference
$\DFMatrixZero$ also gives the background reciprocal bound. Homogeneous projections
contract the norm, so the same response bound makes each diagonal
block invertible. The response does not lower degree, so each block
equation involves only the current and already solved degrees.
Unit-constant background and invertible
diagonal blocks suffice for the finite solve; smallness certifies
invertibility and stability. Finite products and block elimination use
polynomially many arithmetic operations in the matrix dimension.
\end{proof}

\paragraph{Finite block solution.}
The finite solve of \cref{prop:df-finite-coefficient-computation}
starts with scalar correction zero and proceeds in increasing degree
$1\leq\HoeffdingDegree\leq\BoundaryInverseOrder$ by
\[
 (\DFMatrixCorrection)_{\HoeffdingDegree}
 =-\DFMatrixDiagonalBlock_{\HoeffdingDegree\HoeffdingDegree}^{-1}
 \left((\DFMatrixBackgroundVector)_{\HoeffdingDegree}
 +\sum_{\DFMatrixDegreeIndex=1}^{\HoeffdingDegree-1}
   (\DFMatrixLifted)_{\HoeffdingDegree,\DFMatrixDegreeIndex}
   (\DFMatrixCorrection)_{\DFMatrixDegreeIndex}\right).
\]

\subsubsection{Concrete background and local transfer estimates}
\label{sec:df-table-background}

\begin{lemma}[Weighted Hilbert transfer]
\label[lemma]{lem:df-joint-transfer}
A local linear operator consumes $\DFTableConsumedPrimary$ primary
and $\DFTableConsumedBody$ body slots and emits
$\DFTableNewPrimary$ centered primary output profiles. Suppose its raw density operator
has Hilbert norm at most $\CoreHilbertBound$, with the same bound for
Hilbert-valued retained inputs. After lifting the new output
profiles by $\CalibActivation^{-1}\DFFirstResponse^{-1}$ and storing
the resulting densities, its graded operator norm is at most
\[
 \DFProofConstant_{\BoundaryInverseOrder}
 \FockRadius^{-\DFTableConsumedBody}
 \ActivityBodyPool^{\DFTableConsumedBody/2}
 \ActivityPrimaryPool^{(\DFTableConsumedPrimary-\DFTableNewPrimary)/2}
 \CalibActivation^{-\DFTableNewPrimary}\CoreHilbertBound.
\]
The same assertion holds before separate symmetrization.
\end{lemma}
\begin{proof}
The output/input weight ratio in $\DFTableNorm$ gives the radius
and pool factors. Each first inverse costs at most two by
\cref{lem:df-first-inverse}; activation inverses give the remaining
factors. Symmetrization contracts, and retained joint Hilbert
arguments are covered by the assumed local operator bound.
\end{proof}

For the public family, the two new-slot factors are
\[
 \CalibActivation^{-1}\ActivityPrimaryPool^{-1/2}
   =O(\HoeffdingParameterBase\DFPublicScale^{-8}),\qquad
 \FockRadius\ActivityBodyPool^{-1/2}
   =O(\HoeffdingParameterBase^{1/4}\DFPublicScale^{-10}).
\]
Consuming a primary slot costs $O(\DFPublicScale^8)$; consuming
an ordinary slot costs at most $O(\DFPublicScale^{10})$.
These factors also apply to retained joint inputs.

\paragraph{Independent-root background certificates.}

For a probability profile $\CalibProposal$, write
$\DFTableCentering{\CalibProposal}$ for the slotwise projection
subtracting total mass times $\CalibProposal$. A profile certificate
is a specified canonical density coefficient with that leading
evaluation. Primary lifts use the first inverse; safe body seeds
remain body slots without inverting their convolution.

\paragraph{Fringe and translated safe seeds.}
Set $\DFTableSafeHeight=(\LARModulus-2\PaddingSafeMargin)^{-1}$ and
$\DFTableFringe=\DFTableSafeHeight-\DFBodyProfile_0$, using the body
profile of \cref{sec:df-positive-baseline}.
The positive-baseline formula and the pointwise height bounds used
in \cref{thm:df-positive-baseline}, specialized as in
\cref{sec:df-parameter-choice}, give
\begin{equation*}
0\leq\DFTableFringe\leq
 \CalibCompensation\DFBaselinePrimary,
\qquad
\sum\DFTableFringe=O(\LegCutoff/\LARModulus).
\end{equation*}
In the baseline formula, the coefficient of $\DFTableFringe$ is
$\DFFreeCount+\GadgetCount\CalibActivation$. The remaining positive
constant is $\GadgetCount\CalibActivation\DFTableSafeHeight$;
its safe-interval and diffuse-slice losses are each
$O(\HoeffdingParameterBase^{-1})$ relative to this constant by
\cref{sec:df-baseline-proof}. Dividing by the fringe coefficient gives
the displayed bound for sufficiently large parameters.
Since
$\DFBodyProfile_0-\CalibUniform
 = (\sum\DFTableFringe)\CalibUniform-\DFTableFringe$,
\begin{equation*}
\|\DFBodyProfile_0-\CalibUniform\|_{\DFBaselinePrimary^{-1}}
 =O(\HoeffdingParameterBase^{1/2}\DFPublicScale^{-6}).
\end{equation*}
The squared Hilbert estimate uses
$\sum\DFTableFringe^2/\DFBaselinePrimary
 \leq\CalibCompensation\sum\DFTableFringe$.

The safe projection $\DFTableSafeProjection$ keeps safe colors and
replaces every unsafe color by an independent $\CalibSafeUniform$.
For an integer shift $\DFTableScalar$, define cyclic translation of a
mass profile $\CalibProposal$ by
\[
 (\DFTableTranslate_{\DFTableScalar}\CalibProposal)(\CalibColor)
 :=\CalibProposal(\CalibColor-\DFTableScalar).
\]
For $|\DFTableScalar|<2\LegCutoff$, define
\begin{align*}
\DFTableShiftDifference&=\DFTableTranslate_{\DFTableScalar}
                    \CalibSafeUniform-\CalibSafeUniform,&
\DFTableUnsafePart&=\EventIndicator{\HistogramSafeCenters^c}
                    \DFTableTranslate_{\DFTableScalar}\CalibSafeUniform,\\
\DFTableUnsafeMass&=\sum\DFTableUnsafePart,&
\DFTableSafeSeed&=\DFTableShiftDifference-\DFTableUnsafePart
                       +\DFTableUnsafeMass\CalibSafeUniform,\\
\DFTableShiftResidual&=\DFProtectedRootMarginal*\DFTableUnsafePart
                       -\DFTableUnsafeMass\DFBodyProfile_0.
\end{align*}
These give the exact mass-zero identity
\begin{equation*}
\DFProtectedRootMarginal*\DFTableTranslate_{\DFTableScalar}
 \CalibSafeUniform-\CalibUniform
 =\DFBodyProfile_0-\CalibUniform
   +\DFProtectedRootMarginal*\DFTableSafeSeed+\DFTableShiftResidual.
\end{equation*}
Only $O(\LegCutoff)$ boundary colors change, so
$\|\DFTableSafeSeed\|_{\DFBaselineCenter^{-1}}
 =O(\HoeffdingParameterBase\DFPublicScale^{-4})$.
Moreover
$\DFProtectedRootMarginal*\DFTableUnsafePart
 \leq\DFTableFringe$; hence the residual has the Hilbert bound in
the fringe estimate established above. The new-slot costs show that the
certificate $\DFTableTranslationCertificate_{\DFTableScalar}$ defined
by this recipe has norm
$O(\HoeffdingParameterBase^{3/2}\DFPublicScale^{-14})$.
For the public family, $\CalibAnchorWidth\geq5\LegCutoff$ for
$\HoeffdingParameterBase\geq2$. Hence every translated anchor column
used below stays on $\HistogramSafeCenters$ whenever
$|\DFTableScalar|<2\LegCutoff$.
Two further useful safe/profile estimates are
\[
\|\DFTableTranslate_{-\DFTableScalar}\CalibAnchorLaw
             \DFBaselinePrimary-\CalibSafeUniform\|_{\DFBaselineCenter^{-1}}
 =O(\HoeffdingParameterBase\DFPublicScale^{-1}),\qquad
\|\DFBaselinePrimary-\CalibUniform\|_{\DFBaselinePrimary^{-1}}
 =O(\HoeffdingParameterBase^{1/2}\DFPublicScale^{-1}).
\]
For the first, the anchor applied to uniform input differs only on
$O(\CalibAnchorWidth)$ endpoint or midpoint transition colors;
the remaining change is at most
$\DFProofConstant\sqrt{\LARModulus/\CalibAnchorWidth}
 \|\DFBaselinePrimary-\CalibUniform\|_1$ by Minkowski.
Protected translation and safe trimming add
$O(\sqrt{\LegCutoff/\LARModulus})$.
For the second use $\CalibUniform/\DFBaselinePrimary\leq2$ and
$\sum (\DFBaselinePrimary-\CalibUniform)^2/\DFBaselinePrimary
 \leq \DFProofConstant\|\DFBaselinePrimary-\CalibUniform\|_1$.
The first is retained as a safe body seed; the second has a primary
certificate of norm $O(\HoeffdingParameterBase^{3/2}\DFPublicScale^{-9})$.

The first estimate also holds after averaging the protected translate,
so the balancing formula of \cref{sec:df-positive-baseline} gives
\[
\|\DFBaselineCenter-\CalibSafeUniform\|_{\DFBaselineCenter^{-1}}
 =\CalibCompensation
  \|\DFSmoothedAnchor\DFBaselinePrimary-\CalibSafeUniform\|_{\DFBaselineCenter^{-1}}
 =O(\DFPublicScale^{-5}).
\]
Its body certificate costs $O(\HoeffdingParameterBase^{1/2}\DFPublicScale^{-15})$.
Together with the fringe primary certificate this defines a certificate
$\DFTableFreeCertificate$ of norm $O(\HoeffdingParameterBase^{3/2}\DFPublicScale^{-14})$
for $\DFProtectedRootMarginal*\DFBaselineCenter-\CalibUniform$.
The same profile has raw primary-reference norm
$O(\DFPublicScale^{-5})$: convolution from the center
reference is bounded by conditional Jensen and
$\sup(\DFProtectedRootMarginal*\DFBaselineCenter)/\DFBaselinePrimary\leq8$,
and its fringe part is smaller.

\begin{lemma}[Diffuse and marked pair certificates]
\label[lemma]{lem:df-pair-certificates}
At fixed protected leg scalar $\DFTableScalar$, let
$\DFTableDiffusePair_{\DFTableScalar}$ be the actual diffuse pair,
including its random anchor sign, length-$\CalibAnchorWidth$ interval,
and independent public blur. Define
\begin{align*}
\DFTableDiffuseSeed_{\DFTableScalar}
 &=\DFTableCentering{\DFBaselineCenter}^{\otimes2}
   (\DFTableSafeProjection\otimes\DFTableSafeProjection)
       \DFTableDiffusePair_{\DFTableScalar},\\
\DFTableDiffuseResidual_{\DFTableScalar}
 &=\DFTableCentering{\CalibUniform}^{\otimes2}
       \DFTableDiffusePair_{\DFTableScalar}
   -(\DFProtectedRootMarginal*\otimes\DFProtectedRootMarginal*)
       \DFTableDiffuseSeed_{\DFTableScalar}.
\end{align*}
Their profile norms in the respective two-slot references are
$O(\DFPublicScale)$ and
$O(\HoeffdingParameterBase\DFPublicScale^{-2})$.
For $\DFTableAnchorAtScalar=\DFTableTranslate_{-\DFTableScalar}
 \CalibAnchorLaw$, put
\[
\DFTableMarkedPair_{\DFTableScalar}(\CalibColor,\CalibOtherColor)
 =\DFBaselinePrimary(\CalibColor)
   (\DFProtectedRootMarginal*\DFTableAnchorAtScalar_{\CalibColor})
                 (\CalibOtherColor),\quad
\DFTableMarkedSeed_{\DFTableScalar}
 =(\DFTableCentering{\CalibUniform}\otimes
   \DFTableCentering{\DFBaselineCenter})
 [\DFBaselinePrimary(\CalibColor)
            \DFTableAnchorAtScalar_{\CalibColor}(\CalibOtherColor)].
\]
Then $\DFTableMarkedSeed_{\DFTableScalar}$ is supported safely in its
second slot, has norm $O(\DFPublicScale)$ in
$\DFBaselinePrimary^{-1}\otimes\DFBaselineCenter^{-1}$, and
\begin{equation*}
\begin{split}
\DFTableCentering{\CalibUniform}^{\otimes2}
       \DFTableMarkedPair_{\DFTableScalar}
={}&(\LocalIdentityOperator\otimes\DFProtectedRootMarginal*)
       \DFTableMarkedSeed_{\DFTableScalar}\\
 &+(\DFBaselinePrimary-\CalibUniform)\otimes
       (\DFProtectedRootMarginal*\DFBaselineCenter-\CalibUniform).
\end{split}
\end{equation*}
\end{lemma}
\begin{proof}
Conditioning on sign and public blur puts the diffuse pair in a band
of $O(\LARModulus\CalibAnchorWidth)$ cells, each of mass at most
$O((\LARModulus\CalibAnchorWidth)^{-1})$. Its uniform-profile norm is
therefore $O(\sqrt{\LARModulus/\CalibAnchorWidth})$.
Safe projection has column sums one and row sums at most two; its
marginals have height $O(\LARModulus^{-1})$. Projection and centering
are bounded in this profile norm, and
$\DFBaselineCenter\geq\CalibSafeUniform/2$ on safe support. This proves
the safe-seed bound.

Protected translations shift the length-$\CalibAnchorWidth$
sum-offset interval by $O(\LegCutoff)$, costing
$O(\LegCutoff/\CalibAnchorWidth)$ in total variation. At a matched
sum $\LegSliceShift$, the slice
$\LegTrimmedSlice{\LegSliceShift-\CalibRootValue_1-\CalibRootValue_2}
 +\CalibRootValue_1$ and $\LegTrimmedSlice{\LegSliceShift}$
have endpoints within $O(\LegCutoff)$ and lengths
$\Theta(\LARModulus)$. Their uniform laws, including denominator
changes, cost $O(\LegCutoff/\LARModulus)$ to couple.
Safe projection has the same cost by the marginal heights.
Modulo reduction and convolution contract $\ell_1$, while translation
and convolution preserve the height bound. Thus the uncentered
difference from the convolved projected pair has $\ell_1$ norm
$O(\LegCutoff/\CalibAnchorWidth+\LegCutoff/\LARModulus)$ and
maximum entry $O((\LARModulus\CalibAnchorWidth)^{-1})$.
Interpolation, also after averaging sign and blur, gives
\[
 \DFProofConstant\left(
       \frac{\sqrt{\LARModulus\LegCutoff}}{\CalibAnchorWidth}
       +\sqrt{\frac{\LegCutoff}{\CalibAnchorWidth}}\right)
 =O(\HoeffdingParameterBase\DFPublicScale^{-2}).
\]
Since
\[
 \BaseMatrixBodyConvolution\DFTableCentering{\DFBaselineCenter}
 =\DFTableCentering{\BaseMatrixBodyConvolution\DFBaselineCenter}
                  \BaseMatrixBodyConvolution,
\]
the convolved seed is centered at
$\BaseMatrixBodyConvolution\DFBaselineCenter$. Expanding the two
centering differences uses only bounded one-coordinate marginal norms,
so recentering at $\CalibUniform$ costs at most
$\DFProofConstant\|\DFBaselineCenter-\CalibUniform\|_{\CalibUniform^{-1}}
 =O(\HoeffdingParameterBase\DFPublicScale^{-4})$
by convolution contraction and \cref{sec:df-baseline-proof}.
Combining the errors and using $\DFBaselinePrimary\geq\CalibUniform/2$
proves the residual bound.

For the marked seed, the uncentered squared norm is at most
$\DFProofConstant\LARModulus/\CalibAnchorWidth$, by the anchor-column height and
unit column mass. Each centering is bounded in the indicated profile
norm: $\CalibUniform/\DFBaselinePrimary\leq2$ and the second
reference is its own centering measure. This proves its bound.
Convolution commutes with total mass, so applying it to the seed
centers the second output at
$\DFProtectedRootMarginal*\DFBaselineCenter$.
Changing this centering to $\CalibUniform$ gives exactly the product
in the marked mixed identity of \cref{lem:df-pair-certificates}.
\end{proof}

The diffuse safe seed uses two body slots, its residual two primary
slots, and the marked seed one slot of each type. Including all
$\GadgetCount$ legs and the marked activation, their weighted costs are
respectively
\[
 O(\HoeffdingParameterBase\DFPublicScale^{-3}),\qquad
 O(\HoeffdingParameterBase^3\DFPublicScale^{-2}),\qquad
 O(\HoeffdingParameterBase^{1/2}\DFPublicScale^{-1}).
\]
All are $O(\DFPublicScale^{-1/2})$.
The marked product residual is smaller: the primary and free
certificates above give population-weighted cost
$O(\HoeffdingParameterBase^2\DFPublicScale^{-7})$.
For either diffuse marginal $\CalibSliceMean_{\DFTableScalar}$,
use the exact decomposition
\[
\CalibSliceMean_{\DFTableScalar}-\CalibUniform
 =\DFProtectedRootMarginal*
    (\DFTableSafeProjection\CalibSliceMean_{\DFTableScalar}
      -\DFBaselineCenter)
  +(\CalibSliceMean_{\DFTableScalar}
       -\DFProtectedRootMarginal*
          \DFTableSafeProjection\CalibSliceMean_{\DFTableScalar})
  +(\DFProtectedRootMarginal*\DFBaselineCenter-\CalibUniform).
\]
The first parenthesis is a mass-zero safe seed with bounded center
norm, since its marginal height is $O(\LARModulus^{-1})$.
The second is a mass-zero primary residual. Marginalizing the
translation/projection comparison in \cref{lem:df-pair-certificates}
contracts its uniform-profile norm, giving
$O(\HoeffdingParameterBase\DFPublicScale^{-2})$ also in the primary
reference. Lift these two terms by their respective slot maps and use
$\DFTableFreeCertificate$ for the third. This specifies the diffuse
marginal certificate, of norm
$O(\HoeffdingParameterBase^2\DFPublicScale^{-10})$.

\paragraph{One scalar per whole triple.}
At fixed $\DFTableScalar$ the attachment has law
$\DFProtectedRootMarginal*
 \DFTableTranslate_{\DFTableScalar}\CalibSafeUniform$.
Disjoint placement excludes queue injection at this attachment.
The off-body pair and this attachment are conditionally independent
in the independent-root comparison. Their triple is
\[
\CountExpectation_{\DFTableScalar}
 \left[((1-\CalibActivation)\DFTableDiffusePair_{\DFTableScalar}
       +\CalibActivation\DFTableMarkedPair_{\DFTableScalar})
 \otimes(\DFProtectedRootMarginal*
 \DFTableTranslate_{\DFTableScalar}\CalibSafeUniform)
 \right].
\]
The constant activation is justified by independence of the first
leg input from $\DFTableScalar$. Construct quadratic and cubic
certificates by expanding this displayed tensor, substituting the
specified marginal and pair certificates, multiplying at the same
$\DFTableScalar$, and only then averaging that scalar. Denote them
by $\DFTableLegQuadratic_0$ and
$\DFTableLegCubic_0$. They do not depend on the queue scalar.
The pair bounds and attachment certificate give
\[
 \GadgetCount\DFTableNorm(\DFTableLegQuadratic_0)
 =O(\DFPublicScale^{-1/2}),\qquad
 \GadgetCount\DFTableNorm(\DFTableLegCubic_0)
 =O(\DFPublicScale^{-14}).
\]
Indeed, multiplying the attachment certificate by the marked and
diffuse marginal certificates, including activation and summing legs,
costs
$O(\HoeffdingParameterBase^2\DFPublicScale^{-7}
   +\HoeffdingParameterBase^{7/2}\DFPublicScale^{-8})$.
The marked primary gives the first term; the marked companion is
smaller. The third output adds the translation certificate to a pair.

Define the canonical coefficient
$\DFTableTranslationAverage_1
 =\CountExpectation_{\DFTableScalar\sim\LegClippedLaw}
       \DFTableTranslationCertificate_{\DFTableScalar}$.
The subscript on this averaged certificate counts the one scalar
injection; it does not denote translation by the integer $1$.
Choose the linear leg certificate to be
\begin{equation*}
\DFTableLegLinear_0=
 -\frac{\DFFreeCount}{\GadgetCount}\DFTableFreeCertificate
 -\frac{\LARModulus}{\GadgetCount}\DFTableTranslationAverage_1.
\end{equation*}
Its evaluation is the true sum of the three leg marginal deviations:
the averaged mean identity of \cref{thm:df-positive-baseline} gives
$\DFFreeCount\DFTableEvaluation\DFTableFreeCertificate
 +\LARModulus\DFTableEvaluation\DFTableTranslationAverage_1$
plus $\GadgetCount$ times that sum equal to zero.
Since $\DFFreeCount/\GadgetCount=O(\DFPublicScale^4)$ and
$\LARModulus/\GadgetCount=O(\DFPublicScale^{-8})$, this linear
certificate has norm
$O(\HoeffdingParameterBase^{3/2}\DFPublicScale^{-10})$.
With these choices the complete fixed-queue linear coefficient is
\begin{equation}
\label{eq:df-disjoint-table-cancellation}
\begin{split}
\DFTableMeanCertificate_{\DFTableQueue}
 &=\DFFreeCount\DFTableFreeCertificate
   +\LARModulus\DFTableTranslationCertificate_{\DFTableQueue}
   +\GadgetCount\DFTableLegLinear_0\\
 &=\LARModulus
       (\DFTableTranslationCertificate_{\DFTableQueue}
                    -\DFTableTranslationAverage_1).
\end{split}
\end{equation}
This is an equality before evaluation. In the fixed public family,
both $\LegClippedLaw$ and $\QueueClippedLaw$ are the same
$\Gaussian{\ZZ}{\GaussianWidth}$ conditioned on
$|\DFTableScalar|<\LegCutoff=\PaddingChargeCutoff$. Consequently
$\CountExpectation_{\DFTableQueue\sim\QueueClippedLaw}
 \DFTableTranslationCertificate_{\DFTableQueue}
 =\DFTableTranslationAverage_1$ as canonical coefficients and
$\CountExpectation_{\DFTableQueue}\DFTableMeanCertificate_{\DFTableQueue}=0$.
Its uniform norm is
$O(\HoeffdingParameterBase^{3/2}\DFPublicScale^{-6})$.

\begingroup
\mdfsetup{topline=true}
\begin{proposition}[Complete background coefficients]
\label[proposition]{prop:df-actual-background}
There are explicit unit-constant canonical coefficients
$\DFTableBackground_{\DFTableQueue}$ such that
\[
\DFTableEvaluation\DFTableBackground_{\DFTableQueue}
 =[\DFTableIndependentPolynomial_{\DFTableQueue}]_{\leq\BoundaryInverseOrder},
\qquad
\sup_{\DFTableQueue}
 \bigl[\DFTableNorm(\DFTableBackground_{\DFTableQueue}-1)
       +\DFTableNorm(\DFTableBackground_{\DFTableQueue}^{-1}-1)\bigr]
 =O_{\BoundaryInverseOrder}(\HoeffdingParameterBase^{-3}).
\]
Here $\DFTableIndependentPolynomial_{\DFTableQueue}$ is the character
polynomial of the independent-root proposal at the fixed queue value,
with $\DFFreeCount$ free bodies, $\LARModulus$ queue-only bodies, and
$\GadgetCount$ complete leg triples.
\end{proposition}
\endgroup
\begin{proof}
The free population includes the $\GadgetCount$ compensators, each
with an independent control and comparison root before scoring.
Its positive factor multiplies the leg-triple factor; pairing them
for the response leaves the vertex counts unchanged.
Define $\DFTableBackground_{\DFTableQueue}$ by the finite formal product
\[
(1+\DFTableFreeCertificate)^{\DFFreeCount}
(1+\DFTableTranslationCertificate_{\DFTableQueue})^{\LARModulus}
(1+\DFTableLegLinear_0+\DFTableLegQuadratic_0+\DFTableLegCubic_0)
 ^{\GadgetCount}.
\]
Each factor has the specified leading evaluation, so
\cref{lem:df-table-algebra} gives the complete independent-root
polynomial.
Its finite formal logarithm $\DFTableLog_{\DFTableQueue}$ is the sum
of the three role logarithms by \cref{lem:df-table-algebra}. Its degree-one
coefficient is exactly $\DFTableMeanCertificate_{\DFTableQueue}$ by
\eqref{eq:df-disjoint-table-cancellation}. The complete nonconstant leg factor has norm
$O(\HoeffdingParameterBase^{3/2}\DFPublicScale^{-10})$, dominated
by its linear certificate. The free and queue factors have norm
$O(\HoeffdingParameterBase^{3/2}\DFPublicScale^{-14})$.
There are no cross-role products in the logarithm. All powers of order
at least two are therefore bounded by the corresponding squared
norm times a fixed-order constant. Thus
\[
\begin{split}
 \DFTableNorm(\DFTableLog_{\DFTableQueue})
 &\leq\DFProofConstant_{\BoundaryInverseOrder}
 \bigl(\HoeffdingParameterBase^{3/2}\DFPublicScale^{-6}
       +\HoeffdingParameterBase^3\DFPublicScale^{-8}
       +\DFPublicScale^{-1/2}
       +\HoeffdingParameterBase^3\DFPublicScale^{-4}\bigr)\\
 &=O_{\BoundaryInverseOrder}(\DFPublicScale^{-1/2}).
\end{split}
\]
The four terms account respectively for the common-queue linear
residual, higher free and queue powers, leg terms of degree at least
two, and higher powers of the complete leg factor. The finite
exponential identity of \cref{lem:df-table-algebra} gives the same
bound for $\DFTableBackground_{\DFTableQueue}-1$. Since
$\DFPublicScale=\HoeffdingParameterBase^8$, this is
$O_{\BoundaryInverseOrder}(\HoeffdingParameterBase^{-4})$ and hence
implies the stated bound.
The background reciprocal estimate proved in
\cref{prop:df-finite-coefficient-computation} gives it also for
$\DFTableBackground_{\DFTableQueue}^{-1}-1$, uniformly over the queue.
\end{proof}

\subsubsection{Normalized response blocks and protected-cycle defects}
\label{sec:df-table-response}

For a canonical coefficient $\DFTable$, let
$\DFTableSignedPolynomial_{\DFTableQueue}(\DFTable)$ be the
protected-cycle endpoint character polynomial weighted by
$\CoreUStatistic\ActivityCoreLift\DFTable$. This map is linear;
its scalar input one gives the unweighted polynomial
$\DFTableCyclePolynomial_{\DFTableQueue}$.
Use a superscript $0$ on the signed polynomial for independent roots.
Omitting the queue subscript denotes averaging the complete fixed-queue
polynomial under $\QueueClippedLaw$.
All polynomial products, reciprocals, and identities in the remainder
of this appendix are truncated above $\BoundaryInverseOrder$.
Unit-constant reciprocals use the finite formal formula of
\cref{prop:df-finite-coefficient-computation}.
We construct output coefficients as centered signed color profiles to apply
the first inverse, then Fourier-transform them.

\paragraph{One local degree budget.}
Use the local scale $\DFCoarseLocalScale$ from
\cref{lem:df-tagged-local}. Positive paired coefficients and signed
paired numerators of degree $\DFTableNewPrimary$ have retained-Hilbert
norm at most
$\DFProofConstant_{\BoundaryInverseOrder}
 \DFCoarseLocalScale^{\DFTableNewPrimary}$.
The same bound holds for each degree of a finite positive-factor
reciprocal: degrees add in every product of its truncated geometric
series. An ordinary positive body has bounded profile norm, so its
finite reciprocal obeys the same bound because $\DFCoarseLocalScale\geq1$.
All local polynomials retain the complete flag--map pair and
same-source leg scalar. Define the new-primary-slot cost and the
normalized output factor by
\[
 \DFRootNewSlot:=\CalibActivation^{-1}
                     \ActivityPrimaryPool^{-1/2},\qquad
 \DFRootOutputFactor:=\DFRootNewSlot\DFCoarseLocalScale.
\]
For the public family,
$\DFRootNewSlot=O(\HoeffdingParameterBase\DFPublicScale^{-8})$ and
$\DFRootOutputFactor
 =O(\HoeffdingParameterBase^{1/2}\DFPublicScale^{-6})=o(1)$.

\begingroup
\mdfsetup{topline=true}
\begin{lemma}[Independent-root raised response]
\label[lemma]{lem:df-independent-response}
An explicit operator $\DFTableUpper$ on canonical coefficients satisfies
\[
\DFTableSignedPolynomial^0_{\DFTableQueue}(\DFTable)/
 \DFTableIndependentPolynomial_{\DFTableQueue}
 =\DFTableEvaluation(\LocalIdentityOperator+\DFTableUpper)\DFTable,
\qquad
\|\DFTableUpper\|_{\DFTableNorm\to\DFTableNorm}
 =O_{\BoundaryInverseOrder}(\HoeffdingParameterBase^{-3}).
\]
It strictly raises represented degree, creates no new body slot,
and is independent of $\DFTableQueue$.
\end{lemma}
\endgroup
\begin{proof}
Divide each selected group's complete signed polynomial by its
positive polynomial, canceling unselected factors. A primary group
contains its leg triple and distinct compensator; an ordinary group
contains one body output. Independent comparison roots and the
complete conditional flag--map law permit these local operators to
act on arbitrary retained joint Hilbert inputs.

The degree-one primary response is exactly
$\CalibActivation\DFFirstResponse$; represent it by the unchanged
primary slot. A nonleading primary response emits
$\DFTableNewPrimary\geq2$ profiles. The weighted transfer rule of
\cref{lem:df-joint-transfer} and the common degree budget bound its
cost by
\[
 \DFProofConstant_{\BoundaryInverseOrder}
 \ActivityPrimaryPool^{1/2}
             \DFRootOutputFactor^{\DFTableNewPrimary}.
\]
For an ordinary group, retain the leading safe body slot without
inverting its convolution. Each added reciprocal output costs at
most $\DFRootOutputFactor$. Because this factor is eventually at
most one, every nonleading local response gains at most
\[
 \DFProofConstant_{\BoundaryInverseOrder}
 \max\{\ActivityPrimaryPool^{1/2}\DFRootOutputFactor^2,
                              \DFRootOutputFactor\}
 =O_{\BoundaryInverseOrder}(\HoeffdingParameterBase\DFPublicScale^{-4}).
\]
Tensor the local expansions with the exact distinct-index
normalizers and symmetrize. Their leading responses are the identity;
every nonleading fixed-order pattern contains a small factor just
bounded and strictly raises represented degree, creating no body
slot. Extra outputs only multiply by a non-increasing factor. The
number of degree patterns depends on the fixed order alone. This
defines $\DFTableUpper$ and proves its stated bound. Selected groups
avoid the queue, whose unselected factors canceled before expansion,
so the operator is queue-independent.
\end{proof}

\begin{lemma}[Coarse transfer of centered root patterns]
\label[lemma]{lem:df-coarse-root-transfer}
Let the source and public-control reference laws be independent,
and let the centered joint-root density satisfy
$\|\DFCenteredRootDensity_{\DFRootOrder}\|_2
 \leq\DFProofConstant_{\DFRootOrder}\DFRootTagSize^{\DFRootOrder}$
under the product root reference. Let $\DFTable$ be a canonical
public-control kernel and partition the consumed arguments into
disjoint local groups, using each complete control once. Local maps
may be restricted to the centered root and control subspaces. In an
expansion through degree $\BoundaryInverseOrder$, with
$\DFRootOrder\leq\BoundaryInverseOrder$ and the slot factors of
\cref{lem:df-joint-transfer}, each group hit by a root tag emits
$\DFTableNewPrimary\geq1$ profiles.
Let $\DFRootPopulation$ bound
unselected choices per group and $\DFRootInputFactor$ bound a
selected group's consumed-input weight. Assume an unselected
single-output root map has retained-Hilbert norm at most
$\DFRootDeviation$; all other degree-$\DFTableNewPrimary$ numerators
and finite reciprocal coefficients have norm at most
$\DFProofConstant_{\BoundaryInverseOrder}
 \DFCoarseLocalScale^{\DFTableNewPrimary}$.
Assume unhit unselected factors cancel, while unhit selected normalized
responses and the number of incidence and degree patterns are bounded
in terms of the fixed order. Suppose
\[
\begin{gathered}
 \DFRootOutputFactor\leq1,\qquad
 \DFRootTransferBudget:=\DFRootTagSize
 \max\{1,\DFRootPopulation\DFRootNewSlot\DFRootDeviation,
          \DFRootPopulation\DFRootOutputFactor^2,
          \DFRootInputFactor\DFRootOutputFactor\}\leq1.
\end{gathered}
\]
The full response from nonempty root patterns has norm at most
$\DFProofConstant_{\BoundaryInverseOrder}\DFRootTransferBudget$.
\end{lemma}
\begin{proof}
Source independence gives
$\|\DFCenteredRootDensity_{\DFRootOrder}\otimes\DFTable\|_2
 =\|\DFCenteredRootDensity_{\DFRootOrder}\|_2\|\DFTable\|_2$.
Slotwise centering fixes this joint input. Apply the joint-root norm
once and tensor the local maps on the disjoint groups; their
retained-Hilbert bounds apply without factoring the root density.
Reserve one scalar factor $\DFRootTagSize$ for each hit group.
There are no more hit groups than tags, and every remaining tag
factor is at most one since $\DFRootTagSize\leq\DFRootTransferBudget\leq1$.

By \cref{lem:df-joint-transfer}, an unselected hit group costs
$\DFRootPopulation$ times either $\DFRootNewSlot\DFRootDeviation$
for one output or $\DFRootOutputFactor^{\DFTableNewPrimary}$
for at least two outputs. A selected hit group costs
$\DFRootInputFactor\DFRootOutputFactor^{\DFTableNewPrimary}$.
After its reserved tag factor, each is bounded by the common budget
$\DFProofConstant_{\BoundaryInverseOrder}\DFRootTransferBudget$;
extra outputs multiply by $\DFRootOutputFactor\leq1$.
A multiply tagged group still consumes its complete control once.

Every nonempty pattern has a hit group. Multiplying these common
budgets and the bounded unhit responses costs at most
$\DFProofConstant_{\BoundaryInverseOrder}\DFRootTransferBudget$,
since $\DFRootTransferBudget\leq1$.
All population factors are already included; normalized selected-index
averages and symmetrization contract. The remaining incidence and
degree patterns depend only on the fixed order, so summing them
proves the claim.
\end{proof}

\begin{lemma}[Protected-cycle defects on joint input kernels]
\label[lemma]{lem:df-actual-cycle-defect}
There are explicit canonical coefficients
$\DFTableRootDefect_{\DFTableQueue}$ and linear operators
$\DFTableWeightedDefect_{\DFTableQueue}$ satisfying
\begin{align}
\DFTableEvaluation\DFTableRootDefect_{\DFTableQueue}
 &=\DFTableCyclePolynomial_{\DFTableQueue}/
             \DFTableIndependentPolynomial_{\DFTableQueue}-1,\notag\\
\DFTableEvaluation\DFTableWeightedDefect_{\DFTableQueue}\DFTable
 &=\DFTableSignedPolynomial_{\DFTableQueue}(\DFTable)/
             \DFTableIndependentPolynomial_{\DFTableQueue}
       -\DFTableEvaluation(\LocalIdentityOperator+\DFTableUpper)\DFTable,\notag\\
\DFTableWeightedDefect_{\DFTableQueue}1
 &=\DFTableRootDefect_{\DFTableQueue}.
\label{eq:df-actual-cycle-defect-identities}
\end{align}
Uniformly over protected queue values,
\begin{equation*}
\DFTableNorm(\DFTableRootDefect_{\DFTableQueue})
 =O_{\BoundaryInverseOrder}(\HoeffdingParameterBase^{-3}),\qquad
\|\DFTableWeightedDefect_{\DFTableQueue}\|
 =O_{\BoundaryInverseOrder}(\HoeffdingParameterBase^{-3}).
\end{equation*}
The operators do not lower represented degree, and
$\DFTableRootDefect_{\DFTableQueue}^{[0]}
 =\DFTableRootDefect_{\DFTableQueue}^{[1]}=0$ as canonical coefficients.
\end{lemma}
\begin{proof}
Fix a protected queue value $\DFTableQueue$. The two experiments use
the same public controls and local endpoint maps; only the joint law
of the body roots changes. We isolate that change, construct its
canonical coefficients, and then bound their norm. All polynomial
operations below are truncated above $\BoundaryInverseOrder$.

\begin{description}
\item[1. Dependent-root contribution.]
By \cref{lem:df-independent-response}, the required weighted defect
has leading evaluation
\[
 \DFTableEvaluation\DFTableWeightedDefect_{\DFTableQueue}\DFTable
 =\frac{\DFTableSignedPolynomial_{\DFTableQueue}(\DFTable)
       -\DFTableSignedPolynomial^0_{\DFTableQueue}(\DFTable)}
       {\DFTableIndependentPolynomial_{\DFTableQueue}}.
\]
To expand this difference, write the identity in each root coordinate
as the sum of its mean projection and its centering projection under
$\DFProtectedRootMarginal$. Expanding the joint root likelihood gives
the constant term $1$, which produces the independent-root experiment,
and terms centered in nonempty sets of root coordinates. Call those
coordinates \emph{tagged}. A term with $\DFRootOrder$ tags has the
joint density $\DFCenteredRootDensity_{\DFRootOrder}$ of
\cref{thm:df-cycle-density}.

A root enters the local channel only by translating its own body
endpoint. If that endpoint contributes no output to a polynomial
term, the term is independent of the root, and integration against
its centered tag gives zero. Thus every surviving tag requires its
own body output, and at most $\BoundaryInverseOrder$ tags survive.
The cycle-density theorem and the public parameter choice give
\[
 \|\DFCenteredRootDensity_{\DFRootOrder}\|_
 {L^2((\DFProtectedRootMarginal)^{\otimes\DFRootOrder})}
 \leq\DFProofConstant_{\DFRootOrder}\DFRootTagSize^{\DFRootOrder},
 \qquad
 \DFRootTagSize=O(\HoeffdingParameterBase\DFPublicScale^{-10}).
\]

\item[2. Canonical coefficient construction.]
Work first with one summand in the distinct-index control average.
A selected primary group consists of its complete leg triple and
assigned compensator; a selected ordinary group consists of one body.
Call a group \emph{hit} if it contains a tagged root. For each nonempty
root pattern, form the complete local endpoint polynomials, multiplying
the leg and attachment factors at the same leg scalar before averaging.
A paired group uses its complete control once, including when both
its attachment and compensator roots are tagged.

Divide by the independent-root polynomial using its local finite
reciprocals. Unhit unselected factors cancel. Unhit selected groups
retain the normalized response of \cref{lem:df-independent-response},
while hit groups act on the joint root and control kernel.
Center the new output profiles, lift them by
$\CalibActivation^{-1}\DFFirstResponse^{-1}$, and store the resulting
primary densities; retain any unchanged safe body seeds as body slots.
Average with the exact distinct-index normalizers and symmetrize the
two input types. These finite linear operations define
$\DFTableWeightedDefect_{\DFTableQueue}$ on canonical coefficients.
Their evaluation is the difference in Step~1, term by term.
For scalar input $1$, the same construction defines
$\DFTableRootDefect_{\DFTableQueue}
 :=\DFTableWeightedDefect_{\DFTableQueue}1$.
Since the two scalar responses are
$\DFTableCyclePolynomial_{\DFTableQueue}$ and
$\DFTableIndependentPolynomial_{\DFTableQueue}$, this proves all
three response identities.

\item[3. Local and global norm bounds.]
The root density need not factor across groups. Instead, source
independence gives the single joint-input bound
\[
 \|\DFCenteredRootDensity_{\DFRootOrder}\otimes\DFTable\|_2
 =\|\DFCenteredRootDensity_{\DFRootOrder}\|_2\|\DFTable\|_2.
\]
We apply local maps to disjoint groups while retaining all other
arguments in a Hilbert space. The retained-input bounds of
\cref{lem:df-tagged-local} therefore apply to this joint kernel.
Integration over roots uses Cauchy--Schwarz under their probability
reference, so no inverse root-atom probabilities are needed.

The case requiring a sharper local bound is an unselected hit group
with just one output. Its tagged root forces that output to be a body
endpoint. Before the root translation, its profile $\CalibProposal$
is $\DFBaselineCenter$ for a free body or compensator, or a fixed
translate of $\CalibSafeUniform$ for a queue or attachment.
Root centering replaces the translated profile by its difference
from the root average. Translation fixes the uniform law and
preserves its profile norm, while convolution contracts that norm.
Consequently,
\[
 \sup_{\CalibRootValue}
 \|\DFTableTranslate_{\CalibRootValue}\CalibProposal
       -\DFProtectedRootMarginal*\CalibProposal\|_{\CalibUniform^{-1}}
 \leq2\|\CalibProposal-\CalibUniform\|_{\CalibUniform^{-1}}.
\]
Conversion to the primary reference costs at most $\sqrt2$.
Thus \cref{sec:df-baseline-proof,sec:df-parameter-verification}
provide the single-output bound
\[
 \DFRootDeviation:=2\sqrt2\bigl(
 \|\DFBaselineCenter-\CalibUniform\|_{\CalibUniform^{-1}}
 +\|\CalibSafeUniform-\CalibUniform\|_{\CalibUniform^{-1}}\bigr)
 =O(\HoeffdingParameterBase\DFPublicScale^{-4}).
\]
All other local numerators and finite reciprocal coefficients obey
the degree bound
$\DFProofConstant_{\BoundaryInverseOrder}
 \DFCoarseLocalScale^{\DFTableNewPrimary}$ established at the start of
\cref{sec:df-table-response}. This includes ordinary-center scores:
their fixed-root operator norm is $O(1)$ because
$\DFBaselineCenter\leq4/\LARModulus$ and
$\DFBaselinePrimary\geq1/(2\LARModulus)$.

We can now apply \cref{lem:df-coarse-root-transfer}. Unselected hit
groups have at most the body population many choices; selected indices
are already averaged, so they incur only a consumed-input weight.
Accordingly, take
\[
 \DFRootPopulation=\InteriorVertexCount=O(\DFPublicScale^{20}),
 \qquad
 \DFRootInputFactor
 =\max\{\ActivityPrimaryPool^{1/2},
          \ActivityBodyPool^{1/2}/\FockRadius\}
 =O(\DFPublicScale^{10}).
\]
The slot factors in \cref{sec:df-table-response} give the following
three costs, before the root-tag factor:
\[
\begin{aligned}
 \DFRootPopulation\DFRootNewSlot\DFRootDeviation
   &=O(\HoeffdingParameterBase^2\DFPublicScale^8)
     &&\text{(unselected, one output)},\\
 \DFRootPopulation\DFRootOutputFactor^2
   &=O(\HoeffdingParameterBase\DFPublicScale^8)
     &&\text{(unselected, at least two outputs)},\\
 \DFRootInputFactor\DFRootOutputFactor
   &=O(\HoeffdingParameterBase^{1/2}\DFPublicScale^4)
     &&\text{(selected)}.
\end{aligned}
\]
Extra outputs cost at most
$\DFRootOutputFactor=O(\HoeffdingParameterBase^{1/2}
\DFPublicScale^{-6})\leq1$ for sufficiently large parameters.
Every hit group has a tag, so the transfer lemma supplies one factor
$\DFRootTagSize$ to offset its cost. Its common budget is therefore
\[
\begin{split}
 \DFRootTransferBudget
 &=\DFRootTagSize\max\{1,
       \DFRootPopulation\DFRootNewSlot\DFRootDeviation,
       \DFRootPopulation\DFRootOutputFactor^2,
       \DFRootInputFactor\DFRootOutputFactor\}\\
 &=O(\HoeffdingParameterBase^3\DFPublicScale^{-2})
  =O(\HoeffdingParameterBase^{-13}),
\end{split}
\]
using $\DFPublicScale=\HoeffdingParameterBase^8$.
Thus the budget is at most one for sufficiently large parameters.
The lemma sums all nonempty root patterns, including patterns with
several tags in one group, and yields
\[
 \|\DFTableWeightedDefect_{\DFTableQueue}\|
 =O_{\BoundaryInverseOrder}(\HoeffdingParameterBase^{-13}).
\]
Since $\DFTableNorm(1)=1$, the same bound holds for
$\DFTableRootDefect_{\DFTableQueue}$. This is stronger than the
claimed $O_{\BoundaryInverseOrder}(\HoeffdingParameterBase^{-3})$.
All local estimates are uniform in the protected queue value, and
the remaining incidence and degree counts depend only on the fixed
truncation order.

\item[4. Degree assertions.]
For scalar input, \cref{thm:df-cycle-density} gives
$\DFCenteredRootDensity_1=0$, so a nonzero correction has at least
two tags on distinct body endpoints. Both endpoints must contribute
outputs by Step~1, even if they lie in the same local group.
The coefficient lifts preserve these outputs and finite reciprocals
only add degree. Hence the constant and linear canonical coefficients
of $\DFTableRootDefect_{\DFTableQueue}$ vanish.
For a general input kernel, fix the roots and all other controls.
If a selected group contributes no output, integrating its centered
paired or ordinary score gives zero. Every surviving selected group
therefore contributes at least one output. Since input degree counts
selected groups, and subsequent lifts, reciprocals, and symmetrization
do not lower degree, neither defect lowers represented degree.
\end{description}
\end{proof}

\subsubsection{True-cycle response and common-queue assembly}
\label{sec:df-common-queue-inverse}

\paragraph{Assembly of the response.}
\label{sec:df-matrix-ledger}
We combine the background coefficients of
\cref{prop:df-actual-background} with the raised response and joint
root operators of \cref{lem:df-independent-response,lem:df-actual-cycle-defect}.
In the fixed coordinates, let $\AppMatrixQueueBackground$ and
$\AppMatrixQueueDefect$ represent $\DFTableBackground_{\DFTableQueue}$
and $\DFTableRootDefect_{\DFTableQueue}$, respectively; let
$\AppMatrixQueueRoot$ and $\DFMatrixRaising$ represent the operators
$\DFTableWeightedDefect_{\DFTableQueue}$ and $\DFTableUpper$.
Thus $\DFMatrixRaising\DFMatrixZero=0$ and
$\AppMatrixQueueRoot\DFMatrixZero=\AppMatrixQueueDefect$.
Absorb this scalar response into the complete background and normalize once:
\begin{align}
\DFMatrixBackgroundVector
 &=\CountExpectation_{\DFTableQueue}
   [\DFMatrixMultiply(\AppMatrixQueueBackground)
       (\DFMatrixZero+\AppMatrixQueueDefect)],\notag\\
\AppMatrixQueueConnected
 &=\AppMatrixQueueRoot-
   \DFMatrixMultiply(\AppMatrixQueueDefect)
        (\IdentityMatrix+\DFMatrixRaising),\notag\\
\DFMatrixRoot
 &=\DFMatrixBackground^{-1}
   \CountExpectation_{\DFTableQueue}
       [\DFMatrixMultiply(\AppMatrixQueueBackground)\AppMatrixQueueConnected].
\label{eq:df-common-queue-operators}
\end{align}
Here $\DFMatrixBackground=\DFMatrixMultiply(\DFMatrixBackgroundVector)$,
as in \cref{sec:df-matrix-ledger}; the corresponding canonical
coefficient is $\DFTableTotalBackground$.
Every product inside an expectation uses one queue value. Only after
averaging the complete products do we invert the background.
Its constant coordinate is one, so its multiplication inverse exists
by \cref{prop:df-finite-coefficient-computation}.

Convexity, submultiplicativity, and the background and root bounds give
\[
 \DFMatrixNorm{\DFMatrixBackgroundVector-\DFMatrixZero}
 +\sup_{\DFTableQueue}\DFMatrixNorm{\AppMatrixQueueConnected}
 =O_{\BoundaryInverseOrder}(\HoeffdingParameterBase^{-3}).
\]
The same proposition bounds $\DFMatrixBackground^{-1}$ by
$1+O_{\BoundaryInverseOrder}(\HoeffdingParameterBase^{-3})$.
Consequently,
\[
 \DFMatrixNorm{\DFMatrixRoot}
 =O_{\BoundaryInverseOrder}(\HoeffdingParameterBase^{-3}),
 \qquad \DFMatrixRoot\DFMatrixZero=0,
\]
since $\AppMatrixQueueConnected\DFMatrixZero=0$ before evaluation.
For matrices these are induced graded operator norms.

For a fixed queue value, the root identities
\eqref{eq:df-actual-cycle-defect-identities} and multiplicativity of
leading evaluation give the exact physical response
\[
 [\DFTableSignedPolynomial_{\DFTableQueue}(\DFTable)]_{\leq\BoundaryInverseOrder}
 =\DFMatrixLeading\DFMatrixMultiply(\AppMatrixQueueBackground)
   (\IdentityMatrix+\DFMatrixRaising+\AppMatrixQueueRoot)
       \DFMatrixVector.
\]
Substitute $\AppMatrixQueueConnected$ from
\eqref{eq:df-common-queue-operators} and average. Since
$\DFMatrixRaising$ is queue-independent and multiplication matrices
respect the coefficient product, the result is
\[
\begin{split}
 [\DFTableSignedPolynomial(\DFTable)]_{\leq\BoundaryInverseOrder}
 &=\DFMatrixLeading\DFMatrixBackground
       (\IdentityMatrix+\DFMatrixRaising+\DFMatrixRoot)\DFMatrixVector\\
 &=\DFMatrixLeading\DFMatrixLifted\DFMatrixVector,\\
 [\DFTableCyclePolynomial]_{\leq\BoundaryInverseOrder}
 &=\DFMatrixLeading\DFMatrixBackgroundVector.
\end{split}
\]
The scalar-input case gives the second identity. Both use the physical
synthesis, averaged symmetrization, and exact distinct-index normalizers.

\paragraph{Grading and the canonical degree-one term.}
\label{sec:df-canonical-first-degree}
The raw root operator does not lower degree, and its scalar image
$\AppMatrixQueueDefect$ starts in degree two, by
\cref{lem:df-actual-cycle-defect}. Hence
$\AppMatrixQueueConnected$ does not lower degree. Multiplication by
the unit-constant background inverse and averaging preserve this
property, so $\DFMatrixRoot$ does not lower degree either.
Together with the strictly raising $\DFMatrixRaising$, this makes
both the normalized and full response block lower triangular, with
actual diagonal blocks
\[
 \IdentityMatrix+(\DFMatrixRoot)_{\HoeffdingDegree,\HoeffdingDegree}.
\]
These blocks act on the full hybrid degree space, including all
body-to-primary conversions. The protected-root diagonal is retained.

For $\BoundaryInverseOrder\geq1$, the background product in
\cref{prop:df-actual-background} has canonical degree-one coefficient
$\DFTableMeanCertificate_{\DFTableQueue}$.
The coefficient identity \eqref{eq:df-disjoint-table-cancellation}
and the equality of the clipped leg and queue laws give
$\CountExpectation_{\DFTableQueue}
 \DFTableMeanCertificate_{\DFTableQueue}=0$ before evaluation.
Since the root defect starts in degree two, the definition of
$\DFMatrixBackgroundVector$ therefore gives
$(\DFMatrixBackgroundVector)_1=0$ as a canonical coefficient.

\subsubsection{Application to the public correction}
\label{sec:df-graded-application}

\begin{proof}[Proof of \cref{thm:df-actual-table-inverse}]
By \cref{sec:df-common-queue-inverse}, the full response is
$\DFMatrixLifted=\DFMatrixBackground
 (\IdentityMatrix+\DFMatrixRaising+\DFMatrixRoot)$ and
$\DFMatrixLifted\DFMatrixZero=\DFMatrixBackgroundVector$.
Submultiplicativity in \cref{lem:df-table-algebra} and the assembled
bounds give, in the induced graded norm,
\[
\begin{split}
 \DFMatrixNorm{\DFMatrixLifted-\IdentityMatrix}
 &\leq\DFMatrixNorm{\DFMatrixBackgroundVector-\DFMatrixZero}\\
 &\quad+\bigl(1+\DFMatrixNorm{\DFMatrixBackgroundVector-\DFMatrixZero}\bigr)
       \bigl(\DFMatrixNorm{\DFMatrixRaising}+\DFMatrixNorm{\DFMatrixRoot}\bigr)\\
 &=O_{\BoundaryInverseOrder}(\HoeffdingParameterBase^{-3}).
\end{split}
\]
For fixed $\BoundaryInverseOrder$ and sufficiently large
$\HoeffdingParameterBase$, this is at most $1/2$.
Apply \cref{thm:stable-finite-dimensional-calibration} with synthesis
$\DFMatrixPhysical$, evaluation $\DFMatrixLeading$, and target and
reference $\DFMatrixZero$. Its reference-residual bound gives the
unique solution with
\[
 \DFMatrixNorm{\DFMatrixSolution-\DFMatrixZero}
 \leq2\DFMatrixNorm{\DFMatrixBackgroundVector-\DFMatrixZero}
 =O_{\BoundaryInverseOrder}(\HoeffdingParameterBase^{-3}).
\]

The scalar row is the identity, so the correction has scalar zero.
The degree-one equation then has zero right-hand side because
$(\DFMatrixBackgroundVector)_1=0$. Homogeneous projections contract
the norm, so the same small-perturbation bound makes the degree-one
diagonal block invertible and gives $(\DFMatrixCorrection)_1=0$.
Paired lifting preserves canonical centering, so distinct-index
averaging gives mean zero for the correction. Exact calibration yields
\[
 \CountExpectation_{\DFIdealProposal}\DFTableWeight=1,
 \qquad
 [\DFMatrixWeightedPolynomial]_{\leq\BoundaryInverseOrder}
 =\DFMatrixLeading\DFMatrixZero=1.
\]
Independent uniform relabeling makes the law exchangeable.
Each homogeneous character coefficient is then a positive
combinatorial multiple of its support-frequency coefficient, giving
the claimed cancellation. The kernels and inverses are averaged
public quantities, so evaluating the weight reads only the controls
in \cref{sec:df-control-context}.
\end{proof}

%% file: chapters/appendix-direct-fourier-tail.tex
\section{Complete Fourier Tails and Public-Core Moments}
\label{sec:df-tail-appendix}

Use the public family and radius of \cref{sec:df-parameter-choice}.
For a finite polynomial $\FockPolynomial$ of degree at most
$\BlockDim$, write $\FockNorm{\FockPolynomial}{\FockRadius}$ for
the ordered-coefficient norm of \cref{def:df-weighted-character-norm}.
Applied to a character polynomial, this is
$\DFFourierNorm_{\FockRadius}(\SignedEndpointLaw)$.
Symmetrization is the orthogonal projection that averages tensor-slot
permutations. Contractions and conjugation follow
\cref{sec:df-complex-coefficients}.
We apply the bounds below to the physical weight $\DFTableWeight$
and its canonical public-control coefficient $\DFTableCorrection$
from \cref{sec:df-matrix-calibration}.
The real public input coefficients and scores give conjugate-symmetric
character coefficients. Fix a disjoint placement, with the free
population and control pools of
\cref{sec:df-control-context}.
The protected latent cycle, its independent body permutation, and the
single clipped queue scalar are retained under the uncharged experiment
of \cref{sec:df-complete-control}.

\subsection{A mean bound for the full protected cycle}

Write $\CalibRootValue_{\VertexIndex}$ for the protected cyclic
difference at body vertex $\VertexIndex$ after the fixed body
permutation. Expectations in the following lemma average the latent
edge samples of \cref{def:protected-body-cycle}.

\begin{lemma}[Cycle mean exponential moment]
\label[lemma]{lem:df-cycle-mean-mgf}
Fix a disjoint placement, a body permutation sampled independently of
the public core and latent cycle, and the shared queue scalar.
At body vertex $\VertexIndex$, let
$\DFTailBlockLaw_{\VertexIndex}$ be its center profile after averaging
its leg input, if present, but before adding its cycle root.
These profiles are fixed independently of the latent edge samples.
Suppose
\[
 \begin{gathered}
 \|\DFTailBlockLaw_{\VertexIndex}-\CalibUniform\|_{\CalibUniform^{-1}}
       \leq\DFTailLocalDeviation,
 \qquad \FockRadius^2\DFTailLocalDeviation^2=o(1),\\
 \DFTailFluctuation
 =\BlockDim^{-1/2}\sum_{\VertexIndex\in\InteriorVertices}
 \left(\ActivityTranslation{\CalibRootValue_{\VertexIndex}}
                  \DFTailBlockLaw_{\VertexIndex}
       -\CountExpectation
         \ActivityTranslation{\CalibRootValue_{\VertexIndex}}
                  \DFTailBlockLaw_{\VertexIndex}\right).
 \end{gathered}
\]
For every fixed positive $\FockConstant$ and
$\FockRadius=\HoeffdingParameterBase^{1/4}$,
\[
 \CountExpectation\exp\!
 \left(\FockConstant\FockRadius^2
       \|\DFTailFluctuation\|_{\CalibUniform^{-1}}^2\right)=O(1).
\]
The bound is uniform in the fixed data. It also holds for the entire
normalized conditional mean when its root-averaged drift has norm
$O(\FockRadius^{-1})$.
\end{lemma}
\begin{proof}
The latent edge variables are independent samples from
$\DFProtectedEdgeLaw$ by \cref{def:protected-body-cycle}.
Replacing one by an independent copy changes two roots.
Translation fixes $\CalibUniform$
and is an isometry, so each affected summand changes in norm by at most
$2\DFTailLocalDeviation$. Consequently
\[
 \|\DFTailFluctuation-\DFTailFluctuation'\|
 \leq\DFTailDifferenceBound
 :=4\DFTailLocalDeviation/\sqrt{\BlockDim},
 \qquad
 \DFTailVarianceProxy
 :=\InteriorVertexCount\DFTailDifferenceBound^2
 \leq16\DFTailLocalDeviation^2.
\]
All norms in this proof are in the profile Hilbert space.
The case $\DFTailVarianceProxy=0$ is immediate. Otherwise apply the
Efron--Stein inequality
\cite[Proposition~1]{BoucheronEtAl2005Moments} to each orthonormal
coordinate and sum, obtaining
$\CountExpectation\|\DFTailFluctuation\|^2
 \leq\DFTailVarianceProxy/2$.
The scalar norm obeys the same coordinate difference bound, so its
Doob martingale under successive edge exposure has conditional ranges
of length at most $\DFTailDifferenceBound$.
The one-sided bounded-differences inequality
\cite[Theorem~6.7]{McDiarmid1989BoundedDifferences} therefore gives
\[
 \Pr\!\left(\|\DFTailFluctuation\|
    >\sqrt{\DFTailVarianceProxy/2}+\ActivityTailArgument\right)
 \leq e^{-2\ActivityTailArgument^2/\DFTailVarianceProxy}
 \quad(\ActivityTailArgument\geq0).
\]
Tail integration gives, whenever
$0\leq\DFTailLaplace\DFTailVarianceProxy<1$,
\[
 \CountExpectation e^{\DFTailLaplace\|\DFTailFluctuation\|^2}
 \leq
 \frac{e^{\DFTailLaplace\DFTailVarianceProxy}}
      {1-\DFTailLaplace\DFTailVarianceProxy}.
\]
Take $\DFTailLaplace=\FockConstant\FockRadius^2$.
The smallness hypothesis gives an absolute bound.
Add the drift and use
$\|\DFTailVectorFirst+\DFTailVectorSecond\|^2
 \leq2\|\DFTailVectorFirst\|^2+2\|\DFTailVectorSecond\|^2$
to prove the last statement.
\end{proof}

For the proposal in \cref{thm:df-positive-baseline}, the body profiles
before adding cycle roots are
\[
\begin{array}{c|c|c}
 \text{body role}&\text{population}&\DFTailBlockLaw_{\VertexIndex}\\ \hline
 \text{free}&\DFFreeCount&\DFBaselineCenter\\
 \text{attachment}&\GadgetCount&\LegClippedLaw*\CalibSafeUniform\\
 \text{queue}&\LARModulus&
      \ActivityTranslation{\DFTableQueue}\CalibSafeUniform .
\end{array}
\]
The attachment row averages its own leg's second input; the queue row
fixes the common value $\DFTableQueue$.
Translation is an isometry and convolution is a contraction in the
uniform-reference norm. Thus the center estimate established in
\cref{sec:df-parameter-verification} bounds every row by the same
$\DFTailLocalDeviation$, uniformly in the placement.
The off-body coordinates use independent public blurs with the protected
marginal and have no actual-root dependence. Thus the \emph{entire}
histogram mean has exactly the body's root fluctuation.

The root/queue-averaged histogram mean equals
$\BlockDim\CalibUniform$ by \cref{thm:df-positive-baseline}.
For fixed queue value its root-averaged excess is exactly
\[
 \LARModulus\DFProtectedRootMarginal*
 \bigl(\ActivityTranslation{\DFTableQueue}\CalibSafeUniform
                 -\QueueClippedLaw*\CalibSafeUniform\bigr).
\]
Only the $\LARModulus$ queue vertices change when the queue average is
removed; attachments lie outside that set.
Both terms in parentheses are within
$\DFTailLocalDeviation$ of $\CalibUniform$. Hence the normalized
drift is at most
\[
 2\LARModulus\DFTailLocalDeviation/\sqrt{\BlockDim}.
\]
For the selected family, the center deviation is
$\DFTailLocalDeviation=O(\HoeffdingParameterBase\DFPublicScale^{-4})$.
Consequently
\[
 \FockRadius^2\DFTailLocalDeviation^2
 +\frac{\FockRadius^2\LARModulus^2
                     \DFTailLocalDeviation^2}{\BlockDim}
 =O(\HoeffdingParameterBase^{5/2}\DFPublicScale^{-8})=o(1).
\]
This verifies both hypotheses of \cref{lem:df-cycle-mean-mgf} at
every fixed multiple of the radius, uniformly in the common queue value.

\subsection{Gaussian orthogonality and the all-degree product bound}

\begin{lemma}[Complex Gaussian orthogonality]
\label[lemma]{lem:df-gaussian-orthogonality}
Let $\FockPolynomial$ be a polynomial of degree at most $\BlockDim$
with the ordered symmetric coefficient convention of
\cref{def:df-weighted-character-norm}, and let
$\FockGaussian$ have independent circular complex Gaussian coordinates,
with real and imaginary variances $1/2$. Then
\[
 \CountExpectation
 |\FockPolynomial(\FockRadius\FockGaussian/\sqrt{\BlockDim})|^2
 =\FockNorm{\FockPolynomial}{\FockRadius}^2.
\]
The identity also holds for coefficients in a finite-dimensional
Hilbert space, with its squared norm in place of absolute square.
It requires neither an endpoint probability law nor independence
among the endpoints represented by the polynomial.
\end{lemma}
\begin{proof}
For nonnegative integers $\FockQuadraticCount,\FockCubicCount$,
one-variable Gaussian integration gives
\[
 \CountExpectation\FockGaussian_{\DFTailSlot}^{\FockQuadraticCount}
 \overline{\FockGaussian_{\DFTailSlot}}^{\,\FockCubicCount}
 =\EventIndicator{\set{\FockQuadraticCount=\FockCubicCount}}
     \FockQuadraticCount!.
\]
Independence makes distinct monomials orthogonal. Within degree
$\HoeffdingDegree$, grouping ordered tuples with the same multiplicities
contributes exactly $\HoeffdingDegree!$ times the squared Euclidean
norm of the ordered symmetric coefficient tensor. Rescaling by
$\FockRadius/\sqrt{\BlockDim}$ gives the claimed norm at every degree.
Sum over degrees and an orthonormal basis of the coefficient Hilbert
space to obtain both assertions.
\end{proof}

For any homogeneous symmetric tensor of degree $\FockSelectedDegree$,
symmetrization contraction and
$(\FockSelectedDegree+\HoeffdingDegree)!
 \leq2^{\FockSelectedDegree+\HoeffdingDegree}
       \FockSelectedDegree!\HoeffdingDegree!$
give
\begin{equation}
\label{eq:df-finite-multiplication}
 \FockNorm{
   \langle\FockTensor_{\FockSelectedDegree},
          \FockArgument^{\otimes\FockSelectedDegree}\rangle
        \FockPolynomial}{\FockRadius}
 \leq\sqrt{\FockSelectedDegree!}
   \left(\frac{\sqrt2\FockRadius}{\sqrt{\BlockDim}}\right)^{
        \FockSelectedDegree}
   \|\FockTensor_{\FockSelectedDegree}\|_2
   \FockNorm{\FockPolynomial}{\sqrt2\FockRadius}.
\end{equation}
This follows by squaring and summing the degree bounds, so applies to
every finite degree and to Hilbert-valued input coefficient maps.

\begin{lemma}[Finite products with one nonlinear radius enlargement]
\label[lemma]{lem:df-finite-product}
Let $\FockPolynomial$ be a product of blocks of size at most three,
with block polynomials
$1+\FockLinear_{\VertexIndex}
   +\FockNonlinear_{\VertexIndex,2}
   +\FockNonlinear_{\VertexIndex,3}$,
with at most $\GadgetCount$ nonsingleton blocks. Define
\[
\begin{aligned}
 \FockLinearActivity&=\BlockDim^{-1}
                  \sum_{\VertexIndex}\|\FockLinear_{\VertexIndex}\|_2^2,
 &\FockMean&=\BlockDim^{-1/2}
                  \left\|\sum_{\VertexIndex}\FockLinear_{\VertexIndex}\right\|_2,\\
 \FockQuadraticActivity&=\BlockDim^{-1}
                  \sum_{\VertexIndex}\|\FockNonlinear_{\VertexIndex,2}\|_2,
 &\FockCubicActivity&=\BlockDim^{-3/2}
                  \sum_{\VertexIndex}\|\FockNonlinear_{\VertexIndex,3}\|_2.
\end{aligned}
\]
There are absolute positive constants $\FockSmallConstant,\FockConstant$
such that
\[
 \FockRadius^2\FockLinearActivity\leq\FockSmallConstant,\qquad
 \FockRadius^2\FockQuadraticActivity+
  \sqrt{\GadgetCount}\FockRadius^3\FockCubicActivity
       \leq\FockSmallConstant
\]
imply
$\FockNorm{\FockPolynomial}{\FockRadius}^2
 \leq\FockConstant
       e^{\FockConstant\FockRadius^2\FockMean^2}$.
For a mixture, uniform conditional activity bounds and a bounded
expectation of this exponential suffice.
\end{lemma}
\begin{proof}
Apply \cref{lem:df-gaussian-orthogonality} to the full product.
For the product of linear factors use
$|1+\DFTailComplexValue|^2
 \leq\exp(2\operatorname{Re}\DFTailComplexValue+
                               |\DFTailComplexValue|^2)$.
The resulting quadratic matrix has trace
$\FockRadius^2\FockLinearActivity$.
Completing the square in the Gaussian integral gives a determinant
$\det(\LocalIdentityOperator-\FockRadius^2\FockGram)^{-1}$
and a mean factor bounded by
$\exp(2\FockRadius^2\FockMean^2)$,
where
$\FockGram=\BlockDim^{-1}\sum_{\VertexIndex}
 \FockLinear_{\VertexIndex}\FockLinear_{\VertexIndex}^{*}$.
If the trace is at most $1/2$, then
$-\log\det(\LocalIdentityOperator-\FockRadius^2\FockGram)
 \leq2\FockRadius^2\FockLinearActivity$, proving the linear claim.

Expand the product by selecting $\FockQuadraticCount$ quadratic and
$\FockCubicCount$ cubic terms, in disjoint blocks, and put
$\FockSelectedCount=\FockQuadraticCount+\FockCubicCount$.
Omitting their linear factors changes the squared normalized mean to
at most
$2\FockMean^2+2\FockSelectedCount\FockLinearActivity$,
by Cauchy--Schwarz. Thus their remaining linear product costs at most
\[
 \exp\!\left(2\FockSmallConstant+4\FockRadius^2\FockMean^2
                       +4\FockSelectedCount\FockSmallConstant\right)
\]
at radius $\sqrt2\FockRadius$, provided $\FockSmallConstant\leq1/4$.
The factor $e^{2\FockSmallConstant}$ is retained also when no nonlinear
block is selected and $\FockMean=0$.
Multiply \emph{all} selected nonlinear tensors first and apply
\eqref{eq:df-finite-multiplication} once.
Summing their norms and dropping disjointness gives the upper bound
\[
 e^{4\FockSelectedCount\FockSmallConstant}
 2^{(2\FockQuadraticCount+3\FockCubicCount)/2}
 \frac{\sqrt{(2\FockQuadraticCount+3\FockCubicCount)!}}
      {\FockQuadraticCount!\FockCubicCount!}
 (\FockRadius^2\FockQuadraticActivity)^{\FockQuadraticCount}
 (\FockRadius^3\FockCubicActivity)^{\FockCubicCount}
\]
times $\exp(2\FockSmallConstant+4\FockRadius^2\FockMean^2)$.
The factorial quotient is at most
$4^{\FockSelectedCount}(3\GadgetCount)^{\FockCubicCount/2}$:
use
$\sqrt{(2\FockSelectedCount)!}/\FockSelectedCount!
 \leq2^{\FockSelectedCount}$,
$\binom{\FockSelectedCount}{\FockQuadraticCount}
 \leq2^{\FockSelectedCount}$, and bound the remaining
$\FockCubicCount$ factorial factors by
$(3\GadgetCount)^{\FockCubicCount}$ before taking the square root.
For $\FockSmallConstant\leq1/4$, these terms are bounded by the
product of two geometric majorants whose ratios are absolute constant
multiples of
$\FockRadius^2\FockQuadraticActivity$ and
$\sqrt{\GadgetCount}\FockRadius^3\FockCubicActivity$, respectively.
Choose $\FockSmallConstant$ so both ratios are at most $1/2$.
The two geometric sums are then bounded by an absolute constant.
Square the norm bound and enlarge $\FockConstant$ to obtain the claim.
If there are no nonsingleton blocks, the linear estimate applies directly.
The single radius enlargement is independent of the number of blocks
and the total degree.
Jensen in the finite Hilbert direct sum proves the mixture statement.
\end{proof}

\subsection{Local polynomial operators and the weighted Fourier bound}

For the paired lift of \cref{sec:df-first-response}, write
\begin{equation*}
 \DFTailCompensatorScore
 :=\CalibActivation
       \frac{\DFSmoothedAnchor(\DFBaselinePrimary\ActivityCoreScore)}
            {\DFBaselineCenter}
\end{equation*}
for the compensator summand. Its denominator is used only on safe
support. The two public inputs are independent under the proposal.

\begin{lemma}[Local polynomial operators and paired lifts]
\label[lemma]{lem:df-tagged-local}
Use the public family of \cref{sec:df-parameter-choice} and its
positive proposal from \cref{thm:df-positive-baseline}, for all
sufficiently large $\HoeffdingParameterBase$. Conditional on the complete protected
root vector, the body permutation, the placement, and the shared queue,
the unweighted proposal is a product of
$\DFFreeCount+\LARModulus$ singleton laws and $\GadgetCount$
same-source triple laws. A triple contains the two off-body leg
coordinates and its own attachment; the attachment has no queue
shift. Put
\[
 \DFCoarseLocalScale
 :=\sqrt{\frac{\CalibActivation\BlockDim}{\GadgetCount}}.
\]
The positive triple, the positive four-coordinate factor obtained by
adjoining its distinct compensator, and the corresponding signed
paired group induced by a centered unit score in
$L^2(\DFBaselinePrimary)$ have degree-$\HoeffdingDegree$ raw
coefficient operator norm at most
\[
 \FockConstant\DFCoarseLocalScale^{\HoeffdingDegree}
 \qquad(1\leq\HoeffdingDegree\leq4).
\]
The triple has no degree-four term, and the signed paired factor has
zero constant coefficient. An ordinary-center score has only a
degree-one output, with operator norm $O(1)$. The paired lift has
Hilbert operator norm $O(1)$ and evaluation-row norm $O(\LARModulus)$
on its paired reference support.
These raw output bounds hold both in the uniform-reference profile
norm and in the primary-reference profile norm, uniformly in fixed
attachment and compensator roots. All operator statements allow
arbitrary retained Hilbert-valued inputs.
\end{lemma}
\begin{proof}
Each fixed-sign diffuse pair averages a uniform slice of length
$\LARModulus-O(\CalibAnchorWidth+\LegCutoff)$ over an anchor interval
of length $\CalibAnchorWidth$. For a specified pair, its pair sum fixes
at most one anchor value after fixing the blur and leg input.
Its point mass is therefore at most
$\FockConstant/(\LARModulus\CalibAnchorWidth)$, giving pair likelihood
norm at most $\FockConstant\sqrt{\LARModulus/\CalibAnchorWidth}$.
Its marginals have height $O(1/\LARModulus)$.
Convolution with the independent public blur preserves these bounds.

For a marked pair with the same leg scalar fixed, its law is a
translate of a kernel with first marginal $\DFBaselinePrimary$ and
conditional second-coordinate height $1/\CalibAnchorWidth$.
Since $\sum\DFBaselinePrimary^2\leq\|\DFBaselinePrimary\|_\infty$,
Cauchy--Schwarz bounds both the unweighted pair and its operator on
unit primary scores by
\[
 \LARModulus\sqrt{\frac{\|\DFBaselinePrimary\|_\infty}
                             {\CalibAnchorWidth}}.
\]
The analogous primary marginal bound is
$\sqrt{\LARModulus\|\DFBaselinePrimary\|_\infty}$.
Multiply these marked terms by $\CalibActivation$.
For the diffuse part the primary enters only through its anchor sign.
The absolute score coefficient of either sign-conditioned law is at
most $\|\ActivityCoreScore\|_1\leq\|\ActivityCoreScore\|_2$.
This controls that part without division by a small sign probability.

The same-source triple retains the leg scalar in both the companion
and attachment, as in \cref{sec:leg-joint-law}. Conditional on that
scalar, the attachment has a fresh center of height $O(1/\LARModulus)$,
independent of the anchor and public blur. Average the pair over the
anchor and blur, tensor with the attachment law whose
uniform-reference likelihood norm is $O(1)$, and average over the
\emph{same} scalar. This gives the pair bounds for the triple.
The actual root is fixed and only translates the attachment.
Centering output coordinates is orthogonal projection, so it can only
decrease those norms. For the linear coefficient, the other
marginals are bounded by the height estimates.
The same primary height thus controls both the unweighted product
and the score operators.

For the compensating term, the Markov-kernel Cauchy--Schwarz inequality
gives
\[
 \|\DFTailCompensatorScore\|_{L^2(\DFBaselineCenter)}^2
 \leq\FockConstant\CalibActivation\|\ActivityCoreScore\|_2^2.
\]
Here use
$\CalibActivation\DFSmoothedAnchor\DFBaselinePrimary
 \leq\FockConstant/\LARModulus$
and $\DFBaselineCenter\geq\FockConstant^{-1}/\LARModulus$ on its
support.
The compensator singleton's linear deviation is $o(1)$ by
\cref{sec:df-parameter-verification}.
There are exactly $\GadgetCount$ such four-coordinate paired groups,
each containing one triple and a distinct free singleton. The other
free singletons number
$\DFFreeCount-\GadgetCount=\ActivityBodyPool$; the
$\LARModulus$ queue singletons are never controls.
The signed compensator polynomial has zero constant coefficient and
linear norm $O(\sqrt{\CalibActivation})$. Since
$\CalibActivation\leq1$, multiplying it by an unweighted triple
is bounded by the stated score-operator bounds. The signed-primary
triple times the positive compensator obeys the same bounds, since
the compensator has constant coefficient one and linear norm $o(1)$.
This controls every degree through four.

The preceding bounds give linear coefficient cost
$O(1+\CalibActivation
       \sqrt{\LARModulus\|\DFBaselinePrimary\|_\infty})$
and higher-degree cost
$O(\sqrt{\LARModulus/\CalibAnchorWidth}
 (1+\CalibActivation\sqrt{\LARModulus\|\DFBaselinePrimary\|_\infty}))$.
The proposal height bound of \cref{lem:df-symbolic-baseline} gives
\[
 \CalibActivation\sqrt{\LARModulus\|\DFBaselinePrimary\|_\infty}
 \leq\DFCoarseLocalScale.
\]
For the public family,
$\DFCoarseLocalScale\asymp
 \HoeffdingParameterBase^{-1/2}\DFPublicScale^2\geq1$ eventually,
and
$\sqrt{\LARModulus/\CalibAnchorWidth}=O(\DFCoarseLocalScale)$.
Consequently the linear bounds above are
$O(\DFCoarseLocalScale)$, while all higher-degree bounds are
$O(\DFCoarseLocalScale^2)$ and hence
$O(\DFCoarseLocalScale^{\HoeffdingDegree})$ in degree
$\HoeffdingDegree\geq2$.
The positive compensator has constant coefficient one and linear
deviation $o(1)$, so adjoining it preserves this degree envelope.
The bound $\DFBaselinePrimary\geq\CalibUniform/2$ converts a
degree-$\HoeffdingDegree$ uniform-reference profile bound to the
primary reference at cost at most $2^{\HoeffdingDegree/2}$.

The two summands of the paired lift are centered and depend on
independent public inputs. Orthogonality and the compensator bound give
\[
 \|\ActivityCoreLift\ActivityCoreScore\|_2^2
 \leq(1+\FockConstant\CalibActivation)\|\ActivityCoreScore\|_2^2.
\]
Every atom of the paired reference
$\DFBaselinePrimary\otimes\DFBaselineCenter$ on its support has mass
at least $(4\LARModulus^2)^{-1}$. This atom lower bound gives evaluation
operator norm at most $2\LARModulus$ at every complete pair.
Combined with the lift bound, this proves the evaluation-row claim,
including Hilbert-valued retained arguments.
Tensor extension applies to arbitrary joint kernels.
\end{proof}

\begin{theorem}[Complete weighted Fourier bound]
\label{thm:df-weighted-tail}
Fix a positive integer $\HoeffdingOrder$. For all sufficiently large
$\HoeffdingParameterBase$, let $\DFTable$ be a canonical public-control
coefficient with positive degrees at most $\HoeffdingOrder$. Let
$\DFTableNorm$ be its graded weighted $L^2$ norm with body weight
$\FockRadius^{\ActivityBodyDegree}$, where
$\FockRadius=\HoeffdingParameterBase^{1/4}$.
On the protected cycle proposal put
$\FockWeight=1+\CoreUStatistic\ActivityCoreLift\DFTable$.
Then the direct character polynomial of the signed output law
satisfies
\[
 \FockNorm{\FockPolynomial_{\FockWeight}}{\FockRadius}
 \leq\FockConstant_{\HoeffdingOrder}
                   (1+\DFTableNorm(\DFTable)).
\]
The result concerns the signed weight before charge and bounded clipping.
\end{theorem}
\begin{proof}
The character polynomial depends linearly on the weight. We therefore
bound the unweighted proposal and then the contribution of each
arity of $\DFTable$. For a correction term, the selected control
groups affect only a bounded number of endpoints; the remaining
endpoints are handled together by the all-degree product bound.
Throughout the proof, the normalization scale is the full endpoint
count $\BlockDim$, even when some groups are removed.

\emph{The unweighted proposal.}
Fix the disjoint placement, the body permutation, the complete
protected root vector, and the shared queue scalar. By
\cref{lem:df-tagged-local}, the conditional character polynomial
factors into $\GadgetCount=\DFPublicScale^{16}$ triple polynomials and
$\DFFreeCount+\LARModulus=\InteriorVertexCount-\GadgetCount$
singleton polynomials. Each triple keeps the two leg endpoints
together with their own attachment. All queue singletons use the
same fixed translation.

We check the two inputs to \cref{lem:df-finite-product}: small local
activities and an exponential-moment bound for the normalized linear
mean $\FockMean$. Each singleton has deviation
$\DFTailLocalDeviation=O(\HoeffdingParameterBase\DFPublicScale^{-4})$,
while the triple coefficients are bounded by
\cref{lem:df-tagged-local}. Using
$\GadgetCount\DFCoarseLocalScale^2/\BlockDim=\CalibActivation$
in the activity definitions gives
\[
 \begin{aligned}
 \FockRadius^2\FockLinearActivity
   &\leq\FockConstant\FockRadius^2
       (\CalibActivation+\DFTailLocalDeviation^2)
     =O(\HoeffdingParameterBase^{-1/2}),\\
 \FockRadius^2\FockQuadraticActivity
   &\leq\FockConstant\FockRadius^2\CalibActivation
     =O(\HoeffdingParameterBase^{-1/2}),\\
 \sqrt{\GadgetCount}\FockRadius^3\FockCubicActivity
   &\leq\FockConstant(\FockRadius^2\CalibActivation)^{3/2}
     =O(\HoeffdingParameterBase^{-3/4}).
 \end{aligned}
\]
These bounds are uniform in the conditioned data and still tend to
zero at every fixed multiple of $\FockRadius$.
The cycle-mean estimate in \cref{lem:df-cycle-mean-mgf}, with the
fixed-queue drift verified immediately after that lemma, gives
\[
 \mathbb E\exp(\FockConstant\FockRadius^2\FockMean^2)=O(1)
\]
for every fixed positive $\FockConstant$, now averaging the protected
latent cycle while keeping the placement, body permutation, and
shared queue scalar fixed. Thus
\cref{lem:df-finite-product} bounds the conditional product in mean
square over the cycle. Jensen's inequality allows averaging the cycle
and then the body permutation and shared queue scalar, proving the
claim for $\FockWeight=1$.

\emph{The selected control groups.}
For the local estimates, condition again on the root vector, body
permutation, and shared queue scalar.
Fix an arity $(\ActivityPrimaryDegree,\ActivityBodyDegree)$ with
$1\leq\ActivityPrimaryDegree+\ActivityBodyDegree\leq\HoeffdingOrder$,
and one choice of distinct groups in the corresponding U-statistic.
Call these groups tagged. A tagged primary group consists of its
triple and its assigned compensator singleton; a tagged ordinary
group consists of one ordinary-body singleton. These endpoint sets
are disjoint by \cref{sec:df-control-context}.

Canonical centering lets us apply the centered-score operators of
\cref{lem:df-tagged-local} in every input slot. A tagged primary group
has zero constant coefficient and can emit degrees
$1\leq\HoeffdingDegree\leq4$. After the radius and endpoint
normalization, its degree-$\HoeffdingDegree$ operator costs at most
\[
 \FockConstant
 \left(\frac{\FockRadius\DFCoarseLocalScale}{\sqrt{\BlockDim}}\right)^{
                   \HoeffdingDegree}
 =\FockConstant
 \left(\frac{\FockRadius\sqrt{\CalibActivation}}{\sqrt{\GadgetCount}}\right)^{
                   \HoeffdingDegree}
 =O(\ActivityPrimaryPool^{-1/2}).
\]
Here $\ActivityPrimaryPool=\GadgetCount$ and
$\FockRadius\sqrt{\CalibActivation}
=O(\HoeffdingParameterBase^{-1/4})$; the absence of degree zero
ensures at least one factor $\ActivityPrimaryPool^{-1/2}$.
An ordinary group emits only degree one, with raw operator norm
$O(1)$, so its normalized cost is
$O(\FockRadius/\sqrt{\ActivityBodyPool})$, using
$\ActivityBodyPool\asymp\BlockDim$ from
\cref{sec:df-parameter-verification}.

The Hilbert tensor product of these local operators acts on the
joint kernel $\DFTable_{\ActivityPrimaryDegree,\ActivityBodyDegree}$.
Its norm is bounded by the product of the local operator norms,
whether or not the kernel factors across slots. The total tagged
output degree is at most
$4\ActivityPrimaryDegree+\ActivityBodyDegree\leq4\HoeffdingOrder$.
Consequently, summing its degree allocations, symmetrizing the
output tensors, and including the degree factorials in the Fourier
norm cost only a constant depending on $\HoeffdingOrder$.

\emph{Combining tagged and untagged outputs.}
To recover the full correction polynomial, multiply the tagged
output by the unweighted product on the remaining endpoints.
Delete each tagged triple together with its compensator, and each
tagged ordinary singleton. Leave every unselected triple and
singleton as a separate block, so the remaining product still has
blocks of size at most three, as required by
\cref{lem:df-finite-product}.

Deleting blocks decreases all three activities. Since at most
$\HoeffdingOrder$ control groups were selected, it changes the
normalized linear mean by at most
$\FockConstant_{\HoeffdingOrder}
 \DFCoarseLocalScale/\sqrt{\BlockDim}$.
The radius-scaled square of this change is
$O_{\HoeffdingOrder}(\FockRadius^2\CalibActivation/\GadgetCount)=o(1)$,
so the remaining mean has the same bounded exponential moment over
the cycle. The product lemma therefore also controls the remaining
product in mean square.

For each total tagged degree, combine all tagged coefficients into
one tensor and apply \eqref{eq:df-finite-multiplication} once.
This uses the remaining product at radius $\sqrt2\FockRadius$;
the proof of \cref{lem:df-finite-product} uses at most
$2\FockRadius$. Both are fixed multiples covered by the activity
and mean bounds above. Thus the radius does not grow with the
number of untagged blocks.

Average over the complete roots, the shared queue scalar, and the
body permutation using Jensen's inequality. In particular, multiply
all queue factors before averaging their one common scalar. The
resulting norm bound for this specified selection is
\[
 \FockConstant_{\HoeffdingOrder}
 \ActivityPrimaryPool^{-\ActivityPrimaryDegree/2}
 \left(\frac{\FockRadius}{\sqrt{\ActivityBodyPool}}\right)^{
                \ActivityBodyDegree}
 \|\DFTable_{\ActivityPrimaryDegree,\ActivityBodyDegree}\|_2.
\]
The bound is uniform in the fixed disjoint placement, so it also
survives averaging the joint placement law, including the dependence
between attachment and queue positions.

The physical U-statistic is a normalized average over these
distinct-group selections. The triangle inequality preserves the
displayed bound without an additional counting factor. Its
right-hand side is precisely the contribution of this arity to
$\FockConstant_{\HoeffdingOrder}\DFTableNorm(\DFTable)$, by
\cref{sec:df-graded-norm}. Summing over arities and adding the
unweighted bound proves the theorem.
\end{proof}

\subsection{Canonical moment transfer from independent control pools}

Canonical means that integrating any one kernel argument against its
reference law gives zero with all other arguments fixed. The public
control pools are large enough relative to their smallest atoms that
an elementary finite-alphabet estimate suffices.

\begin{lemma}[Finite-alphabet canonical moments]
\label[lemma]{lem:df-rectangular-moments}
Consider two independent pools of positive sizes $\ActivityPrimaryPool$ and
$\ActivityBodyPool$, each consisting of independent identically
distributed variables on a finite space. Let their smallest positive
atom probabilities be $\DFMomentPrimaryAtom$ and $\DFMomentBodyAtom$.
Let $\DFTable$ be a real canonical, separately symmetric kernel of
fixed arities $(\ActivityPrimaryDegree,\ActivityBodyDegree)$.
Suppose the pool sizes are at least their respective arities and,
for some fixed $\DFMomentOccupancy>0$,
\[
 \ActivityPrimaryPool\DFMomentPrimaryAtom\geq\DFMomentOccupancy,
 \qquad
 \ActivityBodyPool\DFMomentBodyAtom\geq\DFMomentOccupancy.
\]
Conditions on a zero-arity pool may be omitted. For every fixed
positive integer $\FockMomentOrder$, the distinct-index average
normalized by
$\binom{\ActivityPrimaryPool}{\ActivityPrimaryDegree}
 \binom{\ActivityBodyPool}{\ActivityBodyDegree}$ over unordered subsets satisfies
\[
 \|\CoreUStatistic\DFTable\|_{2\FockMomentOrder}
 \leq\FockConstant_{\ActivityPrimaryDegree,\ActivityBodyDegree,
                         \FockMomentOrder,\DFMomentOccupancy}
 \ActivityPrimaryPool^{-\ActivityPrimaryDegree/2}
 \ActivityBodyPool^{-\ActivityBodyDegree/2}\|\DFTable\|_2.
\]
The kernel norm uses the product reference law. The constant is
independent of alphabet and pool sizes and of the individual atom
probabilities subject to the displayed lower bounds.
\end{lemma}
\begin{proof}
First consider a single column of $\DFTailSlotSize$ independent samples
$\ProbabilityRandomVariable_{\CoordinateIndex}$ from a finite law with
smallest positive atom $\DFMomentAtom$. For a centered function
$\LocalTestFunction$, put
$\DFMomentAverage\LocalTestFunction
 =\DFTailSlotSize^{-1}\sum_{\CoordinateIndex=1}^{\DFTailSlotSize}
 \LocalTestFunction(\ProbabilityRandomVariable_{\CoordinateIndex})$.
In its $2\FockMomentOrder$th moment expansion, any sample index
occurring once makes the expectation zero. A remaining term with
$\DFMomentBlockCount\leq\FockMomentOrder$ distinct indices has
absolute expectation at most
$\DFMomentAtom^{-(\FockMomentOrder-\DFMomentBlockCount)}
 \|\LocalTestFunction\|_2^{2\FockMomentOrder}$, since, on the reference support,
$\|\LocalTestFunction\|_\infty
 \leq\DFMomentAtom^{-1/2}\|\LocalTestFunction\|_2$.
Each pattern has at most
$\DFTailSlotSize^{\DFMomentBlockCount}$ index assignments, and the
number of patterns depends only on $\FockMomentOrder$. Hence
\[
 \mathbb E|\DFMomentAverage\LocalTestFunction|^{2\FockMomentOrder}
 \leq\FockConstant_{\FockMomentOrder}
 \DFTailSlotSize^{-\FockMomentOrder}
 \|\LocalTestFunction\|_2^{2\FockMomentOrder}
 \sum_{\CoordinateIndex=0}^{\FockMomentOrder-1}
       (\DFTailSlotSize\DFMomentAtom)^{-\CoordinateIndex}.
\]
Thus the column-average operator from centered $L^2$ functions to
$L^{2\FockMomentOrder}$ has norm
$O_{\FockMomentOrder,\DFMomentOccupancy}(\DFTailSlotSize^{-1/2})$
whenever $\DFTailSlotSize\DFMomentAtom\geq\DFMomentOccupancy$.

To tensorize over independent columns, apply the first bound for each
fixed sample array of the second, use Minkowski's inequality
$\|\cdot\|_{L^{2\FockMomentOrder}(L^2)}
 \leq\|\cdot\|_{L^2(L^{2\FockMomentOrder})}$, then apply the second
bound for each fixed reference input of the first.
Iteration multiplies the bounds for any joint kernel, since averaging
other columns preserves centering in the remaining slots.

For each positive arity, divide its pool uniformly into that many
labeled disjoint blocks of equal size, leaving any remainder unused.
Each block size lies between the pool size divided by twice its arity
and the pool size divided by its arity. The atom condition therefore
persists with a lower bound depending only on the fixed arities and
$\DFMomentOccupancy$. Apply the tensorized bound to the rectangular
average selecting one index per block. Its division-average is exactly
the distinct-index average, since every ordered distinct tuple has
the same inclusion probability. Jensen completes the proof.
The zero-total-arity case is immediate.
\end{proof}

\begin{corollary}[Moment transfer to physical corrections]
\label[corollary]{cor:df-weight-moments}
For every fixed positive integer $\FockMomentOrder$ and all sufficiently
large $\HoeffdingParameterBase$, the signed weight of
\cref{thm:df-weighted-tail} satisfies
\[
 \|\FockWeight-1\|_{2\FockMomentOrder}
 \leq\FockConstant_{\HoeffdingOrder,\FockMomentOrder}
             \DFTableNorm(\DFTable).
\]
The estimate holds before charge and clipping, uniformly in the
fixed disjoint placement.
\end{corollary}
\begin{proof}
Take $\HoeffdingParameterBase$ large enough that $\FockRadius\geq1$ and
both control pools have size at least $\HoeffdingOrder$.
At each fixed placement, the uncharged proposal has
$\ActivityPrimaryPool=\GadgetCount$ independent primary--compensator
pairs with law $\DFBaselinePrimary\otimes\DFBaselineCenter$ and an
independent pool of $\ActivityBodyPool$ ordinary centers with law
$\DFBaselineCenter$, by \cref{sec:df-sampling-contract}.
Their smallest positive atoms satisfy
\[
 \DFMomentPrimaryAtom\geq(4\LARModulus^2)^{-1},
 \qquad \DFMomentBodyAtom\geq(2\LARModulus)^{-1}
\]
by \cref{thm:df-positive-baseline}.
The parameter bounds
$\LARModulus^2/\ActivityPrimaryPool=O(1)$ and
$\LARModulus/\ActivityBodyPool=O(\DFPublicScale^{-12})$ from
\cref{sec:df-parameter-verification} supply a common fixed positive
$\DFMomentOccupancy$ in \cref{lem:df-rectangular-moments}.

The paired lift preserves centering in each complete pair and has
Hilbert operator norm $O(1)$ by \cref{lem:df-tagged-local}.
Its tensor extension therefore sends any joint kernel of arities
$(\ActivityPrimaryDegree,\ActivityBodyDegree)$ to a canonical kernel
on these pools with $L^2$ norm at most an order-dependent constant
times $\|\DFTable_{\ActivityPrimaryDegree,\ActivityBodyDegree}\|_2$.
Applying \cref{lem:df-rectangular-moments} to that lifted kernel gives
\[
 \|\CoreUStatistic\ActivityCoreLift
       \DFTable_{\ActivityPrimaryDegree,\ActivityBodyDegree}\|_{2\FockMomentOrder}
 \leq\FockConstant_{\HoeffdingOrder,\FockMomentOrder}
 \ActivityPrimaryPool^{-\ActivityPrimaryDegree/2}
 \ActivityBodyPool^{-\ActivityBodyDegree/2}
 \|\DFTable_{\ActivityPrimaryDegree,\ActivityBodyDegree}\|_2.
\]
Sum over arities and use $\FockRadius\geq1$ to obtain the stated
coefficient-norm bound, uniformly over placements.
The moment order affects only the constant.
\end{proof}

\subsection{Bounds for the calibrated weight and its full Fourier tail}
\label{sec:df-corrected-bounds-proof}

\begin{proof}[Proof of \cref{prop:df-corrected-weight-bounds}]
For the calibrated coefficient,
$\DFTableNorm(\DFTableCorrection)
 =\DFMatrixNorm{\DFMatrixCorrection}
 =O_{\HoeffdingOrder}(\HoeffdingParameterBase^{-3})$.
Substitution into \cref{thm:df-weighted-tail,cor:df-weight-moments}
gives the complete Fourier norm bound and every fixed moment bound.
The analysis radius of \cref{sec:df-parameter-choice} is independent
of either order.
The constants and the sufficiently large parameter threshold are
uniform over placements.

By \cref{thm:df-actual-table-inverse}, the protected endpoint law has
mass one and, after independent uniform final relabeling, is
exchangeable with all nonzero support frequencies through
$\HoeffdingOrder$ canceled. For sufficiently large
$\HoeffdingParameterBase$,
$\DFChargeRadius=\FockRadius/\sqrt e>1$.
Using the Fourier transform of this law and
\cref{lem:df-fourier-isometry} gives
\begin{align*}
 \sum_{\DFFrequency\ne0}|\DFTransform(\DFFrequency)|^2
 &=\sum_{|\operatorname{supp}\DFFrequency|>\HoeffdingOrder}
                          |\DFTransform(\DFFrequency)|^2\\
 &\leq\DFChargeRadius^{-2(\HoeffdingOrder+1)}
       \DFEnergy_{\DFChargeRadius}
           (\SignedEndpointLaw_{\FockWeight_{\HoeffdingOrder}})\\
 &\leq\DFChargeRadius^{-2(\HoeffdingOrder+1)}
       \left(\DFFourierNorm_{\FockRadius}
           (\SignedEndpointLaw_{\FockWeight_{\HoeffdingOrder}})^2-1\right)\\
 &\leq\DFProofConstant_{\HoeffdingOrder}
                   \HoeffdingParameterBase^{-(\HoeffdingOrder+1)/2}.
\end{align*}
The sum includes every remaining frequency, up to full support.
Applying \cref{lem:df-ambient-target} and enlarging the order-dependent
constant gives the stated ambient half-$L^1$ bound.
\end{proof}

%% file: chapters/appendix-direct-fourier-charge.tex
\section{Charge Conditioning and Bounded Positive Sampling}
\label{sec:df-fourier-proofs}

We prove the charge and clipping guarantees of
\cref{sec:df-positive-sampling}, verify the parameters, and establish
the bounded implementation and accuracy of \cref{sec:df-implementation}.
Signed expectations and variation distance follow the conventions of
\cref{sec:df-target}.

\subsection{The public charge test}
\label{sec:df-public-charge-verification}

Each body cycle has integer sum zero. A leg's two off-body residues
have sum $\LARModulus-1+\LegBaselineShift-\LocalSecondError$
modulo $\LARModulus$, and its attachment adds
$\LocalSecondError$. The queue adds $\LARModulus$ times its scalar.
Thus the public charge test of \cref{sec:df-positive-sampling} depends
only on core centers and leg shifts and is equivalent, pointwise in all
remaining randomness, to the auxiliary endpoint lying in
$\ZeroSumEndpointHyperplane$.

\subsection{Clipping and conditioning from ambient error}
\label{sec:df-clipping-conditioning-proof}

\begin{proof}[Proof of \cref{thm:df-positive-conditioning}]
 Fix the placement. Let $\DFClippedEndpointMeasure$ be the
 unnormalized endpoint image of the clipped core measure
 $\DFClippedWeight\DFIdealProposal$. Pushing the signed and clipped core measures
 through the same protected endpoint channel, including final
 relabeling, contracts their $\ell_1$ difference. Thus
 \[
 \|\DFClippedEndpointMeasure-\SignedEndpointLaw\|_1
 \leq\mathbb E|\FockWeight-\DFClippedWeight|
 =\DFClippingError.
 \]
 The triangle inequality gives
 \[
 \|\DFClippedEndpointMeasure-\Unif{\LARField^{\BlockDim}}\|_1
 \leq2\DFAmbientError+\DFClippingError
 =\DFConditioningError.
 \]
 The public charge test is pointwise equivalent to membership in
 $\ZeroSumEndpointHyperplane$, by
 \cref{sec:df-public-charge-verification}. Hence
 $\DFClippedEndpointMeasure(\ZeroSumEndpointHyperplane)
 =\DFPositiveMass$. The ambient uniform law assigns this hyperplane
 mass $1/\LARModulus$, so
 \[
 |\DFPositiveMass-1/\LARModulus|\leq\DFConditioningError,
 \qquad
 \DFPositiveMass\geq\frac1{\LARModulus}-\DFConditioningError
                 \geq\frac1{2\LARModulus}.
 \]
 This is stronger than the asserted mass lower bound. Restriction
 cannot increase $\ell_1$ error. Normalize the restricted clipped
 measure by $\DFPositiveMass$ and the restricted uniform measure
 by $1/\LARModulus$. Adding and subtracting the latter restriction
 divided by $\DFPositiveMass$ gives
 \[
 \begin{aligned}
 \StatisticalDistance{\DFContextPositiveEndpointLaw}
                       {\Unif{\ZeroSumEndpointHyperplane}}
 &\leq
 \frac{\DFConditioningError+
             |\DFPositiveMass-1/\LARModulus|}
      {2\DFPositiveMass}\\
 &\leq\frac{\DFConditioningError}{\DFPositiveMass}
 \leq2\LARModulus\DFConditioningError.
 \end{aligned}
 \]
\end{proof}

\begin{proof}[Proof of \cref{lem:df-positive-repair}]
 Fix the calibration order. For every fixed positive integer
 $\FockMomentOrder$,
 \cref{cor:calibration-bounded-clipping,prop:df-corrected-weight-bounds}
 give
 \[
 \DFClippingError
 \leq\mathbb E|\FockWeight-1|^{2\FockMomentOrder}
 \leq\DFProofConstant_{\HoeffdingOrder,\FockMomentOrder}
          \HoeffdingParameterBase^{-6\FockMomentOrder},
 \]
 uniformly over the disjoint placements. The calibrated weight and
 its proposal do not depend on the moment order. Taking a sufficiently
 large fixed moment order for each desired inverse-polynomial bound
 proves that this same clipping loss is negligible, uniformly over
 placements. Since $\LARModulus=O(\HoeffdingParameterBase^{64})$,
 its charge-amplified loss is negligible as well.
\end{proof}

\subsection{Bounded clipping for a fixed placement}
\label{sec:df-numerical-stability}

\begingroup
\mdfsetup{topline=true}
\begin{lemma}[Bounded rejection and approximation budget]
 \label[lemma]{lem:df-bounded-rejection}
 Under the hypotheses of \cref{thm:df-positive-conditioning}, the ideal
 rule samples the clipped positive core law upon acceptance, with
 acceptance rate at least $1/(8\LARModulus)$. For
 $0<\DFTrialError<1$, a cap
 \[
 \DFTrialCap=
 \left\lceil8\LARModulus\log(1/\DFTrialError)\right\rceil
 \]
 fails with probability at most $\DFTrialError$.

 Let $\DFUnnormalizedError$ bound the $\ell_1$ error in a numerical
 nonnegative charged measure, whose weight also lies in $[0,2]$.
 If $\DFUnnormalizedError\leq1/(8\LARModulus)$, its normalized core law
 is within $4\LARModulus\DFUnnormalizedError$ of the ideal clipped law
 in statistical distance. Its acceptance rate with denominator two
 is at least $1/(16\LARModulus)$, so replacing $8$ by $16$ in the cap
 gives the same failure bound.
 For a proposal at statistical distance at most $\DFProposalError$
 from $\DFIdealProposal$ and a weight satisfying
 $\|\DFApproxWeight-\FockWeight\|_\infty\leq\DFWeightError$,
 clipping and charge restriction permit the choice
 \[
  \DFUnnormalizedError=4\DFProposalError+\DFWeightError.
 \]
\end{lemma}
\endgroup

\begin{proof}[Proof of \cref{lem:df-bounded-rejection}]
 For the fixed placement, the ideal acceptance rate is
 $\AssemblyAcceptanceProbability=\DFPositiveMass/2
 \geq1/(8\LARModulus)$. Independence of the trials gives
 \[
 \Pr[\text{timeout}]
 =(1-\AssemblyAcceptanceProbability)^{\DFTrialCap}
 \leq e^{-\AssemblyAcceptanceProbability\DFTrialCap}.
 \]
 Conditional on acceptance, the core has the clipped charged law,
 proving the ideal cap and output claims.

 Suppose the rational proposal $\DFRationalProposal$ is within
 $\DFProposalError$ of $\DFIdealProposal$ in statistical distance,
 and $\|\DFApproxWeight-\FockWeight\|_\infty\leq\DFWeightError$.
 Clip the rational weight by $\DFClipMap$. Changing the proposal costs
 at most $4\DFProposalError$ in $\ell_1$ because the ideal clipped
 weight lies in $[0,2]$; changing the weight costs at most
 $\DFWeightError$ because clipping is $1$-Lipschitz. Charge restriction
 cannot increase either error, so the unnormalized charged measures
 have $\ell_1$ discrepancy at most
 \[
 \DFUnnormalizedError:=4\DFProposalError+\DFWeightError.
 \]
 Normalize these nonnegative measures as in the proof of
 \cref{thm:df-positive-conditioning}. Its ideal mass bound
 $1/(4\LARModulus)$ gives normalized statistical distance at most
 $4\LARModulus\DFUnnormalizedError$.
 If $\DFUnnormalizedError\leq1/(8\LARModulus)$, the numerical
 mass is at least $1/(8\LARModulus)$, and acceptance with denominator
 two has probability at least $1/(16\LARModulus)$.
 The same geometric estimate proves the numerical cap.
\end{proof}

Retaining the initial placement through retries and timeout fallback
preserves its marginal despite placement-dependent acceptance rates.
Conditional completion follows \cref{sec:df-sampling-contract}.

\subsection{Verification of the public parameter family}
\label{sec:df-parameter-verification}

\paragraph{Geometry and local feasibility.}
Use $\DFPublicScale=\HoeffdingParameterBase^8$ from
\cref{sec:df-parameter-choice}. The public sizes give
\[
 \frac{\LARModulus+2\GadgetCount}{\InteriorVertexCount}
 =O(\DFPublicScale^{-4}),\qquad
 \ActivityPrimaryPool=\DFPublicScale^{16},\qquad
 \ActivityBodyPool\asymp\InteriorVertexCount
       \asymp\BlockDim\asymp\DFPublicScale^{20}.
\]
Thus disjoint placement and distinct compensators exist eventually.
The finite-atom losses are absorbed by the pools:
$\LARModulus^2/\ActivityPrimaryPool=O(1)$ and
$\LARModulus/\ActivityBodyPool=O(\DFPublicScale^{-12})$.
These bounds verify the full paired-alphabet moment estimate in
\cref{cor:df-weight-moments}.
The three source cutoffs sum to the safe-center margin, while
$\HoeffdingParameterBase^2\ll\CalibAnchorWidth=\DFPublicScale^6
 \ll\LARModulus$ verifies the anchor support conditions.

For local affine-map feasibility, the protected input count satisfies
$\LegInputCount=(2\HoeffdingParameterBase^2-1)^2
 \sim4\HoeffdingParameterBase^4$.
Substituting $\LegSumBound=3\DFPublicScale^6$ and
$\LARModulus\geq2^{16}\DFPublicScale^8$ into the three loss
quantities of \cref{sec:leg-sampling-parameters} gives
\[
\begin{aligned}
 \ChartDeletionBound
   &=O(\HoeffdingParameterBase^4\DFPublicScale^{-2})
     =O(\HoeffdingParameterBase^{-12}),\\
 \ChartOperatorBound
   &=O(\HoeffdingParameterBase^8\DFPublicScale^{-2})
     =O(\HoeffdingParameterBase^{-8}),\\
 \ChartLocalPositivity
   &=O(\HoeffdingParameterBase^{12}\DFPublicScale^{-2})
     =O(\HoeffdingParameterBase^{-4}).
\end{aligned}
\]
All three losses tend to zero, so the local feasibility and positivity
conditions hold for all sufficiently large $\HoeffdingParameterBase$.

\paragraph{Proposal and response bounds.}
Substitution in the symbolic estimates of
\cref{sec:df-baseline-proof} gives
\begin{gather*}
 \CalibActivation=\Theta(\HoeffdingParameterBase^{-1}),\qquad
 \CalibCompensation
       =O(\HoeffdingParameterBase^{-1}\DFPublicScale^{-4}),\\
 \DFBaselineLoss+\DFCenterLoss+\DFPrimaryDeviation
       =O(\HoeffdingParameterBase\DFPublicScale^{-2}),\\
 \|\DFBaselinePrimary\|_\infty
       \leq\frac{\BlockDim}
              {\GadgetCount\CalibActivation\LARModulus}
       =O(\HoeffdingParameterBase\DFPublicScale^{-4}),\\
 \CalibActivation\|\DFSmoothedAnchor\DFBaselinePrimary\|_\infty
       =O(\LARModulus^{-1}),\\
 \|\DFBaselineCenter-\CalibUniform\|_{\CalibUniform^{-1}}
 +\|\CalibSafeUniform-\CalibUniform\|_{\CalibUniform^{-1}}
       =O(\HoeffdingParameterBase\DFPublicScale^{-4}).
\end{gather*}
These bounds imply proposal positivity. The first inverse is bounded because
\[
 \CalibActivation^{-1}\sqrt{\CalibAnchorWidth/\LARModulus}
 =O(\HoeffdingParameterBase/\DFPublicScale)=o(1).
\]
The local estimates of \cref{sec:df-actual-tables,sec:df-tail-appendix}
then give an $O_{\BoundaryInverseOrder}(\HoeffdingParameterBase^{-3})$
coefficient correction, the same decay for every fixed physical
moment norm, and a bounded complete Fourier norm. Constants may
depend on the fixed orders. With the common local scale of
\cref{lem:df-tagged-local}, the three product activities at radius
$\FockRadius=\HoeffdingParameterBase^{1/4}$ are
$O(\HoeffdingParameterBase^{-1/2})$,
$O(\HoeffdingParameterBase^{-1/2})$, and
$O(\HoeffdingParameterBase^{-3/4})$, respectively.
They remain small at $2\FockRadius$, as required by
\cref{lem:df-finite-product,thm:df-weighted-tail}.

\paragraph{Source protection and the finite prefix.}
The ratio of each cutoff to $\GaussianWidth$ grows linearly in
$\HoeffdingParameterBase$. Combining the body coupling bound in
\eqref{eq:df-edge-coupling} with the tail estimates in
\cref{lem:joint-source-tails} therefore gives
\[
 \BodyTailProbability+\DFBodyClippingError
       +2\GadgetCount\LegTailProbability+\QueueTailProbability
 \leq \operatorname{poly}(\HoeffdingParameterBase)
                 e^{-\Omega(\HoeffdingParameterBase^2)}.
\]
On any finite prefix excluded by analytic or numerical certification,
use the fallback of \cref{sec:df-certification-fallback}. It is integral
and invertible, and carry-safe on protected inputs. No accuracy guarantee
is asserted on that prefix.

\subsection{Numerical verification and deterministic bounds}
\label{sec:df-numerical-verification}

The sampler uses certified rational intervals and bounded finite-bit
draws. For fixed
$\HoeffdingOrder$, the certificates below prove polynomial
running time, exact safe support, and implementation error at most
$\DFNumericalBudget$.

\begin{algorithm}[H]
 \caption{Bounded public-control matrix sampler}
 \label{alg:df-core-sampler}
 \label{alg:finite-core-inversion}
 \begin{algorithmic}[1]
 \Require Fixed accuracy order $\TransportAccuracyOrder$ and the
 public parameters of \cref{sec:df-parameter-choice}.
 \Ensure Integral affine data independent of the hidden source.
 \State Fix the order and budgets of \cref{sec:df-implementation},
 and the precision, table-size and work caps of
 \cref{sec:df-numerical-verification}.
 \State Draw one disjoint placement $\CalibContext$, pair its labeled
 legs with distinct free compensators, and retain this placement.
 \State Construct the proposal and finite response matrices; solve for
 the signed correction by the finite block computation of
 \cref{prop:df-finite-coefficient-computation}, certifying the proposal,
 inverse gaps, and numerical bounds. If certification fails,
 return the deterministic carry-safe affine pair of
 \cref{sec:df-certification-fallback}.
 \State For at most
 $16\LARModulus\lceil\log_2(5/\DFNumericalBudget)\rceil$
 trials, draw a fresh core at this placement and implement the
 threshold $\EventIndicator{\DFContextChargeEvent}
 \DFClipMap(\DFApproxWeight)/2$ to the certified finite-bit
 precision. Retain the first accepted core.
 \State If no core was accepted, use the charge-correct regular-only
 core of \cref{sec:df-timeout-fallback}.
 \State Draw the complete conditional flag--affine-map pairs at the
 retained core, then independently draw the actual closure Gaussian,
 body permutation, and final global permutation with bounded samplers.
 \State Assemble and return $(\TransportMatrix,\TransportOffset)$
 by \cref{prop:matrix-integer-realization}.
 \end{algorithmic}
\end{algorithm}

\paragraph{Certified precomputation.}
Certification in \cref{alg:df-core-sampler} checks
\cref{lem:df-disjoint-context,thm:df-positive-baseline} before using
population normalizers or calibration inverses. It uses the rational
reference laws below, averages complete fixed-queue expressions before
inversion as in \cref{sec:df-common-queue-inverse}, and checks the
inverse gaps of \cref{prop:df-finite-coefficient-computation}.
The following bounds certify precision, work caps, and fallbacks.

\paragraph{Finite coordinates, matrices, and arithmetic.}
For fixed $\HoeffdingOrder$, the canonical public-control coordinates
of \cref{def:df-control-coordinates}, the multiplication structure
constants, and the normalized response blocks are obtained from the
same finite pattern contractions. Their total stored size satisfies
\[
 \DFTableSize\leq
 (\LARModulus\InteriorVertexCount\LegCutoff\CalibAnchorWidth)^{
                                            \DFProofConstant_{\HoeffdingOrder}}.
\]

Each input or output pattern has $O(\HoeffdingOrder)$ slots.
Enumerate its selected vertices,
their role intersections, and their relative positions on the cycle.
There are $\InteriorVertexCount^{O(\HoeffdingOrder)}$ possibilities.
For each path cluster sum the finitely supported Gaussian masses on
at most $O(\HoeffdingOrder)$ latent edges. This requires
$\LegCutoff^{O(\HoeffdingOrder)}$ operations. Anchors, colors, public
blurs, and the shared queue scalar add factors at most
$(\LARModulus\LegCutoff\CalibAnchorWidth)^{O(\HoeffdingOrder)}$.
The distinct-index weights are ratios of falling factorials.
Multiplication matrices, finite logarithms and reciprocals, and the
block solve in \cref{prop:df-finite-coefficient-computation} use
polynomially many operations in this stored size, retaining the queue
value until its prescribed average. The computation enumerates only
these fixed-order patterns.

Gaussian probabilities are computable reals, so the implementation
uses certified rational approximations. Every nonzero protected edge mass
has negative logarithm $O(\HoeffdingParameterBase^2+\log
\HoeffdingParameterBase)$. A fixed-order joint mass has the same
bound with an order-dependent constant. Source contractions can
therefore use joint masses directly; if likelihood ratios are used,
$O_{\HoeffdingOrder}(\HoeffdingParameterBase^2)$ extra precision bits
suffice to control these possibly exponentially small denominators.
The primary-color and free-center proposal probabilities are at least
$1/(2\LARModulus)$ on their supports.

The diagonal inverse in \cref{prop:df-finite-coefficient-computation}
is bounded in the graded coefficient norm. At each fixed order,
the lower reference atoms and the finite-atom inequality of
\cref{lem:df-norm-comparison} give, for every hybrid type,
\begin{align*}
 \|\DFTable_{\ActivityPrimaryDegree,\ActivityBodyDegree}\|_\infty
 &\leq(2\LARModulus)^{(\ActivityPrimaryDegree+
                              \ActivityBodyDegree)/2}
       \FockRadius^{-\ActivityBodyDegree}\\
 &\quad\cdot
       \ActivityPrimaryPool^{\ActivityPrimaryDegree/2}
       \ActivityBodyPool^{\ActivityBodyDegree/2}
       \DFTableNorm(\DFTable).
\end{align*}
Conversely, a supremum bound controls every product $L^2$ norm.
The coordinate basis changes of \cref{sec:df-table-algebra} and
these explicit population factors are polynomial at fixed order,
so the induced coordinate inverse norm is polynomially bounded.
These factors affect numerical point evaluation; the physical moment
and complete Fourier bounds remain as stated.

The finite-atom bound for a complete paired control, the integrated
paired-lift estimate of \cref{lem:df-tagged-local}, and the
normalized U-statistic average give polynomial uniform propagation
of coefficient errors at fixed order.
Evaluation of
$\DFMatrixPhysical(\CalibCore)\DFMatrixSolution$ enumerates distinct
tuples of at most $\HoeffdingOrder$ controls and the finite anchor
contractions in each paired lift. It takes
$(\BlockDim\LARModulus)^{O(\HoeffdingOrder)}$ arithmetic operations.

\paragraph{Rational approximation with a law-level budget.}
Use outward-rounded rational intervals for Gaussian masses and all
finite operations. Approximate positive masses on their exact
supports and normalize the resulting rational vectors; the primary
and center margins in \cref{thm:df-positive-baseline} permit this while
retaining, for example, lower bounds $1/(4\LARModulus)$ and one quarter
of the safe uniform. Under these rational reference laws, canonicalize
every table by the exact commuting projections
$\LocalIdentityOperator-\mathbb E_{\text{one input slot}}$. The scalar component is
retained separately. Use the same reference colors and ordered basis
indices for the ideal and rational coordinate constructions. Rebuild
the paired lift, response tables, and physical evaluation row using
these same rational laws and normalized rational local kernels.
In particular, let $\DFRationalPrimary,\DFRationalCenter$ be the two
rational control laws, let $\DFRationalAnchor$ be the rational smoothed
anchor kernel, normalized column by column on its exact safe support,
and let $\DFRationalActivation$ approximate the activation. The
rational paired lift is
\[
 (\DFRationalLift\DFDensityInput)(\CalibColor,\CalibOtherColor)
 =\DFDensityInput(\CalibColor)
 -\DFRationalActivation
 \frac{(\DFRationalAnchor\operatorname{diag}(\DFRationalPrimary)
                \DFDensityInput)(\CalibOtherColor)}
      {\DFRationalCenter(\CalibOtherColor)}.
\]
The denominator is positive on this support. Since the anchor columns
sum exactly to one,
\[
 \mathbb E_{\DFRationalPrimary\otimes\DFRationalCenter}
       \DFRationalLift\DFDensityInput
 =(1-\DFRationalActivation)
       \mathbb E_{\DFRationalPrimary}\DFDensityInput=0
\]
for every rationally centered input. Slotwise tensoring and
distinct-index averaging therefore preserve exact centering of the
physical rational correction. Merely projecting the coefficient
tables while retaining an inconsistent lift would not suffice.
The uniform weight error below includes changes to reference laws,
basis conversions, projections, and lifts. Exact Fourier
cancellation is required only for the ideal weight.

Propagate interval errors through the finite arithmetic circuit using
the displayed operand and denominator bounds. For each degree-diagonal
block, let $\DFDiagonalInverseBound$ bound its induced coordinate
$\ell_\infty$ inverse norm, and let $\DFDiagonalEntryError$ bound the
maximum entrywise approximation error. Require
\[
 \DFMatrixDimension\DFDiagonalEntryError
       \leq\frac1{2\DFDiagonalInverseBound}.
\]
The factor $\DFMatrixDimension$ converts entrywise error to operator
error. Apply \cref{thm:stable-finite-dimensional-calibration} after
factoring out the ideal block: the approximate inverse has norm at
most $2\DFDiagonalInverseBound$. The inverse-difference identity
bounds its error by
$2\DFDiagonalInverseBound^2\DFMatrixDimension\DFDiagonalEntryError$.
The polynomial coordinate dimension and inverse bound thus certify an
inverse tolerance for the fixed public family.
The fixed-degree logarithms, multiplication matrices, finite
reciprocals, and graded-inverse products have polynomial total
degree; source ratios have at most
$2^{O_{\HoeffdingOrder}(\HoeffdingParameterBase^2)}$ conditioning.
These rules give an explicit upper bound
\[
 \DFStabilityFactor
 \leq
 2^{\operatorname{poly}_{\HoeffdingOrder}
                           (\HoeffdingParameterBase)}
\]
on the propagated error in any table entry or physical weight.
Compute this bound from the circuit and positive margins. Precision
$O(\log\DFStabilityFactor+
     \log(\DFTableSize/\DFLocalTolerance))$ is polynomial in
$\secpar$ and $\log(1/\DFLocalTolerance)$ at fixed order.
Rational Gaussian approximations can be produced by the convergent
exponential power series with range reduction; its remainder bound
certifies each endpoint interval.

Choose the proposal and weight errors to satisfy
\begin{equation*}
 \DFProposalError\leq
       \frac{\DFNumericalBudget}{320\LARModulus},
 \qquad
 \DFWeightError\leq\frac{\DFNumericalBudget}{80\LARModulus}.
\end{equation*}
A product of at most $O(\BlockDim)$ conditional elementary draws is
within the sum of their uniform conditional errors, which determines
the local table tolerances.
Consequently $\DFUnnormalizedError\leq
\DFNumericalBudget/(40\LARModulus)$. For all sufficiently large
security parameters, uniformly over placements,
\cref{thm:df-positive-conditioning} gives
$\DFPositiveMass\geq1/(4\LARModulus)$.
In this regime, \cref{lem:df-bounded-rejection} bounds the normalized
core error by $\DFNumericalBudget/10$ and certifies the numerical
mass and acceptance bounds for $0<\DFNumericalBudget<1$.

Rational discrete sampling and acceptance are implemented by fixed-bit
inverse-transform sampling. For a finite law of support size
$\DFTableSize$, cumulative dyadic rounding with
$O(\log(\DFTableSize/\DFTolerance))$ bits gives distance at most
$\DFTolerance$, while never selecting a forbidden support point.
The acceptance threshold is
$\EventIndicator{\DFContextChargeEvent}\DFClipMap(\DFApproxWeight)/2$.
Round it within $[0,1]$, keeping it exactly zero outside the public
charge event and whenever the clipped value is zero.
Budget these errors over the predetermined maximum number of draws
and trials. In the same asymptotic regime, exact trials from the
certified rational law satisfy the geometric timeout bound. Couple
these capped trials with their finite-bit implementation: all rounding
effects, including changes to acceptance or timeout, cost at most the
sum of their conditional sampling errors.
Reserve at most $\DFNumericalBudget/5$ for them. Use the integer cap
\[
 \DFTrialCap=16\LARModulus
        \left\lceil\log_2(5/\DFNumericalBudget)\right\rceil.
\]
Compute the ceiling by exact integer comparisons. Since
$\log_2(5/\DFNumericalBudget)\geq\log(5/\DFNumericalBudget)$,
\cref{lem:df-bounded-rejection} bounds timeout by
$\DFNumericalBudget/5$ with
$O(\LARModulus\log(1/\DFNumericalBudget))$ trials.
Reserve at most $\DFNumericalBudget/5$ for the subsequent complete flag--map draws,
closure Gaussian, body permutation, and final relabeling, using
\cref{sec:discrete-leg-injection,sec:discrete-body-injection,sec:bounded-permutation-sampling}.
Reserve a separate $\DFNumericalBudget/10$ for the initial joint
placement permutation. This error is not charge-amplified: both
ideal and approximate algorithms normalize separately for each retained
placement, and total variation contracts under the common conditional
output kernel. Unused budget covers the finite table and interval
endpoints.
This allocation includes charge-normalizer amplification.

\paragraph{Fallbacks and deterministic caps.}
\label{sec:df-certification-fallback}
If certification exceeds the prescribed polynomial work or bit caps,
return the deterministic data
\[
 \begin{aligned}
 \DFFallbackAdjustment
   &:=\CanonicalLift{-\BlockDim\PaddingSafeMargin},\qquad
 (\DFFallbackLoad)_{\VertexIndex}
   &:=\PaddingSafeMargin
       +\EventIndicator{\{\VertexIndex\leq\DFFallbackAdjustment\}},
       \qquad 1\leq\VertexIndex\leq\BlockDim,\\
 \TransportMatrix&:=\IdentityMatrix,\qquad
 \TransportOffset:=\RoundingMatrix^{-1}\DFFallbackLoad.
 \end{aligned}
\]
Since $0\leq\DFFallbackAdjustment<\LARModulus\leq\BlockDim$,
the load sum $\BlockDim\PaddingSafeMargin+\DFFallbackAdjustment$
is divisible by $\LARModulus$; \cref{lem:rounding-image-lattice}
gives an integral offset. The linear part is invertible modulo
$\LARModulus$. Each row of $\RoundingMatrix$ has absolute sum at
most three, and a protected source satisfies
$\|\LWEError\|_\infty\leq\LegCutoff$ by
\cref{def:explicit-rounding-matrix,def:joint-protected-event}.
As $\PaddingSafeMargin=3\LegCutoff$, the signed load obeys
\[
 0\leq(\DFFallbackLoad+\RoundingMatrix\LWEError)_{\VertexIndex}
 \leq2\PaddingSafeMargin+1<\LARModulus.
\]
Its carry is zero; passing to the canonical source lift changes it by
an element of $\LatticeGen{\RoundingMatrix}$, preserving
\cref{lem:affine-transport-endpoint-carry-bound}'s criterion.
Compute the last offset coordinate as
$(\BlockDim\PaddingSafeMargin+\DFFallbackAdjustment)/\LARModulus$
and, for $1\leq\CoordinateIndex<\BlockDim$, set
\[
 (\TransportOffset)_{\CoordinateIndex}
 =\sum_{\VertexIndex=1}^{\CoordinateIndex}
   \left((\DFFallbackLoad)_{\VertexIndex}
    -(\TransportOffset)_{\BlockDim}
       \EventIndicator{\{\VertexIndex\leq\LARModulus\}}\right).
\]
The computation uses public data with polynomial work and bit length
and preserves protected-input carry safety on the excluded prefix.
The asymptotic bounds ensure eventual certification, using interval
inequalities and rational inverse residuals.

\paragraph{Charge-correct timeout.}
\label{sec:df-timeout-fallback}
For rejection timeout in the certified regime,
use a charge-correct regular-only graph core: every body center is
$\PaddingSafeMargin=3\LegCutoff$, every leg shift is $1$, every
marking weight is zero, and the primary target is
$(\LARModulus-1)/2$ with companion
$(\LARModulus+1)/2$. These companion targets remain interior after
every protected second input. The charge is correct because
$\LARModulus$ divides $\InteriorVertexCount$:
\[
 \sum_{\VertexIndex\in\InteriorVertices}\BaselineCenters_{\VertexIndex}
   +\sum_{\GadgetIndex}\LegBaselineShift_{\GadgetIndex}
 =3\InteriorVertexCount\LegCutoff+\GadgetCount
 \equiv\GadgetCount\pmod{\LARModulus}.
\]
The regular complete-map law preserves safe support and nonzero leg
scales. Together with the verified charge constraint,
\cref{prop:matrix-integer-realization} therefore gives integral affine
data and modular invertibility on this fallback.
Its error is charged only to timeout. Complete this core with
independent closure and final permutations; certification failure
instead returns the deterministic affine pair directly.
Both fallbacks use public data only.

\subsection{Completion of fixed-order affine fiber transport}
\label{sec:df-final-aft-proof}

\begin{proof}[Proof of \cref{thm:df-constructive-aft}]
 \Cref{prop:df-corrected-weight-bounds} bounds the ambient error by
 $O_{\HoeffdingOrder}
   (\HoeffdingParameterBase^{-(\HoeffdingOrder+1)/4})$.
 The chosen calibration order and
 $\LARModulus=O(\HoeffdingParameterBase^{64})$ give
 $\LARModulus\DFAmbientError=o(1)$.
 By \cref{lem:df-positive-repair},
 $\LARModulus\DFClippingError=o(1)$ as well, uniformly over placements.
 Thus
 $\LARModulus\DFConditioningError
 =\LARModulus(2\DFAmbientError+\DFClippingError)\leq1/2$
 for all sufficiently large $\HoeffdingParameterBase$.
 \Cref{thm:df-positive-conditioning} then bounds the positive
 protected endpoint error by
 \[
 O_{\HoeffdingOrder}
   (\LARModulus\HoeffdingParameterBase^{-(\HoeffdingOrder+1)/4})
 +2\LARModulus\DFClippingError.
 \]
 For each fixed calibration order,
 \cref{lem:df-positive-repair} makes the second term negligible,
 uniformly over placements.
 Normalization is separate at each disjoint placement; averaging
 adds no error.

 Apply the joint coupling of \cref{sec:protected-comparison-proof}
 to the positive source-independent core. On a match, the pointwise
 criterion of \cref{lem:affine-transport-endpoint-carry-bound}
 identifies the affine output with the canonical pullback of the
 protected endpoint. By \cref{lem:canonical-endpoint-pullback}, the
 fiber error therefore incurs the single additional cost
 \[
 \DFBodyClippingError+2\GadgetCount\LegTailProbability
 +\QueueTailProbability
 \leq\operatorname{poly}(\HoeffdingParameterBase)
                  e^{-\Omega(\HoeffdingParameterBase^2)},
 \]
 as verified in \cref{sec:df-parameter-verification}.
 Absorb this source-protection error and the charge-amplified clipping
 error into the negligible function
 $\DFNegligibleError_{\HoeffdingOrder}(\HoeffdingParameterBase)$.

 The numerical budget of \cref{sec:df-numerical-verification}
 includes charged-core normalization, finite-bit trials, timeout,
 placement, and completion. Its precision and work caps give
 polynomial worst-case time for each fixed order.
 Bertrand's bound gives $\LARModulus=O(\HoeffdingParameterBase^{64})$.
 Find the least prime deterministically by scanning the defining
 interval and testing each integer by trial division. The interval
 length and divisor bounds are polynomial in $\secpar$, so computing
 the public modulus takes polynomial time.
 The calibration order in \cref{sec:df-implementation} satisfies
 $\HoeffdingOrder\geq4\TransportAccuracyOrder+267$, so
 \[
  64-\frac{\HoeffdingOrder+1}{4}
  \leq-\TransportAccuracyOrder-3.
 \]
 Thus the Fourier term is
 $O_{\HoeffdingOrder}(\HoeffdingParameterBase^{-\TransportAccuracyOrder-3})$.
 Together with the negligible remainder and numerical budget, the
 total error is eventually at most $\secpar^{-\TransportAccuracyOrder}$.
\end{proof}

%% file: chapters/appendix-applications-proofs.tex
\section{Proofs for the Applications}
\label{sec:applications-proofs}

We prove the applications of \cref{sec:applications} for the family in
\cref{sec:df-parameter-choice} and
\cref{def:explicit-rounding-matrix}.

\subsection{The Gaussian interface and worst-case reduction}
\label{sec:applications-worst-case-proof}

\begin{proof}[Proof of \cref{cor:applications-worst-case}]
For the fixed public family, $\SecretDim=\secpar$,
$\GaussianWidth=\secpar+2$, and $\LARModulus$ is an odd prime
polynomially bounded in $\SecretDim$.
Set $\LWENoiseRate=\GaussianWidth/(\sqrt{2}\LARModulus)$.
Then $0<\LWENoiseRate<1$ and
$\LWENoiseRate\LARModulus=\GaussianWidth/\sqrt{2}>2\sqrt{\SecretDim}$
for $\secpar\geq3$, as required by
Regev~\cite[full version, Theorem~3.1]{STOC:Regev05}.
Moreover, $\SampleCount$ blocks contain
$\SampleCount\BlockDim$ scalar samples, still polynomially many.
It remains to match the Gaussian and algorithmic interfaces.

Regev's theorem uses continuous Gaussian error on the torus.
Scale each response by $\LARModulus$, apply Gaussian randomized
rounding to $\ZZ$ with width $\GaussianWidth/\sqrt{2}$, and reduce
modulo $\LARModulus$.
By~\cite[full version, Theorem~3.1]{C:Peikert10}, the continuous and
rounding squared widths, both $\GaussianWidth^2/2$, yield error within
$\exp(-\Omega(\GaussianWidth^2))$ in SD of
$\Gaussian{\ZZ}{\GaussianWidth}\bmod\LARModulus$.
Integer-translation equivariance preserves the secret term and sends a
uniform torus response to an exactly uniform residue.
Truncation and polynomial-bit arithmetic give a bounded polynomial-time
implementation with negligible additional error.
Thus rounding supplies the source in \cref{def:gaussian-lwe}, with
negligible loss over polynomially many samples even under QPT
postprocessing, by the contraction bound in \cref{sec:probability-notation}.

Fix a uniform QPT $\LAR$ distinguisher and polynomial bounds on its
sample count and inverse advantage.
\Cref{cor:gaussian-lwe-to-lar-composition} then chooses one fixed
transport accuracy order, without changing the public family, and
provides a discrete Gaussian distinguisher with nonnegligible advantage.
Regev's secret-shift and decision-to-search reductions
\cite[full version, Lemmas~4.1--4.2]{STOC:Regev05} apply to this error law;
their probability estimates use fresh classical tables and independent
QPT executions with measured output bits.
Enumerate all positive table lengths within a computable polynomial
bound. For each length, run the amplified coordinate-recovery
acceptance-gap tests with fixed polynomial time bounds to obtain a
candidate secret or report failure. Undo any secret shift and verify
each candidate against $\secpar$ fresh, independent scalar Gaussian
$\LWE$ samples before accepting it: require every response minus its
predicted inner product to have a centered integer lift of absolute
value at most $\LARModulus/4$.
For the correct candidate, the Gaussian tail bound gives rejection
probability at most
$\secpar\exp(-\Omega(\LARModulus^2/\GaussianWidth^2))$.
For any incorrect candidate, conditioned on its construction, the fresh
residuals are independent and uniform over $\LARField$, so its
acceptance probability is at most
$(1/2+1/\LARModulus)^{\secpar}$.
The length witnessing the distinguishing advantage occurs in the
enumeration and yields the correct candidate with inverse-exponentially
small failure. A union bound over polynomially many candidates preserves
this failure bound for the verified search solver.
Enumeration, amplification, and verification have polynomial overhead
and preserve uniformity even when the sample-count function is not
computable. Feeding the verified search solver rounded samples supplies
the continuous-$\LWE$ solver.

For the coherent decoding call in
\cite[full version, Lemmas~3.11 and~3.14]{STOC:Regev05}, defer measurements
in the amplified uniform QPT decoder, compute the decoder, copy its
output, and uncompute; the promised output is unique.
The coherent error is at most twice the square root of the decoding
failure probability. The Gaussian query state is independent of the
stored Gaussian sample pool, so averaging its amplitude-weighted
failure also over that pool gives negligible error.
Fresh ancillary registers and a hybrid over polynomially many steps
preserve negligible total error with polynomial overhead.
The composition therefore gives uniform QPT algorithms for
$\GapSVP$ and $\SIVP$ with approximation factor
$\widetilde O(\SecretDim/\LWENoiseRate)
=\widetilde O(\SecretDim\LARModulus/\GaussianWidth)$, using only the
distinguisher's guarantees on classical random examples.
Accuracy and amplification choices follow the adversary's polynomial
bounds; if its advantage holds only infinitely often, the algorithms
succeed on the same subsequence of dimensions.
\end{proof}

\subsection{Key sampling, exact marginals, security, and evaluation}
\label{sec:applications-weak-prf-proof}

\begin{proof}[Proof of \cref{prop:applications-tc0}]
For every fixed nonzero key, a uniform $\LARPublicMatrix$ gives
$\LARPublicMatrix\LARSecret\sim\Unif{\LARField^{\BlockDim}}$.
The balanced fibers of \cref{lem:lar-properties} therefore make each
output exactly uniform in $\LARField$.
For bounded-time key generation, sample a key index in
$\set{1,\ldots,\LARModulus^{\SecretDim}-1}$ by standard rejection
sampling capped at $\secpar$ attempts and use its base-$\LARModulus$
expansion as the key; on exhaustion return $\StandardBasisVec_1$.
The resulting law is supported on $\PRFKeySpace$ and is within
$\PRFKeyError:=2^{-\secpar}$ of $\Unif{\PRFKeySpace}$ in SD.
Since the marginal property holds for every key in this support,
it remains exact for the implemented sampler.

For a uniform QPT adversary, let $\WeakPRFAdvantage$ denote its
random-example advantage against a uniformly random function.
Conditioning the $\LAR$ secret to be nonzero costs
$\LARModulus^{-\SecretDim}$, and implementing key sampling costs
$\PRFKeyError$.
Independent uniform labels couple to a random function unless inputs
collide, at cost at most
$\binom{\SampleCount}{2}\LARModulus^{-\BlockDim\SecretDim}$
\cite[Lemma~1]{EPRINT:Shoup04}.
Thus, for a suitable $\LAR$ distinguisher,
\begin{equation*}
 \WeakPRFAdvantage
 \leq\LARDecisionalAdv{\LARDistinguisher}(\secpar,\SampleCount)
      +\LARModulus^{-\SecretDim}+\PRFKeyError
      +\binom{\SampleCount}{2}\LARModulus^{-\BlockDim\SecretDim}.
\end{equation*}
The reduction invokes the adversary once on a classical table and is
uniform QPT; the statistical bounds hold under quantum postprocessing
by the contraction bound in \cref{sec:probability-notation}.
All three extra terms are negligible for every polynomial sample count,
so \cref{cor:applications-worst-case} gives security.

Standard $\TCZero$ circuits for multiplication, iterated addition,
and division~\cite{HesseAllenderBarrington2002} compute
$\LARPublicMatrix\LARSecret\bmod\LARModulus$.
For the matrix in \cref{def:explicit-rounding-matrix}, each local floor
depends on at most three canonical residues and is computable by a
polynomial-size constant-depth lookup circuit.
Summation and modular reduction, adding a sufficiently large public
multiple of $\LARModulus$ to handle negative values, then give the
quotient label.
All intermediate integers have $O(\log\secpar)$ bits, and the
polynomially many operations occur in a constant number of parallel
stages. Since the public parameters and lookup tables are
polynomial-time computable, the resulting joint key-and-input
$\TCZero$ circuits are polynomial-time constructible.
\end{proof}

\subsection{Prediction and regression distinguishers}
\label{sec:applications-learning-proof}

Independent validation turns an inverse-polynomial field-label
prediction advantage over $1/\LARModulus$ into distinguishing advantage
using polynomially many classical examples.

\begin{proof}[Proof of \cref{cor:applications-learning}]
Use the independent-label experiments of \cref{def:lar-distributions},
dividing canonical label representatives by $\LARModulus$.
Train the QPT learner on an initial sample block and measure its
classical hypothesis. Let $\LearningRationalHypothesis$ be the evaluator's
rational approximation to the clipped hypothesis, clipped once more
to $[0,1]$.
Clipping does not increase squared loss, and the evaluator's
$1/1000$ approximation error changes it by at most $1/500$.
Thus, for every nonzero secret, successful training gives risk at most
$1/100+1/500=3/250$.
In the ideal experiment, each validation label is independent of training
and its prediction, even when inputs repeat, so the conditional risk
is at least its variance
$\LearningVariance=(\LARModulus^2-1)/(12\LARModulus^2)\geq2/27$.

Reserve $\LearningValidationCount:=10^4$ fresh samples and let
$\LearningEmpiricalRisk$ be their average squared loss for
$\LearningRationalHypothesis$; accept exactly when
$\LearningEmpiricalRisk\leq1/20$.
Conditioned on the training transcript and measured hypothesis, and
also on the secret in the real experiment, these losses are independent
and lie in $[0,1]$.
The threshold is more than $1/50$ from both risk bounds, so
Hoeffding's inequality~\cite[Theorem~2]{Hoeffding1963} bounds each
one-sided validation failure probability by $1/100$.
Training succeeds with probability at least $2/3$ for every nonzero
secret; the zero-secret event has probability
$\LARModulus^{-\SecretDim}$.
Consequently, the distinguishing advantage is at least
\[
 (1-\LARModulus^{-\SecretDim})\frac23\frac{99}{100}
 -\frac1{100}>\frac12
\]
for all sufficiently large $\secpar$.

The learning model bounds the learner's time on all transcripts,
including unrealizable ideal transcripts, and supplies a uniform
deterministic classical evaluator for every output hypothesis.
Its rational predictions and the validation arithmetic have polynomial
bit length. QPT training followed by this classical validation therefore
gives a uniform QPT distinguisher with time and total sample count
polynomial in $\LearningInputDim$, hence in $\secpar$.
\Cref{cor:applications-worst-case} then gives the hardness consequence
for the same public family.
\end{proof}

\subsection{The two-hidden-layer representation}
\label{sec:applications-relu-proof}

\begin{proof}[Proof of \cref{lem:applications-relu-representation}]
Fix a nonzero key. Integer-lift invariance from
\cref{lem:lar-properties}, with the quotient sum in
\cref{def:explicit-rounding-matrix}, gives
\[
 \LearningIntegerScore_{\LARSecret}(\LARPublicMatrix)
 :=\sum_{\CoordinateIndex=1}^{\BlockDim}
   \left\lfloor
    \frac{(\RoundingMatrix\CanonicalLift{\LARPublicMatrix}
                   \CanonicalLift{\LARSecret})_{\CoordinateIndex}}
         {\LARModulus}\right\rfloor,
 \qquad
 \LearningFunction_{\LARSecret}
 =\frac{\LearningIntegerScore_{\LARSecret}\bmod\LARModulus}{\LARModulus}.
\]
Each numerator is an integer linear form in the input of magnitude
at most $3\SecretDim(\LARModulus-1)^2$.
Put $\LearningRange:=4\SecretDim\LARModulus^2$ and
$\ReLU(\LearningScalar):=\max\{0,\LearningScalar\}$.
For every integer $|\LearningScalar|\leq\LearningRange$, the exact identity
\[
 \left\lfloor\frac{\LearningScalar}{\LARModulus}\right\rfloor
 =-\LearningRange+
   \sum_{\LearningThresholdIndex=-\LearningRange+1}^{\LearningRange}
    \bigl(\ReLU(\LearningScalar-\LARModulus\LearningThresholdIndex+1)
          -\ReLU(\LearningScalar-\LARModulus\LearningThresholdIndex)\bigr)
\]
expresses each floor by one ReLU layer. Indeed, each difference of
ReLUs equals $\EventIndicator{\set{\LearningScalar\geq
\LARModulus\LearningThresholdIndex}}$ on integer inputs. Counting the
thresholds gives $\LearningRange+
\lfloor\LearningScalar/\LARModulus\rfloor$, including negative inputs.
There are $4\LearningRange$ units per row and hence at most
$4\BlockDim\LearningRange$ units in the first hidden layer.
Its affine scalar output defines the extension of
$\LearningIntegerScore_{\LARSecret}$. On the grid this output is an
integer and satisfies
$|\LearningIntegerScore_{\LARSecret}|
\leq\BlockDim\LearningRange$.
Define $\LearningOuterMap$ by linear interpolation of
$\LearningScalar\mapsto(\LearningScalar\bmod\LARModulus)/\LARModulus$
at all integers in this interval, and extend it constantly outside
the interval. It is globally $[0,1]$-valued. To make its network
representation explicit, let $\LearningSplineSlope_{\LearningThresholdIndex}$
be its slope on
$(\LearningThresholdIndex,\LearningThresholdIndex+1)$ for integer
$-\BlockDim\LearningRange\leq\LearningThresholdIndex<\BlockDim\LearningRange$,
and put
$\LearningSplineSlope_{-\BlockDim\LearningRange-1}
=\LearningSplineSlope_{\BlockDim\LearningRange}=0$.
The standard piecewise-linear representation
recalled in~\cite[Remark~2.2]{NeuroIPS:CGKM22} is, in this case,
\[
 \LearningOuterMap(\LearningScalar)
 =\LearningOuterMap(-\BlockDim\LearningRange)
 +\sum_{\LearningThresholdIndex=-\BlockDim\LearningRange}^{\BlockDim\LearningRange}
  (\LearningSplineSlope_{\LearningThresholdIndex}
   -\LearningSplineSlope_{\LearningThresholdIndex-1})
  \ReLU(\LearningScalar-\LearningThresholdIndex).
\]
The slopes have absolute value at most one and denominator
$\LARModulus$. Substituting the first layer's affine output for
$\LearningScalar$ inside these ReLUs gives a second hidden layer
with at most $2\BlockDim\LearningRange+1$ units and an affine output.
The intermediate affine map is absorbed into the second-layer
preactivations, so the composition has exactly two hidden layers.

The total number of hidden units is at most
$6\BlockDim\LearningRange+1
=24\LearningInputDim\LARModulus^2+1=O(\LearningInputDim^3)$.
First-layer input weights have magnitude at most $\LARModulus-1$
and biases at most $\LARModulus\LearningRange+1$.
The intermediate affine output has coefficients in $\set{-1,1}$ and
bias $-\BlockDim\LearningRange$; hence the second-layer weights have
magnitude one and biases at most $2\BlockDim\LearningRange$.
Output weights have magnitude at most two and denominator
$\LARModulus$. Since $\LARModulus\leq\BlockDim\leq\LearningInputDim$,
all parameters have polynomial magnitude and $O(\log\LearningInputDim)$
bit length. Enumerating the units and their connections takes
polynomial time in $\LearningInputDim$.
Finally, the global range of $\LearningOuterMap$ ensures that the
composite extension stays in $[0,1]$ even off the grid.
\end{proof}